\documentclass[reprint,prd,twocolumn,superscriptaddress,nofootinbib]{revtex4-2}   % somehow this puts it on one page 
\usepackage{graphicx}
\usepackage{multirow}
\usepackage{gensymb}
\usepackage{amsmath}
\usepackage{bm}
\usepackage{natbib}
\usepackage{hyperref}
\usepackage{color}  %SdM: allows fancy color names
\usepackage{longtable}
\usepackage{booktabs}
\usepackage{placeins}
\newcommand\aj{{AJ}}%        % Astronomical Journal 
\newcommand\araa{{ARA\&A}}%  % Annual Review of Astron and Astrophys 
\renewcommand\apj{{ApJ}}%    % Astrophysical Journal 
\newcommand\apjl{{ApJL}}%     % Astrophysical Journal, Letters 
\newcommand\apjs{{ApJS}}%    % Astrophysical Journal, Supplement 
\newcommand\aap{{A\&A}}%     % Astronomy and Astrophysics 
\newcommand\aapr{{A\&A~Rev.}}%  % Astronomy and Astrophysics Reviews 
\newcommand\mnras{{MNRAS}}%   % Monthly Notices of the RAS 
\renewcommand\prd{{Phys.~Rev.~D}}% % Physical Review D 
\renewcommand\nat{{Nature}}%  % Nature 
\def\mone{{m}_1}
\def\mtwo{{m}_2}

\def\aone{{\rm a}_1}
\def\atwo{{\rm a}_2}
\def\aonecap{\hat{a_1}}
\def\atwocap{\hat{a_2}}
\def\lcap{\hat{L}}

\def\xeff{\chi_{\rm eff}}

\def\mueff{\mu_{\rm eff}}
\def\nsamp{{\rm N}_{\rm samp}}

\def\nmrg{{\rm N}_{\rm mrg}}

\def\zf{z_{\rm f}}

\def\delage{\Delta t_{\rm age}}
\def\delz{\Delta z}

\def\rate{\mathcal R}

\def\rpess{{\mathcal R}_{-}}

\def\mrgfrac{{\mathcal F}_{\rm mrg}}

\def\clmf{\phi_{\rm CLMF}}
\def\sfh{\phi_{\rm SFH}}
\def\phir{\phi_{\rm r}}
\def\phizz{\phi_{\rm zZ}}

\def\peryg{{\rm~yr}^{-1}{\rm Gpc}^{-3}}

\def\mgclow{M_{\rm GC,l}}
\def\mgchigh{M_{\rm GC,u}}
\def\mcllow{M_{\rm cl,l}}
\def\mclhigh{M_{\rm cl,u}}
\def\zflow{z_{\rm f,l}}
\def\zfhigh{z_{\rm f,u}}
\def\metlow{Z_{\rm l}}
\def\methigh{Z_{\rm u}}
\def\rhlow{r_{\rm h,l}}
\def\rhhigh{r_{\rm h,u}}
\def\permv{{\rm~Mpc}^{-3}}

\def\unif{{\mathcal U}}

\def\porb{P_{\rm orb}}

\def\tend{T_{\rm end}}

\def\mcomp{m_{\rm comp}}

\def\etamrg{\eta_{\rm mrg}}

\def\pastro{p_{\rm astro}}
\def\epsacfe{\epsilon_{\rm CFE}}
\def\psisfr{\psi_{\rm SFR}}
\newcommand{\Ms}{\ensuremath{{\rm M}_{\odot}}}
\newcommand{\Zs}{\ensuremath{{\rm Z}_{\odot}}}
\newcommand{\eg}{{\it e.g.}}
\newcommand{\cf}{{\it c.f.~}}
\newcommand{\ie}{{\it i.e.}}

\newcommand{\beq}{\begin{equation}}
\newcommand{\eeq}{\end{equation}}
\newcommand{\mtot}{\ensuremath{m_{\rm tot}}}
\newcommand{\mzams}{\ensuremath{m_{\rm ZAMS}}}

\newcommand{\kmps}{\ensuremath{{\rm~km~s}^{-1}}}
\newcommand{\peryr}{\ensuremath{{\rm~yr}^{-1}}}

\newcommand{\thub}{\ensuremath{t_{\rm Hubble}}}

\newcommand{\ncl}{\ensuremath{n_{\rm cl}}}
\newcommand{\ngc}{\ensuremath{n_{\rm GC}}}
\newcommand{\rhoclzero}{\ensuremath{\rho_{\rm cl,0}}}
\newcommand{\mcl}{\ensuremath{M_{\rm cl}}}
\newcommand{\mclstar}{\ensuremath{M_{{\rm cl},\ast}}}
\newcommand{\rh}{\ensuremath{r_{\rm h}}}
\newcommand{\rhpst}{\ensuremath{r^\prime_{\rm h,\ast}}}

\newcommand{\tmrg}{\ensuremath{t_{\rm mrg}}}

\newcommand{\tauinsp}{\ensuremath{\tau_{\rm insp}}}

\newcommand{\vej}{\ensuremath{v_{\rm ej}}}

\newcommand{\nbseven}{{\tt NBODY7}}

\newcommand{\nbpp}{{\tt NBODY6++GPU}}

\newcommand{\bse}{{\tt BSE}}

\newcommand{\archain}{{\tt ARCHAIN}}

\newcommand{\bspline}{{\tt B-Spline}}
\newcommand{\spline}{{\tt Spline}}
\newcommand{\bplpp}{{\tt Broken~Power~Law~+~2~peaks}}
\newcommand{\plp}{{\tt Power~Law~+~Peak}}

\newcommand{\fbin}{\ensuremath{f_{\rm bin}}}

\newcommand{\fobin}{\ensuremath{f_{\rm Obin}}}

\newcommand{\mbh}{\ensuremath{m_{\rm BH}}}
\newcommand{\mns}{\ensuremath{m_{\rm NS}}}
\newcommand{\nbh}{\ensuremath{N_{\rm BH}}}

\newcommand{\fmrg}{\ensuremath{f_{\rm mrg}}}
\newcommand{\ftz}{\ensuremath{f_{\rm TZ}}}

\newcommand{\ace}{\ensuremath{\alpha_{\rm CE}}}

\newcommand{\msplit}{\ensuremath{m_{\rm split}}}
\newcommand{\betalow}{\ensuremath{\beta_q^{\rm low}}}
\newcommand{\betahigh}{\ensuremath{\beta_q^{\rm high}}}
\begin{document}

\title[Part VII: comparisons with GWTC-4]
{Stellar-mass black holes in young massive and open stellar clusters – VII. Comparisons with
gravitational-wave events until LVK-O4a and Gaia compact binaries}

\author{Sambaran Banerjee}
\email{banerjee@hiskp.uni-bonn.de; sambaran.banerjee@gmail.com}
\affiliation{Helmholtz-Instituts f\"ur Strahlen- und Kernphysik (HISKP),
Nussallee 14-16, D-53115 Bonn, Germany}

\date{\today}

\begin{abstract}

Gravitational-wave (GW) detections from the LIGO–Virgo–KAGRA (LVK) GW observatories suggest multiple formation channels
contributing to the GW coalescence of compact binaries in the Universe. In this study, I assess the contribution of young massive clusters (YMC)
evolving into old open clusters (OC) -- the YMC/OC channel -- to the GW merger population.
To that end, a homogeneous grid of 90 N-body evolutionary model star clusters, spanning initial masses of $10^4\Ms \leq \mcl(0) \leq 10^5\Ms$,
half-mass radii of $1-3$ pc, and metallicities between $0.0002-0.02$ is computed with the direct, star-by-star, post-Newtonian N-body code $\nbseven$.
The N-body simulations include primordial binaries, delayed stellar remnant masses forming stellar-remnant black holes (BH) and
neutron stars (NS), BH spin prescriptions, and general-relativistic recoil kicks. The model clusters are evolved until they lose most of the retained BHs.
The majority of the GW mergers from the cluster models are dynamically assembled binary black holes (BBH) that merge within their host clusters,
while the rarer escaped mergers are commonly primordially paired. Merger mass ratios reach down to $0.1-0.2$ despite an overall bias
toward nearly symmetric pairs of $q\approx0.8$. The GW merger efficiency varies non-monotonically with cluster mass,
peaking around $\mcl(0)=7.5\times10^4\Ms$ and also for $\mcl(0)\leq3.0\times10^4\Ms$.
The computed mergers reproduce some of the key features of the latest observed
GW event catalogue, including asymmetric low-mass mergers, misaligned events among highly spinning, massive BHs, and
an excess of $30\Ms$ primaries, though they underproduce
$10\Ms$ primaries, hinting at contributions from other channels. The merger rate density from the model grid
at redshift $z=0.2$ accounts for a fiducial 25\%-33\% of the observed rate depending on cosmic metallicity evolution;
it increases with redshift somewhat more steeply than the cosmic star formation rate, and
consistently with the overall growth of the LVK-inferred merger rate. Some of the apparent low-redshift features of the observed BBH mergers'
effective spin distribution, \eg, the positive asymmetry at $z=0$ and the broadening with redshift, are also qualitatively reproduced,
after applying a post-processing prescription to the model mergers.
The model clusters also yield field BH- and NS-main sequence star binaries with parameters consistent with the Gaia-discovered candidates.

\end{abstract}

\maketitle

\section{Introduction}\label{intro}

The known collection of general-relativistic (hereafter GR) compact binary coalescence events in the Universe, as
detected in gravitational waves (hereafter GW), has undergone a major expansion after the LIGO-Virgo-KAGRA (hereafter LVK;
Ref.~\cite{Abbott_GWTC3}) collaboration has just updated their gravitational wave transient catalogue (hereafter GWTC) with
the GW-event candidates from the first half of their fourth observing run, `O4a' \cite{GWTC4a_cat}. 
The current GWTC contains $\approx166$ events that are designated to be astrophysical GW events
(having $p_{\rm astro}>0.5$)\footnote{{\tt https://gwosc.org/eventapi/html/GWTC/}}.
Like the previous GW-event catalogues, the vast majority of the to-date-observed GW events
remain classified as binary stellar-remnant black hole (hereafter BBH) GW mergers.
A range of scenarios have been proposed as potential formation channels for BBH mergers;
see Refs.~\cite{Mandel_2021,Spera_2022} for comprehensive reviews. Such scenarios include
dynamical interactions among stellar-remnant black holes (hereafter BH) in dense star clusters
\cite{Kulkarni_1993,Lee_1995,Benacquista_2013},
isolated evolution of massive-stellar binaries (\eg, \cite{Belczynski_2002}),
interacting BHs in AGN gas disks (\eg, \cite{McKernan_2018,Vaccaro_2024}),
isolated evolution of hierarchical massive-stellar systems (\eg, a massive field triple-star;
\cite{VignaGomez_2021,Stegmann_2022}),
close flyby interactions involving BBHs in the galactic field (\eg, \cite{Michaely_2019}), and cluster and galactic
tides on wide BBHs (\eg, \cite{Hamilton_2019,Stegmann_2024}).
The properties of the observed GW events and of their population would, in turn, serve to
constrain the formation channels, their physics, and the physics of stellar remnant formation.

Indeed, underlying population properties, especially the rather complex, multi-peaked merger primary mass distribution
and the asymmetric effective spin distribution, 
as inferred from the latest ensemble of observed GW events, suggest formation of these events via a range of
mechanisms (or formation channels), rather than through a single dominant channel
(\eg, Refs.~\cite{Zevin_2020,GWTC4a_pop,Banagiri_2025}).
In particular, the latest inferred population distributions point to both the importance of dynamical interactions
(\ie, the `dynamical channel') and isolated binary evolution (\ie, the `isolated-binary channel')
in forming GW sources \cite{GWTC4a_pop}. Therefore, to obtain an
understanding of the underlying distributions of GW events in the Universe and to avoid misleading interpretations, it is
nevertheless crucial to explore individual formation channels thoroughly. A relatively newly conceived but widely discussed
channel is GW merger formation in young massive clusters (hereafter YMC) and moderately
massive open clusters (hereafter OC) \cite{Banerjee_2010,Aarseth_2012,Banerjee_2017,DiCarlo_2019}.
Observed YMCs in the Milky Way (hereafter MW) and local group (hereafter LG) galaxies are typically of $\sim10^4\Ms$ 
and parsecs scale length \cite{PortegiesZwart_2010,Krumholz_2019},
and through dynamical and stellar evolution, they would evolve into moderate mass, extended OCs.
While more massive globular clusters (typically of $\sim10^5\Ms$; hereafter GC) are canonically considered as 
the most prominent source of dynamically assembled BBH mergers \cite{Rodriguez_2021,Ye_2025},
YMCs and OCs, owing to their much shorter relaxation time, low velocity dispersion,
shorter lifespan, and possibly different cosmic formation history, comprise a separate formation channel by their own right
\cite{Santoliquido_2020,Kumamoto_2020,Banerjee_2021,Fragione_2021}.
See also Ref.~\cite{Banerjee_2025} (their Secs.~I and IV) for additional discussions
on the physics of star cluster evolution in the presence of a subpopulation comprising of retained
BHs and neutron stars (hereafter NS). For brevity, this discussion is skipped in this paper.

In this study, this YMC/OC channel of GW source formation
is revisited based on a newly computed grid comprising long-term evolutionary models
of $10^4-10^5\Ms$, pc-scale star clusters. The details of the grid are presented in Sec.~\ref{sims}.
Evolutionary model grids in the YMC/OC regime producing populations of GW mergers
have previously been explored by several groups and authors.
Ref.~\cite{Banerjee_2020c} presented a set of $\approx70$ long-term (evolution time of several to 10 Gyr) direct N-body
evolution (see Sec.~\ref{nbsims}) of model star clusters, of initial masses ranging over $10^4\Ms-10^5\Ms$.
Ref.~\cite{Banerjee_2022} explored a grid of 40 massive (membership $\sim10^5$)
clusters that were evolved for $\approx300$ Myr with direct N-body integration. Ref.~\cite{Barber_2025}
presented a grid of 32 particle-particle-particle-Tree (or `P3Tree') N-body evolutionary model clusters \cite{Wang_2020b}
with initial masses and (half-mass) densities ranging over
$10^4-10^5\Ms$ and $10^3-10^5\Ms{\rm pc}^{-3}$, which were typically evolved up to
1 Gyr (three models were evolved up to 3 Gyr). These authors have additionally evolved two cluster models
of initially $10^6\Ms$ for up to $\approx600$ Myr. Based on a suite of
95 long-term (until 10 Gyr or dissolution) direct N-body simulations of initially
$3\times10^4\Ms$ clusters spanning across a wide range of galactocentric distances (or external
tidal field), Ref.~\cite{Banerjee_2025} addressed the impact of galactic tidal field
on the production of GW sources from YMC/OCs.

Apart from YMC/OC-type clusters, evolutionary grids have been computed for models representing
GCs, with a focus on GW source formation in them.
Ref.~\cite{Kremer_2020} presented a grid of 148 GC models with masses within $10^5\Ms-10^6\Ms$,
which were evolved for 12 Gyr via a Monte Carlo approach. Ref.~\cite{Askar_2016} presented
similar (but more numerous; 2000 models) Monte Carlo GC model grids,
that were evolved up to a similar extent. Ref.~\cite{ArcaSedda_2024a} evolved 
19 GC-like model clusters that are initially of $0.7-5.9\times10^5\Ms$ and extremely dense ($10^4-10^7\Ms{\rm pc}^{-3}$).
With direct N-body integration, they evolved the clusters up to $\lesssim2.4$ Gyr.
On the other hand, by applying direct N-body integration,
Refs.~\cite{DiCarlo_2020,Kumamoto_2020,Rastello_2021} presented extensive sets of evolutionary
models of low mass ($10^2-10^4\Ms$) star clusters.
See Fig.~1 of Ref.~\cite{ArcaSedda_2024a} for a comprehensive overview of the model
cluster suits from various authors. 

Clearly, there exists a dearth of homogeneous, densely-sampled long-term evolutionary model grids of star clusters
that initiate at a mass within $10^4-10^5\Ms$.
This study addesses this deficit by presenting a new evolutionary grid of 90 $10^4-10^5\Ms$ model star clusters
that is homogeneous in the clusters' initial and physical properties, and the clusters are evolved at least
until most of the BHs are depleted from them. Furthermore, the GW merger population
from this grid is compared with the latest GWTC and its inferred population distributions \cite{GWTC4a_cat,GWTC4a_pop}.
Since the model clusters are evolved until the BH population inside
each cluster is processed essentially to exhaustion, no significant additional merger yield would be expected.
The model merger yield is, therefore, not prematurely truncated,
making the model merger population complete.

Since it is likely that the observed GW event population is an outcome of multiple formation
channels (see above), the author does not aim to reproduce the observed population
based on the GW mergers from the model clusters. Rather, the objective here is to see to
what extent the YMC/OC channel contributes to the observed event population.
To date, this is the largest homogeneous grid of GR-merger-forming direct N-body star cluster models in the
$10^4-10^5\Ms$ range that is evolved until the depletion of its BHs,
covering six cluster masses, three cluster sizes, and five metallicities within a single, uniformly constructed
set. In contrast, the earlier model sets of this work-series \cite{Banerjee_2020c,Banerjee_2021} were
developmental and heterogeneous: they mostly employed the `rapid' remnant-mass model, with only a few members
computed with the `delayed' model (see below), and they also mixed different natal-kick prescriptions.
The more recent set of Ref.~\cite{Banerjee_2025}, while homogeneous, is restricted to a single cluster mass of
$3\times10^4\Ms$, being aimed at the effect of the external tidal field.
The present, newly computed grid, in contrast, is homogeneous in prescription choices, systematic in its wide parameter coverage,
and adopts the delayed remnant-mass prescription — the latter significantly affecting BH retention and BBH-merger production (Sec.~\ref{discuss}).
This renders the resulting GW merger
population internally consistent and, simultaneously, spanning across cluster mass, size, and metallicity.
Furthermore, the model merger population is compared against the newly published GWTC-4 and its LVK-inferred population distributions.

This paper is organised as follows. Sec.~\ref{sims} describes the computed N-body star cluster model grid. Sec.~\ref{res}
describes the results of the N-body simulations: Sec.~\ref{evol} discusses the general long-term evolution of the model clusters and Sec.~\ref{mrg}
describes the GW mergers formed from them. Sec.~\ref{gwtc} compares the model mergers with GWTC.
Sec.~\ref{gaia} discusses the formation of field compact-star--normal-star binaries
from the model clusters and compares them with observed Gaia BH-MS and NS-MS binary candidates. Sec.~\ref{discuss}
summarises the results, discusses the current limitations, and suggests potential improvements.

\begin{figure*}
\centering
\includegraphics[width = 0.49\textwidth, angle=0.0]{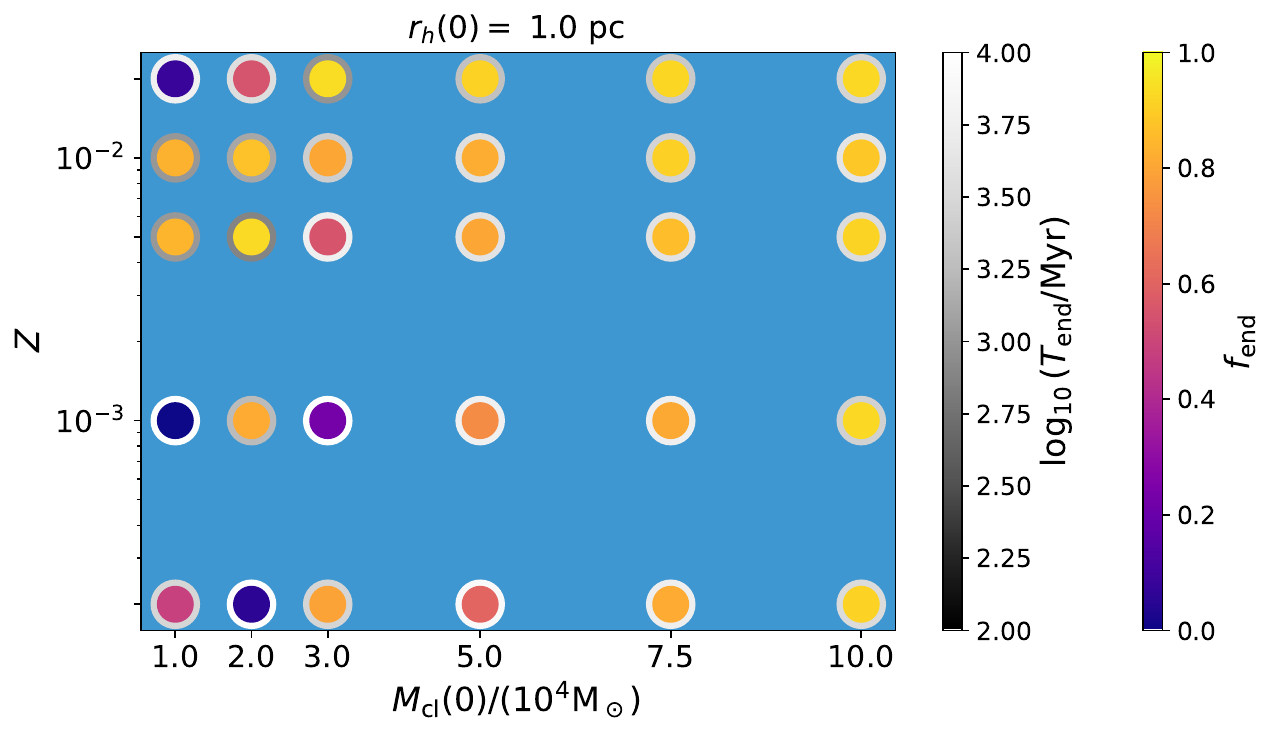}
\includegraphics[width = 0.49\textwidth, angle=0.0]{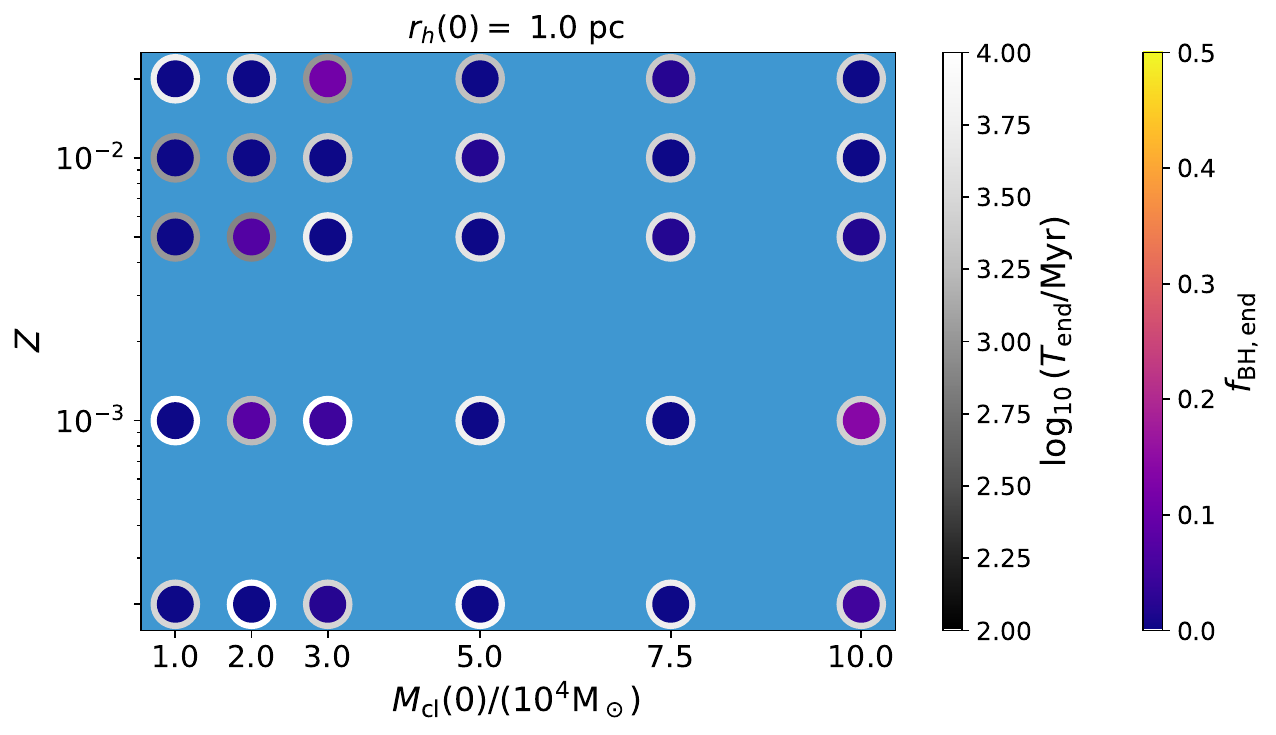}\\
\includegraphics[width = 0.49\textwidth, angle=0.0]{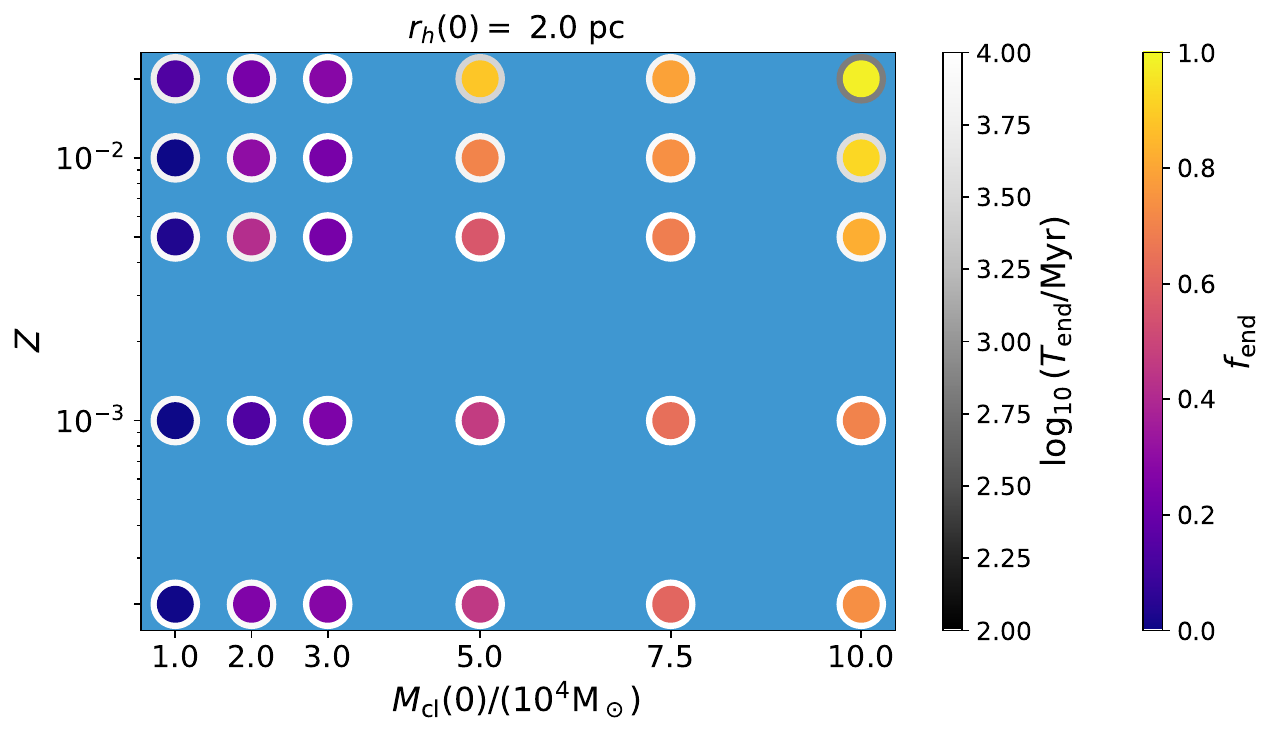}
\includegraphics[width = 0.49\textwidth, angle=0.0]{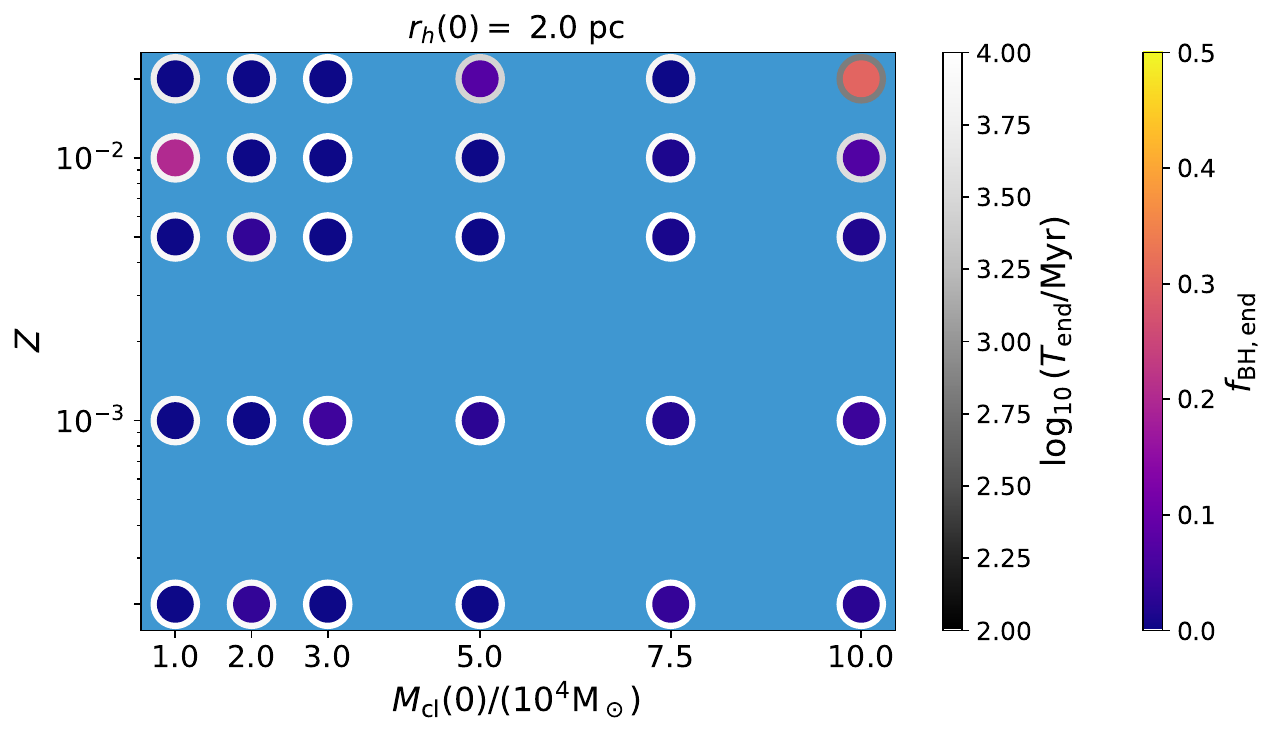}\\
\includegraphics[width = 0.49\textwidth, angle=0.0]{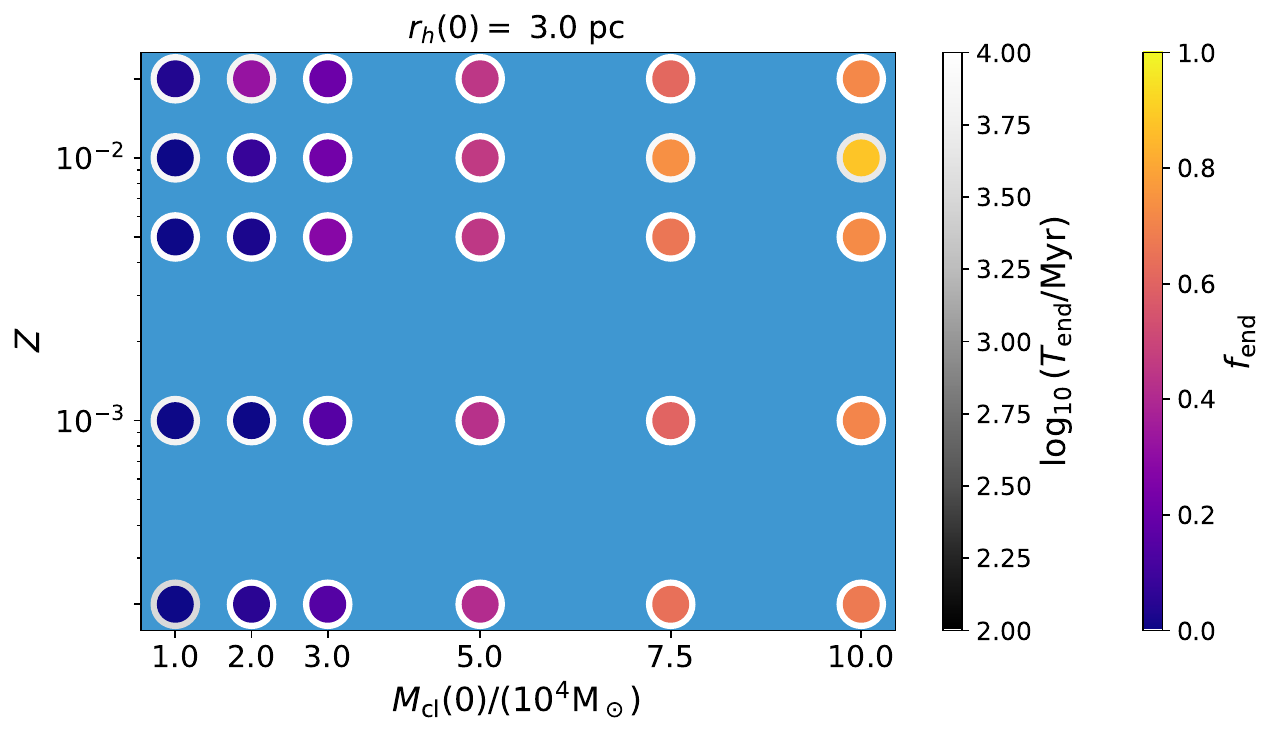}
\includegraphics[width = 0.49\textwidth, angle=0.0]{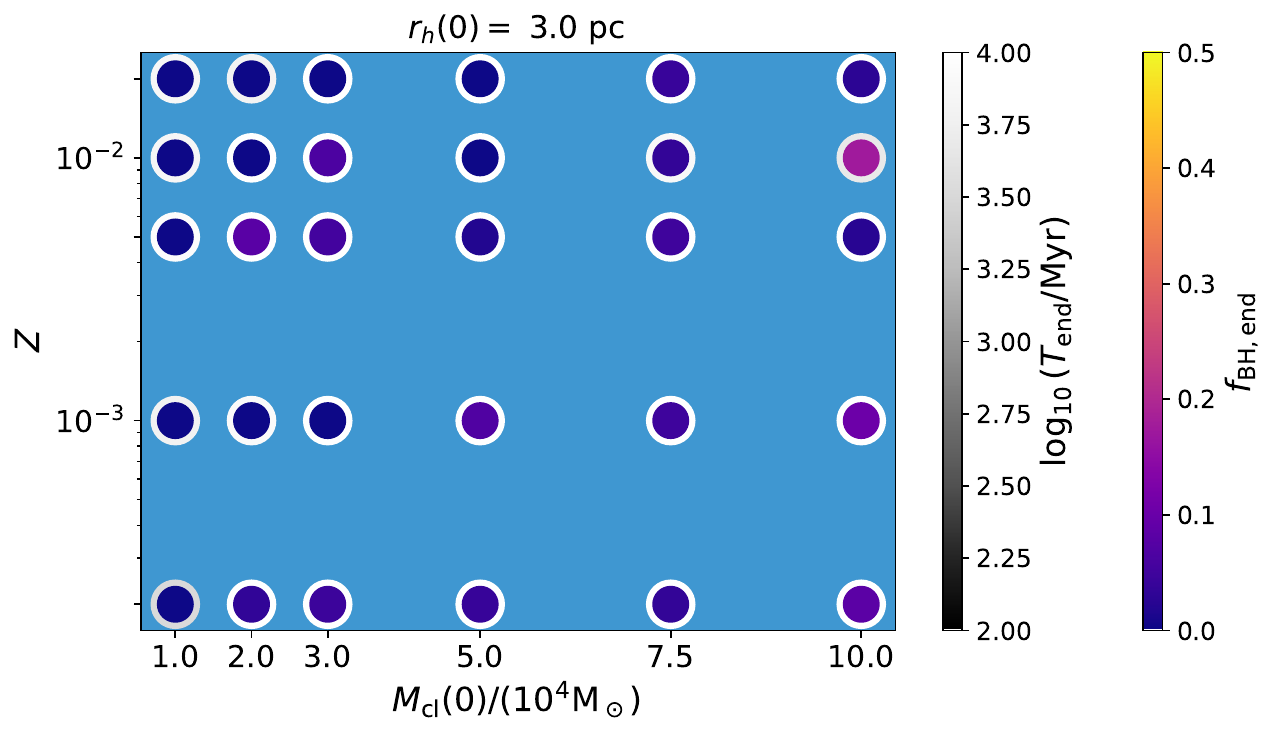}
	\caption{Depiction of the computed grid of 90 evolutionary model star clusters
	spanning across ranges of cluster initial mass, $\mcl(0)$,
	initial half-mass radius, $\rh(0)$, and metallicity, $Z$. The panels in the upper, middle,
	and lower rows correspond to the computed grid points (filled circles) for the
	clusters with $\rh(0)=1$, 2, and 3 pc, respectively. The greyscale color coding
	of the edge of a circle represents the end time in Myr, $\tend$, of the corresponding N-body simulation.
	The circles' fill colour in the left column represents $f_{\rm end}$, which is the number of bound
	cluster members at the end of the simulation relative to the initial cluster membership. 
	The circles' fill colour in the right column represents $f_{\rm BH,end}$, which is the number of
	BHs bound to the cluster at the end of the simulation relative to the peak BH membership
	of the cluster during its evolution. See Table.~\ref{tab:runlist} (Appendix~\ref{runlist}) for more details.}
\label{fig:grid}
\end{figure*}

\section{Modelling star cluster evolution}\label{sims}

This section describes the newly computed grid of evolutionary models of star clusters.

\subsection{The star cluster model grid}\label{grid}

In this study, a grid of star cluster models of initial masses ranging over
$10^4\Ms\leq\mcl(0)\leq10^5\Ms$ is evolved.
The initial density and velocity dispersion profiles of the clusters are taken to be
according to the Plummer model \cite{Plummer_1911,Heggie_2003}. The initial models have masses of
$\mcl(0)=10^4\Ms$, $2\times10^4\Ms$, $3\times10^4\Ms$, $5\times10^4\Ms$, $7.5\times10^4\Ms$, and $10^5\Ms$,
and half-mass radii of $\rh(0)=1$ pc, 2 pc, and 3 pc. Each model cluster
is evolved for the metallicities $Z=0.0002$, 0.001, 0.005, 0.01, and 0.02.
These initial conditions together comprise a grid of 90 evolutionary star cluster models
covering wide ranges of $\mcl(0)$, $\rh(0)$, and $Z$. The computed grid is depicted in Fig.~\ref{fig:grid}.

During its evolution, each model cluster is subjected to a solar-neighbourhood-like external
galactic field by placing the cluster on a circular orbit at a MW galactocentric distance of
$r_g=8.5$ kpc. The MW tidal field is modelled based on an axisymmetric bulge-disk-halo
potential as in Ref.~\cite{Banerjee_2025}. That reference has investigated the impact of galactic tidal
field on GW production from clusters, and has shown that the variation of external field would
only have a moderate impact on GW production, for the cluster mass range considered here.
Therefore, the current model grid complementarily focusses on variation in $\mcl(0)$, $\rh(0)$, and $Z$,
which would allow for constructing a model cluster population that incorporates   
a critical aspect when it comes to physics of star clusters, namely, the cluster's mass. 
In contrast, the model grid of Ref.~\cite{Banerjee_2025} varied $r_g$, $\rh(0)$, and $Z$. 

\subsection{Direct, star-by-star, post-Newtonian N-body simulations}\label{nbsims}

Each initial model in the above-described grid is evolved with the star-by-star, direct, post-Newtonian (hereafter PN)
N-body code $\nbseven$ \cite{Aarseth_2012}. The code integrates the orbit of each member subject to the combined gravitational
field of the rest of the members plus any external gravitational field. The main orbit integrations are carried
out by a fourth-order Hermite integrator. To tackle the time-stepping and stalling issues during
close approaches among the members \cite{vonHoerner_1957}, 
the code does not implement any force-softening or another approximate procedure. Instead, it applies the much more accurate
KS- and Chain-regularisation techniques. The integration is forwarded with individual, grouped time stepping for
all members. The details of all algorithms employed in $\nbseven$ can be found in Ref.~\cite{Aarseth_2003}. In this work,  
an updated version of $\nbseven$ as described in Refs.~\cite{Banerjee_2020} and \cite{Banerjee_2020c} is used.
These updates include implementations of more recent stellar wind mass loss
and remnant-formation prescriptions, and also implementations of numerical relativity
(hereafter NR)-based GR merger recoil and tracking of spins of higher-generation BHs inside the cluster,
at runtime\footnote{This version of $\nbseven$ is publicly available at the following URL:\\
{\tiny https://github.com/sambaranb/updated-nbody7/tree/mpi-parallel}. The {\tt mpi-parallel} branch
contains the standard standalone {\tt X86/CUDA/openmp} variant as well as several under-development variants.}.

Runtime stellar and binary evolution is achieved in $\nbseven$ through its coupling with the rapid,
analytical binary evolution code $\bse$ \cite{Hurley_2000,Hurley_2002}. Runtime PN
evolution of compact binaries and higher order systems is computed by incorporating
the few-body integrator $\archain$ \cite{Mikkola_1999,Mikkola_2008}. In the present computations,
the `delayed' remnant-mass model along with the possibility of
occurrence of pair-instability supernova (hereafter PSN)
and pulsation-pair-instability supernova (hereafter PPSN) are chosen
\cite{Fryer_2012,Belczynski_2016a,Banerjee_2020}.
The delayed remnant-mass prescription allows the formation of stellar remnants
within the so-called `lower mass gap' between $2\Ms-5\Ms$;
in other words, no lower mass gap exists for the chosen delayed remnant-mass prescription.
This choice is consistent with the observations of GW-merger candidates involving such lower-mass-gap
compact objects \cite{Unequal_masss_2020,LowerMassGap2024,GWTC5_pop}. In the present simulations, the canonical
$2.5\Ms$ upper limit for NS mass \cite{Belczynski_2016a} is adopted.
Due to allowing PPSN and PSN, the mass spectrum of BHs derived from single stars and non-interacting binary
members evolving from zero age main sequence exhibits the conventional PSN mass gap over
$\approx40.5\Ms-120\Ms$ \citep{Langer_2007,Woosley_2017} (but see below).
Newly formed NSs and BHs are assigned random, momentum-conserving, fallback-modulated
natal kick as per the recipes in Ref.~\cite{Banerjee_2020}. This causes
BHs of mass $\gtrsim 8\Ms$ and electron capture supernova NSs (hereafter ECS-NS)
to typically retain inside the parent cluster at birth.

Recently, an increasing number of studies are incorporating remnant masses that are based on compactness
of the pre-collapse star's profile (\eg, \cite{Mandel_2020,Mapelli_2026}). Incorporating this in $\nbseven$, however, would require coupling or bridging
the code with a sophisticated interpolator and stellar- and binary-evolution modeller such as
{\tt SEVN} \cite{Iorio_2023} or {\tt POSYDON} \cite{Andrews_2025}.
At present, this is yet to be achieved for classic N-body codes and the compactness-based remnant-mass modelling is
mainly limited to fast binary-evolution codes (see the above references). As such, the presently adopted 
`Fryer-formalism' for remnant masses also comprises analytical fits to detailed hydrodynamical models of stellar collapse
and is applied also in other widely used N-body codes such as $\nbpp$ and {\tt PETAR} \cite{Kamlah_2021,Rastello_2026}.

The stars in the initial models are taken to be at the zero age main sequence (hereafter ZAMS)
and of masses between $0.08\Ms\leq m_\ast\leq150.0\Ms$, which are distributed according
to the Kroupa initial mass function (hereafter IMF) \cite{Kroupa_2001}.
In all initial models, the stars have an overall (see below) primordial-binary fraction of $\fbin=10$\%.  
However, the initial binary fraction of the
O-type stars ($m_\ast\geq16.0\Ms$), which are initially paired only among themselves,
is taken to be $\fobin(0)=100$\%, which is consistent with the observed high binary fraction
among O-stars in young clusters and associations \cite{Sana_2011,Sana_2013,Moe_2017}.
The O-star binaries initially follow the observed orbital-period and eccentricity distributions of Ref.~\cite{Sana_2011}
and a uniform mass-ratio distribution.
The non-O-star primordial binaries initially follow the orbital period
distribution of Ref.~\cite{Duq_1991}, thermal eccentricity distribution \citep{Spitzer_1987}, and a uniform mass ratio distribution.
As explained in Ref.~\cite{Banerjee_2017b}, such a scheme for
primordial binaries provides a reasonable compromise between the economy of computing
and consistencies with observations.

In the present models, small birth spins, as per the `MESA' BH-spin model
of Ref.~\cite{Belczynski_2020}, are assigned to
all BHs derived from single stars or from members of non-mass-transferring or non-interacting binaries.
Such small spin magnitudes correspond to a Kerr parameter (dimensionless spin) of $0.05\lesssim a \lesssim0.15$,
and are caused by dynamo-driven efficient core-to-envelope angular momentum transport in the pre-SN BH progenitor
star \cite{Spruit_2002,Fuller_2019a}. See Appendix~\ref{bhspin} for further details.

In the event of a BH-star merger
(the formation of a BH Thorne--Zytkow object or BH-TZO \cite{TZ_1975}) during a simulation, 
$\ftz=95$\% of the merging star's mass is assumed to be accreted onto the BH.
In runtime star-star collisions, a merged star is formed where an equivalent of $\fmrg=20$\% of the secondary's
mass is assumed to be lost in the merger process.
This assumption is based on outcomes of hydrodynamical studies of
collisions among main sequence (hereafter MS) stars that suggest that $\sim 10$\% of the total mass
is lost during the merger process \citep{Gaburov_2008,Glebbeek_2009}.
In fact, more recent and detailed hydro-simulations of MS-MS mergers suggest
up to only a few percent mass loss \citep{Ryu_2025,Vynatheya_2026}, \ie, the adopted $\fmrg$ serves as an upper limit.
The merged star is rejuvenated according to the recipes implemented in $\bse$ \cite{Hurley_2002}
that assume complete mixing. The extent of mass accretion onto a BH in a BH-star merger is ambiguous and poorly understood. The above
choice of a high $\ftz$ is optimistic and favours the formation of massive BHs, especially those within the canonical PSN mass gap 
between $\approx40\Ms-120\Ms$, even in moderately dense environments.
As demonstrated in Ref.~\cite{Banerjee_2022}, that way the mass range of massive BH formation in YMCs
is consistent with LVK's massive GW events \citep{Abbott_GW190521,Abbott_GWTC3_prop}.
While other channels (\eg, hierarchical BBH mergers in GCs and NSCs \cite{ArcaSedda_2021b}) also have
the potential to form BBH mergers within the PSN gap, in the context of the present study, it would nevertheless
be interesting to allow for adequate room to form PSN-gap events in the computed YMC/OC models.
See Sec.~\ref{GWTC_pop} for further discussions.

The author aimed to evolve each model cluster at least until most of the BHs that were retained
in the cluster at their birth are depleted due to dynamical ejections and GW-merger recoils.
Fig.~\ref{fig:grid} depicts the evolution times and the simulation end states (number fractions of
retained BHs and total membership) for the models corresponding to each grid point.
The relevant end-state values for each run
are quoted in Table~\ref{tab:runlist} (Appendix~\ref{runlist}). As seen, nearly all models are indeed evolved until the cluster
is nearly deprived of BHs. After this, the cluster enters the binary-burning state, often
making the run tedious to continue (see Ref.~\cite{Banerjee_2025} and references therein for a
discussion). Nevertheless, it was possible to evolve $\approx40$ of the model clusters
beyond 10 Gyr. A handful of runs have prematurely crashed in a way that they could not be
recovered for continuation.

\section{Results}\label{res}

Table~\ref{tab:runlist} (Appendix~\ref{runlist}) summarizes the outcomes of each model cluster. Below, the results
are discussed in detail.

\subsection{Star cluster evolution}\label{evol}

\begin{figure}
\centering
\includegraphics[width = 0.49\textwidth, angle=0.0]{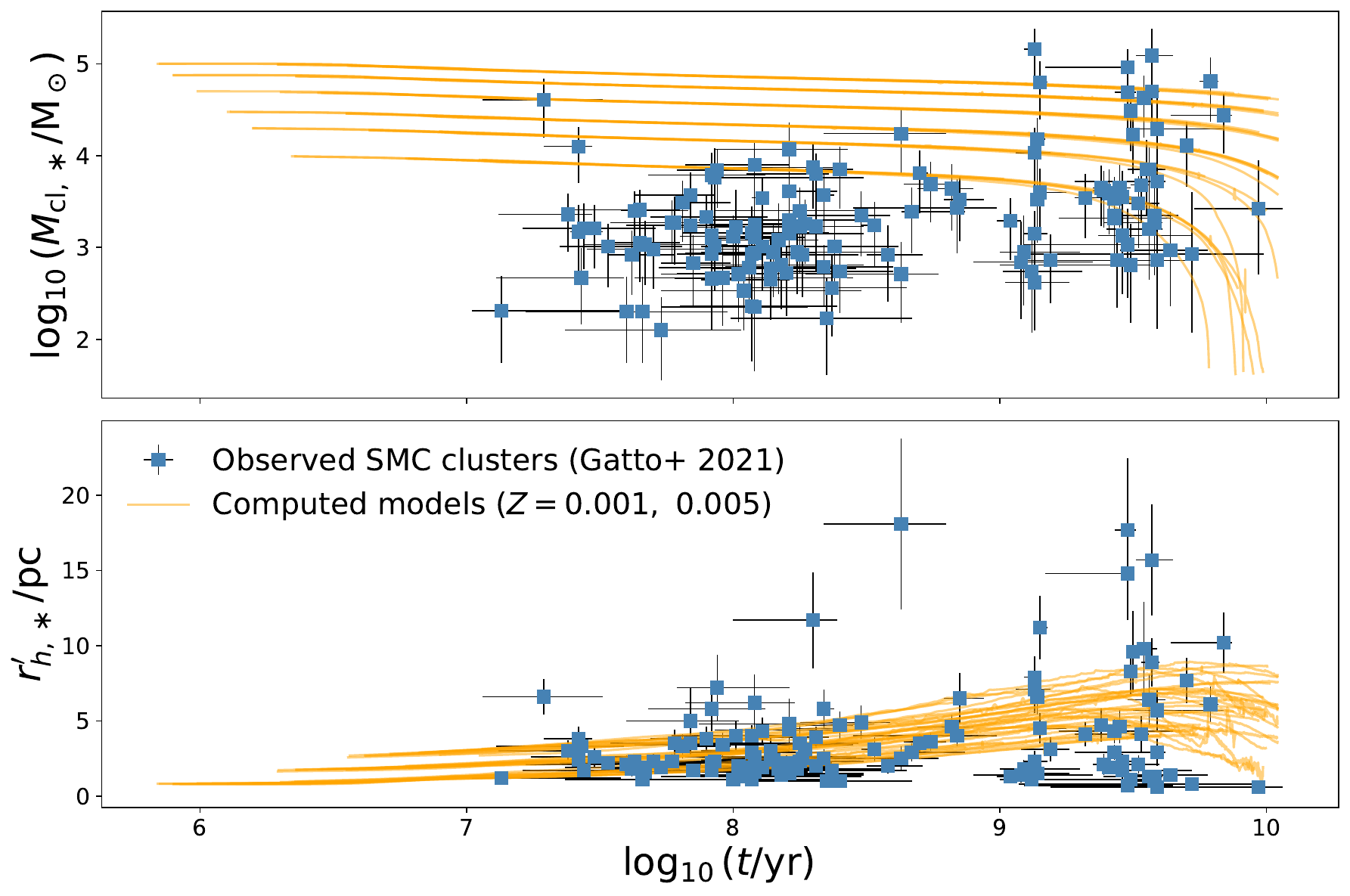}
\includegraphics[width = 0.49\textwidth, angle=0.0]{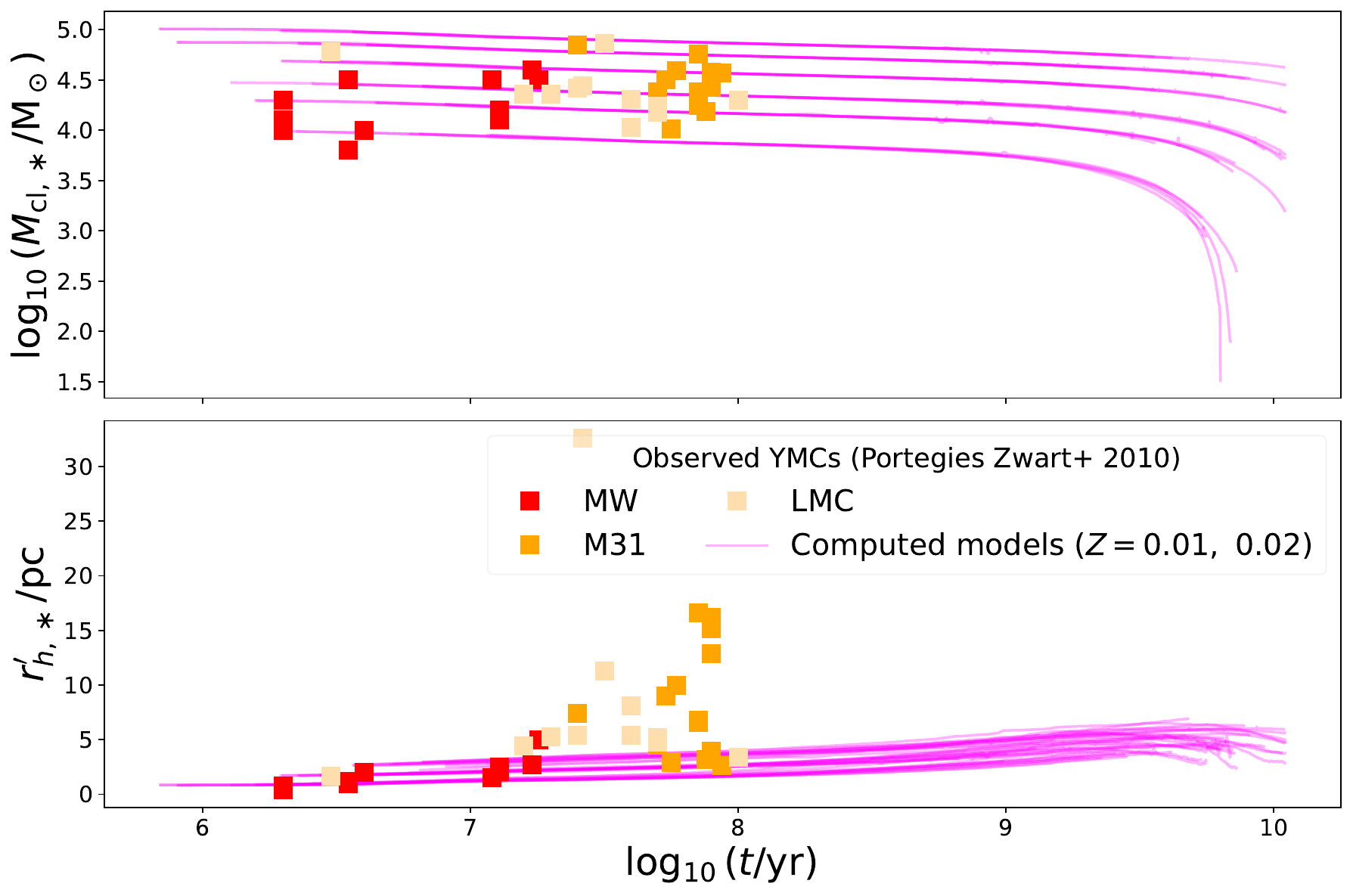}
\caption{Time evolution of the luminous mass (total mass of luminous stars within the
	instantaneous tidal radius), $\mclstar$, and
	size (corresponding projected half-mass radius, averaged over three orthogonal projections), $\rhpst$,
	of the computed model star clusters
	as functions of the cluster-evolutionary age, $t$, at low ($Z=0.001$, 0.005; upper pair of panels)
	and high ($Z=0.01$, 0.02; lower pair of panels) metallicities (solid lines). The masses, sizes, and ages
	of the low-$Z$ (high-$Z$) models are compared with those of the observed star clusters
	across ages in the SMC \cite{Gatto_2021} (MW, M31, and LMC \cite{PortegiesZwart_2010}) (filled symbols).}
\label{fig:smc_lg_comp}
\end{figure}

Due to retention of BHs, the evolution of the model clusters is generally affected by the strong mass segregation
of BHs \cite{Spitzer_1987}, central BH core formation, and the associated balanced evolution --- also termed BH-heating
\cite{Henon_1975,Banerjee_2010,Breen_2013,Antonini_2020}.
For a given initial mass and size of a cluster, a cluster with lower
metallicity generally produces BHs that are more massive and larger in number, due to weaker stellar winds and
shorter stellar lifetimes \cite{Banerjee_2020}. This, in turn, would initially form a more massive and concentrated central
BH core, leading to more energy release in close encounters inside the core, and causing the cluster to expand
to a larger extent. The cluster expansion, in turn,
serves as a `negative feedback' by causing also the BH-core to expand, thereby moderating the
associated dynamical ejections and depletion of the BHs \cite{Breen_2013,Heggie_2014}.
See Sec.~I of Ref.~\cite{Banerjee_2025} and the references therein for a comprehensive discussion
on BH heating. Other initial parameters being identical, a more
massive cluster would retain and dynamically process the BHs for a longer duration, leading
to generally longer delay times of BBH mergers and ejections from them. On the other hand, a more
concentrated cluster would process the BHs faster. See Refs.~\cite{Banerjee_2017,Banerjee_2017b,Kremer_2020}
for detailed discussions and demonstrations of these trends. Fig.~\ref{fig:evol} of Appendix~\ref{more} (this work) comprehensively demonstrates the
evolutionary trends with $Z$, $\mcl(0)$, and $\rh(0)$ by taking advantage of the present complete evolutionary model grid
\footnote{The lines or the boundaries in Figs.~\ref{fig:evol}, \ref{fig:smc_lg_comp},
\ref{fig:zevol} and \ref{fig:xeffz_cn21} are moderately smoothed by applying the Savitzky-Golay-Filter
(with 12 points and polynomial order 3), as available in the {\tt SciPy} module
{\tt scipy.signal.savgol\_filter}.}.

If the BH-progenitor stars of a star cluster are in primordial binaries,
the BHs' birth masses can be affected by the binaries' internal evolutionary processes
(\eg, mass transfer and tidal interaction between binary members) and as well by
stellar collisions triggered by binary-single and binary-binary dynamical encounters.
Furthermore, the internal evolution of a binary itself can be influenced by dynamical encounters.
Accordingly, birth masses of the BHs (and also of the NSs) derived from a dynamically interacting massive binary population
would deviate stochastically from a well-defined initial-mass--remnant-mass relation, as
has been demonstrated in recent studies.
In particular, star-star collision products can produce massive BHs within the canonical PSN mass gap between
$40.5\Ms-120\Ms$ \cite{Spera_2019,Banerjee_2020,Gonzalez_2020,Banerjee_2020c,DiCarlo_2020b,Banerjee_2022,Ballone_2023},
that would typically be unachievable by single stellar evolution beginning from the ZAMS
\footnote{Conventionally used single stellar evolution models, including the one adopted
in this work (Sec.~\ref{nbsims}), result in a PSN mass gap of stellar-remnant BHs. However,
alternative (non-collision-product) single star evolution models exist that allow BH formation within
the conventional PSN mass gap by either modifying/bypassing the occurrence of pair instability
and pulsation pair instability (\eg, Refs.~\cite{Belczynski_2020d,Ziegler_2021})
or modifying the stellar wind \cite{Vink_2021}. Pure isolated evolution of massive binaries
may also result in PSN-gap BHs by, \eg, allowing for super-Eddington accretion \cite{vanSon_2020}.}.
To produce a BH with birth mass within this `forbidden' mass range, the merger needs to take
place between two massive stars with dissimilar evolutionary stages, so that a helium core of mass lower
than the PPSN threshold can acquire an over-massive hydrogen envelope. Dense, dynamically
active environments such as the present cluster models enable collisions and mergers
among a wide range of stars, facilitating the formation of PSN-gap BHs.

Fig.~\ref{fig:smc_lg_comp} shows the time evolution of the total luminous mass, $\mclstar$,
and size, $\rhpst$ (projected half-mass radius of the luminous mass distribution),
of the model clusters at low and high metallicities.
These low-$Z$ (high-$Z$) curves are compared with the masses and sizes of the observed 
star clusters across ages in the SMC \cite{Gatto_2021} (MW, M31, and LMC \cite{PortegiesZwart_2010}).
The $Z$ values of the models presented in Fig.~\ref{fig:smc_lg_comp} are chosen to be similar to the
metallicities of the observed clusters as stated in the corresponding references.
As seen, the computed models represent certain subsets of the observed clusters.
For the observed SMC sample, the computed models mainly cover the relatively massive ($\mclstar \gtrsim 10^3\Ms$),
old (age $t\gtrsim1$ Gyr), and moderately extended
($2{\rm pc} \lesssim \rhpst \lesssim 8{\rm pc}$) clusters. While the SMC clusters' sizes are
also generally covered by the model clusters at younger ages, the present model grid, by construction,
does not contain any cluster over the mass range $10^2\Ms \lesssim \mclstar \lesssim10^3\Ms$ for $t\lesssim300$ Myr,
unlike the observed sample.

For the observed MW-LMC-M31 sample, the masses are well represented
by the model clusters. As for sizes, the model clusters cover up to $\rhpst\lesssim5$ pc - the
highly extended clusters of MW, LMC, and M31 are not covered by the present grid.
Notably, all observed clusters in the MW-LMC-M31 sample are massive ($\mclstar\gtrsim10^4\Ms$) 
and younger than $\approx100$ Myr.
The above model-observation comparison should be considered as preliminary and at a ballpark
level. The observed cluster samples are far from complete. Also, for a detailed comparison,
synthetic observations of the model clusters are necessary, which is beyond the scope of this
study. However, it can be said that the computed models are generally
comparable to the observed massive and compact star clusters of the MW and Local Group galaxies across ages.

\subsection{Gravitational-wave mergers from model clusters}\label{mrg}

\begingroup

\setlength{\tabcolsep}{7.5pt}

\begin{table*}
\centering
\caption{Counts of GW coalescences from the computed model grid (Table~\ref{tab:runlist}, Figure~\ref{fig:grid}).}
\label{tab:mrgcnt}
\begin{tabular}{lrrr}
\toprule
         Merger location  & Dynamically paired & Primordially paired & Total \\
\toprule
 	Inside cluster     &  110               &  3                  & 113   \\
 	Ejected            &  13                &  29                 & 42    \\
 	Total              &  123               &  32                 & 155   \\ 
\bottomrule
\end{tabular}
\end{table*}

\endgroup

\begin{figure*}
\centering
\includegraphics[width = 0.49\textwidth, angle=0.0]{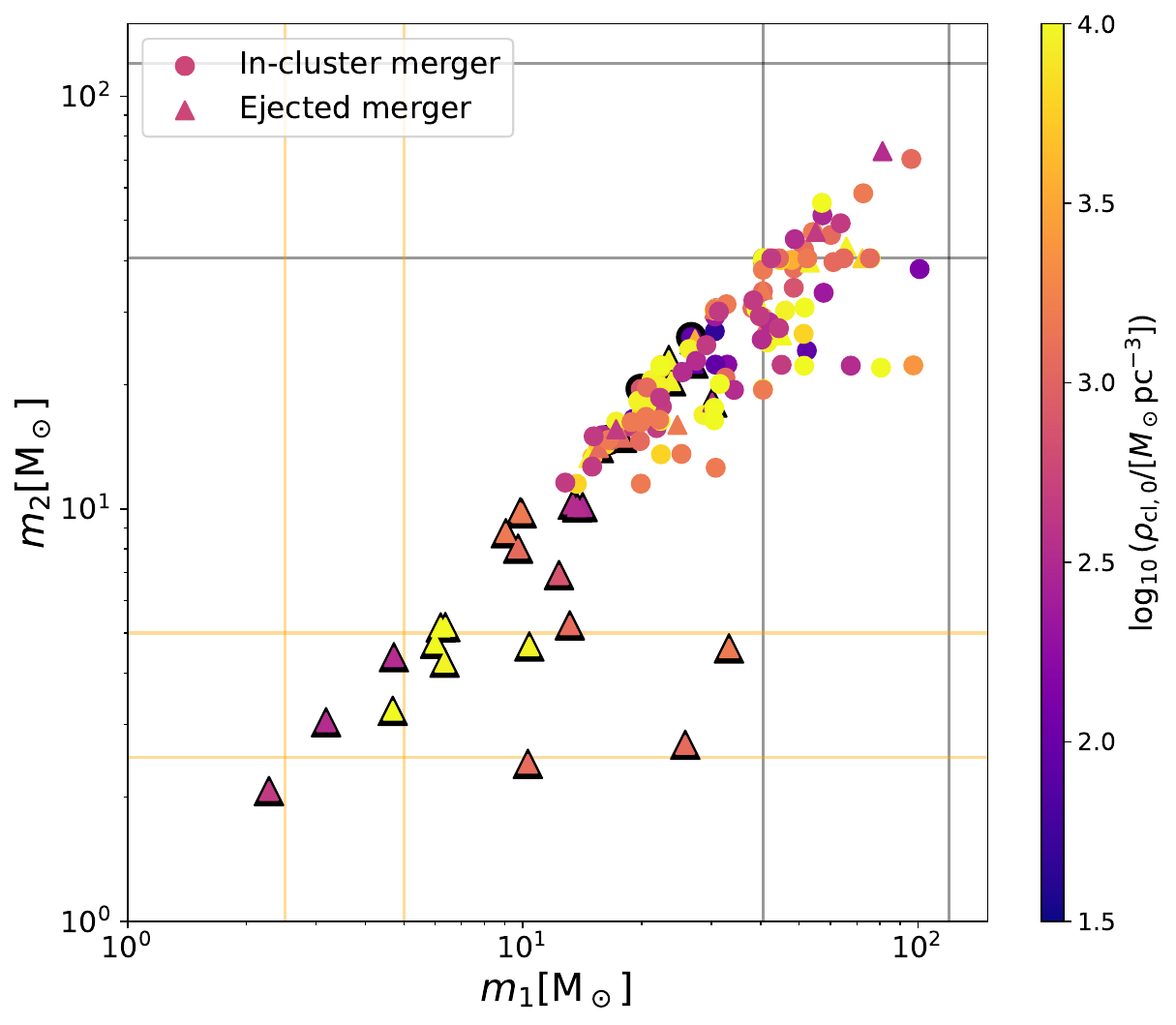}
\includegraphics[width = 0.49\textwidth, angle=0.0]{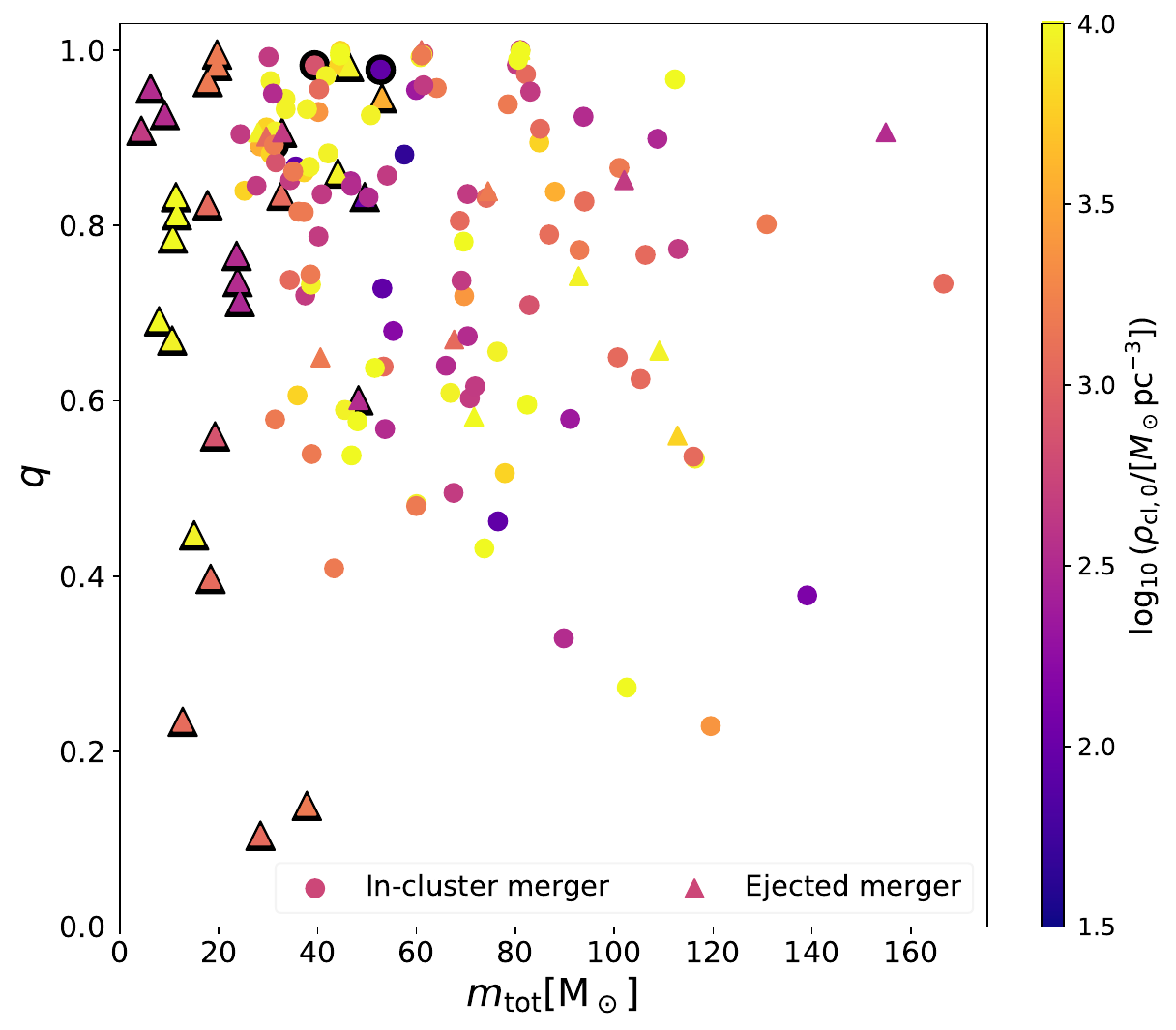}
	\caption{Compilation of GW mergers of compact binaries produced by all the 90 evolutionary model clusters in this work.
	{\bf Left panel:} the filled symbols represent the mergers' primary mass, $\mone$, along the X-axis versus
	the mergers' secondary mass, $\mtwo$, along the Y-axis ($\mone\geq\mtwo$).
	The circles and the triangles mark the in-cluster and the ejected mergers, respectively.
	Those merging binaries that preserve their primordial pairing (\ie, the primordially paired mergers)
	are indicated by an additional thick-lined edge around their respective symbols. The data
	points are colour-coded with respect to the initial half-mass density, $\rhoclzero$, of their respective
	parent cluster (colour bar). The canonical lower or NS-BH mass gap between $2.5\Ms-5.0\Ms$
	and the upper or PSN mas gap between $40.5\Ms-120.0\Ms$ are marked along the axes with the
	horizontal and vertical lines. {\bf Right panel:} the filled symbols represent the mergers'  
	total mass, $\mtot$, along the X-axis versus the mergers' mass ratio, $q\equiv\mtwo/\mone\leq1$,
	along the Y-axis. The meaning of the different symbols and the colour-coding is
	the same as in the left panel. See text for further details.
	}
\label{fig:m1m2}
\end{figure*}

\begin{figure*}
\centering
\includegraphics[width = 0.49\textwidth, angle=0.0]{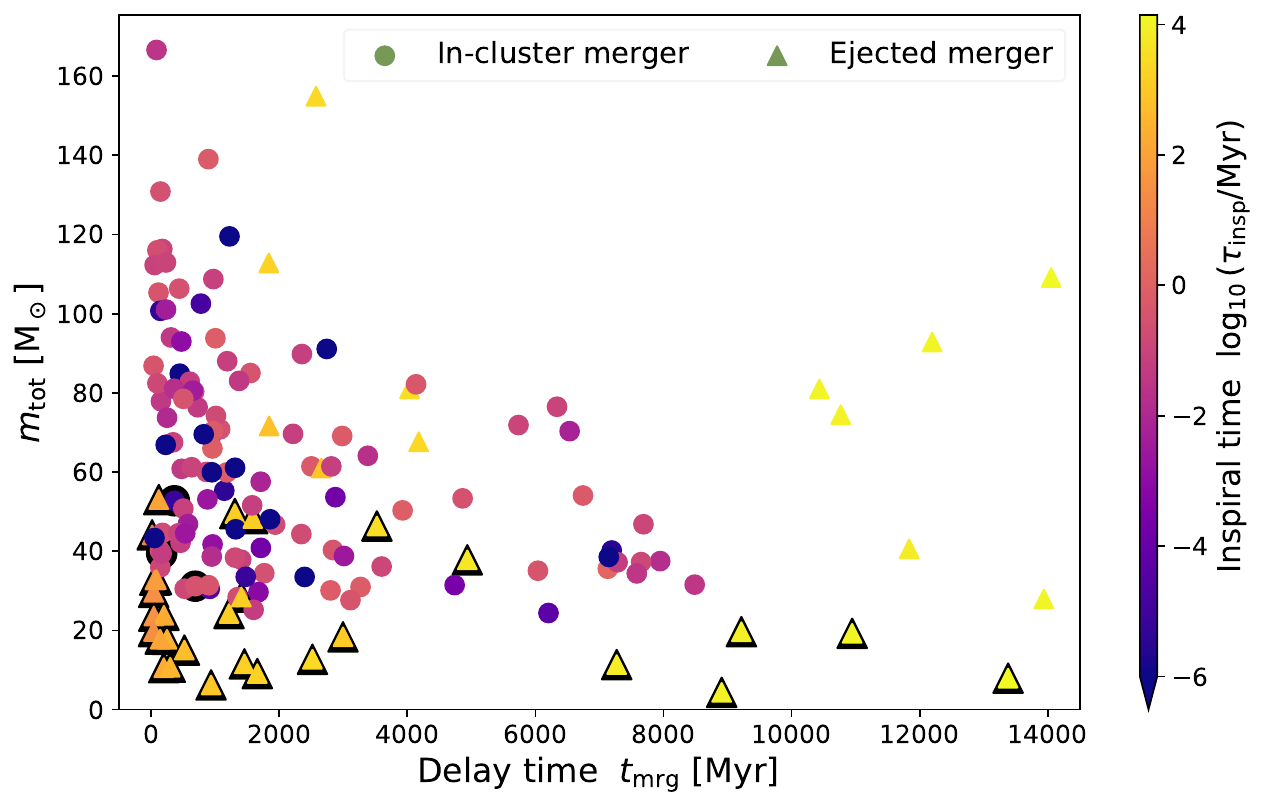}
\includegraphics[width = 0.49\textwidth, angle=0.0]{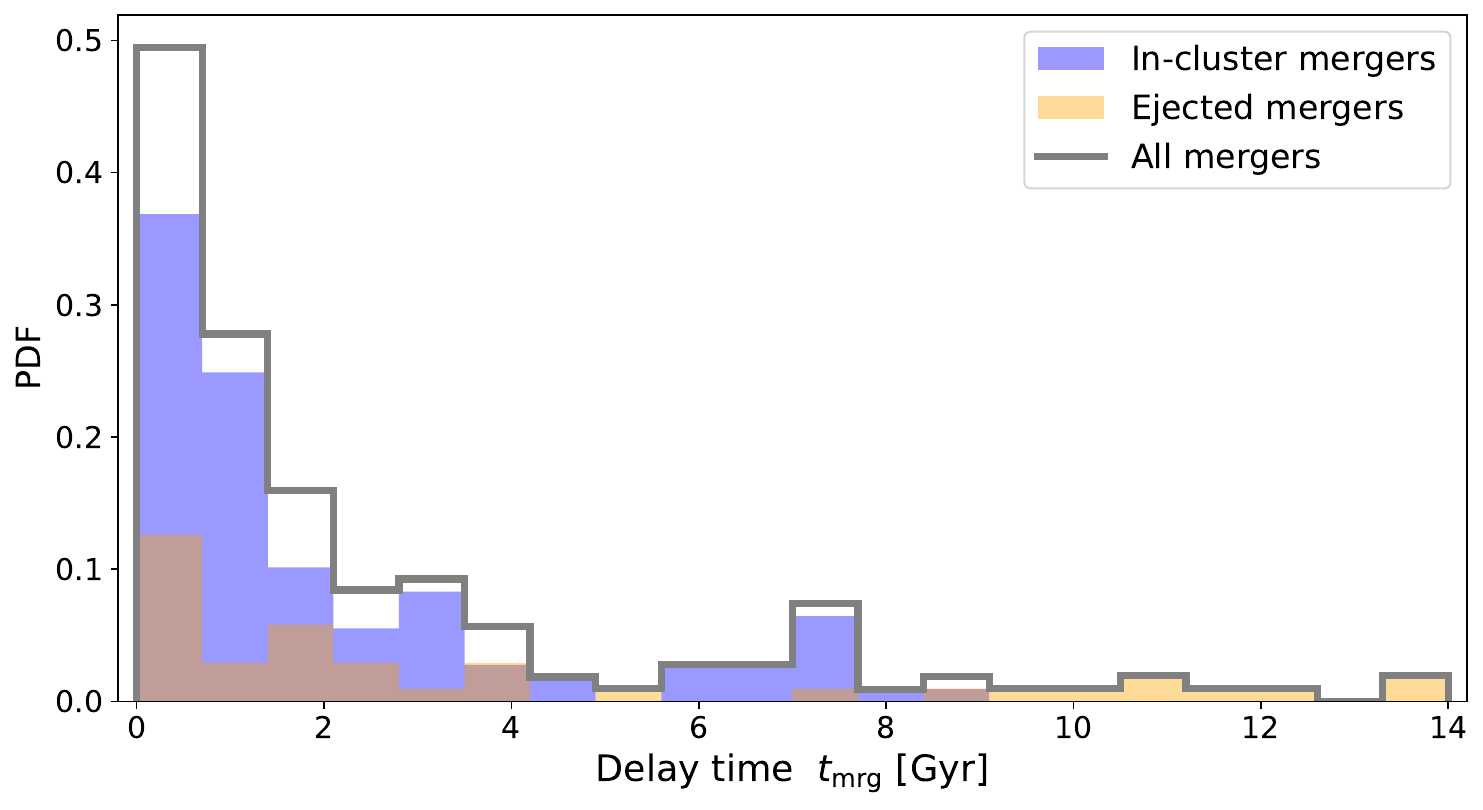}
	\caption{{\bf Left panel:} the filled symbols mark the delay times, $\tmrg$ (X-axis),
	of all compact binary mergers from
	the model-cluster grid versus their corresponding total mass, $\mtot$ (Y-axis). The data points
	are colour-coded according to the mergers' respective inspiral time until merger, $\tauinsp$, since
	their decoupling from the parent cluster (colour bar). For convenient visibility, $\tauinsp$ values below
	$\lesssim 1$ year share the same colour. The meaning of the different symbols is
	the same as in Fig.~\ref{fig:m1m2}. {\bf Right panel:} distribution (probability density function)
	of $\tmrg$ shown for the in-cluster and ejected mergers separately (filled histograms),
	as well as for the combined merger population (empty histogram).
	The combined distribution is normalised to be integrated up to unity. The integrals of the
	in-cluster- and ejected-merger distributions
	are scaled according to the respective sub-population's count relative to the total merger count (so that
	the distributions add up to the normalised combined distribution).}
\label{fig:mtot_tdel}
\end{figure*}

\begin{figure*}
\centering
\includegraphics[width = \textwidth, angle=0.0]{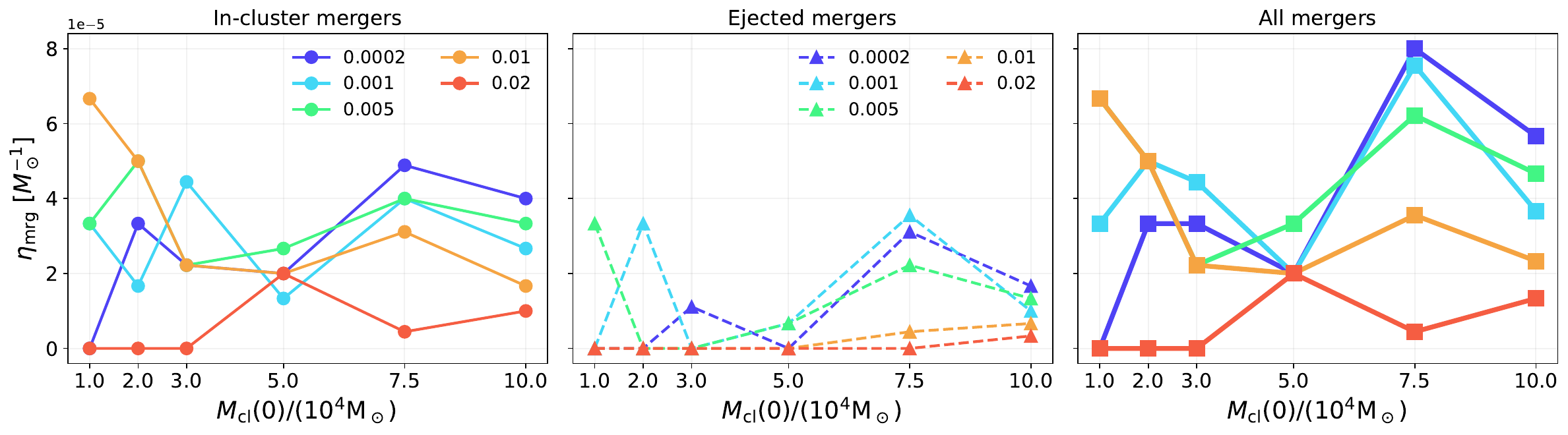}
	\caption{Efficiency per unit initial cluster mass(Y-axis), $\etamrg$, of in-cluster (left panel), ejected (middle),
	and all (right) compact binary mergers as a function of cluster initial mass, $\mcl(0)$ (X-axis),
	and metallicity, $Z$ (legend), as obtained from the computed star cluster model grid in this work.}
\label{fig:etamrg}
\end{figure*}

Due to the formation of a strongly mass-segregated, dense, central BH subpopulation or BH-core, 
these model clusters have the potential to form BBHs through close dynamical interactions,
a fraction of which can undergo GW inspiral and merger.
The key is to boost the BBH's eccentricity to an adequately high value via, \eg, von Zeipel-Kozai-Lidov
mechanism, chaotic triple or higher-multiplicity interactions, or
close flyby encounters \cite{Aarseth_2012,Banerjee_2018,MarinPina_2024}. The GR inspiral and merger
can take place inside the cluster based on the GR coalescence criteria of Ref.~\cite{Aarseth_2012}. Otherwise,
if the BBH is dynamically ejected from the cluster, then, depending on the binary
parameters at the time of the ejection \cite{Peters_1964}, it may merge outside the cluster's tidal radius
within the Hubble time to become a GW event candidate. Due to the presence of massive primordial  
binaries (Sec.~\ref{nbsims}), the pairing of compact binaries (apart from BBHs, also possibly NS-BH and NS-NS
binaries) can happen also due to the evolution of the massive binaries, \ie, the pair maintains
the primordial-binary membership. After formation, such a primordially paired compact binary
can undergo GW coalescence in-cluster or following ejection from the cluster (the ejection being due to either dynamical
encounters or natal kick of the member remnants) maintaining the membership, leading to
a primordially paired GW merger. However, due to the dynamical perturbations of the primordial
binaries or of their derivative compact binaries inside their host clusters,
their binary-evolution pathways are potentially altered. Therefore, the primordially paired
merger population from star clusters does not necessarily represent compact binary GR mergers
from isolated massive binary evolution.

Table~\ref{tab:mrgcnt} quotes the number of in-cluster and ejected GW mergers from all the 90 computed
model clusters, distinguishing between dynamically and primordially paired mergers. The in-cluster mergers are
primarily dynamically paired (97.3\% of them), while the majority (69.0\%) of the ejected mergers are primordially
paired. Overall, the GW mergers from these cluster models are predominantly (79.3\%) dynamically paired.
Also, the majority (72.9\%) of the mergers take place inside the clusters.

Fig.~\ref{fig:m1m2} plots the masses of all the GR mergers from the computed model grid.
The left panel plots the mergers' primary mass, $\mone$, versus secondary mass, $\mtwo$ ($\mone\geq\mtwo$),
and the right panel plots their total mass, $\mtot$, vs mass ratio, $q(\equiv\mtwo/\mone\leq1)$. 
The symbols are colour-coded according to the mergers' parent cluster's initial half-mass density, $\rhoclzero$.
See the figure's caption for the meanings of the symbols and other details.
It is seen that the mergers with $\mtwo\lesssim10\Ms$ are nearly all ejected and primordially paired, whereas
the more massive mergers are more numerous and predominantly in-cluster and dynamically paired. In fact, the
least massive dynamically paired merger has $\mone\approx10\Ms$, $\mtwo\approx10\Ms$.

A number of these GW mergers have one or both of the component masses within the PSN mass gap between $\approx40\Ms-120\Ms$.
In this paper, such mergers are referred to as PSN-gap mergers.
In the current cluster models, these massive BHs are mainly outcomes of star-star and BH-star collisions
(Secs.~\ref{nbsims},\ref{evol}).
While hierarchical GR coalescence is a rather common way to produce BHs within and beyond the PSN mass gap in
massive GCs \cite{Rodriguez_2018,Li_2024b,Ye_2025,Mai_2025} and nuclear star clusters \cite{Chatto_2023,ArcaSedda_2024c,Paiella_2025},
this channel is disfavoured in the present, moderate-mass clusters due to their lower escape
speeds ($\lesssim50\kmps$). The most massive BBH merger obtained from the present computations is of $\mtot=166.6\Ms$. 
The present event set also includes one NS-NS merger,
one NS-BH merger and several mergers involving lower-mass-gap (hereafter LMG)
compact objects (Sec.~\ref{nbsims}), all of which are ejected, primordially paired mergers (see Fig.~\ref{fig:m1m2}).
Due to the adopted delayed remnant mass model, LMG objects form in the model clusters simply due to the single-star
remnant formation process. Due to the adopted NS upper mass limit, all LMG objects are designated
as BH in these computations.

As for the mergers' mass ratio, they can extend to highly asymmetric
values ($q<<1$) although, overall, symmetric mergers ($q\rightarrow1$) are favoured. This
preference arises due to the fact that both massive binary evolution
and BH-core dynamical interactions tend to pair BHs of similar masses \cite{Belczynski_2002,Moody_2009,Banerjee_2017}.
The lower-mass, primordially paired, escaped mergers extend to being
as asymmetric as $q\approx0.1$ (Fig.~\ref{fig:m1m2}), due to the NS-BH and LMG-BH mergers.
For the in-cluster, dynamically paired mergers, $q\gtrsim0.2$. Overall, no clear
trend between the mergers' masses and the parent clusters' initial density is found
in these models (Fig.~\ref{fig:m1m2}) -- it is possible that an underlying trend
is washed away due to the stochastic nature of the dynamical processes and the limited
number of events.

Fig.~\ref{fig:mtot_tdel} (left panel) plots the model mergers' $\mtot$ against their delay time, $\tmrg$,
distinguishing between in-cluster, ejected, primordially paired, and dynamically paired mergers
as in Fig.~\ref{fig:m1m2}. In this study, the delay time of a GW merger (or event) is defined as the time of
the merger since the beginning of the parent model cluster's evolution. 
Here, the data points are colour-coded according to the events' individual
inspiral time until the merger, $\tauinsp$; see the figure's caption for more details. 
For in-cluster mergers, the $\tauinsp$ values are obtained from the compact binary's
orbit-averaged orbital decay due to gravitational radiation \cite{Peters_1964},
beginning from the state when the in-spiralling binary can be considered to be dynamically
decoupled from the rest of the cluster \cite{Aarseth_2012}.
The $\tauinsp$ for ejected mergers are calculated from the binaries' orbit-averaged GW decay
starting from the state at the time of the ejection from the cluster. The $\tmrg$ values are obtained by adding
$\tauinsp$ to the decoupling or the ejection time.

Fig.~\ref{fig:mtot_tdel} (left panel) expectedly shows an overall anti-correlation between
the model events' $\mtot$ and $\tmrg$. Due to shorter mass segregation time \cite{Banerjee_2010}, more massive BHs
inside a cluster are dynamically processed earlier \cite{Chatterjee_2017b,ArcaSedda_2024c}. Furthermore,
more massive BBHs would generally have a shorter $\tauinsp$ \cite{Peters_1964}. These, combined,
give rise to this trend. For all in-cluster mergers, $\tauinsp<1$ Myr (colour coding), which is less
than the typical gravitationally focused binary-single close interaction timescale \cite{Spitzer_1987}
of $\sim 1$ Myr among BHs in such clusters. Only for such small $\tauinsp$ can an in-cluster
binary successfully undergo GW inspiral and coalescence, before the inspiral gets interrupted by an intruder.
See Ref.~\cite{Banerjee_2020c} for further details and discussions (c.f. their Fig.~9).

In contrast, the ejected mergers can have large $\tauinsp$ as they have up to the Hubble time (minus the
ejection time) to coalesce. Fig.~\ref{fig:mtot_tdel} (right panel) demonstrates the contrasting
$\tmrg$ distributions of the in-cluster and ejected mergers as obtained from the present models.
While the in-cluster mergers take place mainly within $\tmrg\lesssim2$ Gyr (although their $\tmrg$
distribution's tail is extended up to $\approx8$ Gyr), the ejected mergers have a much flatter
$\tmrg$ distribution extending up to the Hubble time.

Fig.~\ref{fig:etamrg} plots the GW merger efficiency per unit initial cluster mass, $\etamrg$,
of the model clusters as a function of $\mcl(0)$,
at the different metallicities. In this study, $\etamrg$ of a population of clusters with total
initial mass $M_{\rm cl,tot}(0)$ is defined as
\begin{equation}
\etamrg \equiv \frac{N_{\rm mrg,tot}}{M_{\rm cl,tot}(0)},
\label{eq:etamrg}
\end{equation}
where $N_{\rm mrg,tot}$ is the total number of GR mergers produced by that cluster population.
In Fig.~\ref{fig:etamrg}, the efficiency of in-cluster, ejected, and all mergers are shown separately
(left, middle, and right panels, respectively). All the $\etamrg$s are marginalised with respect
to the clusters' $\rh(0)$.

As seen in Fig.~\ref{fig:etamrg}, $\etamrg$ generally increases with decreasing $Z$ at all $\mcl(0)$
and for both in-cluster and escaped mergers. This is expected since at lower $Z$ cluster-retained BHs are more massive
and numerous, resulting in a denser and longer lasting BH core (Sec.~\ref{evol}; Fig.~\ref{fig:evol}). 
Interestingly, $\etamrg$ does not show a clear monotonic trend with $\mcl(0)$. Both in-cluster and ejected mergers
exhibit a minimum at $\mcl(0)=3\times10^4\Ms$ or $5\times10^4\Ms$ at all $Z$ except at the highest $Z=0.02$;
at $Z=0.02$, $\etamrg$ maximises at $\mcl(0)=5\times10^4\Ms$. Apart from the highest $Z$ (that causes
the lowest BH retention), $\etamrg$ tends to maximise at $\mcl(0)=7.5\times10^4\Ms$ considering all
mergers (Fig.~\ref{fig:etamrg}, right panel). At the lowest $\mcl(0)=1.0\times10^4\Ms$, $\etamrg$ 
attains comparable values at intermediate $Z$s but decreases at the lowest $Z$s. 
While the $\etamrg$ values may have been affected by the limited number of events from
the model grid, qualitatively similar trend with $\mcl(0)$ is exhibited at most metallicities for both
in-cluster and ejected mergers.
This trend can be interpreted as the increasing competition, with increasing cluster mass,
between the formation of a larger number of BBHs (at a given metallicity) and the cluster's
increasing ability to interrupt or ionise them via close dynamical interactions. 
Notably, other studies such as Ref.~\cite{Rastello_2021} have also found non-monotonic behaviour
of $\etamrg$ with $\mcl(0)$. 

Note that the total number of mergers from a cluster, overall, increases with the cluster's mass;
see Table.~\ref{tab:runlist}. However, it is the
merger efficiency that is directly relevant for estimating the merger rate \cite{Rastello_2021,ArcaSedda_2024c}.
In basic terms, the merger rate is
\begin{equation}
R \propto \epsacfe\psisfr\int_{\mcl}\etamrg(\mcl)\clmf(\mcl)d\mcl,
\label{eq:rr}
\end{equation}
where $\psisfr$ is the mass formation rate of stars (typically in $\Ms\peryr\permv$), $\epsacfe$
is the cluster formation efficiency (hereafter CFE) that determines the fraction of star formation
in the form of bound star clusters, and $\clmf(\mcl)$ is the distribution density function
of the clusters' birth masses (typically taken as $\propto\mcl^{-2}$).
The calculation of GR merger rate from the present models is detailed in Appendix~\ref{rate}.

\section{Comparisons with GWTC}\label{gwtc}

\begin{figure*}
\centering
\includegraphics[width = \textwidth, angle=0.0]{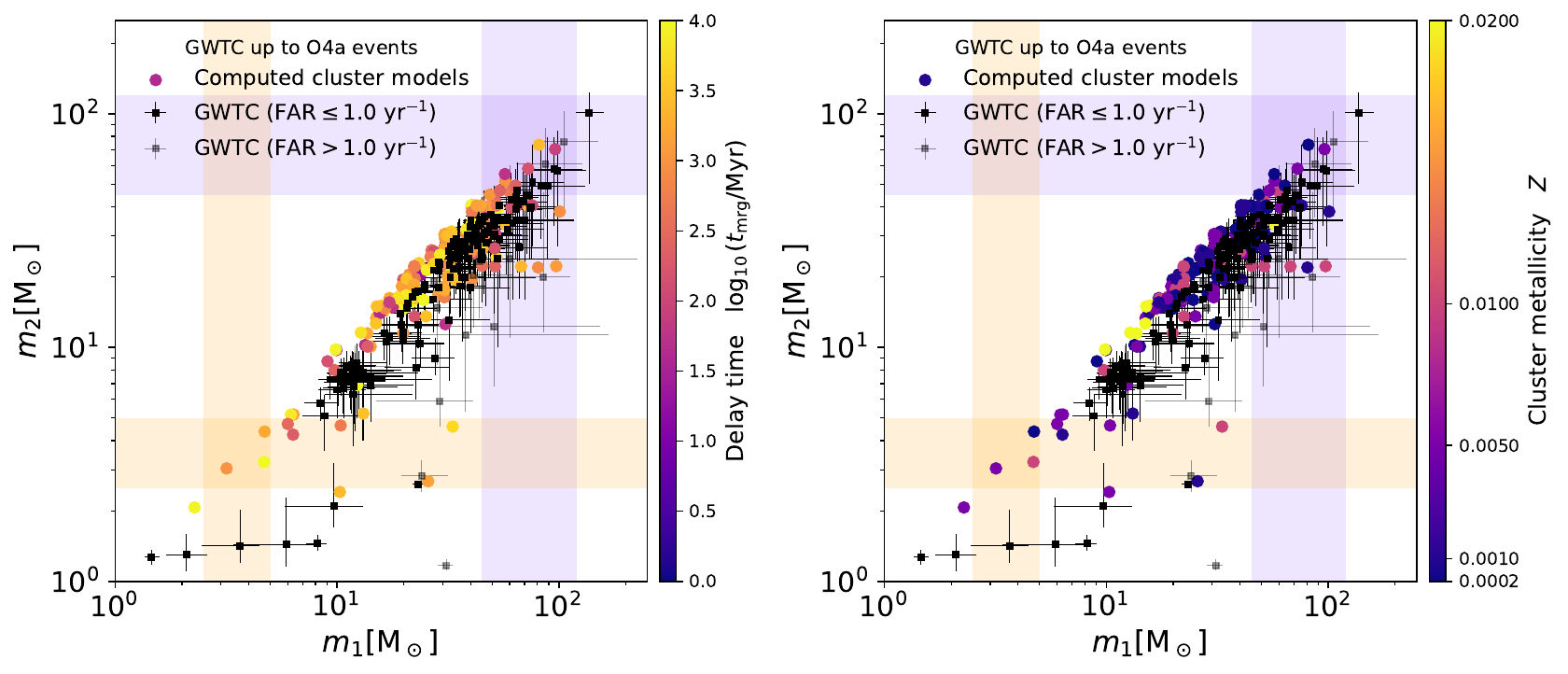}
\includegraphics[width = \textwidth, angle=0.0]{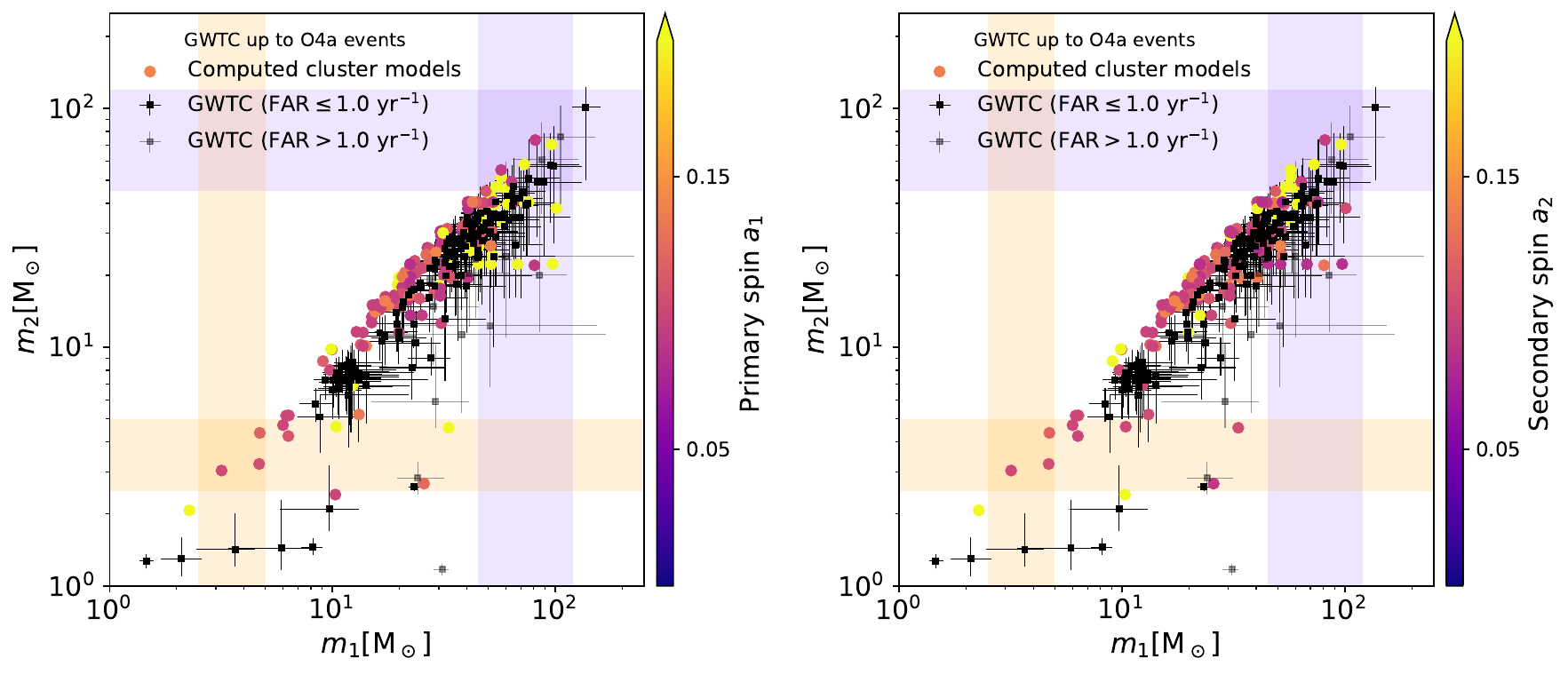}
	\caption{Primary mass-secondary mass ($\mone-\mtwo$; see Fig.~\ref{fig:m1m2}) plot of the GW
	coalescences produced by all the 90 evolutionary model clusters in this work (filled circles). The
	model-merger data points are colour-coded according to the mergers' delay time, $\tmrg$ (upper left panel),
	parent-cluster metallicity, $Z$ (upper right),
	primary-member dimensionless spin (or Kerr parameter), $\aone$ (lower left),
	and secondary-member dimensionless spin, $\atwo$ (lower right).
	On each panel, the canonical NS-BH and PSN mass gaps (see Fig.~\ref{fig:m1m2}) are indicated
	by filled patches. The spin-based colour coding is set such that a BH member with spin $\lesssim0.15$,
	\ie, preserving its natal spin, is represented by orange or a lower-valued colour. The remaining
	saturated-coloured members are candidates of possessing high spins. On each panel, events from
	the up-to-date, publicly available GWTC (the cumulative version) are overlaid for comparison
	(median values and the corresponding 90\% CIs as filled squares and error bars, respectively).
	Among the GWTC events, only the astrophysical candidates, \ie, those with $\pastro>0.5$ \cite{GWTC4a_cat}, are
	plotted that are furthermore distinguished across FAR $=1\peryr$ (legend). No colour coding is applied
	to the GWTC data points. See text for further details.}
\label{fig:m1m2_gwtc}
\end{figure*}

\begin{figure*}
\centering
\includegraphics[width = \textwidth, angle=0.0]{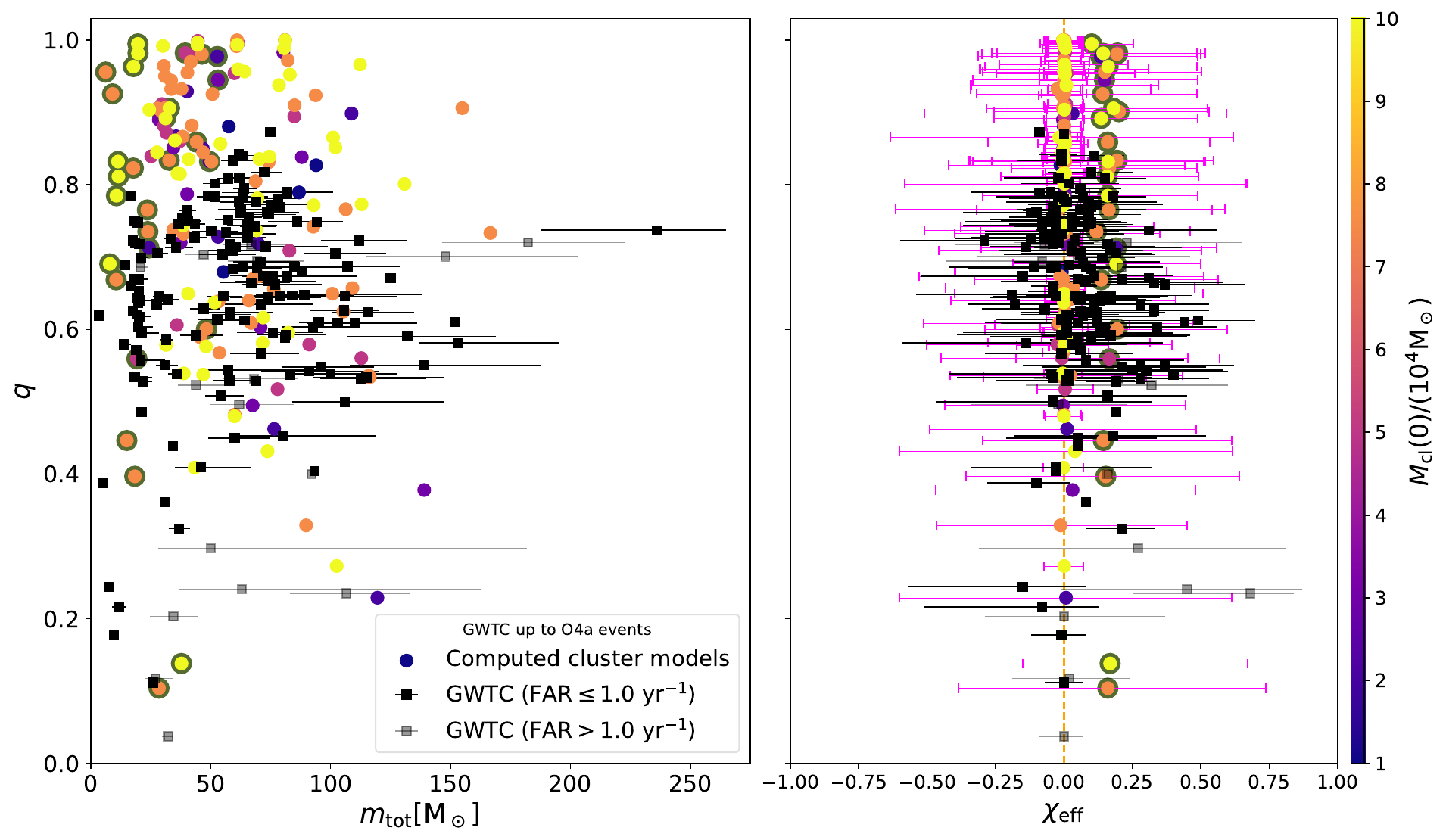}
	\caption{{\bf Filled circles:} The panels show total mass, $\mtot$ (left), and effective spin parameter,
	$\xeff$ (right), versus mass ratio, $q$, of the BBH coalescences produced by all the computed model clusters in this work.
	In both panels, the symbols are colour-coded with the
	initial mass, $\mcl(0)$, of the mergers' respective parent star cluster and
	the primordially paired mergers are marked with an
	additional thick-lined edge around their symbols. In the right panel,
	each symbol and the associated magenta bar represent, respectively, the mean and
	the range (90\% CI) of $\xeff$ values for the possible range of
	spin-orbit orientation of the corresponding merger, after incorporating dynamical tilt (see Appendix~\ref{bhspin}).
	{\bf Filled squares with error bars:} On each panel, events from
	GWTC-4 (cumulative version) are overlaid for comparison.
	The symbols and their error bars are
	plotted in the same way as in Fig.~\ref{fig:m1m2_gwtc}. The public GW event catalogue
	does not explicitly contain the observed GW events' mass ratio values and their uncertainties.
	Therefore, in the upper panels, the GWTC events' median secondary mass to median
	primary mass ratio and no error bars are plotted along the vertical axis. See text for further details.
	}
\label{fig:mt_xeff_q}
\end{figure*}

Fig.~\ref{fig:m1m2_gwtc} plots the model GW mergers' masses and compares them with the latest GWTC, \ie,
the catalogue containing all GW events up to LVK's O4a (Sec.~\ref{intro}). The plotted observed events (filled
squares) are distinguished across false alarm rate (hereafter FAR) $=1\peryr$.
In the four panels, the model events (filled circles)
are colour-coded according to their delay time, primary- and secondary-member dimensionless spin,
$\aone$ and $\atwo$ (Sec.~\ref{nbsims}, Appendix~\ref{bhspin}), and parent-cluster metallicity. It is seen that
the model mergers collectively agree well with the general trend of the observed events,
considering the latter's 90\% credibility intervals (hereafter CI; error bars). Apart from the general trend,
the model events also well reproduce several outlier LVK events, especially those having a companion close to or
within the LMG. This is a consequence of adopting the delayed remnant mass model (Sec.~\ref{nbsims}). 
Several of the (relatively) outlier BBH merger events are also reproduced by the model mergers. 
However, these model events do not reach the component masses of GW231123 -- the to-date most massive observed BBH merger \cite{Abac_2025b},
although the most massive model merger's secondary mass is within the 90\% CI of GW231123's secondary.
Note that to avoid overcrowding, the model mergers in Fig.~\ref{fig:m1m2_gwtc} are not distinguished as in-cluster/ejected or primordially/dynamically paired. However, the panels can be directly mapped onto Fig.~\ref{fig:m1m2} (left panel),
which provides such information (Sec.~\ref{mrg}).

The computed mergers have delay times of $\tmrg\gtrsim100$ Myr (see Sec.~\ref{mrg} for a detailed discussion on $\tmrg$)
and they are produced predominantly by moderate- and low-metallicity clusters ($Z\leq0.01$). The latter is evident also
from Table~\ref{tab:runlist}. As for BH spins (Fig.~\ref{fig:m1m2_gwtc}, lower panels), the PSN-gap model BBH mergers
can have highly spinning or low-spin components. The spin of BBH mergers with components in the PSN gap is of high interest
owing to GW231123 and other massive GW events
\cite{Abac_2025b,Banagiri_2025,Borchers_2025,Kiroglu_2025a,Stegmann_2025,Paiella_2025}.
The present model PSN-gap mergers possess a wide range of spin morphology. They comprise systems with both
components having high spin (like GW231123), a high-spin and a low-spin component, or both components having low spin.
In the current set of model events, all PSN-gap mergers are dynamically paired and the majority of them are in-cluster.

Among the model BBH mergers that have a component within or near the LMG,
highly asymmetric events ($q\lesssim0.5$) are found to be relatively common, with the primary having a high spin.
Qualitatively, such events resemble the observed events GW241011 and GW241110 \cite{Abac_2025a}
\footnote{In this study, the events GW241011 and GW241110 are not plotted in Figs.~\ref{fig:m1m2_gwtc}
and \ref{fig:mt_xeff_q} as they have not yet been included in the cumulative GWTC (as of January 2026). However, they
are included in the full list of $>300$ events (all events, all versions).}. In the present model
event set, all (near) LMG mergers are primordially paired and ejected (see Sec.~\ref{mrg}).
Note that, at present, $\nbseven$ can runtime assign astrophysical-model-based spins only to a fraction of the BHs
that are formed inside the cluster, namely, to those BHs that preserve their low natal spin of $\lesssim0.15$
or that are outcomes of in-cluster BBH mergers (Sec.~\ref{nbsims}).
However, spins of the high-spin BH candidates that are
outcomes of close, internally interacting stellar binaries or that have undergone mass accretion 
have to be reassigned post-run. This procedure is detailed in Appendix~\ref{bhspin}.

Fig.~\ref{fig:mt_xeff_q} plots the model BBH(-only) mergers' $\mtot$ and effective spin parameter,
$\xeff$ (X-axes), versus $q$ (Y-axes) and compares them with the current GWTC.
Here, the model data points corresponding to primordially paired mergers are indicated by a thick edge. 
The model data are colour-coded according to their respective parent clusters' $\mcl(0)$. It is seen that these mergers
well encompass the wide $q$-range of the GWTC events. Asymmetric ($q\lesssim0.5$), dynamically paired
mergers are often ($\approx50$\% of them) produced by lower-mass clusters ($\mcl(0)\leq5\times10^4\Ms$).
This is expected, since due to such clusters' relatively short relaxation and mass-segregation times \cite{Spitzer_1987},
the mass segregation of the BHs in them is more rapid and complete, allowing for dynamical interactions
over a wider BH mass range. See Sec.~\ref{mrg} for more discussions on mass ratios of
these model mergers.

The observable, effective spin parameter \cite{Ajith_2011}, is a measure of the spin-orbit alignment of the
merging binary and is defined as
\begin{equation}
	\xeff = \frac{(\mone\vec\aone + \mtwo\vec\atwo)\cdot\lcap}{\mone+\mtwo}
	= \frac{(\mone\aone\aonecap\cdot\lcap + \mtwo\atwo\atwocap\cdot\lcap)}{\mone+\mtwo}
\label{eq:xeffvec}
\end{equation}
or
\begin{equation}
\xeff = \frac{\mone\aone\cos\theta_1 + \mtwo\atwo\cos\theta_2}{\mone+\mtwo}
	= \frac{\aone\cos\theta_1 +  q\atwo\cos\theta_2}{1 + q}.
\label{eq:xeffdef}
\end{equation}
Here, the merging masses $\mone$, $\mtwo$, have, respectively,
dimensionless spin vectors $\vec\aone=\aonecap\aone$, $\vec\atwo=\atwocap\atwo$.
They make angles $\theta_1$, $\theta_2$, respectively, with the direction of the orbital angular momentum, $\vec L$,
defined by the unit vector $\lcap\equiv{\vec L}/L$
\footnote{In this paper, a vector $\vec A$ has magnitude $A$ and the unit vector in its direction is
denoted by $\hat{A}$.}.
During a GR inspiral, $\xeff$ remains nearly invariant \cite{Yu_2020,Gerosa_2021b}.
Therefore, to a good approximation, one can apply $\theta_1$, $\theta_2$
at the formation of the in-spiralling double compact binary to infer the LVK-measured $\xeff$
at a later time, during the final GW inspiral.

In Fig.~\ref{fig:mt_xeff_q} (right panel), the model $\xeff$ values are plotted by taking
into account the possible range of spin-orbit inclination for both primordially paired and
dynamically paired mergers. For the primordially paired events, the BH spins and the spin-orbit tilts
are adopted from a reference model evolutionary massive binary population that incorporates
formation of BHs with elevated spin from tidally spun-up Wolf-Rayet (hereafter WR) stars.
They are furthermore subjected to additional spin-orbit tilts mimicking experiencing dynamical scatterings
inside their parent cluster.
For the dynamically paired events, an isotropic spin-orbit tilt distribution is always assumed
for both components independently, after reassigning the
spins of the high-spin BH candidates: $(\cos\theta_1,\cos\theta_2) \in \unif(-1,1)$, where $\unif(-1,1)$ is
the uniform distribution over $[-1,1]$. The above procedures are detailed in Appendix~\ref{bhspin}.
In Fig.~\ref{fig:mt_xeff_q}, the random choices of the tilt, azimuth, and spin (when applicable) values are repeated
for 200 times for each model event to determine its sample $\xeff$s.
The corresponding mean and range (90\% CI) of a model event's $\xeff$ are plotted as a filled circle and its error
bar, respectively.

Owing to the preferential spin-orbit alignment even after incorporating
dynamical tilt \cite{Banerjee_2023}, the model primordially paired mergers are
generally biased towards $\xeff>0$. Considering the 90\% CIs, the computed mergers' $\xeff$ generally well
encompass all the observed mergers. However, the model events seemingly do not exhibit any $\xeff-q$
correlation (Fig.~\ref{fig:mt_xeff_q}, right panel). Due to the substantial uncertainty in $\xeff$ measurements,
no significant $\xeff-q$ dependence could be inferred from the to-date-observed GW events \cite{GWTC4a_pop}.

\subsection{Comparisons with GW event population properties}\label{GWTC_pop}

\begin{figure*}
\centering
\includegraphics[width = \textwidth, angle=0.0]{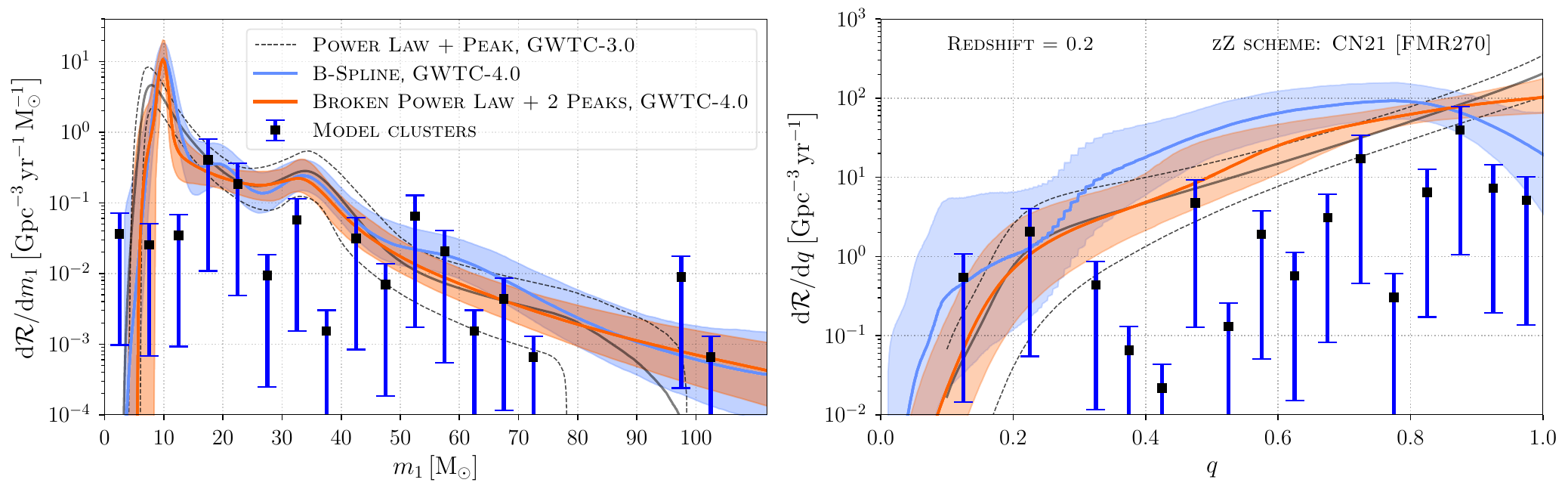}
	\caption{{\bf Black-filled squares with blue bars:}
	differential intrinsic merger rate density distribution at a fiducial redshift of $z=0.2$,
	as derived from the model-merger population synthesis in this work (see Appendix~\ref{rate}).
        At each bin, the bar spans between the standard and the pessimistic rate
	corresponding to that bin, and the filled square lies halfway between the rates.
	The left (right) panel plots the
	differential merger rate density with respect to the merger primary mass (mass ratio),
	$d\rate/d\mone$ ($d\rate/dq$). The model-cluster rate estimates presented here
	correspond to the CN21 redshift-metallicity ($zZ$) scheme (Appendix~\ref{rate}; Fig.~\ref{fig:zZsample}).
	{\bf Background overlays:} corresponding differential intrinsic merger rate
	density distributions at the same fiducial redshift, as inferred by LVK based on the GWTC-3 and GWTC-4 events,
	for different compact-remnant mass models (legend). The solid lines and the shaded areas correspond to
	the median and the 90\% CI of the distributions, respectively. The corresponding LVK limits
	from GWTC-3 are also shown (thin, dashed, black lines). The data and the style
	of the overlays are adopted from the publicly accessible resources corresponding to Ref.~\cite{GWTC4a_pop}
	(see their Figs.~3 and 5) without applying any rescaling or normalisation to the rate values.}
\label{fig:m1_q_dist}
\end{figure*}

To estimate the rate and properties of GW mergers in the Universe with redshift, $z$, due to the YMC/OC channel,
a Model Universe is constructed out of the cluster model grid and its GW mergers. The Model Universe
takes into account observation-based cluster birth mass function, observation-based cosmic star formation rate
(hereafter SFR) and
metallicity evolutions from $z=10$ up to $z=0$, and the standard background cosmological model,
as detailed in Appendix~\ref{rate}. In this section, only the results are discussed.

In Fig.~\ref{fig:m1_q_dist}, the black-filled squares with blue bars show the Model Universe differential intrinsic
merger rate density distribution with respect to primary mass (left panel) and mass ratio (right panel),
at a fiducial redshift of $z=0.2$. At each $\mone$ or $q$ bin, the bar spans between the
standard and the pessimistic rate corresponding to that bin (Appendix~\ref{rate}),
and the filled square lies halfway between the two rates.
For comparison, the corresponding LVK-obtained differential merger rate densities at the same redshift are co-plotted in
the panels' background: here, the $\bspline$ and $\bplpp$ models from GWTC-4 and the
$\plp$ model from GWTC-3 are shown. These background plots are the intact contents of
Figs.~3 and 5 of Ref.~\cite{GWTC4a_pop}.

It is seen that the model $d\rate/d\mone$, overall, accommodates and follows the shape of the continuum structure
of LVK's observed-event-based rate density distributions beyond $\mone\gtrsim15\Ms$. As in the LVK-inferred
$\mone$-distribution, the model distribution exhibits a peak at $\mone\approx30\Ms$\footnote{This feature
holds when the binning is altered.}.
However, the Model Universe
fails to produce a sharp, global peak at $\mone\approx10\Ms$, unlike the observed distribution. Instead, the model
rate density drops for $\mone\lesssim15\Ms$. Such a drop in merger rate for low-mass mergers
has been inferred also in other recent studies, \eg, Ref.~\cite{Antonini_2020b}. The computed models tend to overestimate  
the merger rate for $\mone\gtrsim90\Ms$. 

The model $d\rate/dq$ reasonably accommodates and follows the shape of the LVK's rate density distribution over
the entire range of $q$.
In particular, the model rate density distribution supports a maximum at $q\approx0.8$ as for the observed
$\bspline$ distribution. Note that the model rates presented in Fig.~\ref{fig:m1_q_dist} correspond
to the CN21 redshift-metallicity ($zZ$) dependence -- see Appendix~\ref{rate} for the details. 
Fig.~\ref{fig:m1_q_dist_xtra} (Appendix~\ref{more}) similarly plots the model rate distributions for the other two $zZ$ schemes
considered in this study, namely, CN19 and MF17 (Appendix~\ref{rate}). The above features of the model merger rate distributions
generally remain valid also for these alternative $zZ$ relations. The standard
values of the integrated  model intrinsic merger rate density at $z=0.2$ for the cases
of CN19, CN21, and MF17 $zZ$ dependence are, respectively, $9.9\peryg$, $8.8\peryg$, and $7.3\peryg$.  
This means that GW mergers from the present model clusters account for up to $\approx1/3$rd of the LVK-inferred
merger rate density \cite{GWTC4a_pop}.

Note that the above range of model merger rate density is due to variation
of the cosmic $zZ$ relation alone, which moderately affects the merger rate. There are other quantities
that would also influence the redshift-dependent model merger rate density, $\rate(z)$. In particular, $\rate(z)$ proportionates with the
comoving number density of GCs, $\ngc$ (see Eqn.~\ref{eq:rate}). Another important factor determining $\rate(z)$ is
the power-law index of the cluster birth mass function, $\clmf$, that determines how numerous newborn lower-mass clusters
are relative to GCs --- the steeper $\clmf$ is, the higher would be $\rate(z)$ (see Refs.~\cite{Antonini_2020b,Banerjee_2021} in this context).
The choice of cosmic star formation rate law, $\sfh(z)$, on the other hand, only moderately affects $\rate(z)$ (see, \eg, Ref.~\cite{Banerjee_2021b}).
The model clusters' initial profile, stellar composition, and size distribution would also indirectly affect $\rate(z)$ by partly determining the mixture of
the mergers' delay times. The above-quoted merger rate estimates assume the typical values and functions (Appendix~\ref{rate}) that are widely adopted
in the literature. A detailed study of the uncertainties in the model merger rate is beyond the scope of this study.

The abrupt decay of the model $d\rate/d\mone$ for $\mone\lesssim15\Ms$ is due to the adoption of the
delayed remnant mass model (Sec.~\ref{nbsims}). Owing to a monotonic growth of the SN fallback
fraction in the delayed model beginning from the lowest mass BHs \cite{Fryer_2012} and
the consequent high natal kick for low-mass BHs, the retention
of remnants of $\lesssim10\Ms$ is unlikely in the present model clusters at all metallicities \cite{Banerjee_2020}.
Therefore, dynamical pairing hardly contributes to GW mergers for $\mone\lesssim10\Ms$, wherein
mainly (escaped) primordially paired mergers alone contribute (c.f. Fig.~\ref{fig:m1m2})
\footnote{Although most ECS-NSs retain in the clusters, their mass segregation is
quenched due to BH heating \cite{Banerjee_2017b}. Even in the absence of BH heating,
the NS sub-population in a cluster would not have formed a dynamically
active central core \cite{Samsing_2021}, due to the NSs' much
lower mass. Therefore, the NSs are typically unable to participate in
dynamical pairing unlike the BHs.}.
The $30\Ms$-peak in the model rate distribution is likely caused by the mass distribution
of cluster-retained BHs that possesses such a peak for the delayed remnant model, especially at
low metallicities, combined with the more efficient
mass segregation and dynamical pairing of such BHs \cite{Li_2024a} compared to the lower-mass retained BHs --
see Fig.~8 of Ref.~\cite{Banerjee_2020}.

Of course, the mass distribution of retained BHs is affected in the presence of a massive binary population
due to various binary interactions or binary-mediated close dynamical
interactions (\eg, mass transfer between stellar components, mass accretion,
star-star collisions). Fig.~\ref{fig:remdist} (Appendix~\ref{more}) shows examples of cluster-retained
BH mass distributions in the present models that exhibit prominent peaks at
$\mbh\approx30\Ms$ and $40\Ms$, which are imprinted in the model $d\rate/d\mone$.
The distributions also demonstrate the sharp drop for $\mbh\lesssim15\Ms$ as discussed above.
The maximum of the model $d\rate/dq$ at $q\approx0.8$ is an imprint of the
prominence of BBH mergers around this mass ratio (see, \eg, Fig.~\ref{fig:m1m2}).

So far, the description of the features in the model merger rate density distributions and their comparisons with the LVK merger rate densities
have been qualitative and visual and based on the limited number of model events.
Appendix.~\ref{massmodel} describes a preliminary hierarchical, parametric population inference
based purely on the model mergers from this work --- see the appendix for the details.
The approach is similar to LVK's population
inference, but here a two-regime Power-Law-Multi-Peak primary-mass model separating at $\msplit=45\Ms$ is applied.
Also, independent power-law $q$-distributions across $\msplit$ as well as a global $q$-distribution are explored.
The analysis supports the existence of local peaks at $\mone\approx19\Ms$, $32\Ms$, $40\Ms$, and $52\Ms$
in the model primary-mass distribution (Table~\ref{tab:massmodel}). The parametric fitting also strongly supports a much flatter $q$-distribution
for $\mone\gtrsim45\Ms$ than that below, being consistent with LVK's latest findings \cite{GWTC5_pop}. Evaluation
of Jensen–Shannon (hereafter JS) divergences (Table~\ref{tab:jsd}) supports that the model distribution of massive primaries, for $\mone\gtrsim45\Ms$,
and the model $q$ distribution are consistent
with the form of the corresponding distributions inferred by LVK based on GWTC-4.

Notably, the mass distribution in the high-mass regime, $\mone>\msplit$, could be exaggerated
due to the assumption of a high matter-accretion fraction onto a BH, namely, $\ftz=0.95$ (see Sec.~\ref{nbsims}).
To test the impact of this assumption, the above two-regime hierarchical inference is repeated
by excluding the mass-accreted BHs from the set of model mergers. The re-analysis results
in practically unchanged BBH-merger $\mone$ distribution for $\mone\leq\msplit$. As for $\mone>\msplit$,
although the bulk of the PSN gap remains filled, the high-mass tail truncates at $\approx80\Ms$,
instead of $\approx110\Ms$ as for the full-event-set and LVK's distributions. Also, the support for a flatter $q$-distribution beyond $\msplit$
becomes weaker for the reduced set. That way, the present results, especially those for the high-mass regime,
depend to some extent on the adopted $\ftz$. The present setup produces high-mass $\mone$ and $q$ distributions
consistently with GW observations.

\begin{figure*}
\centering
\includegraphics[width = \textwidth, angle=0.0]{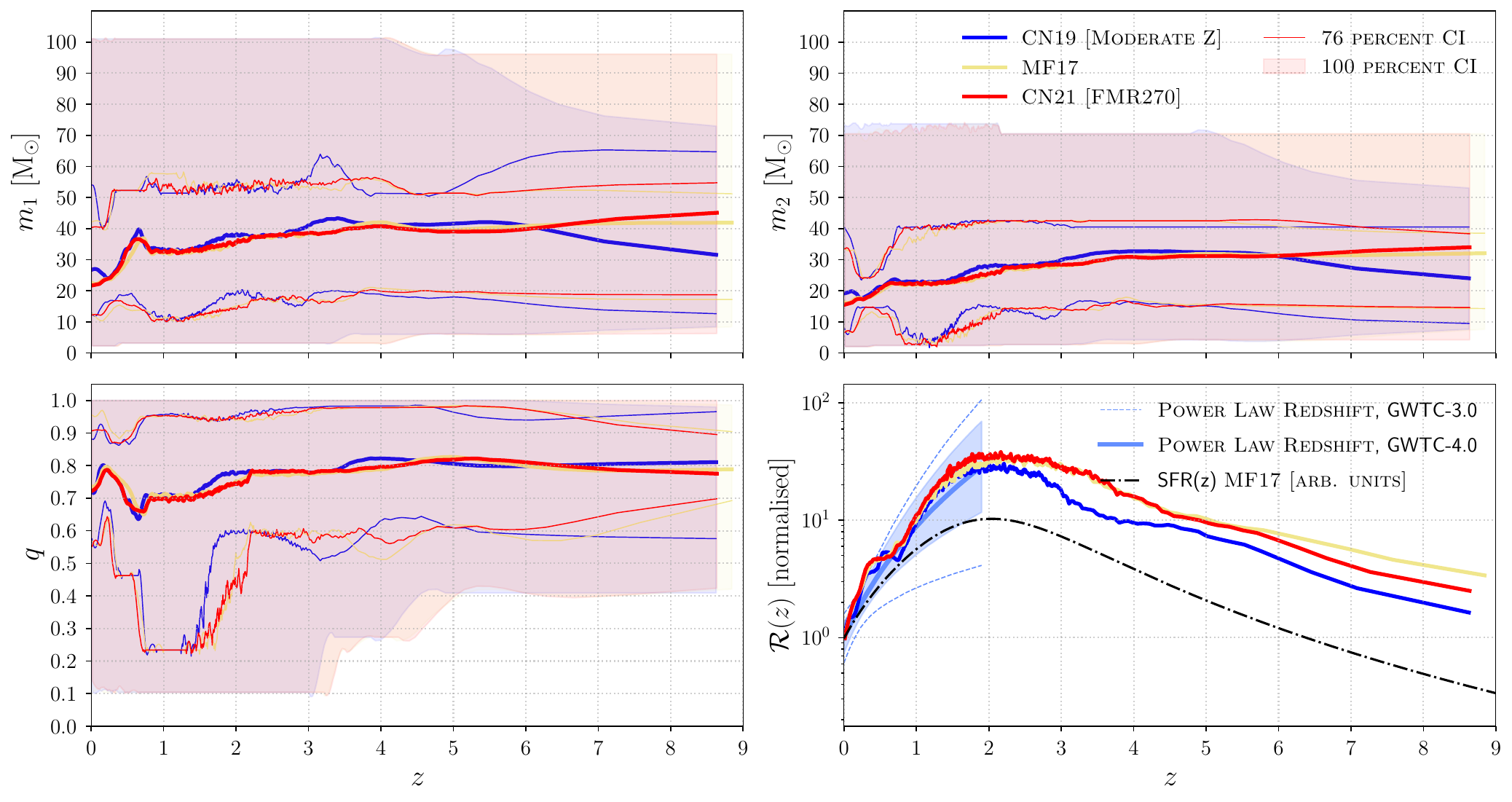}
\caption{Evolution of the distributions of various properties of model mergers with redshift, $z$.
	The upper left, upper right,
	lower left, and lower right panels show the redshift evolution of, respectively, merger primary mass, secondary
	mass, mass ratio, and the intrinsic merger rate density. The evolutions are shown
	for the different $zZ$ schemes considered here, namely,
	MF17, CN19, and CN21 (Appendix~\ref{rate}; Fig.~\ref{fig:zZsample}),
	which are distinguished colour-wise as shown in the legend. In all but the lower right panels,
        for a particular $zZ$ scheme, the thick line traces the mean value,
	the thin lines enclose the 76\% CI, and the shaded area encloses the whole range of
	the distribution at increasing redshifts. {\bf Lower right panel:}
	to facilitate comparison, each $\rate(z)$ curve is normalised with respect to its value at $z=0$.
	The star formation rate density--redshift dependence of MF17 is also shown with an analogous
	normalisation (black dot-dashed curve). Overlaid in the background are the median (thick, solid, sky blue line)
	and 90\% CI (blue-shaded region) of the intrinsic merger rate density evolution as obtained
	by LVK based on GWTC-4, which are, here, analogously scaled. The corresponding LVK limits
	from GWTC-3 are also shown (thin, dashed, sky blue lines). The data and the style
	of the overlays are adopted from the publicly accessible resources corresponding to Ref.~\cite{GWTC4a_pop}
	(see their Fig.~10).}
\label{fig:zevol}
\end{figure*}

\begin{figure*}
\centering
\includegraphics[width = 0.49\textwidth, angle=0.0]{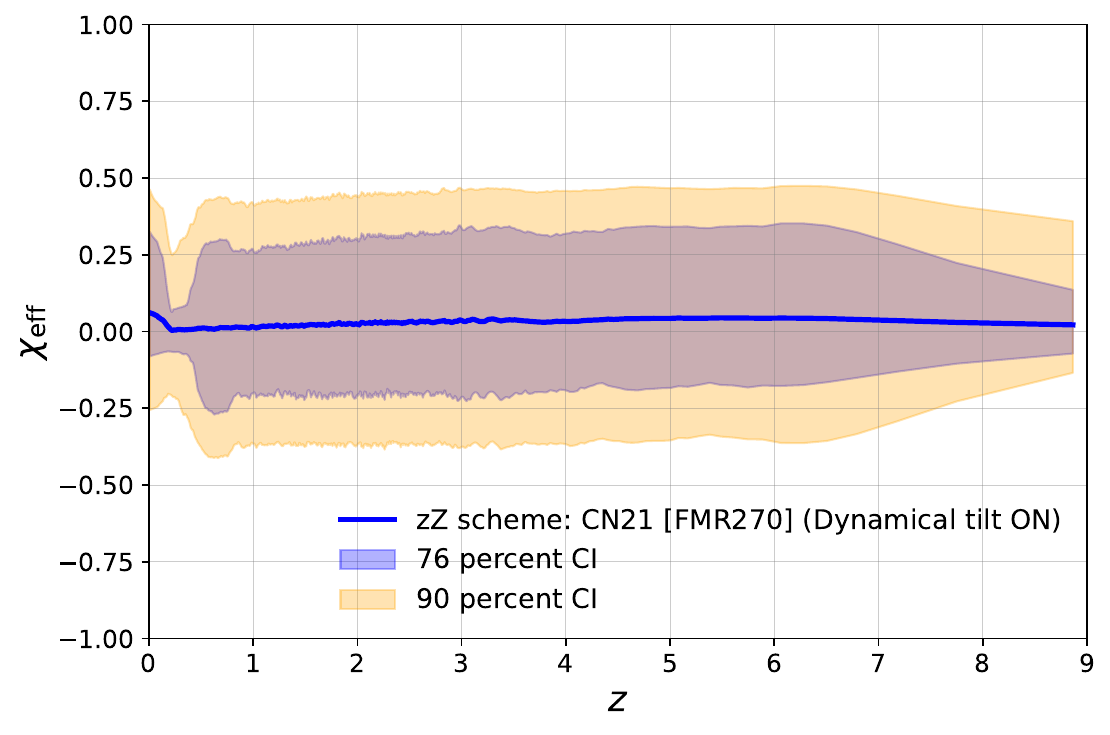}
\includegraphics[width = 0.49\textwidth, angle=0.0]{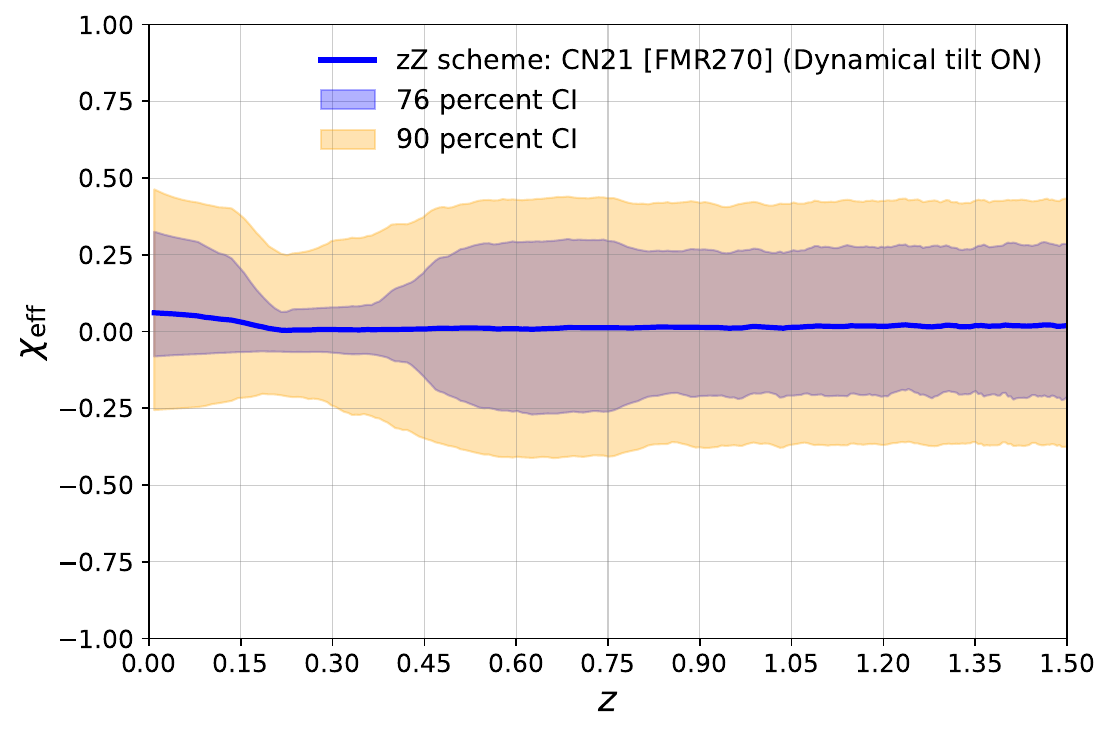}
	\caption{Evolution of the effective spin parameter ($\xeff$) distribution of model BBH mergers with redshift, $z$.
	{\bf Left panel:} the thick, blue line traces the mean and the shaded regions
	enclose the 76\% CI and 90\% CI (legend) of the $\xeff$ distribution at increasing redshifts.
	The evolution shown here corresponds to the CN21 redshift-metallicity ($zZ$) dependence.
	For the case presented in this figure, dynamical tilt is incorporated in evaluating
	the model events' $\xeff$. {\bf Right panel:} the same plot as in the left panel
	except that the X-axis is truncated at $z=1.5$, for better visibility of the
	$\xeff$ evolution at lower redshifts.
	}
\label{fig:xeffz_cn21}
\end{figure*}

\begin{figure*}
\centering
\includegraphics[width = 0.49\textwidth, angle=0.0]{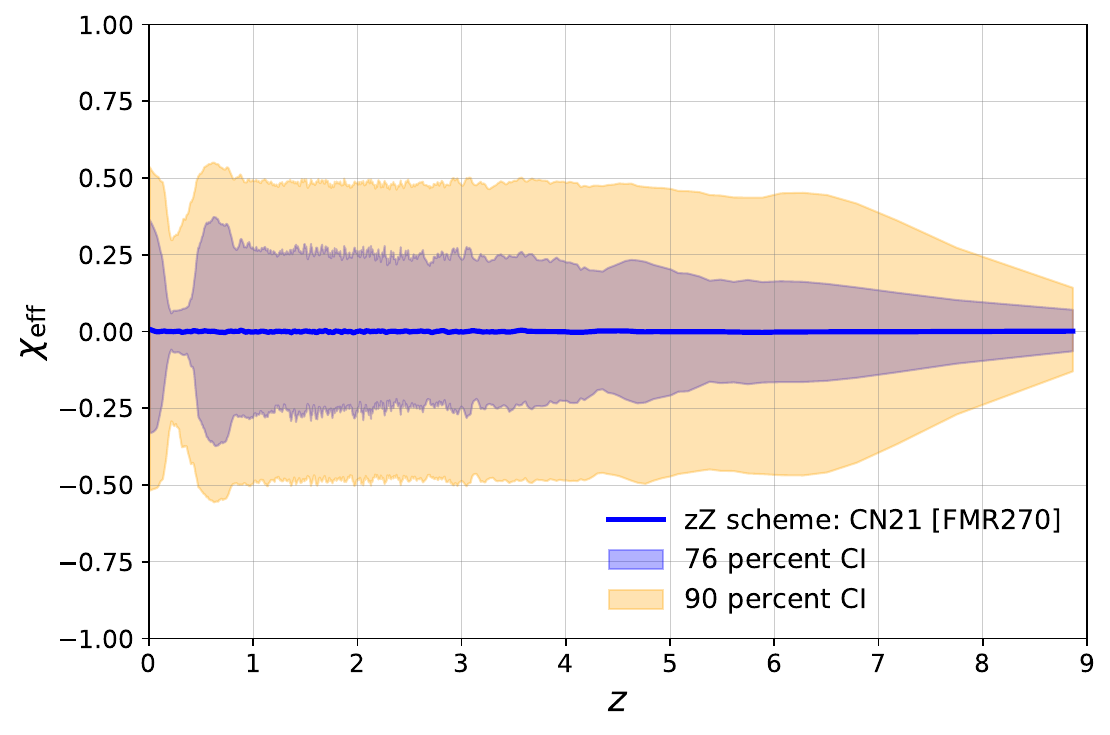}
\includegraphics[width = 0.49\textwidth, angle=0.0]{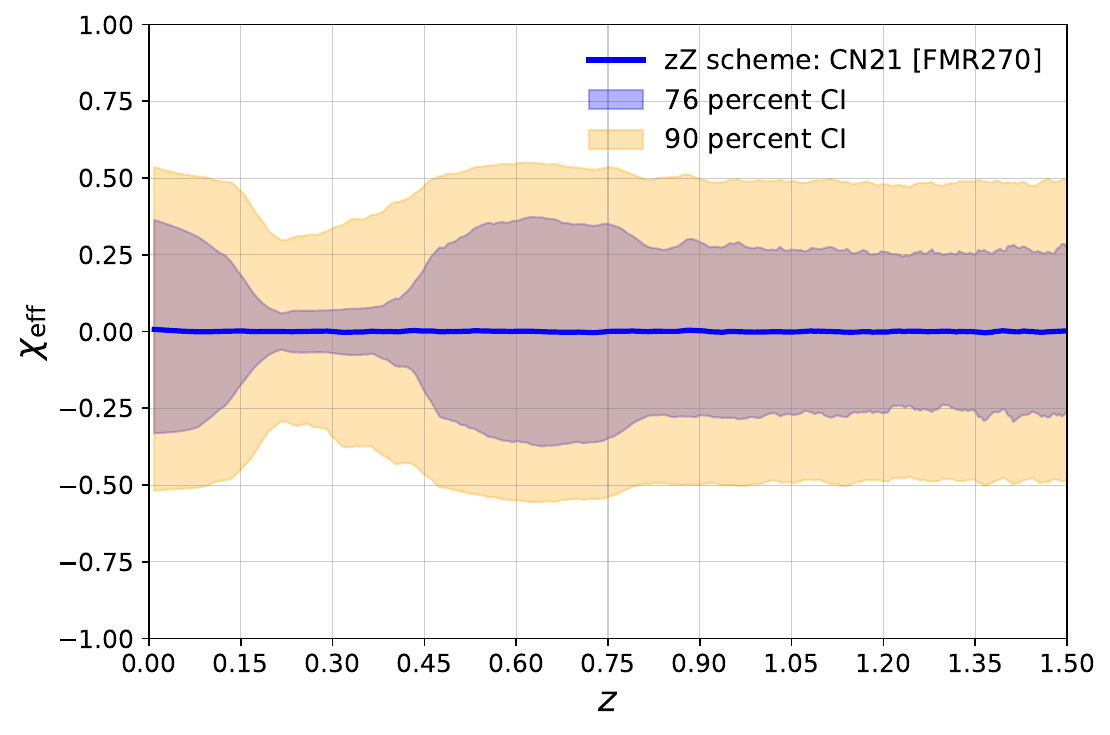}
	\caption{Same description as in Fig.~\ref{fig:xeffz_cn21} applies, except that here the BH spin magnitudes are not
	modified by any post-processing (i.e., are  taken directly from the computed N-body models) and isotropic BH-spin orientation is assumed for all
	merging BBHs.}
\label{fig:xeffz_cn21_direct}
\end{figure*}

\subsection{Evolution with redshift}\label{zevol}

In the context of GW formation channels and GW detection by future-generation detectors \cite{Bailes_2021},
redshift evolution of various GW-merger parameters and the merger rate are of interest. 
Fig.~\ref{fig:zevol} shows the redshift evolution of merger primary mass, secondary mass,
mass ratio, and merger rate as obtained from the Model Universes with the different
$zZ$ relations (Appendix~\ref{rate}). The mean $\mone$, $\mtwo$, and $q$ of the model mergers evolve significantly for
$z\lesssim0.7$, beyond which they evolve much more slowly with $z$. The 76\% CI
(equivalent of the full width at half maximum of a Normal distribution) of the distributions of these
quantities at increasing $z$ exhibit substantially higher variability for $z\lesssim2$ than for beyond this $z$.
The full range of $\mone$ ($\mtwo$) runs from the least massive merging remnant's mass
(a $\approx2.4\Ms$ NS; Fig.~\ref{fig:m1m2}) up to $\approx100\Ms$ ($\approx70\Ms$)
at all $z$ for the MF17 and CN21 $zZ$ cases; for CN19, the upper mass limits decay 
beyond $z\approx5$. In fact, for CN19, the mean $\mone$ and $\mtwo$ also decline
and deviate from the rest of the $zZ$ cases, for $z\gtrsim6$.

These deviations can be explained by the fact that the CN19 $zZ$ dependence is substantially
metal-poorer compared to the CN21 and MF17 counterparts over the entire $0\leq z \leq10$
(Fig.~\ref{fig:zZsample}). Since the primordially paired mergers from the current model
clusters are formed predominantly at low metallicities
with shorter delay times (Figs.~\ref{fig:m1m2}, \ref{fig:mtot_tdel}, \ref{fig:m1m2_gwtc}), the
CN19 Model Universe is richest in primordially paired mergers at all $z$, and, at high $z$,
primordially paired mergers are dominant for CN19. This is demonstrated in
Fig.~\ref{fig:fracz} (Appendix~\ref{more}); see below for further discussions. Since the primordially paired
mergers are generally less massive than the dynamically paired mergers (Fig.~\ref{fig:m1m2}),
their increased fraction at high redshifts causes the merger masses to decline therein,
for the CN19 case.

As for the mass ratio evolution, all the considered $zZ$ schemes produce similar redshift evolutions.
The $q$-distribution widens for $z\lesssim5$ (age of the Universe $\gtrsim1$ Gyr)
and the full $q$-range is achieved
for $z\lesssim3$. This is due to the fact that the most asymmetric model mergers
have delay times $>1$ Gyr (Fig.~\ref{fig:m1m2_gwtc}). Note that in all panels in
Fig.~\ref{fig:zevol}, equal-frequency bins are taken along the redshift axis
so that the evolutions of the means and the CIs are not affected by bin-to-bin
member number variation.

Fig.~\ref{fig:zevol} (lower right panel) plots the evolution of the Model Universe's intrinsic GW merger rate density
with redshift, for the three $zZ$ cases. The LVK-obtained, observed GW event-based  
rate density evolutions are plotted in the background. Here, LVK's power law-based
inferences for GWTC-3 and GWTC-4 \cite{GWTC4a_pop} are shown, since the power-law
model, namely, $\rate(z)=\rate(0)(1+z)^\kappa$, is simpler and is not explicitly
affected by the limited number of detected events from near the instruments' visibility horizon.
Furthermore, the form of the adopted cosmic SFR evolution \cite{Madau_2017}
(Appendix~\ref{rate}) is also shown on the same panel. To aid comparison of the $z$-dependence,
all $\rate(z)$ curves are normalised relative to their respective $\rate(0)$.
(For the GWTC-3 and GWTC-4 curves, the scaling is applied with respect to the respective mean GWTC rates
at $z=0$.)

It is seen that the redshift evolutions of the model merger rates, overall, follow
the GWTC-4 merger rate evolution up to $z\approx2$. The redshift evolutions
for the MF17, CN19, and CN21 merger rate densities are generally similar to each
other and they all achieve their global maxima at $z\approx2$, as for the SFR
evolution. However, the model rates increase with redshift faster than
the SFR, consistently with the GWTC-4 rate.

Fig.~\ref{fig:xeffz_cn21} shows the redshift evolution of the $\xeff$-distribution of BBH mergers
from the CN21 Model Universe. Here, $\xeff$ values of the model mergers
are obtained in the same way as described in Sec.~\ref{gwtc} and Appendix~\ref{bhspin},
\ie, by partially post-processing the BH spins, assigning spin-orbit tilts
based on isolated binary evolution and isotropic orientation distribution,
and incorporating added tilt angles due to close dynamical encounters.
However, since in a Model Universe each event represents a single instance, only a single set of random
draws is applied per model BBH merger. (The present Model Universes contain $\sim 10^6$ events each.)

It is seen that, at $z=0$, the model $\xeff$ distribution is moderately asymmetric and positively biased,
with a positive mean $\xeff$, $\mueff$ (blue solid lines in Fig.~\ref{fig:xeffz_cn21}).
However, $\mueff$ quickly declines to $\approx0$ at $z\approx0.15$ 
and remains vanishing with a gradual increasing trend until $z=1.5$. Over the long redshift range,
$\mueff$ continues to increase until $z\approx6$ and begins to decrease thereafter.
In general, $\mueff\geq0$ and the $\xeff$ distribution is asymmetric and positively biased 
at all $z$, for the model BBH mergers. However, at all $z$, the $\xeff$ distribution
extends significantly also towards $\xeff<0$.

Interestingly, the above low-redshift behaviour of $\mueff$, as obtained from the computed models,
is consistent with LVK's GWTC-4-based inferences of the same. Especially, the small but positive value of
$\mueff$ at $z=0$ and its vanishing at $z\approx0.15$ are supported by the corresponding GWTC-4 $\spline$ inference;
see Fig.~14 of Ref.~\cite{GWTC4a_pop}. The width of the model $\xeff$ distribution continues
to increase up to $z\approx0.75$ as for the GWTC-4 inferences, but after a constriction feature
over $0.15\lesssim z \lesssim 0.3$. The existence of such a constriction feature is marginally
supported by LVK's $\spline$ model for GWTC-4.
The above features in the model $\xeff$ distribution arise from how the fraction of
primordially paired mergers evolves with $z$ for these model mergers, which is shown in Fig.~\ref{fig:fracz} (Appendix~\ref{more}).
The model redshift evolution of $\xeff$ distribution are found to be qualitatively similar for the MF17 and  CN19 $zZ$ relations
considered here.

In Fig.~\ref{fig:xeffz_cn21_direct}, the $\xeff-z$ evolution for the CN21 Model Universe is shown without the BH-spin magnitude
modifications and reference to isolated-binary BBH mergers described in Appendix~\ref{bhspin}. Here, all BH-spin magnitudes
are drawn as-is from the model mergers, which were assigned runtime by the N-body computations (see Appendix~\ref{bhspin}
for the details). Also, for simplicity, all BBH mergers are assigned spin orientations drawn randomly from an isotropic distribution.
As seen, this `raw' $\xeff$ evolution has similar morphological features and $\xeff$-limits as in Fig.~\ref{fig:xeffz_cn21} but is strictly symmetric
around $\xeff=0$, due to the assumption of isotropic spin orientation for all mergers. This demonstrates that the
features in Fig.~\ref{fig:xeffz_cn21} are primarily outcomes of the runtime BH-spin modelling and evolutionary properties of the present
N-body computations --- the post-processing moderately modifies it mainly by adding $z$-dependent asymmetry.

Appendix~\ref{xeffinfer} presents LVK-type parametric inferences on the CN21 Model Universe $\xeff$ distribution
presented in Fig.~\ref{fig:xeffz_cn21} --- all details can be found in the appendix.
Likewise the LVK-inferred observed-event-based $\xeff$ distributions \cite{GWTC4a_pop},
the model events strongly favour a positively skewed $\xeff$ distribution
with a small negative mean  (Table~\ref{tab:xeffpars}) over a symmetric distribution.
Evaluation of JS divergences (Table~\ref{tab:jsd_xeff}) supports that the inferred model skew-normal $\xeff$ distribution
is, overall, consistent with the forms of LVK's $\xeff$ distributions, especially with their $\bspline$ distribution.
However, the sampled model $\xeff$ distribution possesses a core-like structure around $\xeff=0$ (populated mainly by low-spin,
dynamically paired BBH mergers; Fig.~\ref{fig:xeffinfer}) that is not apparent in the LVK-inferred $\xeff$ distributions.

\section{Comparisons with Gaia black holes and neutron stars}\label{gaia}

\begingroup

\setlength{\tabcolsep}{7.5pt}

\begin{table*}
\centering
\caption{Summary of the total formation efficiencies of escaped BH-MS and NS-MS binaries
as obtained from the computed model cluster grid in this study.}
\label{tab:gaiaeff}
\begin{tabular}{ccc}
\toprule
  Binary type &  Low-mass companion ($<3\Ms$) $[\Ms^{-1}]$ & Solar-mass companion ($0.7\Ms-1.1\Ms$)  $[\Ms^{-1}]$ \\
\toprule
  BH-MS       & $6.3\times10^{-6}$     & $7.0\times10^{-7}$ \\
  NS-MS       & $3.0\times10^{-6}$     & $1.2\times10^{-6}$ \\
\bottomrule
\end{tabular}
\end{table*}

\endgroup

\begin{figure*}
\centering
\includegraphics[width = \textwidth, angle=0.0]{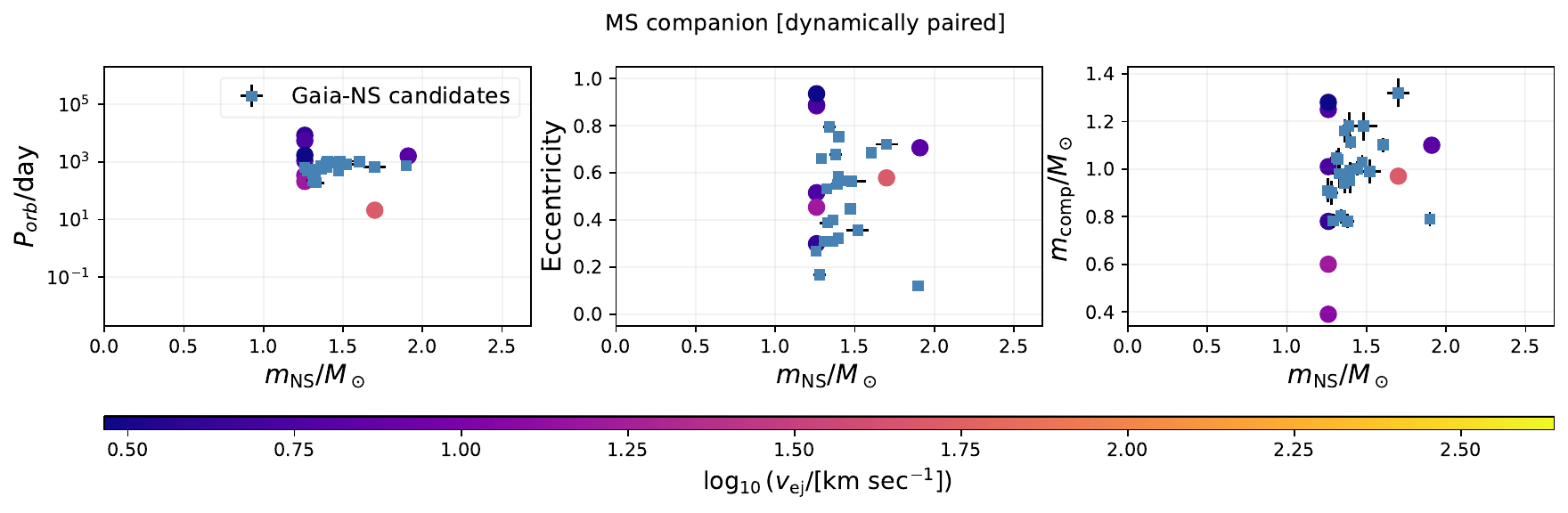}\\
\vspace{-1.3cm}
\includegraphics[width = \textwidth, angle=0.0]{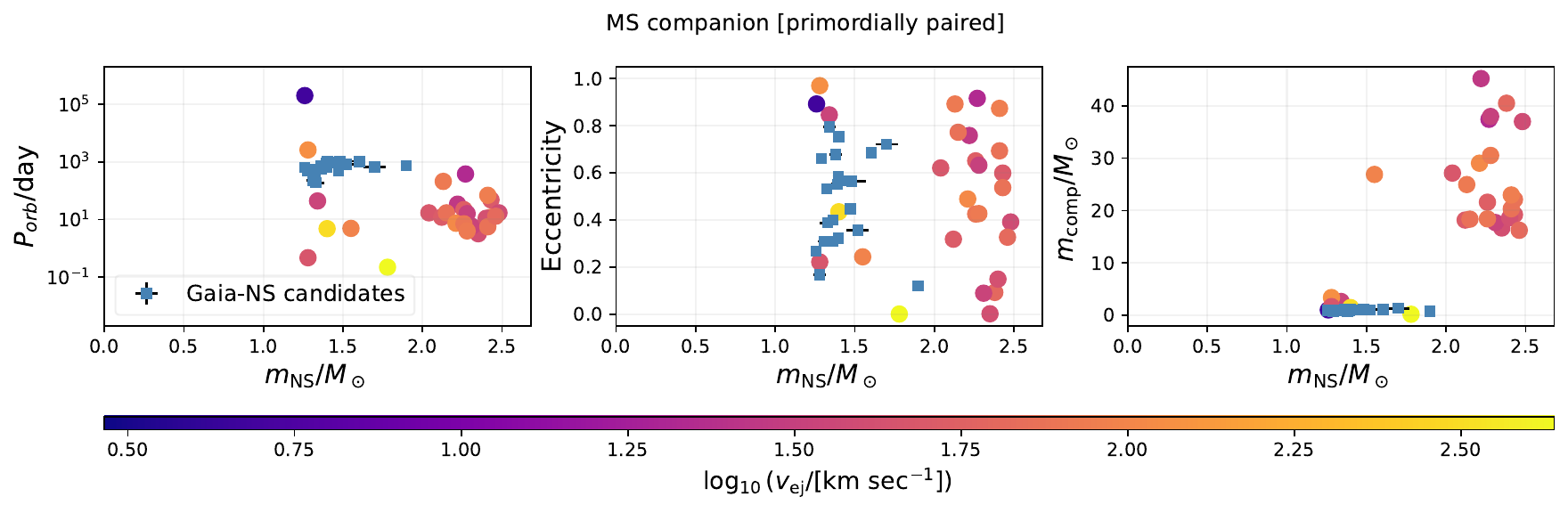}
\caption{Demographics of the NS-MS binaries that have escaped into the galactic field
from the evolutionary model star clusters in this study.
In each row, the filled circles in the panels plot the mass of the
NS member ($\mns$; X-axes) against the orbital period ($\porb$; left panel),
eccentricity (middle), and companion-star mass ($\mcomp$; right)
of these model NS-MS binaries. The binaries are plotted separately depending
on whether they are dynamically or primordially paired, as indicated in each row's title.
The data points are colour-coded according to the velocity, $\vej$,
of the binary's ejection (colour bar).
All plotted values correspond to the event of the binary's crossing of the instantaneous tidal
lobe of its parent cluster. For comparison, the Gaia-observed NS-MS binary candidates
are plotted in each panel (filled squares). For these observed data points, the colour coding is
not followed.}
\label{fig:GaiaNS}
\end{figure*}

Apart from GW observations, detached (\ie, non-mass-transferring), compact-object--normal-star binaries, as discovered
in the Galactic field by the Gaia mission, have the potential to provide clues regarding stellar remnant formation
and the physics of the various channels through which such systems may have formed.  
The current model cluster grid has also produced detached
BH-star and NS-star binaries that got ejected from the cluster and escaped to the
galactic field. The formation mechanisms and the overall properties of these field BH-star binaries are similar
to those found in earlier star cluster computations by the author. Hence, the discussion is not
repeated here and the reader is encouraged to consult Ref.~\cite{Kotko_2024,Banerjee_2025} for the details.
Fig.~\ref{fig:GaiaBH} (Appendix~\ref{more}) shows the ejected black hole--main sequence star (hereafter BH-MS) binaries from
the present model grid, distinguishing between dynamically paired and primordially paired systems and
comparing with the observed, detached BH-MS binary candidates
Gaia-BH1 \citep{Chakrabarti_2023,ElBadry_2023} and Gaia-BH3 \citep{GaiaBH3_2024} in the MW field.

The total formation efficiency\footnote{Defined analogously to Eqn.~\ref{eq:etamrg}.}
of field BH-MS binaries with a low-mass ($<3\Ms$) MS companion from the
present grid is $6.3\times10^{-6}\Ms^{-1}$. Restricting to only solar-mass-like ($0.7\Ms-1.1\Ms$) MS 
companions, the formation efficiency is $7.0\times10^{-7}\Ms^{-1}$. The vast majority (96\%)
of the escaped BH-MS binaries containing a
low-mass MS companion are dynamically paired (see Fig.~\ref{fig:GaiaBH}).

Fig.~\ref{fig:GaiaNS} plots the parameters of the ejected neutron star--main sequence star (hereafter NS-MS) binaries from
the present model grid, distinguishing between dynamically paired and primordially paired systems.
Over-plotted in the panels are the Gaia-detected detached NS-MS binary candidates in the MW field that are reported in
Ref.~\cite{ElBadry_2024}. The dynamically paired field NS-MS binaries from the model clusters well encompass
the Gaia NS binaries in terms of the ranges of NS mass, companion mass, and eccentricity. However, unlike the model
NS-MS binaries, the observed systems have orbital periods $<10^3$ days. It is possible that wider NS-MS
systems have remained undetected by Gaia until now. The primordially paired field NS-MS binaries also reasonably encompass
the Gaia NS systems. However, there is a distinct subpopulation of primordially paired NS-MS systems with NS masses beyond
the Gaia NS masses, of $>2\Ms$, and massive companion masses of $>10\Ms$. However, such massive
NS-MS systems are not expected in the Gaia sample reported in Ref.~\cite{ElBadry_2024}, which authors have
restricted to only systems with low-mass MS companions. Systems with such massive companions are anyway
unlikely to be detected owing to their short lifetimes and poor proper motion of the luminous
member around the binary's barycenter.

Interestingly, the Gaia NS sample's mass distribution is statistically consistent with a narrow
component at $\approx1.3\Ms$ along with an extended component up to $\approx2\Ms$ \cite{Schiebelbein_2025}.
This is indeed supported by these model dynamically paired and primordially paired NS-MS systems
with low-mass components (Fig.~\ref{fig:GaiaNS}). For the dynamically paired systems, the $\approx1.3\Ms$
NSs are ECS-NSs that retain in the clusters due to their small natal kicks (Sec.~\ref{nbsims}).
The total formation efficiency of field NS-MS binaries with a low-mass ($<3\Ms$) MS companion from the
present grid is $3.0\times10^{-6}\Ms^{-1}$. Considering only solar-mass-like ($0.7\Ms-1.1\Ms$) MS 
companions, the formation efficiency is $1.2\times10^{-6}\Ms^{-1}$. A moderate majority (61.5\%)
of the escaped NS-MS binaries containing a low-mass MS companion are dynamically paired (see Fig.~\ref{fig:GaiaNS}).
Table~\ref{tab:gaiaeff} summarises the above-mentioned model formation efficiencies of field
BH-MS and NS-MS binaries. Notably, Ref.~\cite{Tanikawa_2024b} reports an NS-MS formation efficiency
from their N-body model cluster set that is at least an order of magnitude smaller than that obtained here.
This discrepancy can be attributed to much higher cluster masses and longer evolutionary times of the
present models, which allow the cluster-retained NSs to undergo more sustained dynamical interactions.

\section{Summary and discussions}\label{discuss}

The properties of the to-date-observed GW-event candidates by the LVK and of their population suggest
multiple role-playing formation channels for the merging compact binaries in the Universe, rather than a single dominant
channel (Sec.~\ref{intro}). This study surveys the role of young massive clusters evolving into intermediate mass
($\sim 10^3\Ms - 10^4\Ms$), old open clusters, namely, the YMC/OC channel in forming
GW mergers. To that end, a grid of 90 model star clusters of homogeneous properties, over the initial mass range
$10^4\Ms\leq\mcl(0)\leq10^5\Ms$, initial size range $1{\rm~pc}\leq\rh(0)\leq3{\rm~pc}$, and metallicity range
$0.0002\leq Z \leq0.02$ are evolved for long term until most of the BHs are depleted from the clusters
(Sec.~\ref{grid}, Fig.~\ref{fig:grid}, Table~\ref{tab:runlist}). The model-cluster set considered in this study, therefore, serves as a bridge between low-mass
star clusters and massive GCs (Sec.~\ref{intro}). The models incorporate an observationally motivated population of primordial binaries,
the delayed stellar remnant mass model, black hole spins based on stellar evolution models, and in-cluster GR recoil kicks.
They are evolved with a customised version of $\nbseven$, a star-by-star-resolution N-body integrator that
incorporates stellar and binary evolution, stellar remnant formation, PN evolution of compact binaries, and collisions
among stars and stellar remnants (Sec.~\ref{nbsims}). The computed model clusters are generally consistent with the
massive and compact star clusters of the Milky Way and Local Group galaxies across ages (Sec.~\ref{evol}, Fig.~\ref{fig:smc_lg_comp}).
The goal of this work is to assess to what extent the YMC/OC channel can explain the properties of the GW events observed to date,
and whether it can simultaneously explain the compact star--normal star binaries discovered by the Gaia mission.

The main conclusions from this study are as follows:

\begin{itemize}

\item The GW mergers produced by the computed model cluster grid are primarily dynamically paired and take
place inside the clusters (Sec.~\ref{mrg}, Table~\ref{tab:mrgcnt}, Fig.~\ref{fig:m1m2}). On the other hand, among the fewer escaped
mergers, primordially paired systems are common. The vast majority of the model mergers are BBHs. While, overall, the mergers are
biased towards having a symmetric mass ratio, asymmetric mergers are rather common: the dynamically (primordially)
paired mergers extend down to a mass ratio of $q\approx0.2$ (0.1). With increasing primary mass, the model merger set switches from being
predominantly primordially paired and escaped to being predominantly dynamically paired and in-cluster across $\mone\approx10\Ms$.

\item The GW merger efficiency, $\etamrg$, of the model clusters exhibits a non-monotonic behaviour with $\mcl(0)$,
for both in-cluster and ejected events (Sec.~\ref{mrg}, Fig.~\ref{fig:etamrg}). At all metallicities except $Z=0.02$, $\etamrg$
reaches a maximum at $\mcl(0)=7.5\times10^4\Ms$. The $\etamrg$ for $\mcl(0)\leq3.0\times10^4\Ms$ is comparable to
that for $\mcl(0)\geq7.5\times10^4\Ms$, at most metallicities. The values of $\etamrg$ are generally of order $10^{-5}$, which is
consistent with the outcomes of N-body simulations by recent authors (see Ref.~\cite{Banerjee_2025} for a discussion).
Such $\mcl(0)$ dependence of $\etamrg$ suggests that the cluster mass range covered in this study is not exhaustive for GW-forming,
sub-GC-mass star clusters that are gas-free and parsec scale, and the GW-formation threshold of such star clusters
potentially lies at a lower mass. This means that the contributions of sub-GC-mass, parsec-scale, gas-free star clusters
towards GW sources in the Universe as obtained from the current model set are potentially lower limits.

\item The present model cluster grid produces GW mergers that agree with not only the mainstream events of the current GWTC but
also GWTC's outlier low-mass (\eg, highly asymmetric, involving a low-mass BH and a highly spinning primary BH like GW241011/GW241110)
and high-mass (\eg, double PSN-gap, highly spinning BH and misaligned like GW231123) events (Sec.~\ref{gwtc}, Figs.~\ref{fig:m1m2_gwtc},
\ref{fig:mt_xeff_q}). The most massive merger from the present model set is of $\mtot\approx166\Ms$.

\item The GW mergers from the current model grid well accommodate LVK's GWTC-4-based differential merger
rate density beyond $\mone\gtrsim15\Ms$ and exhibit notable features of LVK's mass distribution such as the peak at
$\mone\approx30\Ms$ (Sec.~\ref{GWTC_pop}, Appendix~\ref{rate}, Fig.~\ref{fig:m1_q_dist}). The model mergers also
well accommodate LVK's differential merger rate density w.r.t. mass ratio and reproduce its overall shape,
including the broad peak around $q\approx0.8$ and merger rate density extending
down to $q\approx0.1$. However, the mergers fail to reproduce the sharp global peak of LVK's mass distribution
at $\mone\approx10\Ms$, due to a dearth of events involving $\approx10\Ms$ BHs. This suggests that this feature of the 
GWTC population mass distribution is potentially associated with alternative GW-source formation channels such as
isolated binary evolution, massive GCs, and nuclear clusters.

\item The above features of the model merger rate density distributions hold for largely different cosmic
redshift-metallicity ($zZ$) dependence (Appendix~\ref{rate}; Figs.~\ref{fig:zZsample}, \ref{fig:m1_q_dist_xtra}).
Depending on the $zZ$ relation, the total intrinsic GW merger rate density from
the model cluster grid ranges over $7.3\peryg-9.9\peryg$ (standard, unscaled value) at a fiducial redshift of 0.2 (wherein LVK's GWTC-4-based differential
and total merger rate densities are reported; Sec.~\ref{GWTC_pop}). Hence, YMC/OCs contribute $\approx 25$\% - 33\%
of the total observed merger rate density, as obtained from these cluster models. Note that the above quoted
range in merger rate density is solely due to variation in the redshift-metallicity relation and represents fiducial estimates
based on standard model assumptions (Sec.~\ref{GWTC_pop}, Appendix~\ref{rate}). The range does not represent the full uncertainty in YMC/OC's contribution.

\item The model total merger rate density increases with redshift moderately faster than the cosmic SFR evolution and consistently
with the overall growth rate of GWTC-4 merger rate density, for $z\lesssim2$ (Sec.~\ref{zevol}, Fig.~\ref{fig:zevol}). The mass distribution
of the model mergers evolves at low redshifts, over $z\lesssim0.7$, wherein the mean merger primary and secondary masses
increase with redshift. Beyond this redshift, the mass distributions evolve much more slowly. The model mergers' mass ratio distribution shows
non-monotonic variability up to $z\approx2$, beyond which it becomes approximately steady. The redshift evolutions are, overall,
similar for the different $zZ$ relations considered here.

\item The effective spin parameter ($\xeff$) distribution of the model BBH mergers, as derived via post-processing the mergers (Appendix~\ref{bhspin}),
is generally mildly asymmetric and positively
biased at all redshifts, owing to the presence of primordially paired mergers (Sec.~\ref{zevol}, Fig.~\ref{fig:xeffz_cn21}) that are
partially spin-orbit aligned. Over redshift, the varied mixing fractions of primordially paired and dynamically paired mergers (Fig.~\ref{fig:fracz})
give rise to various features in the model-merger $\xeff$ distribution. Some of these features are reminiscent of features
present in LVK's GWTC-4-based $\xeff$-redshift dependence, \eg, the positive asymmetry and notch near $z\approx0$ and the increasing
width of the $\xeff$ distribution up to  $z\approx0.75$.

\item The model star cluster grid produces BH-MS and NS-MS binaries that escape into the galactic field. A fraction of the field
model BH-MS and NS-MS systems is consistent with the parameters of the BH-MS and NS-MS candidates in the Galactic field as detected
by the Gaia mission (Sec.~\ref{gaia}, Figs.~\ref{fig:GaiaBH}, \ref{fig:GaiaNS}).

\end{itemize}

Notably, the model GW merger rate density obtained here is lower than that estimated for the YMC/OC channel in an earlier study by the author
that applied a similar methodology \cite{Banerjee_2021}. This difference is possibly due to a combination of several aspects that were different
in the earlier work, especially the use of an experimental, heterogeneous grid of model clusters that are biased towards higher cluster mass and
BH retention, and the adoption of the rapid remnant mass model that causes higher BH retentions than the delayed counterpart \cite{Banerjee_2020}.
Furthermore, the present rate estimates focus on a specific redshift, rather than the cosmic comoving volume within an assigned visibility boundary as
was done in the earlier study.

A drawback of the present work is the adoption of (near-) spherical, virialised, gas-free initial star cluster models (Sec.~\ref{sims}). Such initial conditions are
in contrast with highly substructural, non-equilibrium, gas-rich conditions in star-forming regions \cite{Andre_2014,Longmore_2014}. Effectively, the present, idealised
initial condition associates itself with a certain cluster formation efficiency (CFE), $\epsacfe$ (Eqn.~\ref{eq:rr}), \ie,
the fraction involved in assembling gas-free, near-spherical, (typically) parsec-scale, bound young
star clusters that we observe in our and other galaxies, out of the total population of forming stars. The reader is directed to Ref.~\cite{Banerjee_2025}
and references therein for more detailed discussions on this regard, which is omitted here for brevity. Although the scaling procedure for obtaining GW merger
rate densities, as described in Appendix~\ref{rate}, does not explicitly assume a $\epsacfe$, it implicitly assumes that $\epsacfe$ is invariant with
redshift. Furthermore, this study is limited to N-body evolution of only one realisation of the initial cluster per grid point, as is presently
available. Therefore, it does not incorporate or quantify stochastic variations of the simulation outcomes (\eg, the mergers' mass distribution and
cluster-mass-dependent efficiency) due to different random realisations of a particular initial condition.

An important improvement to the current modelling is to implement the formation of tidally spun-up BHs and the assignment of binary-evolution-based
tilts to primordially paired mergers during runtime. This would make the GW-merger treatment in the model clusters more consistent.
For more complete merger rate estimates,
it is important to incorporate star clusters over a wider mass range (\ie, include GCs, nuclear star clusters, and low-mass clusters) and also field binaries.
Comparisons with GW observations need to be expanded by incorporating the latest
of the rapidly evolving GW event catalogue \cite{GWTC5_cat,GWTC5_pop}.
Such endeavours will be undertaken in the near future.

This study highlights the importance of intermediate-mass star clusters, which bridge low-mass clusters
and associations to massive globular clusters, in forming GW mergers in the Universe. The work shows that they would have a role-playing contribution
in shaping the GW-merger population's mass and spin distributions and the GW merger rate.

\section*{Data availability}

The GW merger data, the data corresponding to the BH-star and NS-star binaries from the model clusters,
and the inference scripts and posterior samples of the analyses in the appendices \ref{massmodel} \& \ref{xeffinfer}
are publicly available at this URL:\\
\url{https://doi.org/10.5281/zenodo.22177510} .\\
Additional data will be made available upon reasonable request to the corresponding author.
The version of the $\nbseven$ code used in the study is publicly available at this URL:\\
\url{https://github.com/sambaranb/updated-nbody7/tree/mpi-parallel} .

\begin{acknowledgments}

The author (SB) thanks the referee for constructive comments and criticisms that have helped to
improve the work.
SB thanks Chris Belczynski (late), Aleksandra Olejak, Koushik Sen, Daniel Marin Pina, Iwona Kotko, Fabio Antonini,
Mark Gieles, Norbert Langer for relevant discussions at various occasions. 
SB acknowledges funding for this work by the Deutsche Forschungsgemeinschaft
(DFG; German Research Foundation) through the project ``The dynamics of stellar-mass black holes in
dense stellar systems and their role in gravitational wave generation''
(project number 405620641; PI: S. Banerjee). SB has performed all of the $N$-body simulations reported in this work,
analysed the data, prepared the illustrations, and written the manuscript.
A part of the $N$-body simulations have been carried out on the {\tt gpudyn}-series GPU computing servers containing
NVIDIA Ampere A40 and NVIDIA RTX 2080 GPUs, located at the Argelander-Institut f\"ur Astronomie (AIfA), University of Bonn.
The {\tt gpudyn} servers have been sponsored by the above-mentioned DFG project, HISKP, and the University of Bonn.
A part of the $N$-body simulations have been carried out on the Marvin HPC cluster (A40 GPU nodes) of the University of Bonn.
SB acknowledges the access to the Marvin cluster of the University of Bonn.
SB is thankful to the IT teams of the AIfA, HISKP, and the University of Bonn HPC center for their generous support.
SB thanks the administrators of the AIfA for providing hospitality.
SB acknowledges the use of artificial intelligence (AI) agents in this work for the following purposes:
general proofreading (Apple Intelligence), machine-readable table compilation from
publicly accessible resources for the observed data in Figures~\ref{fig:smc_lg_comp} and \ref{fig:GaiaNS} (ChatGPT-5.2),
bug fixing and optimisations in Python scripts (ChatGPT-5.2/Claude Code), fact-checking and literature search (Scholar AI/Claude Cowork),
parametric inferences in Appendices \ref{massmodel}, \ref{xeffinfer} (Claude Code), composing these appendices,
general data management (Claude Code).

\end{acknowledgments}

\FloatBarrier

\bibliography{bibliography/biblio.bib}

\begin{thebibliography}{164}%
\makeatletter
\providecommand \@ifxundefined [1]{%
 \@ifx{#1\undefined}
}%
\providecommand \@ifnum [1]{%
 \ifnum #1\expandafter \@firstoftwo
 \else \expandafter \@secondoftwo
 \fi
}%
\providecommand \@ifx [1]{%
 \ifx #1\expandafter \@firstoftwo
 \else \expandafter \@secondoftwo
 \fi
}%
\providecommand \natexlab [1]{#1}%
\providecommand \enquote  [1]{``#1''}%
\providecommand \bibnamefont  [1]{#1}%
\providecommand \bibfnamefont [1]{#1}%
\providecommand \citenamefont [1]{#1}%
\providecommand \href@noop [0]{\@secondoftwo}%
\providecommand \href [0]{\begingroup \@sanitize@url \@href}%
\providecommand \@href[1]{\@@startlink{#1}\@@href}%
\providecommand \@@href[1]{\endgroup#1\@@endlink}%
\providecommand \@sanitize@url [0]{\catcode `\\12\catcode `\$12\catcode
  `\&12\catcode `\#12\catcode `\^12\catcode `\_12\catcode `\%12\relax}%
\providecommand \@@startlink[1]{}%
\providecommand \@@endlink[0]{}%
\providecommand \url  [0]{\begingroup\@sanitize@url \@url }%
\providecommand \@url [1]{\endgroup\@href {#1}{\urlprefix }}%
\providecommand \urlprefix  [0]{URL }%
\providecommand \Eprint [0]{\href }%
\providecommand \doibase [0]{https://doi.org/}%
\providecommand \selectlanguage [0]{\@gobble}%
\providecommand \bibinfo  [0]{\@secondoftwo}%
\providecommand \bibfield  [0]{\@secondoftwo}%
\providecommand \translation [1]{[#1]}%
\providecommand \BibitemOpen [0]{}%
\providecommand \bibitemStop [0]{}%
\providecommand \bibitemNoStop [0]{.\EOS\space}%
\providecommand \EOS [0]{\spacefactor3000\relax}%
\providecommand \BibitemShut  [1]{\csname bibitem#1\endcsname}%
\let\auto@bib@innerbib\@empty
%</preamble>
\bibitem [{\citenamefont {{Abbott}}\ \emph
  {et~al.}(2023{\natexlab{a}})\citenamefont {{Abbott}}, \citenamefont
  {{Abbott}}, \citenamefont {{Acernese}}, \citenamefont {{Ackley}},
  \citenamefont {{Adams}}, \citenamefont {{Adhikari}}, \citenamefont
  {{Adhikari}}, \citenamefont {{Adya}},\ and\ \citenamefont
  {et~al.}}]{Abbott_GWTC3}%
  \BibitemOpen
  \bibfield  {author} {\bibinfo {author} {\bibfnamefont {R.}~\bibnamefont
  {{Abbott}}}, \bibinfo {author} {\bibfnamefont {T.~D.}\ \bibnamefont
  {{Abbott}}}, \bibinfo {author} {\bibfnamefont {F.}~\bibnamefont
  {{Acernese}}}, \bibinfo {author} {\bibfnamefont {K.}~\bibnamefont
  {{Ackley}}}, \bibinfo {author} {\bibfnamefont {C.}~\bibnamefont {{Adams}}},
  \bibinfo {author} {\bibfnamefont {N.}~\bibnamefont {{Adhikari}}}, \bibinfo
  {author} {\bibfnamefont {R.~X.}\ \bibnamefont {{Adhikari}}}, \bibinfo
  {author} {\bibfnamefont {V.~B.}\ \bibnamefont {{Adya}}},\ and\ \bibinfo
  {author} {\bibnamefont {et~al.}},\ }\href
  {https://doi.org/10.1103/PhysRevX.13.041039} {\bibfield  {journal} {\bibinfo
  {journal} {Physical Review X}\ }\textbf {\bibinfo {volume} {13}},\ \bibinfo
  {eid} {041039} (\bibinfo {year} {2023}{\natexlab{a}})},\ \Eprint
  {https://arxiv.org/abs/2111.03606} {arXiv:2111.03606 [gr-qc]} \BibitemShut
  {NoStop}%
\bibitem [{\citenamefont {{The LIGO Scientific Collaboration}}\ \emph
  {et~al.}(2025{\natexlab{a}})\citenamefont {{The LIGO Scientific
  Collaboration}}, \citenamefont {{the Virgo Collaboration}}, \citenamefont
  {{the KAGRA Collaboration}}, \citenamefont {{Abac}}, \citenamefont
  {{Abouelfettouh}}, \citenamefont {{Acernese}}, \citenamefont {{Ackley}},
  \citenamefont {{Adamcewicz}}, \citenamefont {{Adhicary}}, \citenamefont
  {{Adhikari}}, \citenamefont {{Adhikari}},\ and\ \citenamefont
  {et~al.}}]{GWTC4a_cat}%
  \BibitemOpen
  \bibfield  {author} {\bibinfo {author} {\bibnamefont {{The LIGO Scientific
  Collaboration}}}, \bibinfo {author} {\bibnamefont {{the Virgo
  Collaboration}}}, \bibinfo {author} {\bibnamefont {{the KAGRA
  Collaboration}}}, \bibinfo {author} {\bibfnamefont {A.~G.}\ \bibnamefont
  {{Abac}}}, \bibinfo {author} {\bibfnamefont {I.}~\bibnamefont
  {{Abouelfettouh}}}, \bibinfo {author} {\bibfnamefont {F.}~\bibnamefont
  {{Acernese}}}, \bibinfo {author} {\bibfnamefont {K.}~\bibnamefont
  {{Ackley}}}, \bibinfo {author} {\bibfnamefont {C.}~\bibnamefont
  {{Adamcewicz}}}, \bibinfo {author} {\bibfnamefont {S.}~\bibnamefont
  {{Adhicary}}}, \bibinfo {author} {\bibfnamefont {D.}~\bibnamefont
  {{Adhikari}}}, \bibinfo {author} {\bibfnamefont {N.}~\bibnamefont
  {{Adhikari}}},\ and\ \bibinfo {author} {\bibnamefont {et~al.}},\ }\href
  {https://doi.org/10.48550/arXiv.2508.18082} {\bibfield  {journal} {\bibinfo
  {journal} {arXiv e-prints}\ ,\ \bibinfo {eid} {arXiv:2508.18082}} (\bibinfo
  {year} {2025}{\natexlab{a}})},\ \Eprint {https://arxiv.org/abs/2508.18082}
  {arXiv:2508.18082 [gr-qc]} \BibitemShut {NoStop}%
\bibitem [{\citenamefont {{Mandel}}\ and\ \citenamefont
  {{Broekgaarden}}(2022)}]{Mandel_2021}%
  \BibitemOpen
  \bibfield  {author} {\bibinfo {author} {\bibfnamefont {I.}~\bibnamefont
  {{Mandel}}}\ and\ \bibinfo {author} {\bibfnamefont {F.~S.}\ \bibnamefont
  {{Broekgaarden}}},\ }\href {https://doi.org/10.1007/s41114-021-00034-3}
  {\bibfield  {journal} {\bibinfo  {journal} {Living Reviews in Relativity}\
  }\textbf {\bibinfo {volume} {25}},\ \bibinfo {eid} {1} (\bibinfo {year}
  {2022})},\ \Eprint {https://arxiv.org/abs/2107.14239} {arXiv:2107.14239
  [astro-ph.HE]} \BibitemShut {NoStop}%
\bibitem [{\citenamefont {{Spera}}\ \emph {et~al.}(2022)\citenamefont
  {{Spera}}, \citenamefont {{Trani}},\ and\ \citenamefont
  {{Mencagli}}}]{Spera_2022}%
  \BibitemOpen
  \bibfield  {author} {\bibinfo {author} {\bibfnamefont {M.}~\bibnamefont
  {{Spera}}}, \bibinfo {author} {\bibfnamefont {A.~A.}\ \bibnamefont
  {{Trani}}},\ and\ \bibinfo {author} {\bibfnamefont {M.}~\bibnamefont
  {{Mencagli}}},\ }\href {https://doi.org/10.3390/galaxies10040076} {\bibfield
  {journal} {\bibinfo  {journal} {Galaxies}\ }\textbf {\bibinfo {volume}
  {10}},\ \bibinfo {pages} {76} (\bibinfo {year} {2022})},\ \Eprint
  {https://arxiv.org/abs/2206.15392} {arXiv:2206.15392 [astro-ph.HE]}
  \BibitemShut {NoStop}%
\bibitem [{\citenamefont {{Kulkarni}}\ \emph {et~al.}(1993)\citenamefont
  {{Kulkarni}}, \citenamefont {{Hut}},\ and\ \citenamefont
  {{McMillan}}}]{Kulkarni_1993}%
  \BibitemOpen
  \bibfield  {author} {\bibinfo {author} {\bibfnamefont {S.~R.}\ \bibnamefont
  {{Kulkarni}}}, \bibinfo {author} {\bibfnamefont {P.}~\bibnamefont {{Hut}}},\
  and\ \bibinfo {author} {\bibfnamefont {S.}~\bibnamefont {{McMillan}}},\
  }\href {https://doi.org/10.1038/364421a0} {\bibfield  {journal} {\bibinfo
  {journal} {\nat}\ }\textbf {\bibinfo {volume} {364}},\ \bibinfo {pages} {421}
  (\bibinfo {year} {1993})}\BibitemShut {NoStop}%
\bibitem [{\citenamefont {{Lee}}(1995)}]{Lee_1995}%
  \BibitemOpen
  \bibfield  {author} {\bibinfo {author} {\bibfnamefont {H.~M.}\ \bibnamefont
  {{Lee}}},\ }\href {https://doi.org/10.1093/mnras/272.3.605} {\bibfield
  {journal} {\bibinfo  {journal} {\mnras}\ }\textbf {\bibinfo {volume} {272}},\
  \bibinfo {pages} {605} (\bibinfo {year} {1995})},\ \Eprint
  {https://arxiv.org/abs/2412.14022} {arXiv:2412.14022 [astro-ph.SR]}
  \BibitemShut {NoStop}%
\bibitem [{\citenamefont {{Benacquista}}\ and\ \citenamefont
  {{Downing}}(2013)}]{Benacquista_2013}%
  \BibitemOpen
  \bibfield  {author} {\bibinfo {author} {\bibfnamefont {M.~J.}\ \bibnamefont
  {{Benacquista}}}\ and\ \bibinfo {author} {\bibfnamefont {J.~M.~B.}\
  \bibnamefont {{Downing}}},\ }\href {https://doi.org/10.12942/lrr-2013-4}
  {\bibfield  {journal} {\bibinfo  {journal} {Living Reviews in Relativity}\
  }\textbf {\bibinfo {volume} {16}},\ \bibinfo {eid} {4} (\bibinfo {year}
  {2013})},\ \Eprint {https://arxiv.org/abs/1110.4423} {arXiv:1110.4423
  [astro-ph.SR]} \BibitemShut {NoStop}%
\bibitem [{\citenamefont {{Belczynski}}\ \emph {et~al.}(2002)\citenamefont
  {{Belczynski}}, \citenamefont {{Kalogera}},\ and\ \citenamefont
  {{Bulik}}}]{Belczynski_2002}%
  \BibitemOpen
  \bibfield  {author} {\bibinfo {author} {\bibfnamefont {K.}~\bibnamefont
  {{Belczynski}}}, \bibinfo {author} {\bibfnamefont {V.}~\bibnamefont
  {{Kalogera}}},\ and\ \bibinfo {author} {\bibfnamefont {T.}~\bibnamefont
  {{Bulik}}},\ }\href {https://doi.org/10.1086/340304} {\bibfield  {journal}
  {\bibinfo  {journal} {\apj}\ }\textbf {\bibinfo {volume} {572}},\ \bibinfo
  {pages} {407} (\bibinfo {year} {2002})},\ \Eprint
  {https://arxiv.org/abs/astro-ph/0111452} {astro-ph/0111452} \BibitemShut
  {NoStop}%
\bibitem [{\citenamefont {{McKernan}}\ \emph {et~al.}(2018)\citenamefont
  {{McKernan}}, \citenamefont {{Ford}}, \citenamefont {{Bellovary}},
  \citenamefont {{Leigh}}, \citenamefont {{Haiman}}, \citenamefont {{Kocsis}},
  \citenamefont {{Lyra}}, \citenamefont {{Mac Low}}, \citenamefont {{Metzger}},
  \citenamefont {{O'Dowd}}, \citenamefont {{Endlich}},\ and\ \citenamefont
  {{Rosen}}}]{McKernan_2018}%
  \BibitemOpen
  \bibfield  {author} {\bibinfo {author} {\bibfnamefont {B.}~\bibnamefont
  {{McKernan}}}, \bibinfo {author} {\bibfnamefont {K.~E.~S.}\ \bibnamefont
  {{Ford}}}, \bibinfo {author} {\bibfnamefont {J.}~\bibnamefont {{Bellovary}}},
  \bibinfo {author} {\bibfnamefont {N.~W.~C.}\ \bibnamefont {{Leigh}}},
  \bibinfo {author} {\bibfnamefont {Z.}~\bibnamefont {{Haiman}}}, \bibinfo
  {author} {\bibfnamefont {B.}~\bibnamefont {{Kocsis}}}, \bibinfo {author}
  {\bibfnamefont {W.}~\bibnamefont {{Lyra}}}, \bibinfo {author} {\bibfnamefont
  {M.~M.}\ \bibnamefont {{Mac Low}}}, \bibinfo {author} {\bibfnamefont
  {B.}~\bibnamefont {{Metzger}}}, \bibinfo {author} {\bibfnamefont
  {M.}~\bibnamefont {{O'Dowd}}}, \bibinfo {author} {\bibfnamefont
  {S.}~\bibnamefont {{Endlich}}},\ and\ \bibinfo {author} {\bibfnamefont
  {D.~J.}\ \bibnamefont {{Rosen}}},\ }\href
  {https://doi.org/10.3847/1538-4357/aadae5} {\bibfield  {journal} {\bibinfo
  {journal} {\apj}\ }\textbf {\bibinfo {volume} {866}},\ \bibinfo {eid} {66}
  (\bibinfo {year} {2018})},\ \Eprint {https://arxiv.org/abs/1702.07818}
  {arXiv:1702.07818 [astro-ph.HE]} \BibitemShut {NoStop}%
\bibitem [{\citenamefont {{Vaccaro}}\ \emph {et~al.}(2024)\citenamefont
  {{Vaccaro}}, \citenamefont {{Mapelli}}, \citenamefont {{P{\'e}rigois}},
  \citenamefont {{Barone}}, \citenamefont {{Artale}}, \citenamefont
  {{Dall'Amico}}, \citenamefont {{Iorio}},\ and\ \citenamefont
  {{Torniamenti}}}]{Vaccaro_2024}%
  \BibitemOpen
  \bibfield  {author} {\bibinfo {author} {\bibfnamefont {M.~P.}\ \bibnamefont
  {{Vaccaro}}}, \bibinfo {author} {\bibfnamefont {M.}~\bibnamefont
  {{Mapelli}}}, \bibinfo {author} {\bibfnamefont {C.}~\bibnamefont
  {{P{\'e}rigois}}}, \bibinfo {author} {\bibfnamefont {D.}~\bibnamefont
  {{Barone}}}, \bibinfo {author} {\bibfnamefont {M.~C.}\ \bibnamefont
  {{Artale}}}, \bibinfo {author} {\bibfnamefont {M.}~\bibnamefont
  {{Dall'Amico}}}, \bibinfo {author} {\bibfnamefont {G.}~\bibnamefont
  {{Iorio}}},\ and\ \bibinfo {author} {\bibfnamefont {S.}~\bibnamefont
  {{Torniamenti}}},\ }\href {https://doi.org/10.1051/0004-6361/202348509}
  {\bibfield  {journal} {\bibinfo  {journal} {\aap}\ }\textbf {\bibinfo
  {volume} {685}},\ \bibinfo {eid} {A51} (\bibinfo {year} {2024})},\ \Eprint
  {https://arxiv.org/abs/2311.18548} {arXiv:2311.18548 [astro-ph.HE]}
  \BibitemShut {NoStop}%
\bibitem [{\citenamefont {{Vigna-G{\'o}mez}}\ \emph {et~al.}(2021)\citenamefont
  {{Vigna-G{\'o}mez}}, \citenamefont {{Toonen}}, \citenamefont
  {{Ramirez-Ruiz}}, \citenamefont {{Leigh}}, \citenamefont {{Riley}},\ and\
  \citenamefont {{Haster}}}]{VignaGomez_2021}%
  \BibitemOpen
  \bibfield  {author} {\bibinfo {author} {\bibfnamefont {A.}~\bibnamefont
  {{Vigna-G{\'o}mez}}}, \bibinfo {author} {\bibfnamefont {S.}~\bibnamefont
  {{Toonen}}}, \bibinfo {author} {\bibfnamefont {E.}~\bibnamefont
  {{Ramirez-Ruiz}}}, \bibinfo {author} {\bibfnamefont {N.~W.~C.}\ \bibnamefont
  {{Leigh}}}, \bibinfo {author} {\bibfnamefont {J.}~\bibnamefont {{Riley}}},\
  and\ \bibinfo {author} {\bibfnamefont {C.-J.}\ \bibnamefont {{Haster}}},\
  }\href {https://doi.org/10.3847/2041-8213/abd5b7} {\bibfield  {journal}
  {\bibinfo  {journal} {\apjl}\ }\textbf {\bibinfo {volume} {907}},\ \bibinfo
  {eid} {L19} (\bibinfo {year} {2021})},\ \Eprint
  {https://arxiv.org/abs/2010.13669} {arXiv:2010.13669 [astro-ph.HE]}
  \BibitemShut {NoStop}%
\bibitem [{\citenamefont {{Stegmann}}\ \emph {et~al.}(2022)\citenamefont
  {{Stegmann}}, \citenamefont {{Antonini}},\ and\ \citenamefont
  {{Moe}}}]{Stegmann_2022}%
  \BibitemOpen
  \bibfield  {author} {\bibinfo {author} {\bibfnamefont {J.}~\bibnamefont
  {{Stegmann}}}, \bibinfo {author} {\bibfnamefont {F.}~\bibnamefont
  {{Antonini}}},\ and\ \bibinfo {author} {\bibfnamefont {M.}~\bibnamefont
  {{Moe}}},\ }\href {https://doi.org/10.1093/mnras/stac2192} {\bibfield
  {journal} {\bibinfo  {journal} {\mnras}\ }\textbf {\bibinfo {volume} {516}},\
  \bibinfo {pages} {1406} (\bibinfo {year} {2022})},\ \Eprint
  {https://arxiv.org/abs/2112.10786} {arXiv:2112.10786 [astro-ph.SR]}
  \BibitemShut {NoStop}%
\bibitem [{\citenamefont {{Michaely}}\ and\ \citenamefont
  {{Perets}}(2019)}]{Michaely_2019}%
  \BibitemOpen
  \bibfield  {author} {\bibinfo {author} {\bibfnamefont {E.}~\bibnamefont
  {{Michaely}}}\ and\ \bibinfo {author} {\bibfnamefont {H.~B.}\ \bibnamefont
  {{Perets}}},\ }\href {https://doi.org/10.3847/2041-8213/ab5b9b} {\bibfield
  {journal} {\bibinfo  {journal} {\apjl}\ }\textbf {\bibinfo {volume} {887}},\
  \bibinfo {eid} {L36} (\bibinfo {year} {2019})},\ \Eprint
  {https://arxiv.org/abs/1902.01864} {arXiv:1902.01864 [astro-ph.SR]}
  \BibitemShut {NoStop}%
\bibitem [{\citenamefont {{Hamilton}}\ and\ \citenamefont
  {{Rafikov}}(2019)}]{Hamilton_2019}%
  \BibitemOpen
  \bibfield  {author} {\bibinfo {author} {\bibfnamefont {C.}~\bibnamefont
  {{Hamilton}}}\ and\ \bibinfo {author} {\bibfnamefont {R.~R.}\ \bibnamefont
  {{Rafikov}}},\ }\href {https://doi.org/10.3847/2041-8213/ab3468} {\bibfield
  {journal} {\bibinfo  {journal} {\apjl}\ }\textbf {\bibinfo {volume} {881}},\
  \bibinfo {eid} {L13} (\bibinfo {year} {2019})},\ \Eprint
  {https://arxiv.org/abs/1907.00994} {arXiv:1907.00994 [astro-ph.GA]}
  \BibitemShut {NoStop}%
\bibitem [{\citenamefont {{Stegmann}}\ \emph {et~al.}(2024)\citenamefont
  {{Stegmann}}, \citenamefont {{Vigna-G{\'o}mez}}, \citenamefont {{Rantala}},
  \citenamefont {{Wagg}}, \citenamefont {{Zwick}}, \citenamefont {{Renzo}},
  \citenamefont {{van Son}}, \citenamefont {{de Mink}},\ and\ \citenamefont
  {{White}}}]{Stegmann_2024}%
  \BibitemOpen
  \bibfield  {author} {\bibinfo {author} {\bibfnamefont {J.}~\bibnamefont
  {{Stegmann}}}, \bibinfo {author} {\bibfnamefont {A.}~\bibnamefont
  {{Vigna-G{\'o}mez}}}, \bibinfo {author} {\bibfnamefont {A.}~\bibnamefont
  {{Rantala}}}, \bibinfo {author} {\bibfnamefont {T.}~\bibnamefont {{Wagg}}},
  \bibinfo {author} {\bibfnamefont {L.}~\bibnamefont {{Zwick}}}, \bibinfo
  {author} {\bibfnamefont {M.}~\bibnamefont {{Renzo}}}, \bibinfo {author}
  {\bibfnamefont {L.~A.~C.}\ \bibnamefont {{van Son}}}, \bibinfo {author}
  {\bibfnamefont {S.~E.}\ \bibnamefont {{de Mink}}},\ and\ \bibinfo {author}
  {\bibfnamefont {S.~D.~M.}\ \bibnamefont {{White}}},\ }\href
  {https://doi.org/10.3847/2041-8213/ad70bb} {\bibfield  {journal} {\bibinfo
  {journal} {\apjl}\ }\textbf {\bibinfo {volume} {972}},\ \bibinfo {eid} {L19}
  (\bibinfo {year} {2024})},\ \Eprint {https://arxiv.org/abs/2405.02912}
  {arXiv:2405.02912 [astro-ph.GA]} \BibitemShut {NoStop}%
\bibitem [{\citenamefont {{Zevin}}\ \emph {et~al.}(2021)\citenamefont
  {{Zevin}}, \citenamefont {{Bavera}}, \citenamefont {{Berry}}, \citenamefont
  {{Kalogera}}, \citenamefont {{Fragos}}, \citenamefont {{Marchant}},
  \citenamefont {{Rodriguez}}, \citenamefont {{Antonini}}, \citenamefont
  {{Holz}},\ and\ \citenamefont {{Pankow}}}]{Zevin_2020}%
  \BibitemOpen
  \bibfield  {author} {\bibinfo {author} {\bibfnamefont {M.}~\bibnamefont
  {{Zevin}}}, \bibinfo {author} {\bibfnamefont {S.~S.}\ \bibnamefont
  {{Bavera}}}, \bibinfo {author} {\bibfnamefont {C.~P.~L.}\ \bibnamefont
  {{Berry}}}, \bibinfo {author} {\bibfnamefont {V.}~\bibnamefont {{Kalogera}}},
  \bibinfo {author} {\bibfnamefont {T.}~\bibnamefont {{Fragos}}}, \bibinfo
  {author} {\bibfnamefont {P.}~\bibnamefont {{Marchant}}}, \bibinfo {author}
  {\bibfnamefont {C.~L.}\ \bibnamefont {{Rodriguez}}}, \bibinfo {author}
  {\bibfnamefont {F.}~\bibnamefont {{Antonini}}}, \bibinfo {author}
  {\bibfnamefont {D.~E.}\ \bibnamefont {{Holz}}},\ and\ \bibinfo {author}
  {\bibfnamefont {C.}~\bibnamefont {{Pankow}}},\ }\href
  {https://doi.org/10.3847/1538-4357/abe40e} {\bibfield  {journal} {\bibinfo
  {journal} {\apj}\ }\textbf {\bibinfo {volume} {910}},\ \bibinfo {eid} {152}
  (\bibinfo {year} {2021})},\ \Eprint {https://arxiv.org/abs/2011.10057}
  {arXiv:2011.10057 [astro-ph.HE]} \BibitemShut {NoStop}%
\bibitem [{\citenamefont {{The LIGO Scientific Collaboration}}\ \emph
  {et~al.}(2025{\natexlab{b}})\citenamefont {{The LIGO Scientific
  Collaboration}}, \citenamefont {{the Virgo Collaboration}}, \citenamefont
  {{the KAGRA Collaboration}}, \citenamefont {{Abac}}, \citenamefont
  {{Abouelfettouh}}, \citenamefont {{Acernese}}, \citenamefont {{Ackley}},
  \citenamefont {{Adamcewicz}}, \citenamefont {{Adhicary}}, \citenamefont
  {{Adhikari}}, \citenamefont {{Adhikari}},\ and\ \citenamefont
  {et~al.}}]{GWTC4a_pop}%
  \BibitemOpen
  \bibfield  {author} {\bibinfo {author} {\bibnamefont {{The LIGO Scientific
  Collaboration}}}, \bibinfo {author} {\bibnamefont {{the Virgo
  Collaboration}}}, \bibinfo {author} {\bibnamefont {{the KAGRA
  Collaboration}}}, \bibinfo {author} {\bibfnamefont {A.~G.}\ \bibnamefont
  {{Abac}}}, \bibinfo {author} {\bibfnamefont {I.}~\bibnamefont
  {{Abouelfettouh}}}, \bibinfo {author} {\bibfnamefont {F.}~\bibnamefont
  {{Acernese}}}, \bibinfo {author} {\bibfnamefont {K.}~\bibnamefont
  {{Ackley}}}, \bibinfo {author} {\bibfnamefont {C.}~\bibnamefont
  {{Adamcewicz}}}, \bibinfo {author} {\bibfnamefont {S.}~\bibnamefont
  {{Adhicary}}}, \bibinfo {author} {\bibfnamefont {D.}~\bibnamefont
  {{Adhikari}}}, \bibinfo {author} {\bibfnamefont {N.}~\bibnamefont
  {{Adhikari}}},\ and\ \bibinfo {author} {\bibnamefont {et~al.}},\ }\href
  {https://doi.org/10.48550/arXiv.2508.18083} {\bibfield  {journal} {\bibinfo
  {journal} {arXiv e-prints}\ ,\ \bibinfo {eid} {arXiv:2508.18083}} (\bibinfo
  {year} {2025}{\natexlab{b}})},\ \Eprint {https://arxiv.org/abs/2508.18083}
  {arXiv:2508.18083 [astro-ph.HE]} \BibitemShut {NoStop}%
\bibitem [{\citenamefont {{Banagiri}}\ \emph {et~al.}(2025)\citenamefont
  {{Banagiri}}, \citenamefont {{Thrane}},\ and\ \citenamefont
  {{Lasky}}}]{Banagiri_2025}%
  \BibitemOpen
  \bibfield  {author} {\bibinfo {author} {\bibfnamefont {S.}~\bibnamefont
  {{Banagiri}}}, \bibinfo {author} {\bibfnamefont {E.}~\bibnamefont
  {{Thrane}}},\ and\ \bibinfo {author} {\bibfnamefont {P.~D.}\ \bibnamefont
  {{Lasky}}},\ }\href {https://doi.org/10.48550/arXiv.2509.15646} {\bibfield
  {journal} {\bibinfo  {journal} {arXiv e-prints}\ ,\ \bibinfo {eid}
  {arXiv:2509.15646}} (\bibinfo {year} {2025})},\ \Eprint
  {https://arxiv.org/abs/2509.15646} {arXiv:2509.15646 [astro-ph.HE]}
  \BibitemShut {NoStop}%
\bibitem [{\citenamefont {{Banerjee}}\ \emph {et~al.}(2010)\citenamefont
  {{Banerjee}}, \citenamefont {{Baumgardt}},\ and\ \citenamefont
  {{Kroupa}}}]{Banerjee_2010}%
  \BibitemOpen
  \bibfield  {author} {\bibinfo {author} {\bibfnamefont {S.}~\bibnamefont
  {{Banerjee}}}, \bibinfo {author} {\bibfnamefont {H.}~\bibnamefont
  {{Baumgardt}}},\ and\ \bibinfo {author} {\bibfnamefont {P.}~\bibnamefont
  {{Kroupa}}},\ }\href {https://doi.org/10.1111/j.1365-2966.2009.15880.x}
  {\bibfield  {journal} {\bibinfo  {journal} {\mnras}\ }\textbf {\bibinfo
  {volume} {402}},\ \bibinfo {pages} {371} (\bibinfo {year} {2010})},\ \Eprint
  {https://arxiv.org/abs/0910.3954} {arXiv:0910.3954 [astro-ph.SR]}
  \BibitemShut {NoStop}%
\bibitem [{\citenamefont {{Aarseth}}(2012)}]{Aarseth_2012}%
  \BibitemOpen
  \bibfield  {author} {\bibinfo {author} {\bibfnamefont {S.~J.}\ \bibnamefont
  {{Aarseth}}},\ }\href {https://doi.org/10.1111/j.1365-2966.2012.20666.x}
  {\bibfield  {journal} {\bibinfo  {journal} {\mnras}\ }\textbf {\bibinfo
  {volume} {422}},\ \bibinfo {pages} {841} (\bibinfo {year} {2012})},\ \Eprint
  {https://arxiv.org/abs/1202.4688} {arXiv:1202.4688 [astro-ph.SR]}
  \BibitemShut {NoStop}%
\bibitem [{\citenamefont {{Banerjee}}(2017)}]{Banerjee_2017}%
  \BibitemOpen
  \bibfield  {author} {\bibinfo {author} {\bibfnamefont {S.}~\bibnamefont
  {{Banerjee}}},\ }\href {https://doi.org/10.1093/mnras/stw3392} {\bibfield
  {journal} {\bibinfo  {journal} {\mnras}\ }\textbf {\bibinfo {volume} {467}},\
  \bibinfo {pages} {524} (\bibinfo {year} {2017})},\ \Eprint
  {https://arxiv.org/abs/1611.09357} {arXiv:1611.09357 [astro-ph.HE]}
  \BibitemShut {NoStop}%
\bibitem [{\citenamefont {{Di Carlo}}\ \emph {et~al.}(2019)\citenamefont {{Di
  Carlo}}, \citenamefont {{Giacobbo}}, \citenamefont {{Mapelli}}, \citenamefont
  {{Pasquato}}, \citenamefont {{Spera}}, \citenamefont {{Wang}},\ and\
  \citenamefont {{Haardt}}}]{DiCarlo_2019}%
  \BibitemOpen
  \bibfield  {author} {\bibinfo {author} {\bibfnamefont {U.~N.}\ \bibnamefont
  {{Di Carlo}}}, \bibinfo {author} {\bibfnamefont {N.}~\bibnamefont
  {{Giacobbo}}}, \bibinfo {author} {\bibfnamefont {M.}~\bibnamefont
  {{Mapelli}}}, \bibinfo {author} {\bibfnamefont {M.}~\bibnamefont
  {{Pasquato}}}, \bibinfo {author} {\bibfnamefont {M.}~\bibnamefont {{Spera}}},
  \bibinfo {author} {\bibfnamefont {L.}~\bibnamefont {{Wang}}},\ and\ \bibinfo
  {author} {\bibfnamefont {F.}~\bibnamefont {{Haardt}}},\ }\href
  {https://doi.org/10.1093/mnras/stz1453} {\bibfield  {journal} {\bibinfo
  {journal} {\mnras}\ }\textbf {\bibinfo {volume} {487}},\ \bibinfo {pages}
  {2947} (\bibinfo {year} {2019})},\ \Eprint {https://arxiv.org/abs/1901.00863}
  {arXiv:1901.00863 [astro-ph.HE]} \BibitemShut {NoStop}%
\bibitem [{\citenamefont {{Portegies Zwart}}\ \emph {et~al.}(2010)\citenamefont
  {{Portegies Zwart}}, \citenamefont {{McMillan}},\ and\ \citenamefont
  {{Gieles}}}]{PortegiesZwart_2010}%
  \BibitemOpen
  \bibfield  {author} {\bibinfo {author} {\bibfnamefont {S.~F.}\ \bibnamefont
  {{Portegies Zwart}}}, \bibinfo {author} {\bibfnamefont {S.~L.~W.}\
  \bibnamefont {{McMillan}}},\ and\ \bibinfo {author} {\bibfnamefont
  {M.}~\bibnamefont {{Gieles}}},\ }\href
  {https://doi.org/10.1146/annurev-astro-081309-130834} {\bibfield  {journal}
  {\bibinfo  {journal} {\araa}\ }\textbf {\bibinfo {volume} {48}},\ \bibinfo
  {pages} {431} (\bibinfo {year} {2010})},\ \Eprint
  {https://arxiv.org/abs/1002.1961} {arXiv:1002.1961} \BibitemShut {NoStop}%
\bibitem [{\citenamefont {{Krumholz}}\ \emph {et~al.}(2019)\citenamefont
  {{Krumholz}}, \citenamefont {{McKee}},\ and\ \citenamefont {{Bland
  -Hawthorn}}}]{Krumholz_2019}%
  \BibitemOpen
  \bibfield  {author} {\bibinfo {author} {\bibfnamefont {M.~R.}\ \bibnamefont
  {{Krumholz}}}, \bibinfo {author} {\bibfnamefont {C.~F.}\ \bibnamefont
  {{McKee}}},\ and\ \bibinfo {author} {\bibfnamefont {J.}~\bibnamefont {{Bland
  -Hawthorn}}},\ }\href {https://doi.org/10.1146/annurev-astro-091918-104430}
  {\bibfield  {journal} {\bibinfo  {journal} {\araa}\ }\textbf {\bibinfo
  {volume} {57}},\ \bibinfo {pages} {227} (\bibinfo {year} {2019})},\ \Eprint
  {https://arxiv.org/abs/1812.01615} {arXiv:1812.01615 [astro-ph.GA]}
  \BibitemShut {NoStop}%
\bibitem [{\citenamefont {{Rodriguez}}\ \emph {et~al.}(2021)\citenamefont
  {{Rodriguez}}, \citenamefont {{Kremer}}, \citenamefont {{Chatterjee}},
  \citenamefont {{Fragione}}, \citenamefont {{Loeb}}, \citenamefont {{Rasio}},
  \citenamefont {{Weatherford}},\ and\ \citenamefont {{Ye}}}]{Rodriguez_2021}%
  \BibitemOpen
  \bibfield  {author} {\bibinfo {author} {\bibfnamefont {C.~L.}\ \bibnamefont
  {{Rodriguez}}}, \bibinfo {author} {\bibfnamefont {K.}~\bibnamefont
  {{Kremer}}}, \bibinfo {author} {\bibfnamefont {S.}~\bibnamefont
  {{Chatterjee}}}, \bibinfo {author} {\bibfnamefont {G.}~\bibnamefont
  {{Fragione}}}, \bibinfo {author} {\bibfnamefont {A.}~\bibnamefont {{Loeb}}},
  \bibinfo {author} {\bibfnamefont {F.~A.}\ \bibnamefont {{Rasio}}}, \bibinfo
  {author} {\bibfnamefont {N.~C.}\ \bibnamefont {{Weatherford}}},\ and\
  \bibinfo {author} {\bibfnamefont {C.~S.}\ \bibnamefont {{Ye}}},\ }\href
  {https://doi.org/10.3847/2515-5172/abdf54} {\bibfield  {journal} {\bibinfo
  {journal} {Research Notes of the American Astronomical Society}\ }\textbf
  {\bibinfo {volume} {5}},\ \bibinfo {eid} {19} (\bibinfo {year} {2021})},\
  \Eprint {https://arxiv.org/abs/2101.07793} {arXiv:2101.07793 [astro-ph.HE]}
  \BibitemShut {NoStop}%
\bibitem [{\citenamefont {{Ye}}\ \emph {et~al.}(2025)\citenamefont {{Ye}},
  \citenamefont {{Fishbach}}, \citenamefont {{Kremer}},\ and\ \citenamefont
  {{Reina-Campos}}}]{Ye_2025}%
  \BibitemOpen
  \bibfield  {author} {\bibinfo {author} {\bibfnamefont {C.~S.}\ \bibnamefont
  {{Ye}}}, \bibinfo {author} {\bibfnamefont {M.}~\bibnamefont {{Fishbach}}},
  \bibinfo {author} {\bibfnamefont {K.}~\bibnamefont {{Kremer}}},\ and\
  \bibinfo {author} {\bibfnamefont {M.}~\bibnamefont {{Reina-Campos}}},\ }\href
  {https://doi.org/10.48550/arXiv.2507.07183} {\bibfield  {journal} {\bibinfo
  {journal} {arXiv e-prints}\ ,\ \bibinfo {eid} {arXiv:2507.07183}} (\bibinfo
  {year} {2025})},\ \Eprint {https://arxiv.org/abs/2507.07183}
  {arXiv:2507.07183 [astro-ph.HE]} \BibitemShut {NoStop}%
\bibitem [{\citenamefont {{Santoliquido}}\ \emph {et~al.}(2020)\citenamefont
  {{Santoliquido}}, \citenamefont {{Mapelli}}, \citenamefont {{Bouffanais}},
  \citenamefont {{Giacobbo}}, \citenamefont {{Di Carlo}}, \citenamefont
  {{Rastello}}, \citenamefont {{Artale}},\ and\ \citenamefont
  {{Ballone}}}]{Santoliquido_2020}%
  \BibitemOpen
  \bibfield  {author} {\bibinfo {author} {\bibfnamefont {F.}~\bibnamefont
  {{Santoliquido}}}, \bibinfo {author} {\bibfnamefont {M.}~\bibnamefont
  {{Mapelli}}}, \bibinfo {author} {\bibfnamefont {Y.}~\bibnamefont
  {{Bouffanais}}}, \bibinfo {author} {\bibfnamefont {N.}~\bibnamefont
  {{Giacobbo}}}, \bibinfo {author} {\bibfnamefont {U.~N.}\ \bibnamefont {{Di
  Carlo}}}, \bibinfo {author} {\bibfnamefont {S.}~\bibnamefont {{Rastello}}},
  \bibinfo {author} {\bibfnamefont {M.~C.}\ \bibnamefont {{Artale}}},\ and\
  \bibinfo {author} {\bibfnamefont {A.}~\bibnamefont {{Ballone}}},\ }\href
  {https://doi.org/10.3847/1538-4357/ab9b78} {\bibfield  {journal} {\bibinfo
  {journal} {\apj}\ }\textbf {\bibinfo {volume} {898}},\ \bibinfo {eid} {152}
  (\bibinfo {year} {2020})},\ \Eprint {https://arxiv.org/abs/2004.09533}
  {arXiv:2004.09533 [astro-ph.HE]} \BibitemShut {NoStop}%
\bibitem [{\citenamefont {{Kumamoto}}\ \emph {et~al.}(2020)\citenamefont
  {{Kumamoto}}, \citenamefont {{Fujii}},\ and\ \citenamefont
  {{Tanikawa}}}]{Kumamoto_2020}%
  \BibitemOpen
  \bibfield  {author} {\bibinfo {author} {\bibfnamefont {J.}~\bibnamefont
  {{Kumamoto}}}, \bibinfo {author} {\bibfnamefont {M.~S.}\ \bibnamefont
  {{Fujii}}},\ and\ \bibinfo {author} {\bibfnamefont {A.}~\bibnamefont
  {{Tanikawa}}},\ }\href {https://doi.org/10.1093/mnras/staa1440} {\bibfield
  {journal} {\bibinfo  {journal} {\mnras}\ }\textbf {\bibinfo {volume} {495}},\
  \bibinfo {pages} {4268} (\bibinfo {year} {2020})},\ \Eprint
  {https://arxiv.org/abs/2001.10690} {arXiv:2001.10690 [astro-ph.HE]}
  \BibitemShut {NoStop}%
\bibitem [{\citenamefont {{Banerjee}}(2021{\natexlab{a}})}]{Banerjee_2021}%
  \BibitemOpen
  \bibfield  {author} {\bibinfo {author} {\bibfnamefont {S.}~\bibnamefont
  {{Banerjee}}},\ }\href {https://doi.org/10.1093/mnras/stab591} {\bibfield
  {journal} {\bibinfo  {journal} {\mnras}\ }\textbf {\bibinfo {volume} {503}},\
  \bibinfo {pages} {3371} (\bibinfo {year} {2021}{\natexlab{a}})},\ \Eprint
  {https://arxiv.org/abs/2011.07000} {arXiv:2011.07000 [astro-ph.HE]}
  \BibitemShut {NoStop}%
\bibitem [{\citenamefont {{Fragione}}\ and\ \citenamefont
  {{Banerjee}}(2021)}]{Fragione_2021}%
  \BibitemOpen
  \bibfield  {author} {\bibinfo {author} {\bibfnamefont {G.}~\bibnamefont
  {{Fragione}}}\ and\ \bibinfo {author} {\bibfnamefont {S.}~\bibnamefont
  {{Banerjee}}},\ }\href {https://doi.org/10.3847/2041-8213/ac00a7} {\bibfield
  {journal} {\bibinfo  {journal} {\apjl}\ }\textbf {\bibinfo {volume} {913}},\
  \bibinfo {eid} {L29} (\bibinfo {year} {2021})},\ \Eprint
  {https://arxiv.org/abs/2103.10447} {arXiv:2103.10447 [astro-ph.HE]}
  \BibitemShut {NoStop}%
\bibitem [{\citenamefont {{Banerjee}}(2025)}]{Banerjee_2025}%
  \BibitemOpen
  \bibfield  {author} {\bibinfo {author} {\bibfnamefont {S.}~\bibnamefont
  {{Banerjee}}},\ }\href {https://doi.org/10.1103/jvb6-3rzz} {\bibfield
  {journal} {\bibinfo  {journal} {\prd}\ }\textbf {\bibinfo {volume} {112}},\
  \bibinfo {eid} {063016} (\bibinfo {year} {2025})},\ \Eprint
  {https://arxiv.org/abs/2505.17780} {arXiv:2505.17780 [astro-ph.GA]}
  \BibitemShut {NoStop}%
\bibitem [{\citenamefont {{Banerjee}}(2021{\natexlab{b}})}]{Banerjee_2020c}%
  \BibitemOpen
  \bibfield  {author} {\bibinfo {author} {\bibfnamefont {S.}~\bibnamefont
  {{Banerjee}}},\ }\href {https://doi.org/10.1093/mnras/staa2392} {\bibfield
  {journal} {\bibinfo  {journal} {\mnras}\ }\textbf {\bibinfo {volume} {500}},\
  \bibinfo {pages} {3002} (\bibinfo {year} {2021}{\natexlab{b}})},\ \Eprint
  {https://arxiv.org/abs/2004.07382} {arXiv:2004.07382 [astro-ph.HE]}
  \BibitemShut {NoStop}%
\bibitem [{\citenamefont {{Banerjee}}(2022{\natexlab{a}})}]{Banerjee_2022}%
  \BibitemOpen
  \bibfield  {author} {\bibinfo {author} {\bibfnamefont {S.}~\bibnamefont
  {{Banerjee}}},\ }\href {https://doi.org/10.1051/0004-6361/202142331}
  {\bibfield  {journal} {\bibinfo  {journal} {\aap}\ }\textbf {\bibinfo
  {volume} {665}},\ \bibinfo {eid} {A20} (\bibinfo {year}
  {2022}{\natexlab{a}})},\ \Eprint {https://arxiv.org/abs/2109.14612}
  {arXiv:2109.14612 [astro-ph.HE]} \BibitemShut {NoStop}%
\bibitem [{\citenamefont {{Barber}}\ and\ \citenamefont
  {{Antonini}}(2025)}]{Barber_2025}%
  \BibitemOpen
  \bibfield  {author} {\bibinfo {author} {\bibfnamefont {J.}~\bibnamefont
  {{Barber}}}\ and\ \bibinfo {author} {\bibfnamefont {F.}~\bibnamefont
  {{Antonini}}},\ }\href {https://doi.org/10.1093/mnras/staf279} {\bibfield
  {journal} {\bibinfo  {journal} {\mnras}\ }\textbf {\bibinfo {volume} {538}},\
  \bibinfo {pages} {639} (\bibinfo {year} {2025})},\ \Eprint
  {https://arxiv.org/abs/2410.03832} {arXiv:2410.03832 [astro-ph.GA]}
  \BibitemShut {NoStop}%
\bibitem [{\citenamefont {{Wang}}\ \emph {et~al.}(2020)\citenamefont {{Wang}},
  \citenamefont {{Iwasawa}}, \citenamefont {{Nitadori}},\ and\ \citenamefont
  {{Makino}}}]{Wang_2020b}%
  \BibitemOpen
  \bibfield  {author} {\bibinfo {author} {\bibfnamefont {L.}~\bibnamefont
  {{Wang}}}, \bibinfo {author} {\bibfnamefont {M.}~\bibnamefont {{Iwasawa}}},
  \bibinfo {author} {\bibfnamefont {K.}~\bibnamefont {{Nitadori}}},\ and\
  \bibinfo {author} {\bibfnamefont {J.}~\bibnamefont {{Makino}}},\ }\href
  {https://doi.org/10.1093/mnras/staa1915} {\bibfield  {journal} {\bibinfo
  {journal} {\mnras}\ }\textbf {\bibinfo {volume} {497}},\ \bibinfo {pages}
  {536} (\bibinfo {year} {2020})},\ \Eprint {https://arxiv.org/abs/2006.16560}
  {arXiv:2006.16560 [astro-ph.IM]} \BibitemShut {NoStop}%
\bibitem [{\citenamefont {{Kremer}}\ \emph {et~al.}(2020)\citenamefont
  {{Kremer}}, \citenamefont {{Ye}}, \citenamefont {{Rui}}, \citenamefont
  {{Weatherford}}, \citenamefont {{Chatterjee}}, \citenamefont {{Fragione}},
  \citenamefont {{Rodriguez}}, \citenamefont {{Spera}},\ and\ \citenamefont
  {{Rasio}}}]{Kremer_2020}%
  \BibitemOpen
  \bibfield  {author} {\bibinfo {author} {\bibfnamefont {K.}~\bibnamefont
  {{Kremer}}}, \bibinfo {author} {\bibfnamefont {C.~S.}\ \bibnamefont {{Ye}}},
  \bibinfo {author} {\bibfnamefont {N.~Z.}\ \bibnamefont {{Rui}}}, \bibinfo
  {author} {\bibfnamefont {N.~C.}\ \bibnamefont {{Weatherford}}}, \bibinfo
  {author} {\bibfnamefont {S.}~\bibnamefont {{Chatterjee}}}, \bibinfo {author}
  {\bibfnamefont {G.}~\bibnamefont {{Fragione}}}, \bibinfo {author}
  {\bibfnamefont {C.~L.}\ \bibnamefont {{Rodriguez}}}, \bibinfo {author}
  {\bibfnamefont {M.}~\bibnamefont {{Spera}}},\ and\ \bibinfo {author}
  {\bibfnamefont {F.~A.}\ \bibnamefont {{Rasio}}},\ }\href
  {https://doi.org/10.3847/1538-4365/ab7919} {\bibfield  {journal} {\bibinfo
  {journal} {\apjs}\ }\textbf {\bibinfo {volume} {247}},\ \bibinfo {eid} {48}
  (\bibinfo {year} {2020})},\ \Eprint {https://arxiv.org/abs/1911.00018}
  {arXiv:1911.00018 [astro-ph.HE]} \BibitemShut {NoStop}%
\bibitem [{\citenamefont {{Askar}}\ \emph {et~al.}(2017)\citenamefont
  {{Askar}}, \citenamefont {{Szkudlarek}}, \citenamefont
  {{Gondek-Rosi{\'n}ska}}, \citenamefont {{Giersz}},\ and\ \citenamefont
  {{Bulik}}}]{Askar_2016}%
  \BibitemOpen
  \bibfield  {author} {\bibinfo {author} {\bibfnamefont {A.}~\bibnamefont
  {{Askar}}}, \bibinfo {author} {\bibfnamefont {M.}~\bibnamefont
  {{Szkudlarek}}}, \bibinfo {author} {\bibfnamefont {D.}~\bibnamefont
  {{Gondek-Rosi{\'n}ska}}}, \bibinfo {author} {\bibfnamefont {M.}~\bibnamefont
  {{Giersz}}},\ and\ \bibinfo {author} {\bibfnamefont {T.}~\bibnamefont
  {{Bulik}}},\ }\href {https://doi.org/10.1093/mnrasl/slw177} {\bibfield
  {journal} {\bibinfo  {journal} {\mnras}\ }\textbf {\bibinfo {volume} {464}},\
  \bibinfo {pages} {L36} (\bibinfo {year} {2017})},\ \Eprint
  {https://arxiv.org/abs/1608.02520} {arXiv:1608.02520 [astro-ph.HE]}
  \BibitemShut {NoStop}%
\bibitem [{\citenamefont {{Arca Sedda}}\ \emph {et~al.}(2024)\citenamefont
  {{Arca Sedda}}, \citenamefont {{Kamlah}}, \citenamefont {{Spurzem}},
  \citenamefont {{Giersz}}, \citenamefont {{Berczik}}, \citenamefont
  {{Rastello}}, \citenamefont {{Iorio}}, \citenamefont {{Mapelli}},
  \citenamefont {{Gatto}},\ and\ \citenamefont {{Grebel}}}]{ArcaSedda_2024a}%
  \BibitemOpen
  \bibfield  {author} {\bibinfo {author} {\bibfnamefont {M.}~\bibnamefont
  {{Arca Sedda}}}, \bibinfo {author} {\bibfnamefont {A.~W.~H.}\ \bibnamefont
  {{Kamlah}}}, \bibinfo {author} {\bibfnamefont {R.}~\bibnamefont {{Spurzem}}},
  \bibinfo {author} {\bibfnamefont {M.}~\bibnamefont {{Giersz}}}, \bibinfo
  {author} {\bibfnamefont {P.}~\bibnamefont {{Berczik}}}, \bibinfo {author}
  {\bibfnamefont {S.}~\bibnamefont {{Rastello}}}, \bibinfo {author}
  {\bibfnamefont {G.}~\bibnamefont {{Iorio}}}, \bibinfo {author} {\bibfnamefont
  {M.}~\bibnamefont {{Mapelli}}}, \bibinfo {author} {\bibfnamefont
  {M.}~\bibnamefont {{Gatto}}},\ and\ \bibinfo {author} {\bibfnamefont {E.~K.}\
  \bibnamefont {{Grebel}}},\ }\href {https://doi.org/10.1093/mnras/stad3952}
  {\bibfield  {journal} {\bibinfo  {journal} {\mnras}\ }\textbf {\bibinfo
  {volume} {528}},\ \bibinfo {pages} {5119} (\bibinfo {year} {2024})},\ \Eprint
  {https://arxiv.org/abs/2307.04805} {arXiv:2307.04805 [astro-ph.GA]}
  \BibitemShut {NoStop}%
\bibitem [{\citenamefont {{Di Carlo}}\ \emph
  {et~al.}(2020{\natexlab{a}})\citenamefont {{Di Carlo}}, \citenamefont
  {{Mapelli}}, \citenamefont {{Giacobbo}}, \citenamefont {{Spera}},
  \citenamefont {{Bouffanais}}, \citenamefont {{Rastello}}, \citenamefont
  {{Santoliquido}}, \citenamefont {{Pasquato}}, \citenamefont {{Ballone}},
  \citenamefont {{Trani}}, \citenamefont {{Torniamenti}},\ and\ \citenamefont
  {{Haardt}}}]{DiCarlo_2020}%
  \BibitemOpen
  \bibfield  {author} {\bibinfo {author} {\bibfnamefont {U.~N.}\ \bibnamefont
  {{Di Carlo}}}, \bibinfo {author} {\bibfnamefont {M.}~\bibnamefont
  {{Mapelli}}}, \bibinfo {author} {\bibfnamefont {N.}~\bibnamefont
  {{Giacobbo}}}, \bibinfo {author} {\bibfnamefont {M.}~\bibnamefont {{Spera}}},
  \bibinfo {author} {\bibfnamefont {Y.}~\bibnamefont {{Bouffanais}}}, \bibinfo
  {author} {\bibfnamefont {S.}~\bibnamefont {{Rastello}}}, \bibinfo {author}
  {\bibfnamefont {F.}~\bibnamefont {{Santoliquido}}}, \bibinfo {author}
  {\bibfnamefont {M.}~\bibnamefont {{Pasquato}}}, \bibinfo {author}
  {\bibfnamefont {A.~r.}\ \bibnamefont {{Ballone}}}, \bibinfo {author}
  {\bibfnamefont {A.~A.}\ \bibnamefont {{Trani}}}, \bibinfo {author}
  {\bibfnamefont {S.}~\bibnamefont {{Torniamenti}}},\ and\ \bibinfo {author}
  {\bibfnamefont {F.}~\bibnamefont {{Haardt}}},\ }\href
  {https://doi.org/10.1093/mnras/staa2286} {\bibfield  {journal} {\bibinfo
  {journal} {\mnras}\ }\textbf {\bibinfo {volume} {498}},\ \bibinfo {pages}
  {495} (\bibinfo {year} {2020}{\natexlab{a}})},\ \Eprint
  {https://arxiv.org/abs/2004.09525} {arXiv:2004.09525 [astro-ph.HE]}
  \BibitemShut {NoStop}%
\bibitem [{\citenamefont {{Rastello}}\ \emph {et~al.}(2021)\citenamefont
  {{Rastello}}, \citenamefont {{Mapelli}}, \citenamefont {{Di Carlo}},
  \citenamefont {{Iorio}}, \citenamefont {{Ballone}}, \citenamefont
  {{Giacobbo}}, \citenamefont {{Santoliquido}},\ and\ \citenamefont
  {{Torniamenti}}}]{Rastello_2021}%
  \BibitemOpen
  \bibfield  {author} {\bibinfo {author} {\bibfnamefont {S.}~\bibnamefont
  {{Rastello}}}, \bibinfo {author} {\bibfnamefont {M.}~\bibnamefont
  {{Mapelli}}}, \bibinfo {author} {\bibfnamefont {U.~N.}\ \bibnamefont {{Di
  Carlo}}}, \bibinfo {author} {\bibfnamefont {G.}~\bibnamefont {{Iorio}}},
  \bibinfo {author} {\bibfnamefont {A.}~\bibnamefont {{Ballone}}}, \bibinfo
  {author} {\bibfnamefont {N.}~\bibnamefont {{Giacobbo}}}, \bibinfo {author}
  {\bibfnamefont {F.}~\bibnamefont {{Santoliquido}}},\ and\ \bibinfo {author}
  {\bibfnamefont {S.}~\bibnamefont {{Torniamenti}}},\ }\href
  {https://doi.org/10.1093/mnras/stab2355} {\bibfield  {journal} {\bibinfo
  {journal} {\mnras}\ }\textbf {\bibinfo {volume} {507}},\ \bibinfo {pages}
  {3612} (\bibinfo {year} {2021})},\ \Eprint {https://arxiv.org/abs/2105.01669}
  {arXiv:2105.01669 [astro-ph.GA]} \BibitemShut {NoStop}%
\bibitem [{\citenamefont {{Plummer}}(1911)}]{Plummer_1911}%
  \BibitemOpen
  \bibfield  {author} {\bibinfo {author} {\bibfnamefont {H.~C.}\ \bibnamefont
  {{Plummer}}},\ }\href {https://doi.org/10.1093/mnras/71.5.460} {\bibfield
  {journal} {\bibinfo  {journal} {\mnras}\ }\textbf {\bibinfo {volume} {71}},\
  \bibinfo {pages} {460} (\bibinfo {year} {1911})}\BibitemShut {NoStop}%
\bibitem [{\citenamefont {{Heggie}}\ and\ \citenamefont
  {{Hut}}(2003)}]{Heggie_2003}%
  \BibitemOpen
  \bibfield  {author} {\bibinfo {author} {\bibfnamefont {D.}~\bibnamefont
  {{Heggie}}}\ and\ \bibinfo {author} {\bibfnamefont {P.}~\bibnamefont
  {{Hut}}},\ }\href@noop {} {\emph {\bibinfo {title} {{The Gravitational
  Million-Body Problem: A Multidisciplinary Approach to Star Cluster
  Dynamics}}}}\ (\bibinfo {year} {2003})\BibitemShut {NoStop}%
\bibitem [{\citenamefont {{von Hoerner}}(1957)}]{vonHoerner_1957}%
  \BibitemOpen
  \bibfield  {author} {\bibinfo {author} {\bibfnamefont {S.}~\bibnamefont {{von
  Hoerner}}},\ }\href {https://doi.org/10.1086/146321} {\bibfield  {journal}
  {\bibinfo  {journal} {\apj}\ }\textbf {\bibinfo {volume} {125}},\ \bibinfo
  {pages} {451} (\bibinfo {year} {1957})}\BibitemShut {NoStop}%
\bibitem [{\citenamefont {{Aarseth}}(2003)}]{Aarseth_2003}%
  \BibitemOpen
  \bibfield  {author} {\bibinfo {author} {\bibfnamefont {S.~J.}\ \bibnamefont
  {{Aarseth}}},\ }\href@noop {} {\emph {\bibinfo {title} {Gravitational N-Body
  Simulations, by Sverre J.~Aarseth, pp.~430.~ISBN 0521432723.~Cambridge, UK:
  Cambridge University Press, November 2003.}}}\ (\bibinfo {year} {2003})\ p.\
  \bibinfo {pages} {430}\BibitemShut {NoStop}%
\bibitem [{\citenamefont {{Banerjee}}\ \emph {et~al.}(2020)\citenamefont
  {{Banerjee}}, \citenamefont {{Belczynski}}, \citenamefont {{Fryer}},
  \citenamefont {{Berczik}}, \citenamefont {{Hurley}}, \citenamefont
  {{Spurzem}},\ and\ \citenamefont {{Wang}}}]{Banerjee_2020}%
  \BibitemOpen
  \bibfield  {author} {\bibinfo {author} {\bibfnamefont {S.}~\bibnamefont
  {{Banerjee}}}, \bibinfo {author} {\bibfnamefont {K.}~\bibnamefont
  {{Belczynski}}}, \bibinfo {author} {\bibfnamefont {C.~L.}\ \bibnamefont
  {{Fryer}}}, \bibinfo {author} {\bibfnamefont {P.}~\bibnamefont {{Berczik}}},
  \bibinfo {author} {\bibfnamefont {J.~R.}\ \bibnamefont {{Hurley}}}, \bibinfo
  {author} {\bibfnamefont {R.}~\bibnamefont {{Spurzem}}},\ and\ \bibinfo
  {author} {\bibfnamefont {L.}~\bibnamefont {{Wang}}},\ }\href
  {https://doi.org/10.1051/0004-6361/201935332} {\bibfield  {journal} {\bibinfo
   {journal} {\aap}\ }\textbf {\bibinfo {volume} {639}},\ \bibinfo {eid} {A41}
  (\bibinfo {year} {2020})},\ \Eprint {https://arxiv.org/abs/1902.07718}
  {arXiv:1902.07718 [astro-ph.SR]} \BibitemShut {NoStop}%
\bibitem [{\citenamefont {Hurley}\ \emph {et~al.}(2000)\citenamefont {Hurley},
  \citenamefont {Pols},\ and\ \citenamefont {Tout}}]{Hurley_2000}%
  \BibitemOpen
  \bibfield  {author} {\bibinfo {author} {\bibfnamefont {J.~R.}\ \bibnamefont
  {Hurley}}, \bibinfo {author} {\bibfnamefont {O.~R.}\ \bibnamefont {Pols}},\
  and\ \bibinfo {author} {\bibfnamefont {C.~A.}\ \bibnamefont {Tout}},\ }\href
  {https://doi.org/10.1046/j.1365-8711.2000.03426.x} {\bibfield  {journal}
  {\bibinfo  {journal} {Monthly Notices of the Royal Astronomical Society}\
  }\textbf {\bibinfo {volume} {315}},\ \bibinfo {pages} {543} (\bibinfo {year}
  {2000})}\BibitemShut {NoStop}%
\bibitem [{\citenamefont {Hurley}\ \emph {et~al.}(2002)\citenamefont {Hurley},
  \citenamefont {Tout},\ and\ \citenamefont {Pols}}]{Hurley_2002}%
  \BibitemOpen
  \bibfield  {author} {\bibinfo {author} {\bibfnamefont {J.~R.}\ \bibnamefont
  {Hurley}}, \bibinfo {author} {\bibfnamefont {C.~A.}\ \bibnamefont {Tout}},\
  and\ \bibinfo {author} {\bibfnamefont {O.~R.}\ \bibnamefont {Pols}},\ }\href
  {https://doi.org/10.1046/j.1365-8711.2002.05038.x} {\bibfield  {journal}
  {\bibinfo  {journal} {Monthly Notices of the Royal Astronomical Society}\
  }\textbf {\bibinfo {volume} {329}},\ \bibinfo {pages} {897} (\bibinfo {year}
  {2002})}\BibitemShut {NoStop}%
\bibitem [{\citenamefont {Mikkola}\ and\ \citenamefont
  {Tanikawa}(1999)}]{Mikkola_1999}%
  \BibitemOpen
  \bibfield  {author} {\bibinfo {author} {\bibfnamefont {S.}~\bibnamefont
  {Mikkola}}\ and\ \bibinfo {author} {\bibfnamefont {K.}~\bibnamefont
  {Tanikawa}},\ }\href {https://doi.org/10.1046/j.1365-8711.1999.02982.x}
  {\bibfield  {journal} {\bibinfo  {journal} {Monthly Notices of the Royal
  Astronomical Society}\ }\textbf {\bibinfo {volume} {310}},\ \bibinfo {pages}
  {745} (\bibinfo {year} {1999})}\BibitemShut {NoStop}%
\bibitem [{\citenamefont {{Mikkola}}\ and\ \citenamefont
  {{Merritt}}(2008)}]{Mikkola_2008}%
  \BibitemOpen
  \bibfield  {author} {\bibinfo {author} {\bibfnamefont {S.}~\bibnamefont
  {{Mikkola}}}\ and\ \bibinfo {author} {\bibfnamefont {D.}~\bibnamefont
  {{Merritt}}},\ }\href {https://doi.org/10.1088/0004-6256/135/6/2398}
  {\bibfield  {journal} {\bibinfo  {journal} {\aj}\ }\textbf {\bibinfo {volume}
  {135}},\ \bibinfo {pages} {2398} (\bibinfo {year} {2008})},\ \Eprint
  {https://arxiv.org/abs/0709.3367} {arXiv:0709.3367} \BibitemShut {NoStop}%
\bibitem [{\citenamefont {{Fryer}}\ \emph {et~al.}(2012)\citenamefont
  {{Fryer}}, \citenamefont {{Belczynski}}, \citenamefont {{Wiktorowicz}},
  \citenamefont {{Dominik}}, \citenamefont {{Kalogera}},\ and\ \citenamefont
  {{Holz}}}]{Fryer_2012}%
  \BibitemOpen
  \bibfield  {author} {\bibinfo {author} {\bibfnamefont {C.~L.}\ \bibnamefont
  {{Fryer}}}, \bibinfo {author} {\bibfnamefont {K.}~\bibnamefont
  {{Belczynski}}}, \bibinfo {author} {\bibfnamefont {G.}~\bibnamefont
  {{Wiktorowicz}}}, \bibinfo {author} {\bibfnamefont {M.}~\bibnamefont
  {{Dominik}}}, \bibinfo {author} {\bibfnamefont {V.}~\bibnamefont
  {{Kalogera}}},\ and\ \bibinfo {author} {\bibfnamefont {D.~E.}\ \bibnamefont
  {{Holz}}},\ }\href {https://doi.org/10.1088/0004-637X/749/1/91} {\bibfield
  {journal} {\bibinfo  {journal} {\apj}\ }\textbf {\bibinfo {volume} {749}},\
  \bibinfo {eid} {91} (\bibinfo {year} {2012})},\ \Eprint
  {https://arxiv.org/abs/1110.1726} {arXiv:1110.1726 [astro-ph.SR]}
  \BibitemShut {NoStop}%
\bibitem [{\citenamefont {{Belczynski}}\ \emph
  {et~al.}(2016{\natexlab{a}})\citenamefont {{Belczynski}}, \citenamefont
  {{Heger}}, \citenamefont {{Gladysz}}, \citenamefont {{Ruiter}}, \citenamefont
  {{Woosley}}, \citenamefont {{Wiktorowicz}}, \citenamefont {{Chen}},
  \citenamefont {{Bulik}}, \citenamefont {{O'Shaughnessy}}, \citenamefont
  {{Holz}}, \citenamefont {{Fryer}},\ and\ \citenamefont
  {{Berti}}}]{Belczynski_2016a}%
  \BibitemOpen
  \bibfield  {author} {\bibinfo {author} {\bibfnamefont {K.}~\bibnamefont
  {{Belczynski}}}, \bibinfo {author} {\bibfnamefont {A.}~\bibnamefont
  {{Heger}}}, \bibinfo {author} {\bibfnamefont {W.}~\bibnamefont {{Gladysz}}},
  \bibinfo {author} {\bibfnamefont {A.~J.}\ \bibnamefont {{Ruiter}}}, \bibinfo
  {author} {\bibfnamefont {S.}~\bibnamefont {{Woosley}}}, \bibinfo {author}
  {\bibfnamefont {G.}~\bibnamefont {{Wiktorowicz}}}, \bibinfo {author}
  {\bibfnamefont {H.-Y.}\ \bibnamefont {{Chen}}}, \bibinfo {author}
  {\bibfnamefont {T.}~\bibnamefont {{Bulik}}}, \bibinfo {author} {\bibfnamefont
  {R.}~\bibnamefont {{O'Shaughnessy}}}, \bibinfo {author} {\bibfnamefont
  {D.~E.}\ \bibnamefont {{Holz}}}, \bibinfo {author} {\bibfnamefont {C.~L.}\
  \bibnamefont {{Fryer}}},\ and\ \bibinfo {author} {\bibfnamefont
  {E.}~\bibnamefont {{Berti}}},\ }\href
  {https://doi.org/10.1051/0004-6361/201628980} {\bibfield  {journal} {\bibinfo
   {journal} {\aap}\ }\textbf {\bibinfo {volume} {594}},\ \bibinfo {eid} {A97}
  (\bibinfo {year} {2016}{\natexlab{a}})},\ \Eprint
  {https://arxiv.org/abs/1607.03116} {arXiv:1607.03116 [astro-ph.HE]}
  \BibitemShut {NoStop}%
\bibitem [{\citenamefont {{LIGO Scientific Collaboration}}\ and\ \citenamefont
  {{Virgo Collaboration}}(2020)}]{Unequal_masss_2020}%
  \BibitemOpen
  \bibfield  {author} {\bibinfo {author} {\bibnamefont {{LIGO Scientific
  Collaboration}}}\ and\ \bibinfo {author} {\bibnamefont {{Virgo
  Collaboration}}},\ }\href {https://doi.org/10.3847/2041-8213/ab960f}
  {\bibfield  {journal} {\bibinfo  {journal} {\apjl}\ }\textbf {\bibinfo
  {volume} {896}},\ \bibinfo {eid} {L44} (\bibinfo {year} {2020})},\ \Eprint
  {https://arxiv.org/abs/2006.12611} {arXiv:2006.12611 [astro-ph.HE]}
  \BibitemShut {NoStop}%
\bibitem [{\citenamefont {{The LIGO Scientific Collaboration}}\ \emph
  {et~al.}(2024)\citenamefont {{The LIGO Scientific Collaboration}},
  \citenamefont {{VIRGO Collaboration}},\ and\ \citenamefont {{Kagra
  Collaboration}}}]{LowerMassGap2024}%
  \BibitemOpen
  \bibfield  {author} {\bibinfo {author} {\bibnamefont {{The LIGO Scientific
  Collaboration}}}, \bibinfo {author} {\bibnamefont {{VIRGO Collaboration}}},\
  and\ \bibinfo {author} {\bibnamefont {{Kagra Collaboration}}},\ }\href
  {https://doi.org/10.3847/2041-8213/ad5beb} {\bibfield  {journal} {\bibinfo
  {journal} {\apjl}\ }\textbf {\bibinfo {volume} {970}},\ \bibinfo {eid} {L34}
  (\bibinfo {year} {2024})},\ \Eprint {https://arxiv.org/abs/2404.04248}
  {arXiv:2404.04248 [astro-ph.HE]} \BibitemShut {NoStop}%
\bibitem [{\citenamefont {{The LIGO Scientific Collaboration}}\ \emph
  {et~al.}(2026{\natexlab{a}})\citenamefont {{The LIGO Scientific
  Collaboration}}, \citenamefont {{the Virgo Collaboration}}, \citenamefont
  {{the KAGRA Collaboration}}, \citenamefont {{Abac}}, \citenamefont {{Abe}},
  \citenamefont {{Abouelfettouh}}, \citenamefont {{Acernese}},\ and\
  \citenamefont {{Ackley}}}]{GWTC5_pop}%
  \BibitemOpen
  \bibfield  {author} {\bibinfo {author} {\bibnamefont {{The LIGO Scientific
  Collaboration}}}, \bibinfo {author} {\bibnamefont {{the Virgo
  Collaboration}}}, \bibinfo {author} {\bibnamefont {{the KAGRA
  Collaboration}}}, \bibinfo {author} {\bibfnamefont {A.~G.}\ \bibnamefont
  {{Abac}}}, \bibinfo {author} {\bibfnamefont {A.}~\bibnamefont {{Abe}}},
  \bibinfo {author} {\bibfnamefont {I.}~\bibnamefont {{Abouelfettouh}}},
  \bibinfo {author} {\bibfnamefont {F.}~\bibnamefont {{Acernese}}},\ and\
  \bibinfo {author} {\bibfnamefont {K.}~\bibnamefont {{Ackley}}},\ }\href
  {https://doi.org/10.48550/arXiv.2605.27226} {\bibfield  {journal} {\bibinfo
  {journal} {arXiv e-prints}\ ,\ \bibinfo {eid} {arXiv:2605.27226}} (\bibinfo
  {year} {2026}{\natexlab{a}})},\ \Eprint {https://arxiv.org/abs/2605.27226}
  {arXiv:2605.27226 [astro-ph.HE]} \BibitemShut {NoStop}%
\bibitem [{\citenamefont {{Langer}}\ \emph {et~al.}(2007)\citenamefont
  {{Langer}}, \citenamefont {{Norman}}, \citenamefont {{de Koter}},
  \citenamefont {{Vink}}, \citenamefont {{Cantiello}},\ and\ \citenamefont
  {{Yoon}}}]{Langer_2007}%
  \BibitemOpen
  \bibfield  {author} {\bibinfo {author} {\bibfnamefont {N.}~\bibnamefont
  {{Langer}}}, \bibinfo {author} {\bibfnamefont {C.~A.}\ \bibnamefont
  {{Norman}}}, \bibinfo {author} {\bibfnamefont {A.}~\bibnamefont {{de
  Koter}}}, \bibinfo {author} {\bibfnamefont {J.~S.}\ \bibnamefont {{Vink}}},
  \bibinfo {author} {\bibfnamefont {M.}~\bibnamefont {{Cantiello}}},\ and\
  \bibinfo {author} {\bibfnamefont {S.~C.}\ \bibnamefont {{Yoon}}},\ }\href
  {https://doi.org/10.1051/0004-6361:20078482} {\bibfield  {journal} {\bibinfo
  {journal} {\aap}\ }\textbf {\bibinfo {volume} {475}},\ \bibinfo {pages} {L19}
  (\bibinfo {year} {2007})},\ \Eprint {https://arxiv.org/abs/0708.1970}
  {arXiv:0708.1970 [astro-ph]} \BibitemShut {NoStop}%
\bibitem [{\citenamefont {{Woosley}}(2017)}]{Woosley_2017}%
  \BibitemOpen
  \bibfield  {author} {\bibinfo {author} {\bibfnamefont {S.~E.}\ \bibnamefont
  {{Woosley}}},\ }\href {https://doi.org/10.3847/1538-4357/836/2/244}
  {\bibfield  {journal} {\bibinfo  {journal} {\apj}\ }\textbf {\bibinfo
  {volume} {836}},\ \bibinfo {eid} {244} (\bibinfo {year} {2017})},\ \Eprint
  {https://arxiv.org/abs/1608.08939} {arXiv:1608.08939 [astro-ph.HE]}
  \BibitemShut {NoStop}%
\bibitem [{\citenamefont {{Mandel}}\ and\ \citenamefont
  {{M{\"u}ller}}(2020)}]{Mandel_2020}%
  \BibitemOpen
  \bibfield  {author} {\bibinfo {author} {\bibfnamefont {I.}~\bibnamefont
  {{Mandel}}}\ and\ \bibinfo {author} {\bibfnamefont {B.}~\bibnamefont
  {{M{\"u}ller}}},\ }\href {https://doi.org/10.1093/mnras/staa3043} {\bibfield
  {journal} {\bibinfo  {journal} {\mnras}\ }\textbf {\bibinfo {volume} {499}},\
  \bibinfo {pages} {3214} (\bibinfo {year} {2020})},\ \Eprint
  {https://arxiv.org/abs/2006.08360} {arXiv:2006.08360 [astro-ph.HE]}
  \BibitemShut {NoStop}%
\bibitem [{\citenamefont {{Mapelli}}\ \emph {et~al.}(2026)\citenamefont
  {{Mapelli}}, \citenamefont {{Sgalletta}}, \citenamefont {{M{\"u}ller-Horn}},
  \citenamefont {{Iorio}}, \citenamefont {{Rinaldi}}, \citenamefont {{Burt}},
  \citenamefont {{Mar{\'\i}n Pina}},\ and\ \citenamefont
  {{Romagnolo}}}]{Mapelli_2026}%
  \BibitemOpen
  \bibfield  {author} {\bibinfo {author} {\bibfnamefont {M.}~\bibnamefont
  {{Mapelli}}}, \bibinfo {author} {\bibfnamefont {C.}~\bibnamefont
  {{Sgalletta}}}, \bibinfo {author} {\bibfnamefont {J.}~\bibnamefont
  {{M{\"u}ller-Horn}}}, \bibinfo {author} {\bibfnamefont {G.}~\bibnamefont
  {{Iorio}}}, \bibinfo {author} {\bibfnamefont {S.}~\bibnamefont {{Rinaldi}}},
  \bibinfo {author} {\bibfnamefont {C.}~\bibnamefont {{Burt}}}, \bibinfo
  {author} {\bibfnamefont {D.}~\bibnamefont {{Mar{\'\i}n Pina}}},\ and\
  \bibinfo {author} {\bibfnamefont {A.}~\bibnamefont {{Romagnolo}}},\ }\href
  {https://doi.org/10.48550/arXiv.2604.12839} {\bibfield  {journal} {\bibinfo
  {journal} {arXiv e-prints}\ ,\ \bibinfo {eid} {arXiv:2604.12839}} (\bibinfo
  {year} {2026})},\ \Eprint {https://arxiv.org/abs/2604.12839}
  {arXiv:2604.12839 [astro-ph.HE]} \BibitemShut {NoStop}%
\bibitem [{\citenamefont {{Iorio}}\ \emph {et~al.}(2023)\citenamefont
  {{Iorio}}, \citenamefont {{Mapelli}}, \citenamefont {{Costa}}, \citenamefont
  {{Spera}}, \citenamefont {{Escobar}}, \citenamefont {{Sgalletta}},
  \citenamefont {{Trani}}, \citenamefont {{Korb}}, \citenamefont
  {{Santoliquido}}, \citenamefont {{Dall'Amico}}, \citenamefont {{Gaspari}},\
  and\ \citenamefont {{Bressan}}}]{Iorio_2023}%
  \BibitemOpen
  \bibfield  {author} {\bibinfo {author} {\bibfnamefont {G.}~\bibnamefont
  {{Iorio}}}, \bibinfo {author} {\bibfnamefont {M.}~\bibnamefont {{Mapelli}}},
  \bibinfo {author} {\bibfnamefont {G.}~\bibnamefont {{Costa}}}, \bibinfo
  {author} {\bibfnamefont {M.}~\bibnamefont {{Spera}}}, \bibinfo {author}
  {\bibfnamefont {G.~J.}\ \bibnamefont {{Escobar}}}, \bibinfo {author}
  {\bibfnamefont {C.}~\bibnamefont {{Sgalletta}}}, \bibinfo {author}
  {\bibfnamefont {A.~A.}\ \bibnamefont {{Trani}}}, \bibinfo {author}
  {\bibfnamefont {E.}~\bibnamefont {{Korb}}}, \bibinfo {author} {\bibfnamefont
  {F.}~\bibnamefont {{Santoliquido}}}, \bibinfo {author} {\bibfnamefont
  {M.}~\bibnamefont {{Dall'Amico}}}, \bibinfo {author} {\bibfnamefont
  {N.}~\bibnamefont {{Gaspari}}},\ and\ \bibinfo {author} {\bibfnamefont
  {A.}~\bibnamefont {{Bressan}}},\ }\href
  {https://doi.org/10.1093/mnras/stad1630} {\bibfield  {journal} {\bibinfo
  {journal} {\mnras}\ }\textbf {\bibinfo {volume} {524}},\ \bibinfo {pages}
  {426} (\bibinfo {year} {2023})},\ \Eprint {https://arxiv.org/abs/2211.11774}
  {arXiv:2211.11774 [astro-ph.HE]} \BibitemShut {NoStop}%
\bibitem [{\citenamefont {{Andrews}}\ \emph {et~al.}(2025)\citenamefont
  {{Andrews}}, \citenamefont {{Bavera}}, \citenamefont {{Briel}}, \citenamefont
  {{Chattaraj}}, \citenamefont {{Dotter}}, \citenamefont {{Fragos}},
  \citenamefont {{Gallegos-Garcia}}, \citenamefont {{Gossage}}, \citenamefont
  {{Kalogera}}, \citenamefont {{Kasdagli}}, \citenamefont {{Katsaggelos}},
  \citenamefont {{Kimball}}, \citenamefont {{Kovlakas}}, \citenamefont
  {{Kruckow}}, \citenamefont {{Liotine}}, \citenamefont {{Misra}},
  \citenamefont {{Rocha}}, \citenamefont {{Souropanis}}, \citenamefont
  {{Srivastava}}, \citenamefont {{Sun}}, \citenamefont {{Teng}}, \citenamefont
  {{Xing}}, \citenamefont {{Zapartas}},\ and\ \citenamefont
  {{Zevin}}}]{Andrews_2025}%
  \BibitemOpen
  \bibfield  {author} {\bibinfo {author} {\bibfnamefont {J.~J.}\ \bibnamefont
  {{Andrews}}}, \bibinfo {author} {\bibfnamefont {S.~S.}\ \bibnamefont
  {{Bavera}}}, \bibinfo {author} {\bibfnamefont {M.}~\bibnamefont {{Briel}}},
  \bibinfo {author} {\bibfnamefont {A.}~\bibnamefont {{Chattaraj}}}, \bibinfo
  {author} {\bibfnamefont {A.}~\bibnamefont {{Dotter}}}, \bibinfo {author}
  {\bibfnamefont {T.}~\bibnamefont {{Fragos}}}, \bibinfo {author}
  {\bibfnamefont {M.}~\bibnamefont {{Gallegos-Garcia}}}, \bibinfo {author}
  {\bibfnamefont {S.}~\bibnamefont {{Gossage}}}, \bibinfo {author}
  {\bibfnamefont {V.}~\bibnamefont {{Kalogera}}}, \bibinfo {author}
  {\bibfnamefont {E.}~\bibnamefont {{Kasdagli}}}, \bibinfo {author}
  {\bibfnamefont {A.}~\bibnamefont {{Katsaggelos}}}, \bibinfo {author}
  {\bibfnamefont {C.}~\bibnamefont {{Kimball}}}, \bibinfo {author}
  {\bibfnamefont {K.}~\bibnamefont {{Kovlakas}}}, \bibinfo {author}
  {\bibfnamefont {M.~U.}\ \bibnamefont {{Kruckow}}}, \bibinfo {author}
  {\bibfnamefont {C.}~\bibnamefont {{Liotine}}}, \bibinfo {author}
  {\bibfnamefont {D.}~\bibnamefont {{Misra}}}, \bibinfo {author} {\bibfnamefont
  {K.~A.}\ \bibnamefont {{Rocha}}}, \bibinfo {author} {\bibfnamefont
  {D.}~\bibnamefont {{Souropanis}}}, \bibinfo {author} {\bibfnamefont {P.~M.}\
  \bibnamefont {{Srivastava}}}, \bibinfo {author} {\bibfnamefont
  {M.}~\bibnamefont {{Sun}}}, \bibinfo {author} {\bibfnamefont
  {E.}~\bibnamefont {{Teng}}}, \bibinfo {author} {\bibfnamefont
  {Z.}~\bibnamefont {{Xing}}}, \bibinfo {author} {\bibfnamefont
  {E.}~\bibnamefont {{Zapartas}}},\ and\ \bibinfo {author} {\bibfnamefont
  {M.}~\bibnamefont {{Zevin}}},\ }\href
  {https://doi.org/10.3847/1538-4365/adfb78} {\bibfield  {journal} {\bibinfo
  {journal} {\apjs}\ }\textbf {\bibinfo {volume} {281}},\ \bibinfo {eid} {3}
  (\bibinfo {year} {2025})},\ \Eprint {https://arxiv.org/abs/2411.02376}
  {arXiv:2411.02376 [astro-ph.GA]} \BibitemShut {NoStop}%
\bibitem [{\citenamefont {{Kamlah}}\ \emph {et~al.}(2022)\citenamefont
  {{Kamlah}}, \citenamefont {{Leveque}}, \citenamefont {{Spurzem}},
  \citenamefont {{Arca Sedda}}, \citenamefont {{Askar}}, \citenamefont
  {{Banerjee}}, \citenamefont {{Berczik}}, \citenamefont {{Giersz}},
  \citenamefont {{Hurley}}, \citenamefont {{Belloni}}, \citenamefont
  {{K{\"u}hmichel}},\ and\ \citenamefont {{Wang}}}]{Kamlah_2021}%
  \BibitemOpen
  \bibfield  {author} {\bibinfo {author} {\bibfnamefont {A.~W.~H.}\
  \bibnamefont {{Kamlah}}}, \bibinfo {author} {\bibfnamefont {A.}~\bibnamefont
  {{Leveque}}}, \bibinfo {author} {\bibfnamefont {R.}~\bibnamefont
  {{Spurzem}}}, \bibinfo {author} {\bibfnamefont {M.}~\bibnamefont {{Arca
  Sedda}}}, \bibinfo {author} {\bibfnamefont {A.}~\bibnamefont {{Askar}}},
  \bibinfo {author} {\bibfnamefont {S.}~\bibnamefont {{Banerjee}}}, \bibinfo
  {author} {\bibfnamefont {P.}~\bibnamefont {{Berczik}}}, \bibinfo {author}
  {\bibfnamefont {M.}~\bibnamefont {{Giersz}}}, \bibinfo {author}
  {\bibfnamefont {J.}~\bibnamefont {{Hurley}}}, \bibinfo {author}
  {\bibfnamefont {D.}~\bibnamefont {{Belloni}}}, \bibinfo {author}
  {\bibfnamefont {L.}~\bibnamefont {{K{\"u}hmichel}}},\ and\ \bibinfo {author}
  {\bibfnamefont {L.}~\bibnamefont {{Wang}}},\ }\href
  {https://doi.org/10.1093/mnras/stab3748} {\bibfield  {journal} {\bibinfo
  {journal} {\mnras}\ }\textbf {\bibinfo {volume} {511}},\ \bibinfo {pages}
  {4060} (\bibinfo {year} {2022})},\ \Eprint {https://arxiv.org/abs/2105.08067}
  {arXiv:2105.08067 [astro-ph.GA]} \BibitemShut {NoStop}%
\bibitem [{\citenamefont {{Rastello}}\ \emph {et~al.}(2026)\citenamefont
  {{Rastello}}, \citenamefont {{Iorio}}, \citenamefont {{Gieles}},\ and\
  \citenamefont {{Wang}}}]{Rastello_2026}%
  \BibitemOpen
  \bibfield  {author} {\bibinfo {author} {\bibfnamefont {S.}~\bibnamefont
  {{Rastello}}}, \bibinfo {author} {\bibfnamefont {G.}~\bibnamefont {{Iorio}}},
  \bibinfo {author} {\bibfnamefont {M.}~\bibnamefont {{Gieles}}},\ and\
  \bibinfo {author} {\bibfnamefont {L.}~\bibnamefont {{Wang}}},\ }\href
  {https://doi.org/10.1051/0004-6361/202556781} {\bibfield  {journal} {\bibinfo
   {journal} {\aap}\ }\textbf {\bibinfo {volume} {707}},\ \bibinfo {eid} {A217}
  (\bibinfo {year} {2026})},\ \Eprint {https://arxiv.org/abs/2509.07067}
  {arXiv:2509.07067 [astro-ph.HE]} \BibitemShut {NoStop}%
\bibitem [{\citenamefont {{Kroupa}}(2001)}]{Kroupa_2001}%
  \BibitemOpen
  \bibfield  {author} {\bibinfo {author} {\bibfnamefont {P.}~\bibnamefont
  {{Kroupa}}},\ }\href {https://doi.org/10.1046/j.1365-8711.2001.04022.x}
  {\bibfield  {journal} {\bibinfo  {journal} {\mnras}\ }\textbf {\bibinfo
  {volume} {322}},\ \bibinfo {pages} {231} (\bibinfo {year} {2001})},\ \Eprint
  {https://arxiv.org/abs/astro-ph/0009005} {astro-ph/0009005} \BibitemShut
  {NoStop}%
\bibitem [{\citenamefont {{Sana}}\ and\ \citenamefont
  {{Evans}}(2011)}]{Sana_2011}%
  \BibitemOpen
  \bibfield  {author} {\bibinfo {author} {\bibfnamefont {H.}~\bibnamefont
  {{Sana}}}\ and\ \bibinfo {author} {\bibfnamefont {C.~J.}\ \bibnamefont
  {{Evans}}},\ }in\ \href {https://doi.org/10.1017/S1743921311011124} {\emph
  {\bibinfo {booktitle} {Active OB Stars: Structure, Evolution, Mass Loss, and
  Critical Limits}}},\ \bibinfo {series} {IAU Symposium}, Vol.\ \bibinfo
  {volume} {272},\ \bibinfo {editor} {edited by\ \bibinfo {editor}
  {\bibfnamefont {C.}~\bibnamefont {{Neiner}}}, \bibinfo {editor}
  {\bibfnamefont {G.}~\bibnamefont {{Wade}}}, \bibinfo {editor} {\bibfnamefont
  {G.}~\bibnamefont {{Meynet}}},\ and\ \bibinfo {editor} {\bibfnamefont
  {G.}~\bibnamefont {{Peters}}}}\ (\bibinfo {year} {2011})\ pp.\ \bibinfo
  {pages} {474--485},\ \Eprint {https://arxiv.org/abs/1009.4197}
  {arXiv:1009.4197 [astro-ph.SR]} \BibitemShut {NoStop}%
\bibitem [{\citenamefont {{Sana}}\ \emph {et~al.}(2013)\citenamefont {{Sana}},
  \citenamefont {{de Koter}}, \citenamefont {{de Mink}}, \citenamefont
  {{Dunstall}}, \citenamefont {{Evans}}, \citenamefont {{H{\'e}nault-Brunet}},
  \citenamefont {{Ma{\'{\i}}z Apell{\'a}niz}}, \citenamefont
  {{Ram{\'{\i}}rez-Agudelo}}, \citenamefont {{Taylor}}, \citenamefont
  {{Walborn}}, \citenamefont {{Clark}}, \citenamefont {{Crowther}},
  \citenamefont {{Herrero}}, \citenamefont {{Gieles}}, \citenamefont
  {{Langer}}, \citenamefont {{Lennon}},\ and\ \citenamefont
  {{Vink}}}]{Sana_2013}%
  \BibitemOpen
  \bibfield  {author} {\bibinfo {author} {\bibfnamefont {H.}~\bibnamefont
  {{Sana}}}, \bibinfo {author} {\bibfnamefont {A.}~\bibnamefont {{de Koter}}},
  \bibinfo {author} {\bibfnamefont {S.~E.}\ \bibnamefont {{de Mink}}}, \bibinfo
  {author} {\bibfnamefont {P.~R.}\ \bibnamefont {{Dunstall}}}, \bibinfo
  {author} {\bibfnamefont {C.~J.}\ \bibnamefont {{Evans}}}, \bibinfo {author}
  {\bibfnamefont {V.}~\bibnamefont {{H{\'e}nault-Brunet}}}, \bibinfo {author}
  {\bibfnamefont {J.}~\bibnamefont {{Ma{\'{\i}}z Apell{\'a}niz}}}, \bibinfo
  {author} {\bibfnamefont {O.~H.}\ \bibnamefont {{Ram{\'{\i}}rez-Agudelo}}},
  \bibinfo {author} {\bibfnamefont {W.~D.}\ \bibnamefont {{Taylor}}}, \bibinfo
  {author} {\bibfnamefont {N.~R.}\ \bibnamefont {{Walborn}}}, \bibinfo {author}
  {\bibfnamefont {J.~S.}\ \bibnamefont {{Clark}}}, \bibinfo {author}
  {\bibfnamefont {P.~A.}\ \bibnamefont {{Crowther}}}, \bibinfo {author}
  {\bibfnamefont {A.}~\bibnamefont {{Herrero}}}, \bibinfo {author}
  {\bibfnamefont {M.}~\bibnamefont {{Gieles}}}, \bibinfo {author}
  {\bibfnamefont {N.}~\bibnamefont {{Langer}}}, \bibinfo {author}
  {\bibfnamefont {D.~J.}\ \bibnamefont {{Lennon}}},\ and\ \bibinfo {author}
  {\bibfnamefont {J.~S.}\ \bibnamefont {{Vink}}},\ }\href
  {https://doi.org/10.1051/0004-6361/201219621} {\bibfield  {journal} {\bibinfo
   {journal} {\aap}\ }\textbf {\bibinfo {volume} {550}},\ \bibinfo {eid} {A107}
  (\bibinfo {year} {2013})},\ \Eprint {https://arxiv.org/abs/1209.4638}
  {arXiv:1209.4638 [astro-ph.SR]} \BibitemShut {NoStop}%
\bibitem [{\citenamefont {{Moe}}\ and\ \citenamefont {{Di
  Stefano}}(2017)}]{Moe_2017}%
  \BibitemOpen
  \bibfield  {author} {\bibinfo {author} {\bibfnamefont {M.}~\bibnamefont
  {{Moe}}}\ and\ \bibinfo {author} {\bibfnamefont {R.}~\bibnamefont {{Di
  Stefano}}},\ }\href {https://doi.org/10.3847/1538-4365/aa6fb6} {\bibfield
  {journal} {\bibinfo  {journal} {\apjs}\ }\textbf {\bibinfo {volume} {230}},\
  \bibinfo {eid} {15} (\bibinfo {year} {2017})},\ \Eprint
  {https://arxiv.org/abs/1606.05347} {arXiv:1606.05347 [astro-ph.SR]}
  \BibitemShut {NoStop}%
\bibitem [{\citenamefont {{Duquennoy}}\ and\ \citenamefont
  {{Mayor}}(1991)}]{Duq_1991}%
  \BibitemOpen
  \bibfield  {author} {\bibinfo {author} {\bibfnamefont {A.}~\bibnamefont
  {{Duquennoy}}}\ and\ \bibinfo {author} {\bibfnamefont {M.}~\bibnamefont
  {{Mayor}}},\ }\href@noop {} {\bibfield  {journal} {\bibinfo  {journal}
  {\aap}\ }\textbf {\bibinfo {volume} {248}},\ \bibinfo {pages} {485} (\bibinfo
  {year} {1991})}\BibitemShut {NoStop}%
\bibitem [{\citenamefont {{Spitzer}}(1987)}]{Spitzer_1987}%
  \BibitemOpen
  \bibfield  {author} {\bibinfo {author} {\bibfnamefont {L.}~\bibnamefont
  {{Spitzer}}},\ }\href@noop {} {\emph {\bibinfo {title} {Princeton, NJ,
  Princeton University Press, 1987, 191 p.}}}\ (\bibinfo {year}
  {1987})\BibitemShut {NoStop}%
\bibitem [{\citenamefont {{Banerjee}}(2018{\natexlab{a}})}]{Banerjee_2017b}%
  \BibitemOpen
  \bibfield  {author} {\bibinfo {author} {\bibfnamefont {S.}~\bibnamefont
  {{Banerjee}}},\ }\href {https://doi.org/10.1093/mnras/stx2347} {\bibfield
  {journal} {\bibinfo  {journal} {\mnras}\ }\textbf {\bibinfo {volume} {473}},\
  \bibinfo {pages} {909} (\bibinfo {year} {2018}{\natexlab{a}})},\ \Eprint
  {https://arxiv.org/abs/1707.00922} {arXiv:1707.00922 [astro-ph.HE]}
  \BibitemShut {NoStop}%
\bibitem [{\citenamefont {{Belczynski}}\ \emph {et~al.}(2020)\citenamefont
  {{Belczynski}}, \citenamefont {{Klencki}}, \citenamefont {{Fields}},
  \citenamefont {{Olejak}}, \citenamefont {{Berti}}, \citenamefont {{Meynet}},
  \citenamefont {{Fryer}}, \citenamefont {{Holz}}, \citenamefont
  {{O'Shaughnessy}}, \citenamefont {{Brown}}, \citenamefont {{Bulik}},
  \citenamefont {{Leung}}, \citenamefont {{Nomoto}}, \citenamefont {{Madau}},
  \citenamefont {{Hirschi}}, \citenamefont {{Kaiser}}, \citenamefont {{Jones}},
  \citenamefont {{Mondal}}, \citenamefont {{Chruslinska}}, \citenamefont
  {{Drozda}}, \citenamefont {{Gerosa}}, \citenamefont {{Doctor}}, \citenamefont
  {{Giersz}}, \citenamefont {{Ekstrom}}, \citenamefont {{Georgy}},
  \citenamefont {{Askar}}, \citenamefont {{Baibhav}}, \citenamefont
  {{Wysocki}}, \citenamefont {{Natan}}, \citenamefont {{Farr}}, \citenamefont
  {{Wiktorowicz}}, \citenamefont {{Coleman Miller}}, \citenamefont {{Farr}},\
  and\ \citenamefont {{Lasota}}}]{Belczynski_2020}%
  \BibitemOpen
  \bibfield  {author} {\bibinfo {author} {\bibfnamefont {K.}~\bibnamefont
  {{Belczynski}}}, \bibinfo {author} {\bibfnamefont {J.}~\bibnamefont
  {{Klencki}}}, \bibinfo {author} {\bibfnamefont {C.~E.}\ \bibnamefont
  {{Fields}}}, \bibinfo {author} {\bibfnamefont {A.}~\bibnamefont {{Olejak}}},
  \bibinfo {author} {\bibfnamefont {E.}~\bibnamefont {{Berti}}}, \bibinfo
  {author} {\bibfnamefont {G.}~\bibnamefont {{Meynet}}}, \bibinfo {author}
  {\bibfnamefont {C.~L.}\ \bibnamefont {{Fryer}}}, \bibinfo {author}
  {\bibfnamefont {D.~E.}\ \bibnamefont {{Holz}}}, \bibinfo {author}
  {\bibfnamefont {R.}~\bibnamefont {{O'Shaughnessy}}}, \bibinfo {author}
  {\bibfnamefont {D.~A.}\ \bibnamefont {{Brown}}}, \bibinfo {author}
  {\bibfnamefont {T.}~\bibnamefont {{Bulik}}}, \bibinfo {author} {\bibfnamefont
  {S.~C.}\ \bibnamefont {{Leung}}}, \bibinfo {author} {\bibfnamefont
  {K.}~\bibnamefont {{Nomoto}}}, \bibinfo {author} {\bibfnamefont
  {P.}~\bibnamefont {{Madau}}}, \bibinfo {author} {\bibfnamefont
  {R.}~\bibnamefont {{Hirschi}}}, \bibinfo {author} {\bibfnamefont
  {E.}~\bibnamefont {{Kaiser}}}, \bibinfo {author} {\bibfnamefont
  {S.}~\bibnamefont {{Jones}}}, \bibinfo {author} {\bibfnamefont
  {S.}~\bibnamefont {{Mondal}}}, \bibinfo {author} {\bibfnamefont
  {M.}~\bibnamefont {{Chruslinska}}}, \bibinfo {author} {\bibfnamefont
  {P.}~\bibnamefont {{Drozda}}}, \bibinfo {author} {\bibfnamefont
  {D.}~\bibnamefont {{Gerosa}}}, \bibinfo {author} {\bibfnamefont
  {Z.}~\bibnamefont {{Doctor}}}, \bibinfo {author} {\bibfnamefont
  {M.}~\bibnamefont {{Giersz}}}, \bibinfo {author} {\bibfnamefont
  {S.}~\bibnamefont {{Ekstrom}}}, \bibinfo {author} {\bibfnamefont
  {C.}~\bibnamefont {{Georgy}}}, \bibinfo {author} {\bibfnamefont
  {A.}~\bibnamefont {{Askar}}}, \bibinfo {author} {\bibfnamefont
  {V.}~\bibnamefont {{Baibhav}}}, \bibinfo {author} {\bibfnamefont
  {D.}~\bibnamefont {{Wysocki}}}, \bibinfo {author} {\bibfnamefont
  {T.}~\bibnamefont {{Natan}}}, \bibinfo {author} {\bibfnamefont {W.~M.}\
  \bibnamefont {{Farr}}}, \bibinfo {author} {\bibfnamefont {G.}~\bibnamefont
  {{Wiktorowicz}}}, \bibinfo {author} {\bibfnamefont {M.}~\bibnamefont
  {{Coleman Miller}}}, \bibinfo {author} {\bibfnamefont {B.}~\bibnamefont
  {{Farr}}},\ and\ \bibinfo {author} {\bibfnamefont {J.~P.}\ \bibnamefont
  {{Lasota}}},\ }\href {https://doi.org/10.1051/0004-6361/201936528} {\bibfield
   {journal} {\bibinfo  {journal} {\aap}\ }\textbf {\bibinfo {volume} {636}},\
  \bibinfo {eid} {A104} (\bibinfo {year} {2020})},\ \Eprint
  {https://arxiv.org/abs/1706.07053} {arXiv:1706.07053 [astro-ph.HE]}
  \BibitemShut {NoStop}%
\bibitem [{\citenamefont {{Spruit}}(2002)}]{Spruit_2002}%
  \BibitemOpen
  \bibfield  {author} {\bibinfo {author} {\bibfnamefont {H.~C.}\ \bibnamefont
  {{Spruit}}},\ }\href {https://doi.org/10.1051/0004-6361:20011465} {\bibfield
  {journal} {\bibinfo  {journal} {\aap}\ }\textbf {\bibinfo {volume} {381}},\
  \bibinfo {pages} {923} (\bibinfo {year} {2002})},\ \Eprint
  {https://arxiv.org/abs/astro-ph/0108207} {astro-ph/0108207} \BibitemShut
  {NoStop}%
\bibitem [{\citenamefont {{Fuller}}\ and\ \citenamefont
  {{Ma}}(2019)}]{Fuller_2019a}%
  \BibitemOpen
  \bibfield  {author} {\bibinfo {author} {\bibfnamefont {J.}~\bibnamefont
  {{Fuller}}}\ and\ \bibinfo {author} {\bibfnamefont {L.}~\bibnamefont
  {{Ma}}},\ }\href {https://doi.org/10.3847/2041-8213/ab339b} {\bibfield
  {journal} {\bibinfo  {journal} {\apjl}\ }\textbf {\bibinfo {volume} {881}},\
  \bibinfo {eid} {L1} (\bibinfo {year} {2019})},\ \Eprint
  {https://arxiv.org/abs/1907.03714} {arXiv:1907.03714 [astro-ph.SR]}
  \BibitemShut {NoStop}%
\bibitem [{\citenamefont {{Thorne}}\ and\ \citenamefont
  {{Zytkow}}(1975)}]{TZ_1975}%
  \BibitemOpen
  \bibfield  {author} {\bibinfo {author} {\bibfnamefont {K.~S.}\ \bibnamefont
  {{Thorne}}}\ and\ \bibinfo {author} {\bibfnamefont {A.~N.}\ \bibnamefont
  {{Zytkow}}},\ }\href {https://doi.org/10.1086/181839} {\bibfield  {journal}
  {\bibinfo  {journal} {\apjl}\ }\textbf {\bibinfo {volume} {199}},\ \bibinfo
  {pages} {L19} (\bibinfo {year} {1975})}\BibitemShut {NoStop}%
\bibitem [{\citenamefont {{Gaburov}}\ \emph {et~al.}(2008)\citenamefont
  {{Gaburov}}, \citenamefont {{Lombardi}},\ and\ \citenamefont {{Portegies
  Zwart}}}]{Gaburov_2008}%
  \BibitemOpen
  \bibfield  {author} {\bibinfo {author} {\bibfnamefont {E.}~\bibnamefont
  {{Gaburov}}}, \bibinfo {author} {\bibfnamefont {J.~C.}\ \bibnamefont
  {{Lombardi}}},\ and\ \bibinfo {author} {\bibfnamefont {S.}~\bibnamefont
  {{Portegies Zwart}}},\ }\href
  {https://doi.org/10.1111/j.1745-3933.2007.00399.x} {\bibfield  {journal}
  {\bibinfo  {journal} {\mnras}\ }\textbf {\bibinfo {volume} {383}},\ \bibinfo
  {pages} {L5} (\bibinfo {year} {2008})},\ \Eprint
  {https://arxiv.org/abs/0707.3021} {arXiv:0707.3021 [astro-ph]} \BibitemShut
  {NoStop}%
\bibitem [{\citenamefont {{Glebbeek}}\ \emph {et~al.}(2009)\citenamefont
  {{Glebbeek}}, \citenamefont {{Gaburov}}, \citenamefont {{de Mink}},
  \citenamefont {{Pols}},\ and\ \citenamefont {{Portegies
  Zwart}}}]{Glebbeek_2009}%
  \BibitemOpen
  \bibfield  {author} {\bibinfo {author} {\bibfnamefont {E.}~\bibnamefont
  {{Glebbeek}}}, \bibinfo {author} {\bibfnamefont {E.}~\bibnamefont
  {{Gaburov}}}, \bibinfo {author} {\bibfnamefont {S.~E.}\ \bibnamefont {{de
  Mink}}}, \bibinfo {author} {\bibfnamefont {O.~R.}\ \bibnamefont {{Pols}}},\
  and\ \bibinfo {author} {\bibfnamefont {S.~F.}\ \bibnamefont {{Portegies
  Zwart}}},\ }\href {https://doi.org/10.1051/0004-6361/200810425} {\bibfield
  {journal} {\bibinfo  {journal} {\aap}\ }\textbf {\bibinfo {volume} {497}},\
  \bibinfo {pages} {255} (\bibinfo {year} {2009})},\ \Eprint
  {https://arxiv.org/abs/0902.1753} {arXiv:0902.1753 [astro-ph.SR]}
  \BibitemShut {NoStop}%
\bibitem [{\citenamefont {{Ryu}}\ \emph {et~al.}(2025)\citenamefont {{Ryu}},
  \citenamefont {{Sills}}, \citenamefont {{Pakmor}}, \citenamefont {{de
  Mink}},\ and\ \citenamefont {{Mathieu}}}]{Ryu_2025}%
  \BibitemOpen
  \bibfield  {author} {\bibinfo {author} {\bibfnamefont {T.}~\bibnamefont
  {{Ryu}}}, \bibinfo {author} {\bibfnamefont {A.}~\bibnamefont {{Sills}}},
  \bibinfo {author} {\bibfnamefont {R.}~\bibnamefont {{Pakmor}}}, \bibinfo
  {author} {\bibfnamefont {S.}~\bibnamefont {{de Mink}}},\ and\ \bibinfo
  {author} {\bibfnamefont {R.}~\bibnamefont {{Mathieu}}},\ }\href
  {https://doi.org/10.3847/2041-8213/adaf94} {\bibfield  {journal} {\bibinfo
  {journal} {\apjl}\ }\textbf {\bibinfo {volume} {980}},\ \bibinfo {eid} {L38}
  (\bibinfo {year} {2025})},\ \Eprint {https://arxiv.org/abs/2410.00148}
  {arXiv:2410.00148 [astro-ph.SR]} \BibitemShut {NoStop}%
\bibitem [{\citenamefont {{Vynatheya}}\ \emph {et~al.}(2026)\citenamefont
  {{Vynatheya}}, \citenamefont {{Ryu}}, \citenamefont {{Wang}}, \citenamefont
  {{Sills}},\ and\ \citenamefont {{Pakmor}}}]{Vynatheya_2026}%
  \BibitemOpen
  \bibfield  {author} {\bibinfo {author} {\bibfnamefont {P.}~\bibnamefont
  {{Vynatheya}}}, \bibinfo {author} {\bibfnamefont {T.}~\bibnamefont {{Ryu}}},
  \bibinfo {author} {\bibfnamefont {C.}~\bibnamefont {{Wang}}}, \bibinfo
  {author} {\bibfnamefont {A.}~\bibnamefont {{Sills}}},\ and\ \bibinfo {author}
  {\bibfnamefont {R.}~\bibnamefont {{Pakmor}}},\ }\href
  {https://doi.org/10.3847/1538-4357/ae40ba} {\bibfield  {journal} {\bibinfo
  {journal} {\apj}\ }\textbf {\bibinfo {volume} {999}},\ \bibinfo {eid} {64}
  (\bibinfo {year} {2026})},\ \Eprint {https://arxiv.org/abs/2510.13736}
  {arXiv:2510.13736 [astro-ph.SR]} \BibitemShut {NoStop}%
\bibitem [{\citenamefont {Abbott}\ \emph {et~al.}(2020)\citenamefont {Abbott},
  \citenamefont {Abbott}, \citenamefont {Abraham}, \citenamefont {Acernese},
  \citenamefont {Ackley}, \citenamefont {Adams}, \citenamefont {Adhikari},
  \citenamefont {Adya},\ and\ \citenamefont {et~al.}}]{Abbott_GW190521}%
  \BibitemOpen
  \bibfield  {author} {\bibinfo {author} {\bibfnamefont {R.}~\bibnamefont
  {Abbott}}, \bibinfo {author} {\bibfnamefont {T.~D.}\ \bibnamefont {Abbott}},
  \bibinfo {author} {\bibfnamefont {S.}~\bibnamefont {Abraham}}, \bibinfo
  {author} {\bibfnamefont {F.}~\bibnamefont {Acernese}}, \bibinfo {author}
  {\bibfnamefont {K.}~\bibnamefont {Ackley}}, \bibinfo {author} {\bibfnamefont
  {C.}~\bibnamefont {Adams}}, \bibinfo {author} {\bibfnamefont {R.~X.}\
  \bibnamefont {Adhikari}}, \bibinfo {author} {\bibfnamefont {V.~B.}\
  \bibnamefont {Adya}},\ and\ \bibinfo {author} {\bibnamefont {et~al.}}
  (\bibinfo {collaboration} {LIGO Scientific Collaboration and Virgo
  Collaboration}),\ }\href {https://doi.org/10.1103/PhysRevLett.125.101102}
  {\bibfield  {journal} {\bibinfo  {journal} {Phys. Rev. Lett.}\ }\textbf
  {\bibinfo {volume} {125}},\ \bibinfo {pages} {101102} (\bibinfo {year}
  {2020})}\BibitemShut {NoStop}%
\bibitem [{\citenamefont {{Abbott}}\ \emph
  {et~al.}(2023{\natexlab{b}})\citenamefont {{Abbott}}, \citenamefont
  {{Abbott}}, \citenamefont {{Acernese}}, \citenamefont {{Ackley}},
  \citenamefont {{Adams}}, \citenamefont {{Adhikari}}, \citenamefont
  {{Adhikari}}, \citenamefont {{Adya}},\ and\ \citenamefont
  {et~al.}}]{Abbott_GWTC3_prop}%
  \BibitemOpen
  \bibfield  {author} {\bibinfo {author} {\bibfnamefont {R.}~\bibnamefont
  {{Abbott}}}, \bibinfo {author} {\bibfnamefont {T.~D.}\ \bibnamefont
  {{Abbott}}}, \bibinfo {author} {\bibfnamefont {F.}~\bibnamefont
  {{Acernese}}}, \bibinfo {author} {\bibfnamefont {K.}~\bibnamefont
  {{Ackley}}}, \bibinfo {author} {\bibfnamefont {C.}~\bibnamefont {{Adams}}},
  \bibinfo {author} {\bibfnamefont {N.}~\bibnamefont {{Adhikari}}}, \bibinfo
  {author} {\bibfnamefont {R.~X.}\ \bibnamefont {{Adhikari}}}, \bibinfo
  {author} {\bibfnamefont {V.~B.}\ \bibnamefont {{Adya}}},\ and\ \bibinfo
  {author} {\bibnamefont {et~al.}},\ }\href
  {https://doi.org/10.1103/PhysRevX.13.011048} {\bibfield  {journal} {\bibinfo
  {journal} {Physical Review X}\ }\textbf {\bibinfo {volume} {13}},\ \bibinfo
  {eid} {011048} (\bibinfo {year} {2023}{\natexlab{b}})},\ \Eprint
  {https://arxiv.org/abs/2111.03634} {arXiv:2111.03634 [astro-ph.HE]}
  \BibitemShut {NoStop}%
\bibitem [{\citenamefont {{Arca Sedda}}\ \emph {et~al.}(2023)\citenamefont
  {{Arca Sedda}}, \citenamefont {{Mapelli}}, \citenamefont {{Benacquista}},\
  and\ \citenamefont {{Spera}}}]{ArcaSedda_2021b}%
  \BibitemOpen
  \bibfield  {author} {\bibinfo {author} {\bibfnamefont {M.}~\bibnamefont
  {{Arca Sedda}}}, \bibinfo {author} {\bibfnamefont {M.}~\bibnamefont
  {{Mapelli}}}, \bibinfo {author} {\bibfnamefont {M.}~\bibnamefont
  {{Benacquista}}},\ and\ \bibinfo {author} {\bibfnamefont {M.}~\bibnamefont
  {{Spera}}},\ }\href {https://doi.org/10.1093/mnras/stad331} {\bibfield
  {journal} {\bibinfo  {journal} {\mnras}\ }\textbf {\bibinfo {volume} {520}},\
  \bibinfo {pages} {5259} (\bibinfo {year} {2023})},\ \Eprint
  {https://arxiv.org/abs/2109.12119} {arXiv:2109.12119 [astro-ph.GA]}
  \BibitemShut {NoStop}%
\bibitem [{\citenamefont {{Gatto}}\ \emph {et~al.}(2021)\citenamefont
  {{Gatto}}, \citenamefont {{Ripepi}}, \citenamefont {{Bellazzini}},
  \citenamefont {{Tosi}}, \citenamefont {{Cignoni}}, \citenamefont {{Tortora}},
  \citenamefont {{Leccia}}, \citenamefont {{Clementini}}, \citenamefont
  {{Grebel}}, \citenamefont {{Longo}}, \citenamefont {{Marconi}},\ and\
  \citenamefont {{Musella}}}]{Gatto_2021}%
  \BibitemOpen
  \bibfield  {author} {\bibinfo {author} {\bibfnamefont {M.}~\bibnamefont
  {{Gatto}}}, \bibinfo {author} {\bibfnamefont {V.}~\bibnamefont {{Ripepi}}},
  \bibinfo {author} {\bibfnamefont {M.}~\bibnamefont {{Bellazzini}}}, \bibinfo
  {author} {\bibfnamefont {M.}~\bibnamefont {{Tosi}}}, \bibinfo {author}
  {\bibfnamefont {M.}~\bibnamefont {{Cignoni}}}, \bibinfo {author}
  {\bibfnamefont {C.}~\bibnamefont {{Tortora}}}, \bibinfo {author}
  {\bibfnamefont {S.}~\bibnamefont {{Leccia}}}, \bibinfo {author}
  {\bibfnamefont {G.}~\bibnamefont {{Clementini}}}, \bibinfo {author}
  {\bibfnamefont {E.~K.}\ \bibnamefont {{Grebel}}}, \bibinfo {author}
  {\bibfnamefont {G.}~\bibnamefont {{Longo}}}, \bibinfo {author} {\bibfnamefont
  {M.}~\bibnamefont {{Marconi}}},\ and\ \bibinfo {author} {\bibfnamefont
  {I.}~\bibnamefont {{Musella}}},\ }\href
  {https://doi.org/10.1093/mnras/stab2297} {\bibfield  {journal} {\bibinfo
  {journal} {\mnras}\ }\textbf {\bibinfo {volume} {507}},\ \bibinfo {pages}
  {3312} (\bibinfo {year} {2021})},\ \Eprint {https://arxiv.org/abs/2108.02791}
  {arXiv:2108.02791 [astro-ph.GA]} \BibitemShut {NoStop}%
\bibitem [{\citenamefont {{H{\'e}non}}(1975)}]{Henon_1975}%
  \BibitemOpen
  \bibfield  {author} {\bibinfo {author} {\bibfnamefont {M.}~\bibnamefont
  {{H{\'e}non}}},\ }in\ \href@noop {} {\emph {\bibinfo {booktitle} {Dynamics of
  the Solar Systems}}},\ \bibinfo {series} {IAU Symposium}, Vol.~\bibinfo
  {volume} {69},\ \bibinfo {editor} {edited by\ \bibinfo {editor}
  {\bibfnamefont {A.}~\bibnamefont {{Hayli}}}}\ (\bibinfo {year} {1975})\ p.\
  \bibinfo {pages} {133}\BibitemShut {NoStop}%
\bibitem [{\citenamefont {Breen}\ and\ \citenamefont
  {Heggie}(2013)}]{Breen_2013}%
  \BibitemOpen
  \bibfield  {author} {\bibinfo {author} {\bibfnamefont {P.~G.}\ \bibnamefont
  {Breen}}\ and\ \bibinfo {author} {\bibfnamefont {D.~C.}\ \bibnamefont
  {Heggie}},\ }\href {https://doi.org/10.1093/mnras/stt628} {\bibfield
  {journal} {\bibinfo  {journal} {Monthly Notices of the Royal Astronomical
  Society}\ }\textbf {\bibinfo {volume} {432}},\ \bibinfo {pages} {2779}
  (\bibinfo {year} {2013})}\BibitemShut {NoStop}%
\bibitem [{\citenamefont {{Antonini}}\ and\ \citenamefont
  {{Gieles}}(2020{\natexlab{a}})}]{Antonini_2020}%
  \BibitemOpen
  \bibfield  {author} {\bibinfo {author} {\bibfnamefont {F.}~\bibnamefont
  {{Antonini}}}\ and\ \bibinfo {author} {\bibfnamefont {M.}~\bibnamefont
  {{Gieles}}},\ }\href {https://doi.org/10.1093/mnras/stz3584} {\bibfield
  {journal} {\bibinfo  {journal} {\mnras}\ }\textbf {\bibinfo {volume} {492}},\
  \bibinfo {pages} {2936} (\bibinfo {year} {2020}{\natexlab{a}})},\ \Eprint
  {https://arxiv.org/abs/1906.11855} {arXiv:1906.11855 [astro-ph.HE]}
  \BibitemShut {NoStop}%
\bibitem [{\citenamefont {{Heggie}}\ and\ \citenamefont
  {{Giersz}}(2014)}]{Heggie_2014}%
  \BibitemOpen
  \bibfield  {author} {\bibinfo {author} {\bibfnamefont {D.~C.}\ \bibnamefont
  {{Heggie}}}\ and\ \bibinfo {author} {\bibfnamefont {M.}~\bibnamefont
  {{Giersz}}},\ }\href {https://doi.org/10.1093/mnras/stu102} {\bibfield
  {journal} {\bibinfo  {journal} {\mnras}\ }\textbf {\bibinfo {volume} {439}},\
  \bibinfo {pages} {2459} (\bibinfo {year} {2014})},\ \Eprint
  {https://arxiv.org/abs/1401.3657} {arXiv:1401.3657 [astro-ph.GA]}
  \BibitemShut {NoStop}%
\bibitem [{\citenamefont {{Spera}}\ \emph {et~al.}(2019)\citenamefont
  {{Spera}}, \citenamefont {{Mapelli}}, \citenamefont {{Giacobbo}},
  \citenamefont {{Trani}}, \citenamefont {{Bressan}},\ and\ \citenamefont
  {{Costa}}}]{Spera_2019}%
  \BibitemOpen
  \bibfield  {author} {\bibinfo {author} {\bibfnamefont {M.}~\bibnamefont
  {{Spera}}}, \bibinfo {author} {\bibfnamefont {M.}~\bibnamefont {{Mapelli}}},
  \bibinfo {author} {\bibfnamefont {N.}~\bibnamefont {{Giacobbo}}}, \bibinfo
  {author} {\bibfnamefont {A.~A.}\ \bibnamefont {{Trani}}}, \bibinfo {author}
  {\bibfnamefont {A.}~\bibnamefont {{Bressan}}},\ and\ \bibinfo {author}
  {\bibfnamefont {G.}~\bibnamefont {{Costa}}},\ }\href
  {https://doi.org/10.1093/mnras/stz359} {\bibfield  {journal} {\bibinfo
  {journal} {\mnras}\ }\textbf {\bibinfo {volume} {485}},\ \bibinfo {pages}
  {889} (\bibinfo {year} {2019})},\ \Eprint {https://arxiv.org/abs/1809.04605}
  {arXiv:1809.04605 [astro-ph.HE]} \BibitemShut {NoStop}%
\bibitem [{\citenamefont {{Gonz{\'a}lez}}\ \emph {et~al.}(2021)\citenamefont
  {{Gonz{\'a}lez}}, \citenamefont {{Kremer}}, \citenamefont {{Chatterjee}},
  \citenamefont {{Fragione}}, \citenamefont {{Rodriguez}}, \citenamefont
  {{Weatherford}}, \citenamefont {{Ye}},\ and\ \citenamefont
  {{Rasio}}}]{Gonzalez_2020}%
  \BibitemOpen
  \bibfield  {author} {\bibinfo {author} {\bibfnamefont {E.}~\bibnamefont
  {{Gonz{\'a}lez}}}, \bibinfo {author} {\bibfnamefont {K.}~\bibnamefont
  {{Kremer}}}, \bibinfo {author} {\bibfnamefont {S.}~\bibnamefont
  {{Chatterjee}}}, \bibinfo {author} {\bibfnamefont {G.}~\bibnamefont
  {{Fragione}}}, \bibinfo {author} {\bibfnamefont {C.~L.}\ \bibnamefont
  {{Rodriguez}}}, \bibinfo {author} {\bibfnamefont {N.~C.}\ \bibnamefont
  {{Weatherford}}}, \bibinfo {author} {\bibfnamefont {C.~S.}\ \bibnamefont
  {{Ye}}},\ and\ \bibinfo {author} {\bibfnamefont {F.~A.}\ \bibnamefont
  {{Rasio}}},\ }\href {https://doi.org/10.3847/2041-8213/abdf5b} {\bibfield
  {journal} {\bibinfo  {journal} {\apjl}\ }\textbf {\bibinfo {volume} {908}},\
  \bibinfo {eid} {L29} (\bibinfo {year} {2021})},\ \Eprint
  {https://arxiv.org/abs/2012.10497} {arXiv:2012.10497 [astro-ph.HE]}
  \BibitemShut {NoStop}%
\bibitem [{\citenamefont {{Di Carlo}}\ \emph
  {et~al.}(2020{\natexlab{b}})\citenamefont {{Di Carlo}}, \citenamefont
  {{Mapelli}}, \citenamefont {{Bouffanais}}, \citenamefont {{Giacobbo}},
  \citenamefont {{Santoliquido}}, \citenamefont {{Bressan}}, \citenamefont
  {{Spera}},\ and\ \citenamefont {{Haardt}}}]{DiCarlo_2020b}%
  \BibitemOpen
  \bibfield  {author} {\bibinfo {author} {\bibfnamefont {U.~N.}\ \bibnamefont
  {{Di Carlo}}}, \bibinfo {author} {\bibfnamefont {M.}~\bibnamefont
  {{Mapelli}}}, \bibinfo {author} {\bibfnamefont {Y.}~\bibnamefont
  {{Bouffanais}}}, \bibinfo {author} {\bibfnamefont {N.}~\bibnamefont
  {{Giacobbo}}}, \bibinfo {author} {\bibfnamefont {F.}~\bibnamefont
  {{Santoliquido}}}, \bibinfo {author} {\bibfnamefont {A.}~\bibnamefont
  {{Bressan}}}, \bibinfo {author} {\bibfnamefont {M.}~\bibnamefont {{Spera}}},\
  and\ \bibinfo {author} {\bibfnamefont {F.}~\bibnamefont {{Haardt}}},\ }\href
  {https://doi.org/10.1093/mnras/staa1997} {\bibfield  {journal} {\bibinfo
  {journal} {\mnras}\ }\textbf {\bibinfo {volume} {497}},\ \bibinfo {pages}
  {1043} (\bibinfo {year} {2020}{\natexlab{b}})},\ \Eprint
  {https://arxiv.org/abs/1911.01434} {arXiv:1911.01434 [astro-ph.HE]}
  \BibitemShut {NoStop}%
\bibitem [{\citenamefont {{Ballone}}\ \emph {et~al.}(2023)\citenamefont
  {{Ballone}}, \citenamefont {{Costa}}, \citenamefont {{Mapelli}},
  \citenamefont {{MacLeod}}, \citenamefont {{Torniamenti}},\ and\ \citenamefont
  {{Pacheco-Arias}}}]{Ballone_2023}%
  \BibitemOpen
  \bibfield  {author} {\bibinfo {author} {\bibfnamefont {A.}~\bibnamefont
  {{Ballone}}}, \bibinfo {author} {\bibfnamefont {G.}~\bibnamefont {{Costa}}},
  \bibinfo {author} {\bibfnamefont {M.}~\bibnamefont {{Mapelli}}}, \bibinfo
  {author} {\bibfnamefont {M.}~\bibnamefont {{MacLeod}}}, \bibinfo {author}
  {\bibfnamefont {S.}~\bibnamefont {{Torniamenti}}},\ and\ \bibinfo {author}
  {\bibfnamefont {J.~M.}\ \bibnamefont {{Pacheco-Arias}}},\ }\href
  {https://doi.org/10.1093/mnras/stac3752} {\bibfield  {journal} {\bibinfo
  {journal} {\mnras}\ }\textbf {\bibinfo {volume} {519}},\ \bibinfo {pages}
  {5191} (\bibinfo {year} {2023})},\ \Eprint {https://arxiv.org/abs/2204.03493}
  {arXiv:2204.03493 [astro-ph.SR]} \BibitemShut {NoStop}%
\bibitem [{\citenamefont {{Belczynski}}(2020)}]{Belczynski_2020d}%
  \BibitemOpen
  \bibfield  {author} {\bibinfo {author} {\bibfnamefont {K.}~\bibnamefont
  {{Belczynski}}},\ }\href {https://doi.org/10.3847/2041-8213/abcbf1}
  {\bibfield  {journal} {\bibinfo  {journal} {\apjl}\ }\textbf {\bibinfo
  {volume} {905}},\ \bibinfo {eid} {L15} (\bibinfo {year} {2020})},\ \Eprint
  {https://arxiv.org/abs/2009.13526} {arXiv:2009.13526 [astro-ph.HE]}
  \BibitemShut {NoStop}%
\bibitem [{\citenamefont {{Ziegler}}\ and\ \citenamefont
  {{Freese}}(2021)}]{Ziegler_2021}%
  \BibitemOpen
  \bibfield  {author} {\bibinfo {author} {\bibfnamefont {J.}~\bibnamefont
  {{Ziegler}}}\ and\ \bibinfo {author} {\bibfnamefont {K.}~\bibnamefont
  {{Freese}}},\ }\href {https://doi.org/10.1103/PhysRevD.104.043015} {\bibfield
   {journal} {\bibinfo  {journal} {\prd}\ }\textbf {\bibinfo {volume} {104}},\
  \bibinfo {eid} {043015} (\bibinfo {year} {2021})},\ \Eprint
  {https://arxiv.org/abs/2010.00254} {arXiv:2010.00254 [astro-ph.HE]}
  \BibitemShut {NoStop}%
\bibitem [{\citenamefont {{Vink}}\ \emph {et~al.}(2021)\citenamefont {{Vink}},
  \citenamefont {{Higgins}}, \citenamefont {{Sander}},\ and\ \citenamefont
  {{Sabhahit}}}]{Vink_2021}%
  \BibitemOpen
  \bibfield  {author} {\bibinfo {author} {\bibfnamefont {J.~S.}\ \bibnamefont
  {{Vink}}}, \bibinfo {author} {\bibfnamefont {E.~R.}\ \bibnamefont
  {{Higgins}}}, \bibinfo {author} {\bibfnamefont {A.~A.~C.}\ \bibnamefont
  {{Sander}}},\ and\ \bibinfo {author} {\bibfnamefont {G.~N.}\ \bibnamefont
  {{Sabhahit}}},\ }\href {https://doi.org/10.1093/mnras/stab842} {\bibfield
  {journal} {\bibinfo  {journal} {\mnras}\ }\textbf {\bibinfo {volume} {504}},\
  \bibinfo {pages} {146} (\bibinfo {year} {2021})},\ \Eprint
  {https://arxiv.org/abs/2010.11730} {arXiv:2010.11730 [astro-ph.HE]}
  \BibitemShut {NoStop}%
\bibitem [{\citenamefont {{van Son}}\ \emph {et~al.}(2020)\citenamefont {{van
  Son}}, \citenamefont {{De Mink}}, \citenamefont {{Broekgaarden}},
  \citenamefont {{Renzo}}, \citenamefont {{Justham}}, \citenamefont
  {{Laplace}}, \citenamefont {{Mor{\'a}n-Fraile}}, \citenamefont {{Hendriks}},\
  and\ \citenamefont {{Farmer}}}]{vanSon_2020}%
  \BibitemOpen
  \bibfield  {author} {\bibinfo {author} {\bibfnamefont {L.~A.~C.}\
  \bibnamefont {{van Son}}}, \bibinfo {author} {\bibfnamefont {S.~E.}\
  \bibnamefont {{De Mink}}}, \bibinfo {author} {\bibfnamefont {F.~S.}\
  \bibnamefont {{Broekgaarden}}}, \bibinfo {author} {\bibfnamefont
  {M.}~\bibnamefont {{Renzo}}}, \bibinfo {author} {\bibfnamefont
  {S.}~\bibnamefont {{Justham}}}, \bibinfo {author} {\bibfnamefont
  {E.}~\bibnamefont {{Laplace}}}, \bibinfo {author} {\bibfnamefont
  {J.}~\bibnamefont {{Mor{\'a}n-Fraile}}}, \bibinfo {author} {\bibfnamefont
  {D.~D.}\ \bibnamefont {{Hendriks}}},\ and\ \bibinfo {author} {\bibfnamefont
  {R.}~\bibnamefont {{Farmer}}},\ }\href
  {https://doi.org/10.3847/1538-4357/ab9809} {\bibfield  {journal} {\bibinfo
  {journal} {\apj}\ }\textbf {\bibinfo {volume} {897}},\ \bibinfo {eid} {100}
  (\bibinfo {year} {2020})},\ \Eprint {https://arxiv.org/abs/2004.05187}
  {arXiv:2004.05187 [astro-ph.HE]} \BibitemShut {NoStop}%
\bibitem [{\citenamefont {{Banerjee}}(2018{\natexlab{b}})}]{Banerjee_2018}%
  \BibitemOpen
  \bibfield  {author} {\bibinfo {author} {\bibfnamefont {S.}~\bibnamefont
  {{Banerjee}}},\ }\href {https://doi.org/10.1093/mnras/sty2608} {\bibfield
  {journal} {\bibinfo  {journal} {\mnras}\ }\textbf {\bibinfo {volume} {481}},\
  \bibinfo {pages} {5123} (\bibinfo {year} {2018}{\natexlab{b}})},\ \Eprint
  {https://arxiv.org/abs/1805.06466} {arXiv:1805.06466 [astro-ph.HE]}
  \BibitemShut {NoStop}%
\bibitem [{\citenamefont {{Mar{\'\i}n Pina}}\ and\ \citenamefont
  {{Gieles}}(2024)}]{MarinPina_2024}%
  \BibitemOpen
  \bibfield  {author} {\bibinfo {author} {\bibfnamefont {D.}~\bibnamefont
  {{Mar{\'\i}n Pina}}}\ and\ \bibinfo {author} {\bibfnamefont {M.}~\bibnamefont
  {{Gieles}}},\ }\href {https://doi.org/10.1093/mnras/stad3777} {\bibfield
  {journal} {\bibinfo  {journal} {\mnras}\ }\textbf {\bibinfo {volume} {527}},\
  \bibinfo {pages} {8369} (\bibinfo {year} {2024})},\ \Eprint
  {https://arxiv.org/abs/2308.10318} {arXiv:2308.10318 [astro-ph.GA]}
  \BibitemShut {NoStop}%
\bibitem [{\citenamefont {{Peters}}(1964)}]{Peters_1964}%
  \BibitemOpen
  \bibfield  {author} {\bibinfo {author} {\bibfnamefont {P.~C.}\ \bibnamefont
  {{Peters}}},\ }\href {https://doi.org/10.1103/PhysRev.136.B1224} {\bibfield
  {journal} {\bibinfo  {journal} {Physical Review}\ }\textbf {\bibinfo {volume}
  {136}},\ \bibinfo {pages} {1224} (\bibinfo {year} {1964})}\BibitemShut
  {NoStop}%
\bibitem [{\citenamefont {Rodriguez}\ \emph {et~al.}(2018)\citenamefont
  {Rodriguez}, \citenamefont {Amaro-Seoane}, \citenamefont {Chatterjee},\ and\
  \citenamefont {Rasio}}]{Rodriguez_2018}%
  \BibitemOpen
  \bibfield  {author} {\bibinfo {author} {\bibfnamefont {C.~L.}\ \bibnamefont
  {Rodriguez}}, \bibinfo {author} {\bibfnamefont {P.}~\bibnamefont
  {Amaro-Seoane}}, \bibinfo {author} {\bibfnamefont {S.}~\bibnamefont
  {Chatterjee}},\ and\ \bibinfo {author} {\bibfnamefont {F.~A.}\ \bibnamefont
  {Rasio}},\ }\href {https://doi.org/10.1103/PhysRevLett.120.151101} {\bibfield
   {journal} {\bibinfo  {journal} {Phys. Rev. Lett.}\ }\textbf {\bibinfo
  {volume} {120}},\ \bibinfo {pages} {151101} (\bibinfo {year}
  {2018})}\BibitemShut {NoStop}%
\bibitem [{\citenamefont {Li}\ \emph {et~al.}(2024{\natexlab{a}})\citenamefont
  {Li}, \citenamefont {Wang}, \citenamefont {Tang},\ and\ \citenamefont
  {Fan}}]{Li_2024b}%
  \BibitemOpen
  \bibfield  {author} {\bibinfo {author} {\bibfnamefont {Y.-J.}\ \bibnamefont
  {Li}}, \bibinfo {author} {\bibfnamefont {Y.-Z.}\ \bibnamefont {Wang}},
  \bibinfo {author} {\bibfnamefont {S.-P.}\ \bibnamefont {Tang}},\ and\
  \bibinfo {author} {\bibfnamefont {Y.-Z.}\ \bibnamefont {Fan}},\ }\href
  {https://doi.org/10.1103/PhysRevLett.133.051401} {\bibfield  {journal}
  {\bibinfo  {journal} {Phys. Rev. Lett.}\ }\textbf {\bibinfo {volume} {133}},\
  \bibinfo {pages} {051401} (\bibinfo {year} {2024}{\natexlab{a}})},\ \Eprint
  {https://arxiv.org/abs/2303.02973} {arXiv:2303.02973 [astro-ph.HE]}
  \BibitemShut {NoStop}%
\bibitem [{\citenamefont {{Mai}}\ \emph {et~al.}(2025)\citenamefont {{Mai}},
  \citenamefont {{Kremer}},\ and\ \citenamefont {{Kiroglu}}}]{Mai_2025}%
  \BibitemOpen
  \bibfield  {author} {\bibinfo {author} {\bibfnamefont {A.}~\bibnamefont
  {{Mai}}}, \bibinfo {author} {\bibfnamefont {K.}~\bibnamefont {{Kremer}}},\
  and\ \bibinfo {author} {\bibfnamefont {F.}~\bibnamefont {{Kiroglu}}},\ }\href
  {https://doi.org/10.48550/arXiv.2510.21916} {\bibfield  {journal} {\bibinfo
  {journal} {arXiv e-prints}\ ,\ \bibinfo {eid} {arXiv:2510.21916}} (\bibinfo
  {year} {2025})},\ \Eprint {https://arxiv.org/abs/2510.21916}
  {arXiv:2510.21916 [astro-ph.GA]} \BibitemShut {NoStop}%
\bibitem [{\citenamefont {{Chattopadhyay}}\ \emph {et~al.}(2023)\citenamefont
  {{Chattopadhyay}}, \citenamefont {{Stegmann}}, \citenamefont {{Antonini}},
  \citenamefont {{Barber}},\ and\ \citenamefont {{Romero-Shaw}}}]{Chatto_2023}%
  \BibitemOpen
  \bibfield  {author} {\bibinfo {author} {\bibfnamefont {D.}~\bibnamefont
  {{Chattopadhyay}}}, \bibinfo {author} {\bibfnamefont {J.}~\bibnamefont
  {{Stegmann}}}, \bibinfo {author} {\bibfnamefont {F.}~\bibnamefont
  {{Antonini}}}, \bibinfo {author} {\bibfnamefont {J.}~\bibnamefont
  {{Barber}}},\ and\ \bibinfo {author} {\bibfnamefont {I.~M.}\ \bibnamefont
  {{Romero-Shaw}}},\ }\href {https://doi.org/10.1093/mnras/stad3048} {\bibfield
   {journal} {\bibinfo  {journal} {\mnras}\ }\textbf {\bibinfo {volume}
  {526}},\ \bibinfo {pages} {4908} (\bibinfo {year} {2023})},\ \Eprint
  {https://arxiv.org/abs/2308.10884} {arXiv:2308.10884 [astro-ph.HE]}
  \BibitemShut {NoStop}%
\bibitem [{\citenamefont {{Arca sedda}}\ \emph {et~al.}(2024)\citenamefont
  {{Arca sedda}}, \citenamefont {{Kamlah}}, \citenamefont {{Spurzem}},
  \citenamefont {{Rizzuto}}, \citenamefont {{Giersz}}, \citenamefont {{Naab}},\
  and\ \citenamefont {{Berczik}}}]{ArcaSedda_2024c}%
  \BibitemOpen
  \bibfield  {author} {\bibinfo {author} {\bibfnamefont {M.}~\bibnamefont
  {{Arca sedda}}}, \bibinfo {author} {\bibfnamefont {A.~W.~H.}\ \bibnamefont
  {{Kamlah}}}, \bibinfo {author} {\bibfnamefont {R.}~\bibnamefont {{Spurzem}}},
  \bibinfo {author} {\bibfnamefont {F.~P.}\ \bibnamefont {{Rizzuto}}}, \bibinfo
  {author} {\bibfnamefont {M.}~\bibnamefont {{Giersz}}}, \bibinfo {author}
  {\bibfnamefont {T.}~\bibnamefont {{Naab}}},\ and\ \bibinfo {author}
  {\bibfnamefont {P.}~\bibnamefont {{Berczik}}},\ }\href
  {https://doi.org/10.1093/mnras/stad3951} {\bibfield  {journal} {\bibinfo
  {journal} {\mnras}\ }\textbf {\bibinfo {volume} {528}},\ \bibinfo {pages}
  {5140} (\bibinfo {year} {2024})},\ \Eprint {https://arxiv.org/abs/2307.04807}
  {arXiv:2307.04807 [astro-ph.HE]} \BibitemShut {NoStop}%
\bibitem [{\citenamefont {Paiella}\ \emph {et~al.}(2025)\citenamefont
  {Paiella}, \citenamefont {Ugolini}, \citenamefont {Spera}, \citenamefont
  {Branchesi},\ and\ \citenamefont {Arca~Sedda}}]{Paiella_2025}%
  \BibitemOpen
  \bibfield  {author} {\bibinfo {author} {\bibfnamefont {L.}~\bibnamefont
  {Paiella}}, \bibinfo {author} {\bibfnamefont {C.}~\bibnamefont {Ugolini}},
  \bibinfo {author} {\bibfnamefont {M.}~\bibnamefont {Spera}}, \bibinfo
  {author} {\bibfnamefont {M.}~\bibnamefont {Branchesi}},\ and\ \bibinfo
  {author} {\bibfnamefont {M.}~\bibnamefont {Arca~Sedda}},\ }\href
  {https://doi.org/10.3847/2041-8213/ae1447} {\bibfield  {journal} {\bibinfo
  {journal} {\apjl}\ }\textbf {\bibinfo {volume} {994}},\ \bibinfo {pages}
  {L54} (\bibinfo {year} {2025})}\BibitemShut {NoStop}%
\bibitem [{\citenamefont {{Moody}}\ and\ \citenamefont
  {{Sigurdsson}}(2009)}]{Moody_2009}%
  \BibitemOpen
  \bibfield  {author} {\bibinfo {author} {\bibfnamefont {K.}~\bibnamefont
  {{Moody}}}\ and\ \bibinfo {author} {\bibfnamefont {S.}~\bibnamefont
  {{Sigurdsson}}},\ }\href {https://doi.org/10.1088/0004-637X/690/2/1370}
  {\bibfield  {journal} {\bibinfo  {journal} {\apj}\ }\textbf {\bibinfo
  {volume} {690}},\ \bibinfo {pages} {1370} (\bibinfo {year} {2009})},\ \Eprint
  {https://arxiv.org/abs/0809.1617} {arXiv:0809.1617 [astro-ph]} \BibitemShut
  {NoStop}%
\bibitem [{\citenamefont {{Chatterjee}}\ \emph {et~al.}(2017)\citenamefont
  {{Chatterjee}}, \citenamefont {{Rodriguez}}, \citenamefont {{Kalogera}},\
  and\ \citenamefont {{Rasio}}}]{Chatterjee_2017b}%
  \BibitemOpen
  \bibfield  {author} {\bibinfo {author} {\bibfnamefont {S.}~\bibnamefont
  {{Chatterjee}}}, \bibinfo {author} {\bibfnamefont {C.~L.}\ \bibnamefont
  {{Rodriguez}}}, \bibinfo {author} {\bibfnamefont {V.}~\bibnamefont
  {{Kalogera}}},\ and\ \bibinfo {author} {\bibfnamefont {F.~A.}\ \bibnamefont
  {{Rasio}}},\ }\href {https://doi.org/10.3847/2041-8213/aa5caa} {\bibfield
  {journal} {\bibinfo  {journal} {\apjl}\ }\textbf {\bibinfo {volume} {836}},\
  \bibinfo {eid} {L26} (\bibinfo {year} {2017})},\ \Eprint
  {https://arxiv.org/abs/1609.06689} {arXiv:1609.06689} \BibitemShut {NoStop}%
\bibitem [{\citenamefont {{Abac}}\ \emph {et~al.}(2025)\citenamefont {{Abac}},
  \citenamefont {{Abouelfettouh}}, \citenamefont {{Acernese}}, \citenamefont
  {{Ackley}}, \citenamefont {{Adamcewicz}}, \citenamefont {{Adhicary}},
  \citenamefont {{Adhikari}}, \citenamefont {{Adhikari}},\ and\ \citenamefont
  {et~al.}}]{Abac_2025b}%
  \BibitemOpen
  \bibfield  {author} {\bibinfo {author} {\bibfnamefont {A.~G.}\ \bibnamefont
  {{Abac}}}, \bibinfo {author} {\bibfnamefont {I.}~\bibnamefont
  {{Abouelfettouh}}}, \bibinfo {author} {\bibfnamefont {F.}~\bibnamefont
  {{Acernese}}}, \bibinfo {author} {\bibfnamefont {K.}~\bibnamefont
  {{Ackley}}}, \bibinfo {author} {\bibfnamefont {C.}~\bibnamefont
  {{Adamcewicz}}}, \bibinfo {author} {\bibfnamefont {S.}~\bibnamefont
  {{Adhicary}}}, \bibinfo {author} {\bibfnamefont {D.}~\bibnamefont
  {{Adhikari}}}, \bibinfo {author} {\bibfnamefont {N.}~\bibnamefont
  {{Adhikari}}},\ and\ \bibinfo {author} {\bibnamefont {et~al.}},\ }\href
  {https://doi.org/10.3847/2041-8213/ae0c9c} {\bibfield  {journal} {\bibinfo
  {journal} {\apjl}\ }\textbf {\bibinfo {volume} {993}},\ \bibinfo {eid} {L25}
  (\bibinfo {year} {2025})},\ \Eprint {https://arxiv.org/abs/2507.08219}
  {arXiv:2507.08219 [astro-ph.HE]} \BibitemShut {NoStop}%
\bibitem [{\citenamefont {{Borchers}}\ \emph {et~al.}(2025)\citenamefont
  {{Borchers}}, \citenamefont {{Ye}},\ and\ \citenamefont
  {{Fishbach}}}]{Borchers_2025}%
  \BibitemOpen
  \bibfield  {author} {\bibinfo {author} {\bibfnamefont {A.}~\bibnamefont
  {{Borchers}}}, \bibinfo {author} {\bibfnamefont {C.~S.}\ \bibnamefont
  {{Ye}}},\ and\ \bibinfo {author} {\bibfnamefont {M.}~\bibnamefont
  {{Fishbach}}},\ }\href {https://doi.org/10.3847/1538-4357/addec6} {\bibfield
  {journal} {\bibinfo  {journal} {\apj}\ }\textbf {\bibinfo {volume} {987}},\
  \bibinfo {eid} {146} (\bibinfo {year} {2025})},\ \Eprint
  {https://arxiv.org/abs/2503.21278} {arXiv:2503.21278 [astro-ph.HE]}
  \BibitemShut {NoStop}%
\bibitem [{\citenamefont {{K{\i}ro{\u{g}}lu}}\ \emph
  {et~al.}(2025{\natexlab{a}})\citenamefont {{K{\i}ro{\u{g}}lu}}, \citenamefont
  {{Kremer}},\ and\ \citenamefont {{Rasio}}}]{Kiroglu_2025a}%
  \BibitemOpen
  \bibfield  {author} {\bibinfo {author} {\bibfnamefont {F.}~\bibnamefont
  {{K{\i}ro{\u{g}}lu}}}, \bibinfo {author} {\bibfnamefont {K.}~\bibnamefont
  {{Kremer}}},\ and\ \bibinfo {author} {\bibfnamefont {F.~A.}\ \bibnamefont
  {{Rasio}}},\ }\href {https://doi.org/10.48550/arXiv.2509.05415} {\bibfield
  {journal} {\bibinfo  {journal} {arXiv e-prints}\ ,\ \bibinfo {eid}
  {arXiv:2509.05415}} (\bibinfo {year} {2025}{\natexlab{a}})},\ \Eprint
  {https://arxiv.org/abs/2509.05415} {arXiv:2509.05415 [astro-ph.HE]}
  \BibitemShut {NoStop}%
\bibitem [{\citenamefont {{Stegmann}}\ \emph {et~al.}(2025)\citenamefont
  {{Stegmann}}, \citenamefont {{Olejak}},\ and\ \citenamefont {{de
  Mink}}}]{Stegmann_2025}%
  \BibitemOpen
  \bibfield  {author} {\bibinfo {author} {\bibfnamefont {J.}~\bibnamefont
  {{Stegmann}}}, \bibinfo {author} {\bibfnamefont {A.}~\bibnamefont
  {{Olejak}}},\ and\ \bibinfo {author} {\bibfnamefont {S.~E.}\ \bibnamefont
  {{de Mink}}},\ }\href {https://doi.org/10.3847/2041-8213/ae0e5f} {\bibfield
  {journal} {\bibinfo  {journal} {\apjl}\ }\textbf {\bibinfo {volume} {992}},\
  \bibinfo {eid} {L26} (\bibinfo {year} {2025})},\ \Eprint
  {https://arxiv.org/abs/2507.15967} {arXiv:2507.15967 [astro-ph.HE]}
  \BibitemShut {NoStop}%
\bibitem [{\citenamefont {{The LIGO Scientific Collaboration}}\ \emph
  {et~al.}(2025{\natexlab{c}})\citenamefont {{The LIGO Scientific
  Collaboration}}, \citenamefont {{the Virgo Collaboration}}, \citenamefont
  {{the KAGRA Collaboration}}, \citenamefont {{Abac}}, \citenamefont
  {{Abouelfettouh}}, \citenamefont {{Acernese}}, \citenamefont {{Ackley}},
  \citenamefont {{Adamcewicz}}, \citenamefont {{Adhicary}}, \citenamefont
  {{Adhikari}}, \citenamefont {{Adhikari}},\ and\ \citenamefont
  {et~al.}}]{Abac_2025a}%
  \BibitemOpen
  \bibfield  {author} {\bibinfo {author} {\bibnamefont {{The LIGO Scientific
  Collaboration}}}, \bibinfo {author} {\bibnamefont {{the Virgo
  Collaboration}}}, \bibinfo {author} {\bibnamefont {{the KAGRA
  Collaboration}}}, \bibinfo {author} {\bibfnamefont {A.~G.}\ \bibnamefont
  {{Abac}}}, \bibinfo {author} {\bibfnamefont {I.}~\bibnamefont
  {{Abouelfettouh}}}, \bibinfo {author} {\bibfnamefont {F.}~\bibnamefont
  {{Acernese}}}, \bibinfo {author} {\bibfnamefont {K.}~\bibnamefont
  {{Ackley}}}, \bibinfo {author} {\bibfnamefont {C.}~\bibnamefont
  {{Adamcewicz}}}, \bibinfo {author} {\bibfnamefont {S.}~\bibnamefont
  {{Adhicary}}}, \bibinfo {author} {\bibfnamefont {D.}~\bibnamefont
  {{Adhikari}}}, \bibinfo {author} {\bibfnamefont {N.}~\bibnamefont
  {{Adhikari}}},\ and\ \bibinfo {author} {\bibnamefont {et~al.}},\ }\href
  {https://doi.org/10.3847/2041-8213/ae0d54} {\bibfield  {journal} {\bibinfo
  {journal} {The Astrophysical Journal Letters}\ }\textbf {\bibinfo {volume}
  {993}},\ \bibinfo {pages} {L21} (\bibinfo {year}
  {2025}{\natexlab{c}})}\BibitemShut {NoStop}%
\bibitem [{\citenamefont {Ajith}\ \emph {et~al.}(2011)\citenamefont {Ajith},
  \citenamefont {Hannam}, \citenamefont {Husa}, \citenamefont {Chen},
  \citenamefont {Br\"ugmann}, \citenamefont {Dorband}, \citenamefont
  {M\"uller}, \citenamefont {Ohme}, \citenamefont {Pollney}, \citenamefont
  {Reisswig}, \citenamefont {Santamar\'{\i}a},\ and\ \citenamefont
  {Seiler}}]{Ajith_2011}%
  \BibitemOpen
  \bibfield  {author} {\bibinfo {author} {\bibfnamefont {P.}~\bibnamefont
  {Ajith}}, \bibinfo {author} {\bibfnamefont {M.}~\bibnamefont {Hannam}},
  \bibinfo {author} {\bibfnamefont {S.}~\bibnamefont {Husa}}, \bibinfo {author}
  {\bibfnamefont {Y.}~\bibnamefont {Chen}}, \bibinfo {author} {\bibfnamefont
  {B.}~\bibnamefont {Br\"ugmann}}, \bibinfo {author} {\bibfnamefont
  {N.}~\bibnamefont {Dorband}}, \bibinfo {author} {\bibfnamefont
  {D.}~\bibnamefont {M\"uller}}, \bibinfo {author} {\bibfnamefont
  {F.}~\bibnamefont {Ohme}}, \bibinfo {author} {\bibfnamefont {D.}~\bibnamefont
  {Pollney}}, \bibinfo {author} {\bibfnamefont {C.}~\bibnamefont {Reisswig}},
  \bibinfo {author} {\bibfnamefont {L.}~\bibnamefont {Santamar\'{\i}a}},\ and\
  \bibinfo {author} {\bibfnamefont {J.}~\bibnamefont {Seiler}},\ }\href
  {https://doi.org/10.1103/PhysRevLett.106.241101} {\bibfield  {journal}
  {\bibinfo  {journal} {Phys. Rev. Lett.}\ }\textbf {\bibinfo {volume} {106}},\
  \bibinfo {pages} {241101} (\bibinfo {year} {2011})}\BibitemShut {NoStop}%
\bibitem [{\citenamefont {{Yu}}\ \emph {et~al.}(2020)\citenamefont {{Yu}},
  \citenamefont {{Ma}}, \citenamefont {{Giesler}},\ and\ \citenamefont
  {{Chen}}}]{Yu_2020}%
  \BibitemOpen
  \bibfield  {author} {\bibinfo {author} {\bibfnamefont {H.}~\bibnamefont
  {{Yu}}}, \bibinfo {author} {\bibfnamefont {S.}~\bibnamefont {{Ma}}}, \bibinfo
  {author} {\bibfnamefont {M.}~\bibnamefont {{Giesler}}},\ and\ \bibinfo
  {author} {\bibfnamefont {Y.}~\bibnamefont {{Chen}}},\ }\href
  {https://doi.org/10.1103/PhysRevD.102.123009} {\bibfield  {journal} {\bibinfo
   {journal} {\prd}\ }\textbf {\bibinfo {volume} {102}},\ \bibinfo {eid}
  {123009} (\bibinfo {year} {2020})},\ \Eprint
  {https://arxiv.org/abs/2007.12978} {arXiv:2007.12978 [gr-qc]} \BibitemShut
  {NoStop}%
\bibitem [{\citenamefont {Gerosa}\ \emph {et~al.}(2021)\citenamefont {Gerosa},
  \citenamefont {Mould}, \citenamefont {Gangardt}, \citenamefont {Schmidt},
  \citenamefont {Pratten},\ and\ \citenamefont {Thomas}}]{Gerosa_2021b}%
  \BibitemOpen
  \bibfield  {author} {\bibinfo {author} {\bibfnamefont {D.}~\bibnamefont
  {Gerosa}}, \bibinfo {author} {\bibfnamefont {M.}~\bibnamefont {Mould}},
  \bibinfo {author} {\bibfnamefont {D.}~\bibnamefont {Gangardt}}, \bibinfo
  {author} {\bibfnamefont {P.}~\bibnamefont {Schmidt}}, \bibinfo {author}
  {\bibfnamefont {G.}~\bibnamefont {Pratten}},\ and\ \bibinfo {author}
  {\bibfnamefont {L.~M.}\ \bibnamefont {Thomas}},\ }\href
  {https://doi.org/10.1103/PhysRevD.103.064067} {\bibfield  {journal} {\bibinfo
   {journal} {Phys. Rev. D}\ }\textbf {\bibinfo {volume} {103}},\ \bibinfo
  {pages} {064067} (\bibinfo {year} {2021})}\BibitemShut {NoStop}%
\bibitem [{\citenamefont {{Banerjee}}\ \emph {et~al.}(2023)\citenamefont
  {{Banerjee}}, \citenamefont {{Olejak}},\ and\ \citenamefont
  {{Belczynski}}}]{Banerjee_2023}%
  \BibitemOpen
  \bibfield  {author} {\bibinfo {author} {\bibfnamefont {S.}~\bibnamefont
  {{Banerjee}}}, \bibinfo {author} {\bibfnamefont {A.}~\bibnamefont
  {{Olejak}}},\ and\ \bibinfo {author} {\bibfnamefont {K.}~\bibnamefont
  {{Belczynski}}},\ }\href {https://doi.org/10.3847/1538-4357/acdd59}
  {\bibfield  {journal} {\bibinfo  {journal} {\apj}\ }\textbf {\bibinfo
  {volume} {953}},\ \bibinfo {eid} {80} (\bibinfo {year} {2023})},\ \Eprint
  {https://arxiv.org/abs/2302.10851} {arXiv:2302.10851 [astro-ph.HE]}
  \BibitemShut {NoStop}%
\bibitem [{\citenamefont {{Antonini}}\ and\ \citenamefont
  {{Gieles}}(2020{\natexlab{b}})}]{Antonini_2020b}%
  \BibitemOpen
  \bibfield  {author} {\bibinfo {author} {\bibfnamefont {F.}~\bibnamefont
  {{Antonini}}}\ and\ \bibinfo {author} {\bibfnamefont {M.}~\bibnamefont
  {{Gieles}}},\ }\href {https://doi.org/10.1103/PhysRevD.102.123016} {\bibfield
   {journal} {\bibinfo  {journal} {\prd}\ }\textbf {\bibinfo {volume} {102}},\
  \bibinfo {eid} {123016} (\bibinfo {year} {2020}{\natexlab{b}})},\ \Eprint
  {https://arxiv.org/abs/2009.01861} {arXiv:2009.01861 [astro-ph.HE]}
  \BibitemShut {NoStop}%
\bibitem [{\citenamefont {{Banerjee}}(2022{\natexlab{b}})}]{Banerjee_2021b}%
  \BibitemOpen
  \bibfield  {author} {\bibinfo {author} {\bibfnamefont {S.}~\bibnamefont
  {{Banerjee}}},\ }\href {https://doi.org/10.1103/PhysRevD.105.023004}
  {\bibfield  {journal} {\bibinfo  {journal} {\prd}\ }\textbf {\bibinfo
  {volume} {105}},\ \bibinfo {eid} {023004} (\bibinfo {year}
  {2022}{\natexlab{b}})},\ \Eprint {https://arxiv.org/abs/2108.04250}
  {arXiv:2108.04250 [astro-ph.HE]} \BibitemShut {NoStop}%
\bibitem [{\citenamefont {{Samsing}}\ and\ \citenamefont
  {{Hotokezaka}}(2021)}]{Samsing_2021}%
  \BibitemOpen
  \bibfield  {author} {\bibinfo {author} {\bibfnamefont {J.}~\bibnamefont
  {{Samsing}}}\ and\ \bibinfo {author} {\bibfnamefont {K.}~\bibnamefont
  {{Hotokezaka}}},\ }\href {https://doi.org/10.3847/1538-4357/ac2b27}
  {\bibfield  {journal} {\bibinfo  {journal} {\apj}\ }\textbf {\bibinfo
  {volume} {923}},\ \bibinfo {eid} {126} (\bibinfo {year} {2021})},\ \Eprint
  {https://arxiv.org/abs/2006.09744} {arXiv:2006.09744 [astro-ph.HE]}
  \BibitemShut {NoStop}%
\bibitem [{\citenamefont {Li}\ \emph {et~al.}(2024{\natexlab{b}})\citenamefont
  {Li}, \citenamefont {Tang}, \citenamefont {Gao}, \citenamefont {Wu},\ and\
  \citenamefont {Wang}}]{Li_2024a}%
  \BibitemOpen
  \bibfield  {author} {\bibinfo {author} {\bibfnamefont {Y.-J.}\ \bibnamefont
  {Li}}, \bibinfo {author} {\bibfnamefont {S.-P.}\ \bibnamefont {Tang}},
  \bibinfo {author} {\bibfnamefont {S.-J.}\ \bibnamefont {Gao}}, \bibinfo
  {author} {\bibfnamefont {D.-C.}\ \bibnamefont {Wu}},\ and\ \bibinfo {author}
  {\bibfnamefont {Y.-Z.}\ \bibnamefont {Wang}},\ }\href
  {https://doi.org/10.3847/1538-4357/ad83b5} {\bibfield  {journal} {\bibinfo
  {journal} {Astrophys. J.}\ }\textbf {\bibinfo {volume} {977}},\ \bibinfo
  {pages} {67} (\bibinfo {year} {2024}{\natexlab{b}})},\ \Eprint
  {https://arxiv.org/abs/2404.09668} {arXiv:2404.09668 [astro-ph.HE]}
  \BibitemShut {NoStop}%
\bibitem [{\citenamefont {{Bailes}}\ \emph {et~al.}(2021)\citenamefont
  {{Bailes}}, \citenamefont {{Berger}}, \citenamefont {{Brady}}, \citenamefont
  {{Branchesi}}, \citenamefont {{Danzmann}}, \citenamefont {{Evans}},
  \citenamefont {{Holley-Bockelmann}}, \citenamefont {{Iyer}},\ and\
  \citenamefont {et.al.}}]{Bailes_2021}%
  \BibitemOpen
  \bibfield  {author} {\bibinfo {author} {\bibfnamefont {M.}~\bibnamefont
  {{Bailes}}}, \bibinfo {author} {\bibfnamefont {B.~K.}\ \bibnamefont
  {{Berger}}}, \bibinfo {author} {\bibfnamefont {P.~R.}\ \bibnamefont
  {{Brady}}}, \bibinfo {author} {\bibfnamefont {M.}~\bibnamefont
  {{Branchesi}}}, \bibinfo {author} {\bibfnamefont {K.}~\bibnamefont
  {{Danzmann}}}, \bibinfo {author} {\bibfnamefont {M.}~\bibnamefont {{Evans}}},
  \bibinfo {author} {\bibfnamefont {K.}~\bibnamefont {{Holley-Bockelmann}}},
  \bibinfo {author} {\bibfnamefont {B.~R.}\ \bibnamefont {{Iyer}}},\ and\
  \bibinfo {author} {\bibnamefont {et.al.}},\ }\href
  {https://doi.org/10.1038/s42254-021-00303-8} {\bibfield  {journal} {\bibinfo
  {journal} {Nature Reviews Physics}\ }\textbf {\bibinfo {volume} {3}},\
  \bibinfo {pages} {344} (\bibinfo {year} {2021})}\BibitemShut {NoStop}%
\bibitem [{\citenamefont {{Madau}}\ and\ \citenamefont
  {{Fragos}}(2017)}]{Madau_2017}%
  \BibitemOpen
  \bibfield  {author} {\bibinfo {author} {\bibfnamefont {P.}~\bibnamefont
  {{Madau}}}\ and\ \bibinfo {author} {\bibfnamefont {T.}~\bibnamefont
  {{Fragos}}},\ }\href {https://doi.org/10.3847/1538-4357/aa6af9} {\bibfield
  {journal} {\bibinfo  {journal} {\apj}\ }\textbf {\bibinfo {volume} {840}},\
  \bibinfo {eid} {39} (\bibinfo {year} {2017})},\ \Eprint
  {https://arxiv.org/abs/1606.07887} {arXiv:1606.07887 [astro-ph.GA]}
  \BibitemShut {NoStop}%
\bibitem [{\citenamefont {{Kotko}}\ \emph {et~al.}(2024)\citenamefont
  {{Kotko}}, \citenamefont {{Banerjee}},\ and\ \citenamefont
  {{Belczynski}}}]{Kotko_2024}%
  \BibitemOpen
  \bibfield  {author} {\bibinfo {author} {\bibfnamefont {I.}~\bibnamefont
  {{Kotko}}}, \bibinfo {author} {\bibfnamefont {S.}~\bibnamefont
  {{Banerjee}}},\ and\ \bibinfo {author} {\bibfnamefont {K.}~\bibnamefont
  {{Belczynski}}},\ }\href {https://doi.org/10.1093/mnras/stae2591} {\bibfield
  {journal} {\bibinfo  {journal} {\mnras}\ }\textbf {\bibinfo {volume} {535}},\
  \bibinfo {pages} {3577} (\bibinfo {year} {2024})},\ \Eprint
  {https://arxiv.org/abs/2403.13579} {arXiv:2403.13579 [astro-ph.SR]}
  \BibitemShut {NoStop}%
\bibitem [{\citenamefont {{Chakrabarti}}\ \emph {et~al.}(2023)\citenamefont
  {{Chakrabarti}}, \citenamefont {{Simon}}, \citenamefont {{Craig}},
  \citenamefont {{Reggiani}}, \citenamefont {{Brandt}}, \citenamefont
  {{Guhathakurta}}, \citenamefont {{Dalba}}, \citenamefont {{Kirby}},
  \citenamefont {{Chang}}, \citenamefont {{Hey}}, \citenamefont {{Savino}},
  \citenamefont {{Geha}},\ and\ \citenamefont {{Thompson}}}]{Chakrabarti_2023}%
  \BibitemOpen
  \bibfield  {author} {\bibinfo {author} {\bibfnamefont {S.}~\bibnamefont
  {{Chakrabarti}}}, \bibinfo {author} {\bibfnamefont {J.~D.}\ \bibnamefont
  {{Simon}}}, \bibinfo {author} {\bibfnamefont {P.~A.}\ \bibnamefont
  {{Craig}}}, \bibinfo {author} {\bibfnamefont {H.}~\bibnamefont {{Reggiani}}},
  \bibinfo {author} {\bibfnamefont {T.~D.}\ \bibnamefont {{Brandt}}}, \bibinfo
  {author} {\bibfnamefont {P.}~\bibnamefont {{Guhathakurta}}}, \bibinfo
  {author} {\bibfnamefont {P.~A.}\ \bibnamefont {{Dalba}}}, \bibinfo {author}
  {\bibfnamefont {E.~N.}\ \bibnamefont {{Kirby}}}, \bibinfo {author}
  {\bibfnamefont {P.}~\bibnamefont {{Chang}}}, \bibinfo {author} {\bibfnamefont
  {D.~R.}\ \bibnamefont {{Hey}}}, \bibinfo {author} {\bibfnamefont
  {A.}~\bibnamefont {{Savino}}}, \bibinfo {author} {\bibfnamefont
  {M.}~\bibnamefont {{Geha}}},\ and\ \bibinfo {author} {\bibfnamefont {I.~B.}\
  \bibnamefont {{Thompson}}},\ }\href
  {https://doi.org/10.3847/1538-3881/accf21} {\bibfield  {journal} {\bibinfo
  {journal} {\aj}\ }\textbf {\bibinfo {volume} {166}},\ \bibinfo {eid} {6}
  (\bibinfo {year} {2023})},\ \Eprint {https://arxiv.org/abs/2210.05003}
  {arXiv:2210.05003 [astro-ph.GA]} \BibitemShut {NoStop}%
\bibitem [{\citenamefont {{El-Badry}}\ \emph {et~al.}(2023)\citenamefont
  {{El-Badry}}, \citenamefont {{Rix}}, \citenamefont {{Quataert}},
  \citenamefont {{Howard}}, \citenamefont {{Isaacson}}, \citenamefont
  {{Fuller}}, \citenamefont {{Hawkins}}, \citenamefont {{Breivik}},
  \citenamefont {{Wong}}, \citenamefont {{Rodriguez}}, \citenamefont
  {{Conroy}}, \citenamefont {{Shahaf}}, \citenamefont {{Mazeh}}, \citenamefont
  {{Arenou}}, \citenamefont {{Burdge}}, \citenamefont {{Bashi}}, \citenamefont
  {{Faigler}}, \citenamefont {{Weisz}}, \citenamefont {{Seeburger}},
  \citenamefont {{Almada Monter}},\ and\ \citenamefont
  {{Wojno}}}]{ElBadry_2023}%
  \BibitemOpen
  \bibfield  {author} {\bibinfo {author} {\bibfnamefont {K.}~\bibnamefont
  {{El-Badry}}}, \bibinfo {author} {\bibfnamefont {H.-W.}\ \bibnamefont
  {{Rix}}}, \bibinfo {author} {\bibfnamefont {E.}~\bibnamefont {{Quataert}}},
  \bibinfo {author} {\bibfnamefont {A.~W.}\ \bibnamefont {{Howard}}}, \bibinfo
  {author} {\bibfnamefont {H.}~\bibnamefont {{Isaacson}}}, \bibinfo {author}
  {\bibfnamefont {J.}~\bibnamefont {{Fuller}}}, \bibinfo {author}
  {\bibfnamefont {K.}~\bibnamefont {{Hawkins}}}, \bibinfo {author}
  {\bibfnamefont {K.}~\bibnamefont {{Breivik}}}, \bibinfo {author}
  {\bibfnamefont {K.~W.~K.}\ \bibnamefont {{Wong}}}, \bibinfo {author}
  {\bibfnamefont {A.~C.}\ \bibnamefont {{Rodriguez}}}, \bibinfo {author}
  {\bibfnamefont {C.}~\bibnamefont {{Conroy}}}, \bibinfo {author}
  {\bibfnamefont {S.}~\bibnamefont {{Shahaf}}}, \bibinfo {author}
  {\bibfnamefont {T.}~\bibnamefont {{Mazeh}}}, \bibinfo {author} {\bibfnamefont
  {F.}~\bibnamefont {{Arenou}}}, \bibinfo {author} {\bibfnamefont {K.~B.}\
  \bibnamefont {{Burdge}}}, \bibinfo {author} {\bibfnamefont {D.}~\bibnamefont
  {{Bashi}}}, \bibinfo {author} {\bibfnamefont {S.}~\bibnamefont {{Faigler}}},
  \bibinfo {author} {\bibfnamefont {D.~R.}\ \bibnamefont {{Weisz}}}, \bibinfo
  {author} {\bibfnamefont {R.}~\bibnamefont {{Seeburger}}}, \bibinfo {author}
  {\bibfnamefont {S.}~\bibnamefont {{Almada Monter}}},\ and\ \bibinfo {author}
  {\bibfnamefont {J.}~\bibnamefont {{Wojno}}},\ }\href
  {https://doi.org/10.1093/mnras/stac3140} {\bibfield  {journal} {\bibinfo
  {journal} {\mnras}\ }\textbf {\bibinfo {volume} {518}},\ \bibinfo {pages}
  {1057} (\bibinfo {year} {2023})},\ \Eprint {https://arxiv.org/abs/2209.06833}
  {arXiv:2209.06833 [astro-ph.SR]} \BibitemShut {NoStop}%
\bibitem [{\citenamefont {{Gaia Collaboration}}\ \emph
  {et~al.}(2024)\citenamefont {{Gaia Collaboration}}, \citenamefont
  {{Panuzzo}}, \citenamefont {{Mazeh}}, \citenamefont {{Arenou}}, \citenamefont
  {{Holl}}, \citenamefont {{Caffau}}, \citenamefont {{Jorissen}}, \citenamefont
  {{Babusiaux}}, \citenamefont {{Gavras}},\ and\ \citenamefont
  {et~al.}}]{GaiaBH3_2024}%
  \BibitemOpen
  \bibfield  {author} {\bibinfo {author} {\bibnamefont {{Gaia Collaboration}}},
  \bibinfo {author} {\bibfnamefont {P.}~\bibnamefont {{Panuzzo}}}, \bibinfo
  {author} {\bibfnamefont {T.}~\bibnamefont {{Mazeh}}}, \bibinfo {author}
  {\bibfnamefont {F.}~\bibnamefont {{Arenou}}}, \bibinfo {author}
  {\bibfnamefont {B.}~\bibnamefont {{Holl}}}, \bibinfo {author} {\bibfnamefont
  {E.}~\bibnamefont {{Caffau}}}, \bibinfo {author} {\bibfnamefont
  {A.}~\bibnamefont {{Jorissen}}}, \bibinfo {author} {\bibfnamefont
  {C.}~\bibnamefont {{Babusiaux}}}, \bibinfo {author} {\bibfnamefont
  {P.}~\bibnamefont {{Gavras}}},\ and\ \bibinfo {author} {\bibnamefont
  {et~al.}},\ }\href {https://doi.org/10.1051/0004-6361/202449763} {\bibfield
  {journal} {\bibinfo  {journal} {\aap}\ }\textbf {\bibinfo {volume} {686}},\
  \bibinfo {eid} {L2} (\bibinfo {year} {2024})},\ \Eprint
  {https://arxiv.org/abs/2404.10486} {arXiv:2404.10486 [astro-ph.GA]}
  \BibitemShut {NoStop}%
\bibitem [{\citenamefont {{El-Badry}}\ \emph {et~al.}(2024)\citenamefont
  {{El-Badry}}, \citenamefont {{Rix}}, \citenamefont {{Latham}}, \citenamefont
  {{Shahaf}}, \citenamefont {{Mazeh}}, \citenamefont {{Bieryla}}, \citenamefont
  {{Buchhave}}, \citenamefont {{Andrae}}, \citenamefont {{Yamaguchi}},
  \citenamefont {{Isaacson}}, \citenamefont {{Howard}}, \citenamefont
  {{Savino}},\ and\ \citenamefont {{Ilyin}}}]{ElBadry_2024}%
  \BibitemOpen
  \bibfield  {author} {\bibinfo {author} {\bibfnamefont {K.}~\bibnamefont
  {{El-Badry}}}, \bibinfo {author} {\bibfnamefont {H.-W.}\ \bibnamefont
  {{Rix}}}, \bibinfo {author} {\bibfnamefont {D.~W.}\ \bibnamefont {{Latham}}},
  \bibinfo {author} {\bibfnamefont {S.}~\bibnamefont {{Shahaf}}}, \bibinfo
  {author} {\bibfnamefont {T.}~\bibnamefont {{Mazeh}}}, \bibinfo {author}
  {\bibfnamefont {A.}~\bibnamefont {{Bieryla}}}, \bibinfo {author}
  {\bibfnamefont {L.~A.}\ \bibnamefont {{Buchhave}}}, \bibinfo {author}
  {\bibfnamefont {R.}~\bibnamefont {{Andrae}}}, \bibinfo {author}
  {\bibfnamefont {N.}~\bibnamefont {{Yamaguchi}}}, \bibinfo {author}
  {\bibfnamefont {H.}~\bibnamefont {{Isaacson}}}, \bibinfo {author}
  {\bibfnamefont {A.~W.}\ \bibnamefont {{Howard}}}, \bibinfo {author}
  {\bibfnamefont {A.}~\bibnamefont {{Savino}}},\ and\ \bibinfo {author}
  {\bibfnamefont {I.~V.}\ \bibnamefont {{Ilyin}}},\ }\href
  {https://doi.org/10.33232/001c.121261} {\bibfield  {journal} {\bibinfo
  {journal} {The Open Journal of Astrophysics}\ }\textbf {\bibinfo {volume}
  {7}},\ \bibinfo {eid} {58} (\bibinfo {year} {2024})},\ \Eprint
  {https://arxiv.org/abs/2405.00089} {arXiv:2405.00089 [astro-ph.SR]}
  \BibitemShut {NoStop}%
\bibitem [{\citenamefont {{Schiebelbein-Zwack}}\ \emph
  {et~al.}(2025)\citenamefont {{Schiebelbein-Zwack}}, \citenamefont {{van
  Son}}, \citenamefont {{Fishbach}},\ and\ \citenamefont
  {{Farr}}}]{Schiebelbein_2025}%
  \BibitemOpen
  \bibfield  {author} {\bibinfo {author} {\bibfnamefont {A.}~\bibnamefont
  {{Schiebelbein-Zwack}}}, \bibinfo {author} {\bibfnamefont {L.~A.~C.}\
  \bibnamefont {{van Son}}}, \bibinfo {author} {\bibfnamefont {M.}~\bibnamefont
  {{Fishbach}}},\ and\ \bibinfo {author} {\bibfnamefont {W.~M.}\ \bibnamefont
  {{Farr}}},\ }\href {https://doi.org/10.48550/arXiv.2511.07393} {\bibfield
  {journal} {\bibinfo  {journal} {arXiv e-prints}\ ,\ \bibinfo {eid}
  {arXiv:2511.07393}} (\bibinfo {year} {2025})},\ \Eprint
  {https://arxiv.org/abs/2511.07393} {arXiv:2511.07393 [astro-ph.SR]}
  \BibitemShut {NoStop}%
\bibitem [{\citenamefont {{Tanikawa}}\ \emph {et~al.}(2024)\citenamefont
  {{Tanikawa}}, \citenamefont {{Wang}},\ and\ \citenamefont
  {{Fujii}}}]{Tanikawa_2024b}%
  \BibitemOpen
  \bibfield  {author} {\bibinfo {author} {\bibfnamefont {A.}~\bibnamefont
  {{Tanikawa}}}, \bibinfo {author} {\bibfnamefont {L.}~\bibnamefont {{Wang}}},\
  and\ \bibinfo {author} {\bibfnamefont {M.~S.}\ \bibnamefont {{Fujii}}},\
  }\href {https://doi.org/10.33232/001c.117684} {\bibfield  {journal} {\bibinfo
   {journal} {The Open Journal of Astrophysics}\ }\textbf {\bibinfo {volume}
  {7}},\ \bibinfo {eid} {39} (\bibinfo {year} {2024})},\ \Eprint
  {https://arxiv.org/abs/2404.01731} {arXiv:2404.01731 [astro-ph.SR]}
  \BibitemShut {NoStop}%
\bibitem [{\citenamefont {{Andr{\'e}}}\ \emph {et~al.}(2014)\citenamefont
  {{Andr{\'e}}}, \citenamefont {{Di Francesco}}, \citenamefont
  {{Ward-Thompson}}, \citenamefont {{Inutsuka}}, \citenamefont {{Pudritz}},\
  and\ \citenamefont {{Pineda}}}]{Andre_2014}%
  \BibitemOpen
  \bibfield  {author} {\bibinfo {author} {\bibfnamefont {P.}~\bibnamefont
  {{Andr{\'e}}}}, \bibinfo {author} {\bibfnamefont {J.}~\bibnamefont {{Di
  Francesco}}}, \bibinfo {author} {\bibfnamefont {D.}~\bibnamefont
  {{Ward-Thompson}}}, \bibinfo {author} {\bibfnamefont {S.~I.}\ \bibnamefont
  {{Inutsuka}}}, \bibinfo {author} {\bibfnamefont {R.~E.}\ \bibnamefont
  {{Pudritz}}},\ and\ \bibinfo {author} {\bibfnamefont {J.~E.}\ \bibnamefont
  {{Pineda}}},\ }in\ \href
  {https://doi.org/10.2458/azu_uapress_9780816531240-ch002} {\emph {\bibinfo
  {booktitle} {Protostars and Planets VI}}},\ \bibinfo {editor} {edited by\
  \bibinfo {editor} {\bibfnamefont {H.}~\bibnamefont {{Beuther}}}, \bibinfo
  {editor} {\bibfnamefont {R.~S.}\ \bibnamefont {{Klessen}}}, \bibinfo {editor}
  {\bibfnamefont {C.~P.}\ \bibnamefont {{Dullemond}}},\ and\ \bibinfo {editor}
  {\bibfnamefont {T.}~\bibnamefont {{Henning}}}}\ (\bibinfo {year} {2014})\
  pp.\ \bibinfo {pages} {27--51},\ \Eprint {https://arxiv.org/abs/1312.6232}
  {arXiv:1312.6232 [astro-ph.GA]} \BibitemShut {NoStop}%
\bibitem [{\citenamefont {{Longmore}}\ \emph {et~al.}(2014)\citenamefont
  {{Longmore}}, \citenamefont {{Kruijssen}}, \citenamefont {{Bastian}},
  \citenamefont {{Bally}}, \citenamefont {{Rathborne}}, \citenamefont
  {{Testi}}, \citenamefont {{Stolte}}, \citenamefont {{Dale}}, \citenamefont
  {{Bressert}},\ and\ \citenamefont {{Alves}}}]{Longmore_2014}%
  \BibitemOpen
  \bibfield  {author} {\bibinfo {author} {\bibfnamefont {S.~N.}\ \bibnamefont
  {{Longmore}}}, \bibinfo {author} {\bibfnamefont {J.~M.~D.}\ \bibnamefont
  {{Kruijssen}}}, \bibinfo {author} {\bibfnamefont {N.}~\bibnamefont
  {{Bastian}}}, \bibinfo {author} {\bibfnamefont {J.}~\bibnamefont {{Bally}}},
  \bibinfo {author} {\bibfnamefont {J.}~\bibnamefont {{Rathborne}}}, \bibinfo
  {author} {\bibfnamefont {L.}~\bibnamefont {{Testi}}}, \bibinfo {author}
  {\bibfnamefont {A.}~\bibnamefont {{Stolte}}}, \bibinfo {author}
  {\bibfnamefont {J.}~\bibnamefont {{Dale}}}, \bibinfo {author} {\bibfnamefont
  {E.}~\bibnamefont {{Bressert}}},\ and\ \bibinfo {author} {\bibfnamefont
  {J.}~\bibnamefont {{Alves}}},\ }\href
  {https://doi.org/10.2458/azu_uapress_9780816531240-ch013} {\bibfield
  {journal} {\bibinfo  {journal} {Protostars and Planets VI}\ ,\ \bibinfo
  {pages} {291}} (\bibinfo {year} {2014})},\ \Eprint
  {https://arxiv.org/abs/1401.4175} {arXiv:1401.4175} \BibitemShut {NoStop}%
\bibitem [{\citenamefont {{The LIGO Scientific Collaboration}}\ \emph
  {et~al.}(2026{\natexlab{b}})\citenamefont {{The LIGO Scientific
  Collaboration}}, \citenamefont {{the Virgo Collaboration}}, \citenamefont
  {{the KAGRA Collaboration}}, \citenamefont {{Abac}}, \citenamefont {{Abe}},
  \citenamefont {{Abouelfettouh}}, \citenamefont {{Acernese}},\ and\
  \citenamefont {{Ackley}}}]{GWTC5_cat}%
  \BibitemOpen
  \bibfield  {author} {\bibinfo {author} {\bibnamefont {{The LIGO Scientific
  Collaboration}}}, \bibinfo {author} {\bibnamefont {{the Virgo
  Collaboration}}}, \bibinfo {author} {\bibnamefont {{the KAGRA
  Collaboration}}}, \bibinfo {author} {\bibfnamefont {A.~G.}\ \bibnamefont
  {{Abac}}}, \bibinfo {author} {\bibfnamefont {A.}~\bibnamefont {{Abe}}},
  \bibinfo {author} {\bibfnamefont {I.}~\bibnamefont {{Abouelfettouh}}},
  \bibinfo {author} {\bibfnamefont {F.}~\bibnamefont {{Acernese}}},\ and\
  \bibinfo {author} {\bibfnamefont {K.}~\bibnamefont {{Ackley}}},\ }\href
  {https://doi.org/10.48550/arXiv.2605.27225} {\bibfield  {journal} {\bibinfo
  {journal} {arXiv e-prints}\ ,\ \bibinfo {eid} {arXiv:2605.27225}} (\bibinfo
  {year} {2026}{\natexlab{b}})},\ \Eprint {https://arxiv.org/abs/2605.27225}
  {arXiv:2605.27225 [gr-qc]} \BibitemShut {NoStop}%
\bibitem [{\citenamefont {{Chruslinska}}\ and\ \citenamefont
  {{Nelemans}}(2019)}]{Chruslinska_2019}%
  \BibitemOpen
  \bibfield  {author} {\bibinfo {author} {\bibfnamefont {M.}~\bibnamefont
  {{Chruslinska}}}\ and\ \bibinfo {author} {\bibfnamefont {G.}~\bibnamefont
  {{Nelemans}}},\ }\href {https://doi.org/10.1093/mnras/stz2057} {\bibfield
  {journal} {\bibinfo  {journal} {\mnras}\ }\textbf {\bibinfo {volume} {488}},\
  \bibinfo {pages} {5300} (\bibinfo {year} {2019})},\ \Eprint
  {https://arxiv.org/abs/1907.11243} {arXiv:1907.11243 [astro-ph.GA]}
  \BibitemShut {NoStop}%
\bibitem [{\citenamefont {{Chru{\'s}li{\'n}ska}}\ \emph
  {et~al.}(2021)\citenamefont {{Chru{\'s}li{\'n}ska}}, \citenamefont
  {{Nelemans}}, \citenamefont {{Boco}},\ and\ \citenamefont
  {{Lapi}}}]{Chruslinska_2021}%
  \BibitemOpen
  \bibfield  {author} {\bibinfo {author} {\bibfnamefont {M.}~\bibnamefont
  {{Chru{\'s}li{\'n}ska}}}, \bibinfo {author} {\bibfnamefont {G.}~\bibnamefont
  {{Nelemans}}}, \bibinfo {author} {\bibfnamefont {L.}~\bibnamefont {{Boco}}},\
  and\ \bibinfo {author} {\bibfnamefont {A.}~\bibnamefont {{Lapi}}},\ }\href
  {https://doi.org/10.1093/mnras/stab2690} {\bibfield  {journal} {\bibinfo
  {journal} {\mnras}\ }\textbf {\bibinfo {volume} {508}},\ \bibinfo {pages}
  {4994} (\bibinfo {year} {2021})},\ \Eprint {https://arxiv.org/abs/2109.06187}
  {arXiv:2109.06187 [astro-ph.GA]} \BibitemShut {NoStop}%
\bibitem [{\citenamefont {{Peebles}}(1993)}]{Peebles_1993}%
  \BibitemOpen
  \bibfield  {author} {\bibinfo {author} {\bibfnamefont {P.~J.~E.}\
  \bibnamefont {{Peebles}}},\ }\href@noop {} {\emph {\bibinfo {title}
  {{Principles of Physical Cosmology}}}}\ (\bibinfo {year} {1993})\BibitemShut
  {NoStop}%
\bibitem [{\citenamefont {{Narlikar}}(2002)}]{Narlikar_2002}%
  \BibitemOpen
  \bibfield  {author} {\bibinfo {author} {\bibfnamefont {J.~V.}\ \bibnamefont
  {{Narlikar}}},\ }\href@noop {} {\emph {\bibinfo {title} {{An introduction to
  cosmology}}}}\ (\bibinfo {year} {2002})\BibitemShut {NoStop}%
\bibitem [{\citenamefont {{Planck Collaboration}}\ \emph
  {et~al.}(2020)\citenamefont {{Planck Collaboration}}, \citenamefont
  {{Aghanim}}, \citenamefont {{Akrami}}, \citenamefont {{Ashdown}},
  \citenamefont {{Aumont}}, \citenamefont {{Baccigalupi}}, \citenamefont
  {{Ballardini}}, \citenamefont {{Banday}}, \citenamefont {{Barreiro}},\ and\
  \citenamefont {et~al.}}]{Planck_2018}%
  \BibitemOpen
  \bibfield  {author} {\bibinfo {author} {\bibnamefont {{Planck
  Collaboration}}}, \bibinfo {author} {\bibfnamefont {N.}~\bibnamefont
  {{Aghanim}}}, \bibinfo {author} {\bibfnamefont {Y.}~\bibnamefont {{Akrami}}},
  \bibinfo {author} {\bibfnamefont {M.}~\bibnamefont {{Ashdown}}}, \bibinfo
  {author} {\bibfnamefont {J.}~\bibnamefont {{Aumont}}}, \bibinfo {author}
  {\bibfnamefont {C.}~\bibnamefont {{Baccigalupi}}}, \bibinfo {author}
  {\bibfnamefont {M.}~\bibnamefont {{Ballardini}}}, \bibinfo {author}
  {\bibfnamefont {A.~J.}\ \bibnamefont {{Banday}}}, \bibinfo {author}
  {\bibfnamefont {R.~B.}\ \bibnamefont {{Barreiro}}},\ and\ \bibinfo {author}
  {\bibnamefont {et~al.}},\ }\href
  {https://doi.org/10.1051/0004-6361/201833910} {\bibfield  {journal} {\bibinfo
   {journal} {\aap}\ }\textbf {\bibinfo {volume} {641}},\ \bibinfo {eid} {A6}
  (\bibinfo {year} {2020})},\ \Eprint {https://arxiv.org/abs/1807.06209}
  {arXiv:1807.06209 [astro-ph.CO]} \BibitemShut {NoStop}%
\bibitem [{\citenamefont {{Gieles}}\ \emph {et~al.}(2006)\citenamefont
  {{Gieles}}, \citenamefont {{Larsen}}, \citenamefont {{Bastian}},\ and\
  \citenamefont {{Stein}}}]{Gieles_2006b}%
  \BibitemOpen
  \bibfield  {author} {\bibinfo {author} {\bibfnamefont {M.}~\bibnamefont
  {{Gieles}}}, \bibinfo {author} {\bibfnamefont {S.~S.}\ \bibnamefont
  {{Larsen}}}, \bibinfo {author} {\bibfnamefont {N.}~\bibnamefont
  {{Bastian}}},\ and\ \bibinfo {author} {\bibfnamefont {I.~T.}\ \bibnamefont
  {{Stein}}},\ }\href {https://doi.org/10.1051/0004-6361:20053589} {\bibfield
  {journal} {\bibinfo  {journal} {\aap}\ }\textbf {\bibinfo {volume} {450}},\
  \bibinfo {pages} {129} (\bibinfo {year} {2006})},\ \Eprint
  {https://arxiv.org/abs/astro-ph/0512297} {astro-ph/0512297} \BibitemShut
  {NoStop}%
\bibitem [{\citenamefont {Larsen}(2009)}]{Larsen_2009}%
  \BibitemOpen
  \bibfield  {author} {\bibinfo {author} {\bibfnamefont {S.~S.}\ \bibnamefont
  {Larsen}},\ }\href {https://doi.org/10.1051/0004-6361:200811212} {\bibfield
  {journal} {\bibinfo  {journal} {Astronomy and Astrophysics}\ }\textbf
  {\bibinfo {volume} {494}},\ \bibinfo {pages} {539} (\bibinfo {year}
  {2009})}\BibitemShut {NoStop}%
\bibitem [{\citenamefont {{Bastian}}\ \emph {et~al.}(2012)\citenamefont
  {{Bastian}}, \citenamefont {{Adamo}}, \citenamefont {{Gieles}}, \citenamefont
  {{Silva-Villa}}, \citenamefont {{Lamers}}, \citenamefont {{Larsen}},
  \citenamefont {{Smith}}, \citenamefont {{Konstantopoulos}},\ and\
  \citenamefont {{Zackrisson}}}]{Bastian_2012}%
  \BibitemOpen
  \bibfield  {author} {\bibinfo {author} {\bibfnamefont {N.}~\bibnamefont
  {{Bastian}}}, \bibinfo {author} {\bibfnamefont {A.}~\bibnamefont {{Adamo}}},
  \bibinfo {author} {\bibfnamefont {M.}~\bibnamefont {{Gieles}}}, \bibinfo
  {author} {\bibfnamefont {E.}~\bibnamefont {{Silva-Villa}}}, \bibinfo {author}
  {\bibfnamefont {H.~J.~G.~L.~M.}\ \bibnamefont {{Lamers}}}, \bibinfo {author}
  {\bibfnamefont {S.~S.}\ \bibnamefont {{Larsen}}}, \bibinfo {author}
  {\bibfnamefont {L.~J.}\ \bibnamefont {{Smith}}}, \bibinfo {author}
  {\bibfnamefont {I.~S.}\ \bibnamefont {{Konstantopoulos}}},\ and\ \bibinfo
  {author} {\bibfnamefont {E.}~\bibnamefont {{Zackrisson}}},\ }\href
  {https://doi.org/10.1111/j.1365-2966.2011.19909.x} {\bibfield  {journal}
  {\bibinfo  {journal} {\mnras}\ }\textbf {\bibinfo {volume} {419}},\ \bibinfo
  {pages} {2606} (\bibinfo {year} {2012})},\ \Eprint
  {https://arxiv.org/abs/1109.6015} {arXiv:1109.6015 [astro-ph.CO]}
  \BibitemShut {NoStop}%
\bibitem [{\citenamefont {{Belczynski}}\ \emph
  {et~al.}(2016{\natexlab{b}})\citenamefont {{Belczynski}}, \citenamefont
  {{Holz}}, \citenamefont {{Bulik}},\ and\ \citenamefont
  {{O'Shaughnessy}}}]{Belczynski_2016}%
  \BibitemOpen
  \bibfield  {author} {\bibinfo {author} {\bibfnamefont {K.}~\bibnamefont
  {{Belczynski}}}, \bibinfo {author} {\bibfnamefont {D.~E.}\ \bibnamefont
  {{Holz}}}, \bibinfo {author} {\bibfnamefont {T.}~\bibnamefont {{Bulik}}},\
  and\ \bibinfo {author} {\bibfnamefont {R.}~\bibnamefont {{O'Shaughnessy}}},\
  }\href {https://doi.org/10.1038/nature18322} {\bibfield  {journal} {\bibinfo
  {journal} {\nat}\ }\textbf {\bibinfo {volume} {534}},\ \bibinfo {pages} {512}
  (\bibinfo {year} {2016}{\natexlab{b}})},\ \Eprint
  {https://arxiv.org/abs/1602.04531} {arXiv:1602.04531 [astro-ph.HE]}
  \BibitemShut {NoStop}%
\bibitem [{\citenamefont {{Rafelski}}\ \emph {et~al.}(2012)\citenamefont
  {{Rafelski}}, \citenamefont {{Wolfe}}, \citenamefont {{Prochaska}},
  \citenamefont {{Neeleman}},\ and\ \citenamefont {{Mendez}}}]{Rafelski_2012}%
  \BibitemOpen
  \bibfield  {author} {\bibinfo {author} {\bibfnamefont {M.}~\bibnamefont
  {{Rafelski}}}, \bibinfo {author} {\bibfnamefont {A.~M.}\ \bibnamefont
  {{Wolfe}}}, \bibinfo {author} {\bibfnamefont {J.~X.}\ \bibnamefont
  {{Prochaska}}}, \bibinfo {author} {\bibfnamefont {M.}~\bibnamefont
  {{Neeleman}}},\ and\ \bibinfo {author} {\bibfnamefont {A.~J.}\ \bibnamefont
  {{Mendez}}},\ }\href {https://doi.org/10.1088/0004-637X/755/2/89} {\bibfield
  {journal} {\bibinfo  {journal} {\apj}\ }\textbf {\bibinfo {volume} {755}},\
  \bibinfo {eid} {89} (\bibinfo {year} {2012})},\ \Eprint
  {https://arxiv.org/abs/1205.5047} {arXiv:1205.5047 [astro-ph.CO]}
  \BibitemShut {NoStop}%
\bibitem [{\citenamefont {{Chru{\'s}li{\'n}ska}}(2022)}]{Chruslinska_2022}%
  \BibitemOpen
  \bibfield  {author} {\bibinfo {author} {\bibfnamefont {M.}~\bibnamefont
  {{Chru{\'s}li{\'n}ska}}},\ }\href@noop {} {\bibfield  {journal} {\bibinfo
  {journal} {arXiv e-prints}\ ,\ \bibinfo {eid} {arXiv:2206.10622}} (\bibinfo
  {year} {2022})},\ \Eprint {https://arxiv.org/abs/2206.10622}
  {arXiv:2206.10622 [astro-ph.GA]} \BibitemShut {NoStop}%
\bibitem [{\citenamefont {{Kremer}}\ \emph {et~al.}(2025)\citenamefont
  {{Kremer}}, \citenamefont {{Weatherford}}, \citenamefont {{Hopkins}},
  \citenamefont {{Rui}},\ and\ \citenamefont {{Ye}}}]{Kremer_2025}%
  \BibitemOpen
  \bibfield  {author} {\bibinfo {author} {\bibfnamefont {K.}~\bibnamefont
  {{Kremer}}}, \bibinfo {author} {\bibfnamefont {N.~C.}\ \bibnamefont
  {{Weatherford}}}, \bibinfo {author} {\bibfnamefont {P.~F.}\ \bibnamefont
  {{Hopkins}}}, \bibinfo {author} {\bibfnamefont {N.~Z.}\ \bibnamefont
  {{Rui}}},\ and\ \bibinfo {author} {\bibfnamefont {C.~S.}\ \bibnamefont
  {{Ye}}},\ }\href {https://doi.org/10.3847/2041-8213/ae1233} {\bibfield
  {journal} {\bibinfo  {journal} {\apjl}\ }\textbf {\bibinfo {volume} {993}},\
  \bibinfo {eid} {L34} (\bibinfo {year} {2025})},\ \Eprint
  {https://arxiv.org/abs/2510.11787} {arXiv:2510.11787 [astro-ph.GA]}
  \BibitemShut {NoStop}%
\bibitem [{\citenamefont {{El-Badry}}\ \emph {et~al.}(2019)\citenamefont
  {{El-Badry}}, \citenamefont {{Quataert}}, \citenamefont {{Weisz}},
  \citenamefont {{Choksi}},\ and\ \citenamefont
  {{Boylan-Kolchin}}}]{ElBadry_2019b}%
  \BibitemOpen
  \bibfield  {author} {\bibinfo {author} {\bibfnamefont {K.}~\bibnamefont
  {{El-Badry}}}, \bibinfo {author} {\bibfnamefont {E.}~\bibnamefont
  {{Quataert}}}, \bibinfo {author} {\bibfnamefont {D.~R.}\ \bibnamefont
  {{Weisz}}}, \bibinfo {author} {\bibfnamefont {N.}~\bibnamefont {{Choksi}}},\
  and\ \bibinfo {author} {\bibfnamefont {M.}~\bibnamefont {{Boylan-Kolchin}}},\
  }\href {https://doi.org/10.1093/mnras/sty3007} {\bibfield  {journal}
  {\bibinfo  {journal} {\mnras}\ }\textbf {\bibinfo {volume} {482}},\ \bibinfo
  {pages} {4528} (\bibinfo {year} {2019})},\ \Eprint
  {https://arxiv.org/abs/1805.03652} {arXiv:1805.03652 [astro-ph.GA]}
  \BibitemShut {NoStop}%
\bibitem [{\citenamefont {{Portegies Zwart}}\ and\ \citenamefont
  {{McMillan}}(2000)}]{PortegiesZwart_2000}%
  \BibitemOpen
  \bibfield  {author} {\bibinfo {author} {\bibfnamefont {S.~F.}\ \bibnamefont
  {{Portegies Zwart}}}\ and\ \bibinfo {author} {\bibfnamefont {S.~L.~W.}\
  \bibnamefont {{McMillan}}},\ }\href {https://doi.org/10.1086/312422}
  {\bibfield  {journal} {\bibinfo  {journal} {\apjl}\ }\textbf {\bibinfo
  {volume} {528}},\ \bibinfo {pages} {L17} (\bibinfo {year} {2000})},\ \Eprint
  {https://arxiv.org/abs/astro-ph/9910061} {astro-ph/9910061} \BibitemShut
  {NoStop}%
\bibitem [{\citenamefont {Rodriguez}\ \emph {et~al.}(2015)\citenamefont
  {Rodriguez}, \citenamefont {Morscher}, \citenamefont {Pattabiraman},
  \citenamefont {Chatterjee}, \citenamefont {Haster},\ and\ \citenamefont
  {Rasio}}]{Rodriguez_2015}%
  \BibitemOpen
  \bibfield  {author} {\bibinfo {author} {\bibfnamefont {C.~L.}\ \bibnamefont
  {Rodriguez}}, \bibinfo {author} {\bibfnamefont {M.}~\bibnamefont {Morscher}},
  \bibinfo {author} {\bibfnamefont {B.}~\bibnamefont {Pattabiraman}}, \bibinfo
  {author} {\bibfnamefont {S.}~\bibnamefont {Chatterjee}}, \bibinfo {author}
  {\bibfnamefont {C.-J.}\ \bibnamefont {Haster}},\ and\ \bibinfo {author}
  {\bibfnamefont {F.~A.}\ \bibnamefont {Rasio}},\ }\bibfield  {journal}
  {\bibinfo  {journal} {Phys. Rev. Lett.}\ }\textbf {\bibinfo {volume} {115}},\
  \href {https://doi.org/10.1103/physrevlett.115.051101}
  {10.1103/physrevlett.115.051101} (\bibinfo {year} {2015})\BibitemShut
  {NoStop}%
\bibitem [{\citenamefont {Kerr}(1963)}]{Kerr_1963}%
  \BibitemOpen
  \bibfield  {author} {\bibinfo {author} {\bibfnamefont {R.~P.}\ \bibnamefont
  {Kerr}},\ }\href {https://doi.org/10.1103/PhysRevLett.11.237} {\bibfield
  {journal} {\bibinfo  {journal} {Phys. Rev. Lett.}\ }\textbf {\bibinfo
  {volume} {11}},\ \bibinfo {pages} {237} (\bibinfo {year} {1963})}\BibitemShut
  {NoStop}%
\bibitem [{\citenamefont {{Callister}}\ and\ \citenamefont
  {{Farr}}(2024)}]{Callister_2024a}%
  \BibitemOpen
  \bibfield  {author} {\bibinfo {author} {\bibfnamefont {T.~A.}\ \bibnamefont
  {{Callister}}}\ and\ \bibinfo {author} {\bibfnamefont {W.~M.}\ \bibnamefont
  {{Farr}}},\ }\href {https://doi.org/10.1103/PhysRevX.14.021005} {\bibfield
  {journal} {\bibinfo  {journal} {Physical Review X}\ }\textbf {\bibinfo
  {volume} {14}},\ \bibinfo {eid} {021005} (\bibinfo {year} {2024})},\ \Eprint
  {https://arxiv.org/abs/2302.07289} {arXiv:2302.07289 [astro-ph.HE]}
  \BibitemShut {NoStop}%
\bibitem [{\citenamefont {{Qin}}\ \emph {et~al.}(2018)\citenamefont {{Qin}},
  \citenamefont {{Fragos}}, \citenamefont {{Meynet}}, \citenamefont
  {{Andrews}}, \citenamefont {{S{\o}rensen}},\ and\ \citenamefont
  {{Song}}}]{Qin_2018}%
  \BibitemOpen
  \bibfield  {author} {\bibinfo {author} {\bibfnamefont {Y.}~\bibnamefont
  {{Qin}}}, \bibinfo {author} {\bibfnamefont {T.}~\bibnamefont {{Fragos}}},
  \bibinfo {author} {\bibfnamefont {G.}~\bibnamefont {{Meynet}}}, \bibinfo
  {author} {\bibfnamefont {J.}~\bibnamefont {{Andrews}}}, \bibinfo {author}
  {\bibfnamefont {M.}~\bibnamefont {{S{\o}rensen}}},\ and\ \bibinfo {author}
  {\bibfnamefont {H.~F.}\ \bibnamefont {{Song}}},\ }\href
  {https://doi.org/10.1051/0004-6361/201832839} {\bibfield  {journal} {\bibinfo
   {journal} {\aap}\ }\textbf {\bibinfo {volume} {616}},\ \bibinfo {eid} {A28}
  (\bibinfo {year} {2018})},\ \Eprint {https://arxiv.org/abs/1802.05738}
  {arXiv:1802.05738 [astro-ph.SR]} \BibitemShut {NoStop}%
\bibitem [{\citenamefont {{Bavera}}\ \emph {et~al.}(2020)\citenamefont
  {{Bavera}}, \citenamefont {{Fragos}}, \citenamefont {{Qin}}, \citenamefont
  {{Zapartas}}, \citenamefont {{Neijssel}}, \citenamefont {{Mandel}},
  \citenamefont {{Batta}}, \citenamefont {{Gaebel}}, \citenamefont
  {{Kimball}},\ and\ \citenamefont {{Stevenson}}}]{Bavera_2020b}%
  \BibitemOpen
  \bibfield  {author} {\bibinfo {author} {\bibfnamefont {S.~S.}\ \bibnamefont
  {{Bavera}}}, \bibinfo {author} {\bibfnamefont {T.}~\bibnamefont {{Fragos}}},
  \bibinfo {author} {\bibfnamefont {Y.}~\bibnamefont {{Qin}}}, \bibinfo
  {author} {\bibfnamefont {E.}~\bibnamefont {{Zapartas}}}, \bibinfo {author}
  {\bibfnamefont {C.~J.}\ \bibnamefont {{Neijssel}}}, \bibinfo {author}
  {\bibfnamefont {I.}~\bibnamefont {{Mandel}}}, \bibinfo {author}
  {\bibfnamefont {A.}~\bibnamefont {{Batta}}}, \bibinfo {author} {\bibfnamefont
  {S.~M.}\ \bibnamefont {{Gaebel}}}, \bibinfo {author} {\bibfnamefont
  {C.}~\bibnamefont {{Kimball}}},\ and\ \bibinfo {author} {\bibfnamefont
  {S.}~\bibnamefont {{Stevenson}}},\ }\href
  {https://doi.org/10.1051/0004-6361/201936204} {\bibfield  {journal} {\bibinfo
   {journal} {\aap}\ }\textbf {\bibinfo {volume} {635}},\ \bibinfo {eid} {A97}
  (\bibinfo {year} {2020})},\ \Eprint {https://arxiv.org/abs/1906.12257}
  {arXiv:1906.12257 [astro-ph.HE]} \BibitemShut {NoStop}%
\bibitem [{\citenamefont {{Qin}}\ \emph {et~al.}(2019)\citenamefont {{Qin}},
  \citenamefont {{Marchant}}, \citenamefont {{Fragos}}, \citenamefont
  {{Meynet}},\ and\ \citenamefont {{Kalogera}}}]{Qin_2019}%
  \BibitemOpen
  \bibfield  {author} {\bibinfo {author} {\bibfnamefont {Y.}~\bibnamefont
  {{Qin}}}, \bibinfo {author} {\bibfnamefont {P.}~\bibnamefont {{Marchant}}},
  \bibinfo {author} {\bibfnamefont {T.}~\bibnamefont {{Fragos}}}, \bibinfo
  {author} {\bibfnamefont {G.}~\bibnamefont {{Meynet}}},\ and\ \bibinfo
  {author} {\bibfnamefont {V.}~\bibnamefont {{Kalogera}}},\ }\href
  {https://doi.org/10.3847/2041-8213/aaf97b} {\bibfield  {journal} {\bibinfo
  {journal} {\apjl}\ }\textbf {\bibinfo {volume} {870}},\ \bibinfo {eid} {L18}
  (\bibinfo {year} {2019})},\ \Eprint {https://arxiv.org/abs/1810.13016}
  {arXiv:1810.13016 [astro-ph.SR]} \BibitemShut {NoStop}%
\bibitem [{\citenamefont {{K{\i}ro{\u{g}}lu}}\ \emph
  {et~al.}(2025{\natexlab{b}})\citenamefont {{K{\i}ro{\u{g}}lu}}, \citenamefont
  {{Lombardi}}, \citenamefont {{Kremer}}, \citenamefont {{Vanderzyden}},\ and\
  \citenamefont {{Rasio}}}]{Kiroglu_2025b}%
  \BibitemOpen
  \bibfield  {author} {\bibinfo {author} {\bibfnamefont {F.}~\bibnamefont
  {{K{\i}ro{\u{g}}lu}}}, \bibinfo {author} {\bibfnamefont {J.~C.}\ \bibnamefont
  {{Lombardi}}}, \bibinfo {author} {\bibfnamefont {K.}~\bibnamefont
  {{Kremer}}}, \bibinfo {author} {\bibfnamefont {H.~D.}\ \bibnamefont
  {{Vanderzyden}}},\ and\ \bibinfo {author} {\bibfnamefont {F.~A.}\
  \bibnamefont {{Rasio}}},\ }\href {https://doi.org/10.3847/2041-8213/adc263}
  {\bibfield  {journal} {\bibinfo  {journal} {\apjl}\ }\textbf {\bibinfo
  {volume} {983}},\ \bibinfo {eid} {L9} (\bibinfo {year}
  {2025}{\natexlab{b}})},\ \Eprint {https://arxiv.org/abs/2501.09068}
  {arXiv:2501.09068 [astro-ph.HE]} \BibitemShut {NoStop}%
\bibitem [{\citenamefont {{Baker}}\ \emph {et~al.}(2008)\citenamefont
  {{Baker}}, \citenamefont {{Boggs}}, \citenamefont {{Centrella}},
  \citenamefont {{Kelly}}, \citenamefont {{McWilliams}}, \citenamefont
  {{Miller}},\ and\ \citenamefont {{van Meter}}}]{Baker_2008}%
  \BibitemOpen
  \bibfield  {author} {\bibinfo {author} {\bibfnamefont {J.~G.}\ \bibnamefont
  {{Baker}}}, \bibinfo {author} {\bibfnamefont {W.~D.}\ \bibnamefont
  {{Boggs}}}, \bibinfo {author} {\bibfnamefont {J.}~\bibnamefont
  {{Centrella}}}, \bibinfo {author} {\bibfnamefont {B.~J.}\ \bibnamefont
  {{Kelly}}}, \bibinfo {author} {\bibfnamefont {S.~T.}\ \bibnamefont
  {{McWilliams}}}, \bibinfo {author} {\bibfnamefont {M.~C.}\ \bibnamefont
  {{Miller}}},\ and\ \bibinfo {author} {\bibfnamefont {J.~R.}\ \bibnamefont
  {{van Meter}}},\ }\href {https://doi.org/10.1086/590927} {\bibfield
  {journal} {\bibinfo  {journal} {\apjl}\ }\textbf {\bibinfo {volume} {682}},\
  \bibinfo {pages} {L29} (\bibinfo {year} {2008})},\ \Eprint
  {https://arxiv.org/abs/0802.0416} {arXiv:0802.0416 [astro-ph]} \BibitemShut
  {NoStop}%
\bibitem [{\citenamefont {{Hofmann}}\ \emph {et~al.}(2016)\citenamefont
  {{Hofmann}}, \citenamefont {{Barausse}},\ and\ \citenamefont
  {{Rezzolla}}}]{Hofmann_2016}%
  \BibitemOpen
  \bibfield  {author} {\bibinfo {author} {\bibfnamefont {F.}~\bibnamefont
  {{Hofmann}}}, \bibinfo {author} {\bibfnamefont {E.}~\bibnamefont
  {{Barausse}}},\ and\ \bibinfo {author} {\bibfnamefont {L.}~\bibnamefont
  {{Rezzolla}}},\ }\href {https://doi.org/10.3847/2041-8205/825/2/L19}
  {\bibfield  {journal} {\bibinfo  {journal} {\apjl}\ }\textbf {\bibinfo
  {volume} {825}},\ \bibinfo {eid} {L19} (\bibinfo {year} {2016})},\ \Eprint
  {https://arxiv.org/abs/1605.01938} {arXiv:1605.01938 [gr-qc]} \BibitemShut
  {NoStop}%
\bibitem [{\citenamefont {Jim\'enez-Forteza}\ \emph {et~al.}(2017)\citenamefont
  {Jim\'enez-Forteza}, \citenamefont {Keitel}, \citenamefont {Husa},
  \citenamefont {Hannam}, \citenamefont {Khan},\ and\ \citenamefont
  {P\"urrer}}]{JimenezForteza_2017}%
  \BibitemOpen
  \bibfield  {author} {\bibinfo {author} {\bibfnamefont {X.}~\bibnamefont
  {Jim\'enez-Forteza}}, \bibinfo {author} {\bibfnamefont {D.}~\bibnamefont
  {Keitel}}, \bibinfo {author} {\bibfnamefont {S.}~\bibnamefont {Husa}},
  \bibinfo {author} {\bibfnamefont {M.}~\bibnamefont {Hannam}}, \bibinfo
  {author} {\bibfnamefont {S.}~\bibnamefont {Khan}},\ and\ \bibinfo {author}
  {\bibfnamefont {M.}~\bibnamefont {P\"urrer}},\ }\href
  {https://doi.org/10.1103/PhysRevD.95.064024} {\bibfield  {journal} {\bibinfo
  {journal} {Phys. Rev. D}\ }\textbf {\bibinfo {volume} {95}},\ \bibinfo
  {pages} {064024} (\bibinfo {year} {2017})}\BibitemShut {NoStop}%
\bibitem [{\citenamefont {{Lousto}}\ \emph {et~al.}(2012)\citenamefont
  {{Lousto}}, \citenamefont {{Zlochower}}, \citenamefont {{Dotti}},\ and\
  \citenamefont {{Volonteri}}}]{Lousto_2012}%
  \BibitemOpen
  \bibfield  {author} {\bibinfo {author} {\bibfnamefont {C.~O.}\ \bibnamefont
  {{Lousto}}}, \bibinfo {author} {\bibfnamefont {Y.}~\bibnamefont
  {{Zlochower}}}, \bibinfo {author} {\bibfnamefont {M.}~\bibnamefont
  {{Dotti}}},\ and\ \bibinfo {author} {\bibfnamefont {M.}~\bibnamefont
  {{Volonteri}}},\ }\href {https://doi.org/10.1103/PhysRevD.85.084015}
  {\bibfield  {journal} {\bibinfo  {journal} {\prd}\ }\textbf {\bibinfo
  {volume} {85}},\ \bibinfo {eid} {084015} (\bibinfo {year} {2012})},\ \Eprint
  {https://arxiv.org/abs/1201.1923} {arXiv:1201.1923 [gr-qc]} \BibitemShut
  {NoStop}%
\bibitem [{\citenamefont {{Berti}}\ and\ \citenamefont
  {{Volonteri}}(2008)}]{Berti_2008}%
  \BibitemOpen
  \bibfield  {author} {\bibinfo {author} {\bibfnamefont {E.}~\bibnamefont
  {{Berti}}}\ and\ \bibinfo {author} {\bibfnamefont {M.}~\bibnamefont
  {{Volonteri}}},\ }\href {https://doi.org/10.1086/590379} {\bibfield
  {journal} {\bibinfo  {journal} {\apj}\ }\textbf {\bibinfo {volume} {684}},\
  \bibinfo {pages} {822} (\bibinfo {year} {2008})},\ \Eprint
  {https://arxiv.org/abs/0802.0025} {arXiv:0802.0025 [astro-ph]} \BibitemShut
  {NoStop}%
\bibitem [{\citenamefont {{Perego}}\ \emph {et~al.}(2009)\citenamefont
  {{Perego}}, \citenamefont {{Dotti}}, \citenamefont {{Colpi}},\ and\
  \citenamefont {{Volonteri}}}]{Perego_2009}%
  \BibitemOpen
  \bibfield  {author} {\bibinfo {author} {\bibfnamefont {A.}~\bibnamefont
  {{Perego}}}, \bibinfo {author} {\bibfnamefont {M.}~\bibnamefont {{Dotti}}},
  \bibinfo {author} {\bibfnamefont {M.}~\bibnamefont {{Colpi}}},\ and\ \bibinfo
  {author} {\bibfnamefont {M.}~\bibnamefont {{Volonteri}}},\ }\href
  {https://doi.org/10.1111/j.1365-2966.2009.15427.x} {\bibfield  {journal}
  {\bibinfo  {journal} {\mnras}\ }\textbf {\bibinfo {volume} {399}},\ \bibinfo
  {pages} {2249} (\bibinfo {year} {2009})},\ \Eprint
  {https://arxiv.org/abs/0907.3742} {arXiv:0907.3742 [astro-ph.CO]}
  \BibitemShut {NoStop}%
\bibitem [{\citenamefont {{Cenci}}\ \emph {et~al.}(2021)\citenamefont
  {{Cenci}}, \citenamefont {{Sala}}, \citenamefont {{Lupi}}, \citenamefont
  {{Capelo}},\ and\ \citenamefont {{Dotti}}}]{Cenci_2021}%
  \BibitemOpen
  \bibfield  {author} {\bibinfo {author} {\bibfnamefont {E.}~\bibnamefont
  {{Cenci}}}, \bibinfo {author} {\bibfnamefont {L.}~\bibnamefont {{Sala}}},
  \bibinfo {author} {\bibfnamefont {A.}~\bibnamefont {{Lupi}}}, \bibinfo
  {author} {\bibfnamefont {P.~R.}\ \bibnamefont {{Capelo}}},\ and\ \bibinfo
  {author} {\bibfnamefont {M.}~\bibnamefont {{Dotti}}},\ }\href
  {https://doi.org/10.1093/mnras/staa3449} {\bibfield  {journal} {\bibinfo
  {journal} {\mnras}\ }\textbf {\bibinfo {volume} {500}},\ \bibinfo {pages}
  {3719} (\bibinfo {year} {2021})},\ \Eprint {https://arxiv.org/abs/2011.06596}
  {arXiv:2011.06596 [astro-ph.GA]} \BibitemShut {NoStop}%
\bibitem [{\citenamefont {{Banerjee}}\ and\ \citenamefont
  {{Olejak}}(2024)}]{Banerjee_2024}%
  \BibitemOpen
  \bibfield  {author} {\bibinfo {author} {\bibfnamefont {S.}~\bibnamefont
  {{Banerjee}}}\ and\ \bibinfo {author} {\bibfnamefont {A.}~\bibnamefont
  {{Olejak}}},\ }\href {https://doi.org/10.48550/arXiv.2411.15112} {\bibfield
  {journal} {\bibinfo  {journal} {arXiv e-prints}\ ,\ \bibinfo {eid}
  {arXiv:2411.15112}} (\bibinfo {year} {2024})},\ \Eprint
  {https://arxiv.org/abs/2411.15112} {arXiv:2411.15112 [astro-ph.HE]}
  \BibitemShut {NoStop}%
\bibitem [{\citenamefont {{Ivanova}}\ \emph {et~al.}(2013)\citenamefont
  {{Ivanova}}, \citenamefont {{Justham}}, \citenamefont {{Chen}}, \citenamefont
  {{De Marco}}, \citenamefont {{Fryer}}, \citenamefont {{Gaburov}},
  \citenamefont {{Ge}}, \citenamefont {{Glebbeek}}, \citenamefont {{Han}},
  \citenamefont {{Li}}, \citenamefont {{Lu}}, \citenamefont {{Marsh}},
  \citenamefont {{Podsiadlowski}}, \citenamefont {{Potter}}, \citenamefont
  {{Soker}}, \citenamefont {{Taam}}, \citenamefont {{Tauris}}, \citenamefont
  {{van den Heuvel}},\ and\ \citenamefont {{Webbink}}}]{Ivanova_2013}%
  \BibitemOpen
  \bibfield  {author} {\bibinfo {author} {\bibfnamefont {N.}~\bibnamefont
  {{Ivanova}}}, \bibinfo {author} {\bibfnamefont {S.}~\bibnamefont
  {{Justham}}}, \bibinfo {author} {\bibfnamefont {X.}~\bibnamefont {{Chen}}},
  \bibinfo {author} {\bibfnamefont {O.}~\bibnamefont {{De Marco}}}, \bibinfo
  {author} {\bibfnamefont {C.~L.}\ \bibnamefont {{Fryer}}}, \bibinfo {author}
  {\bibfnamefont {E.}~\bibnamefont {{Gaburov}}}, \bibinfo {author}
  {\bibfnamefont {H.}~\bibnamefont {{Ge}}}, \bibinfo {author} {\bibfnamefont
  {E.}~\bibnamefont {{Glebbeek}}}, \bibinfo {author} {\bibfnamefont
  {Z.}~\bibnamefont {{Han}}}, \bibinfo {author} {\bibfnamefont {X.~D.}\
  \bibnamefont {{Li}}}, \bibinfo {author} {\bibfnamefont {G.}~\bibnamefont
  {{Lu}}}, \bibinfo {author} {\bibfnamefont {T.}~\bibnamefont {{Marsh}}},
  \bibinfo {author} {\bibfnamefont {P.}~\bibnamefont {{Podsiadlowski}}},
  \bibinfo {author} {\bibfnamefont {A.}~\bibnamefont {{Potter}}}, \bibinfo
  {author} {\bibfnamefont {N.}~\bibnamefont {{Soker}}}, \bibinfo {author}
  {\bibfnamefont {R.}~\bibnamefont {{Taam}}}, \bibinfo {author} {\bibfnamefont
  {T.~M.}\ \bibnamefont {{Tauris}}}, \bibinfo {author} {\bibfnamefont
  {E.~P.~J.}\ \bibnamefont {{van den Heuvel}}},\ and\ \bibinfo {author}
  {\bibfnamefont {R.~F.}\ \bibnamefont {{Webbink}}},\ }\href
  {https://doi.org/10.1007/s00159-013-0059-2} {\bibfield  {journal} {\bibinfo
  {journal} {\aapr}\ }\textbf {\bibinfo {volume} {21}},\ \bibinfo {eid} {59}
  (\bibinfo {year} {2013})},\ \Eprint {https://arxiv.org/abs/1209.4302}
  {arXiv:1209.4302 [astro-ph.HE]} \BibitemShut {NoStop}%
\bibitem [{\citenamefont {{Trani}}\ \emph {et~al.}(2021)\citenamefont
  {{Trani}}, \citenamefont {{Tanikawa}}, \citenamefont {{Fujii}}, \citenamefont
  {{Leigh}},\ and\ \citenamefont {{Kumamoto}}}]{Trani_2021}%
  \BibitemOpen
  \bibfield  {author} {\bibinfo {author} {\bibfnamefont {A.~A.}\ \bibnamefont
  {{Trani}}}, \bibinfo {author} {\bibfnamefont {A.}~\bibnamefont {{Tanikawa}}},
  \bibinfo {author} {\bibfnamefont {M.~S.}\ \bibnamefont {{Fujii}}}, \bibinfo
  {author} {\bibfnamefont {N.~W.~C.}\ \bibnamefont {{Leigh}}},\ and\ \bibinfo
  {author} {\bibfnamefont {J.}~\bibnamefont {{Kumamoto}}},\ }\href
  {https://doi.org/10.1093/mnras/stab967} {\bibfield  {journal} {\bibinfo
  {journal} {\mnras}\ }\textbf {\bibinfo {volume} {504}},\ \bibinfo {pages}
  {910} (\bibinfo {year} {2021})},\ \Eprint {https://arxiv.org/abs/2102.01689}
  {arXiv:2102.01689 [astro-ph.HE]} \BibitemShut {NoStop}%
\bibitem [{\citenamefont {{Speagle}}(2020)}]{Speagle_2020}%
  \BibitemOpen
  \bibfield  {author} {\bibinfo {author} {\bibfnamefont {J.~S.}\ \bibnamefont
  {{Speagle}}},\ }\href {https://doi.org/10.1093/mnras/staa278} {\bibfield
  {journal} {\bibinfo  {journal} {\mnras}\ }\textbf {\bibinfo {volume} {493}},\
  \bibinfo {pages} {3132} (\bibinfo {year} {2020})},\ \Eprint
  {https://arxiv.org/abs/1904.02180} {arXiv:1904.02180 [astro-ph.IM]}
  \BibitemShut {NoStop}%
\bibitem [{\citenamefont {{Skilling}}(2006)}]{Skilling_2006}%
  \BibitemOpen
  \bibfield  {author} {\bibinfo {author} {\bibfnamefont {J.}~\bibnamefont
  {{Skilling}}},\ }\href {https://doi.org/10.1214/06-BA127} {\bibfield
  {journal} {\bibinfo  {journal} {Bayesian Analysis}\ }\textbf {\bibinfo
  {volume} {1}},\ \bibinfo {pages} {833} (\bibinfo {year} {2006})}\BibitemShut
  {NoStop}%
\bibitem [{\citenamefont {{Ashton}}\ \emph {et~al.}(2019)\citenamefont
  {{Ashton}}, \citenamefont {{H{\"u}bner}}, \citenamefont {{Lasky}},
  \citenamefont {{Talbot}}, \citenamefont {{Ackley}}, \citenamefont
  {{Biscoveanu}}, \citenamefont {{Chu}}, \citenamefont {{Divakarla}},
  \citenamefont {{Easter}}, \citenamefont {{Goncharov}},\ and\ \citenamefont
  {et~al.}}]{Ashton_2019}%
  \BibitemOpen
  \bibfield  {author} {\bibinfo {author} {\bibfnamefont {G.}~\bibnamefont
  {{Ashton}}}, \bibinfo {author} {\bibfnamefont {M.}~\bibnamefont
  {{H{\"u}bner}}}, \bibinfo {author} {\bibfnamefont {P.~D.}\ \bibnamefont
  {{Lasky}}}, \bibinfo {author} {\bibfnamefont {C.}~\bibnamefont {{Talbot}}},
  \bibinfo {author} {\bibfnamefont {K.}~\bibnamefont {{Ackley}}}, \bibinfo
  {author} {\bibfnamefont {S.}~\bibnamefont {{Biscoveanu}}}, \bibinfo {author}
  {\bibfnamefont {Q.}~\bibnamefont {{Chu}}}, \bibinfo {author} {\bibfnamefont
  {A.}~\bibnamefont {{Divakarla}}}, \bibinfo {author} {\bibfnamefont {P.~J.}\
  \bibnamefont {{Easter}}}, \bibinfo {author} {\bibfnamefont {B.}~\bibnamefont
  {{Goncharov}}},\ and\ \bibinfo {author} {\bibnamefont {et~al.}},\ }\href
  {https://doi.org/10.3847/1538-4365/ab06fc} {\bibfield  {journal} {\bibinfo
  {journal} {\apjs}\ }\textbf {\bibinfo {volume} {241}},\ \bibinfo {eid} {27}
  (\bibinfo {year} {2019})},\ \Eprint {https://arxiv.org/abs/1811.02042}
  {arXiv:1811.02042 [astro-ph.IM]} \BibitemShut {NoStop}%
\bibitem [{\citenamefont {{Lin}}(1991)}]{Lin_1991}%
  \BibitemOpen
  \bibfield  {author} {\bibinfo {author} {\bibfnamefont {J.}~\bibnamefont
  {{Lin}}},\ }\href {https://doi.org/10.1109/18.61115} {\bibfield  {journal}
  {\bibinfo  {journal} {IEEE Transactions on Information Theory}\ }\textbf
  {\bibinfo {volume} {37}},\ \bibinfo {pages} {145} (\bibinfo {year}
  {1991})}\BibitemShut {NoStop}%
\end{thebibliography}%

\FloatBarrier

\appendix

\onecolumngrid
\section{Computed model clusters}\label{runlist}

\begingroup

\setlength{\tabcolsep}{7.0pt}
\renewcommand{\arraystretch}{1.3}
\LTcapwidth=\linewidth

\begin{longtable}{cclcccclccc}
\caption{Summary of the evolutionary star cluster model grid computed in this work and their GR-merger outcomes.
	All the 90 models are evolved with the direct N-body integration code $\nbseven$ (Sec.~\ref{nbsims}). The model
	clusters initiate their evolution with a Plummer profile of total mass $\mcl(0)$ and half-mass radius $\rh(0)$.
	The initial models have a primordial binary fraction of 100\% for stars with zero age main sequence mass 
	of $\mzams\geq16\Ms$. For the rest of the members, an initial primordial binary fraction of 10\% is adopted.  
	All clusters orbit in a solar-neighbourhood-like external field. See Sec.~\ref{nbsims} for further details on initial conditions.
	The columns from left to right quote, for each model cluster, the initial mass, $\mcl(0)$, initial membership, $N(0)$,
	initial half-mass radius, $r_h(0)$, metallicity, $Z$, maximum number of bound BH members during cluster evolution,
	$N_{\rm BH, max}$, end time of the simulation, $t_{\rm end}$,
	total membership at the end of the simulation, $N_{\rm end}$,
	BH membership at the end of the simulation, $N_{\rm BH, end}$, 
	number of in-cluster mergers, $N_{\rm mrg,in}$, and number of ejected mergers, $N_{\rm mrg,esc}$.
	}
\label{tab:runlist} \\
\toprule
Model nr. & $M_{\rm cl}(0) [M_\odot]$ & $N(0)$ & $r_h(0) [{\rm pc}]$ & $Z$ & $N_{\rm BH, max}$ & $t_{\rm end} [{\rm Gyr}]$ & $N_{\rm end}$ & $N_{\rm BH, end}$ & $N_{\rm mrg,in}$ & $N_{\rm mrg,esc}$ \\
\midrule
\endfirsthead
\toprule
Model nr. & $M_{\rm cl}(0) [M_\odot]$ & $N(0)$ & $r_h(0) [{\rm pc}]$ & $Z$ & $N_{\rm BH, max}$ & $t_{\rm end} [{\rm Gyr}]$ & $N_{\rm end}$ & $N_{\rm BH, end}$ & $N_{\rm mrg,in}$ & $N_{\rm mrg,esc}$ \\
\midrule
\endhead
\midrule
\multicolumn{11}{r}{Continued on next page} \\
\midrule
\endfoot
\bottomrule
\endlastfoot
1 & 10000.0 & 17329 & 1.0 & 0.0002 & 12 & 3.0 & 8290 & 0 & 0 & 0 \\
2 & 10000.0 & 17329 & 1.0 & 0.0010 & 16 & 9.7 & 51 & 0 & 1 & 0 \\
3 & 10000.0 & 17329 & 1.0 & 0.0050 & 12 & 1.0 & 14510 & 0 & 0 & 1 \\
4 & 10000.0 & 17329 & 1.0 & 0.0100 & 10 & 1.0 & 14408 & 0 & 1 & 0 \\
5 & 10000.0 & 17329 & 1.0 & 0.0200 & 6 & 5.6 & 1357 & 0 & 0 & 0 \\
6 & 10000.0 & 17329 & 2.0 & 0.0002 & 18 & 9.2 & 96 & 0 & 0 & 0 \\
7 & 10000.0 & 17329 & 2.0 & 0.0010 & 11 & 7.7 & 52 & 0 & 0 & 0 \\
8 & 10000.0 & 17329 & 2.0 & 0.0050 & 10 & 7.8 & 492 & 0 & 0 & 0 \\
9 & 10000.0 & 17329 & 2.0 & 0.0100 & 10 & 6.3 & 53 & 2 & 1 & 0 \\
10 & 10000.0 & 17329 & 2.0 & 0.0200 & 5 & 5.3 & 2345 & 0 & 0 & 0 \\
11 & 10000.0 & 17329 & 3.0 & 0.0002 & 17 & 3.2 & 55 & 0 & 0 & 0 \\
12 & 10000.0 & 17329 & 3.0 & 0.0010 & 15 & 6.1 & 57 & 0 & 0 & 0 \\
13 & 10000.0 & 17329 & 3.0 & 0.0050 & 9 & 8.9 & 51 & 0 & 1 & 0 \\
14 & 10000.0 & 17329 & 3.0 & 0.0100 & 11 & 6.9 & 2 & 0 & 0 & 0 \\
15 & 10000.0 & 17329 & 3.0 & 0.0200 & 5 & 7.3 & 554 & 0 & 0 & 0 \\
16 & 20000.1 & 34425 & 1.0 & 0.0002 & 35 & 11.0 & 1783 & 0 & 1 & 0 \\
17 & 20000.1 & 34426 & 1.0 & 0.0010 & 26 & 1.7 & 27936 & 2 & 0 & 0 \\
18 & 20000.1 & 34426 & 1.0 & 0.0050 & 28 & 0.7 & 32000 & 2 & 0 & 0 \\
19 & 20000.1 & 34424 & 1.0 & 0.0100 & 18 & 1.3 & 29860 & 0 & 2 & 0 \\
20 & 20000.1 & 34426 & 1.0 & 0.0200 & 12 & 3.6 & 18702 & 0 & 0 & 0 \\
21 & 20000.1 & 34426 & 2.0 & 0.0002 & 30 & 7.9 & 8810 & 1 & 1 & 0 \\
22 & 20000.1 & 34426 & 2.0 & 0.0010 & 28 & 9.9 & 4697 & 0 & 0 & 1 \\
23 & 20000.1 & 34425 & 2.0 & 0.0050 & 30 & 5.8 & 14157 & 1 & 1 & 0 \\
24 & 20000.1 & 34426 & 2.0 & 0.0100 & 20 & 7.1 & 10313 & 0 & 0 & 0 \\
25 & 20000.1 & 34426 & 2.0 & 0.0200 & 9 & 7.0 & 8091 & 0 & 0 & 0 \\
26 & 20000.1 & 34426 & 3.0 & 0.0002 & 31 & 11.0 & 1626 & 1 & 0 & 0 \\
27 & 20000.1 & 34426 & 3.0 & 0.0010 & 28 & 8.2 & 61 & 0 & 1 & 1 \\
28 & 20000.1 & 34426 & 3.0 & 0.0050 & 25 & 11.0 & 754 & 2 & 2 & 0 \\
29 & 20000.1 & 34425 & 3.0 & 0.0100 & 17 & 11.0 & 2631 & 0 & 1 & 0 \\
30 & 20000.1 & 34426 & 3.0 & 0.0200 & 12 & 6.2 & 10933 & 0 & 0 & 0 \\
31 & 30000.1 & 51509 & 1.0 & 0.0002 & 48 & 2.9 & 40723 & 1 & 0 & 1 \\
32 & 30000.1 & 51509 & 1.0 & 0.0010 & 44 & 11.0 & 11703 & 2 & 1 & 0 \\
33 & 30000.1 & 51509 & 1.0 & 0.0050 & 38 & 5.5 & 28333 & 0 & 1 & 0 \\
34 & 30000.1 & 51509 & 1.0 & 0.0100 & 31 & 2.5 & 41210 & 0 & 0 & 0 \\
35 & 30000.1 & 51509 & 1.0 & 0.0200 & 18 & 1.0 & 48273 & 2 & 0 & 0 \\
36 & 30000.1 & 51509 & 2.0 & 0.0002 & 48 & 10.3 & 13886 & 0 & 2 & 0 \\
37 & 30000.1 & 51509 & 2.0 & 0.0010 & 42 & 11.0 & 12727 & 2 & 2 & 0 \\
38 & 30000.1 & 51509 & 2.0 & 0.0050 & 31 & 11.0 & 11933 & 0 & 1 & 0 \\
39 & 30000.1 & 51509 & 2.0 & 0.0100 & 27 & 11.0 & 12164 & 0 & 2 & 0 \\
40 & 30000.1 & 51509 & 2.0 & 0.0200 & 16 & 9.2 & 14111 & 0 & 0 & 0 \\
41 & 30000.1 & 51508 & 3.0 & 0.0002 & 45 & 11.0 & 7501 & 2 & 0 & 0 \\
42 & 30000.1 & 51508 & 3.0 & 0.0010 & 41 & 11.0 & 7837 & 0 & 1 & 0 \\
43 & 30000.1 & 51508 & 3.0 & 0.0050 & 38 & 10.1 & 14168 & 2 & 0 & 0 \\
44 & 30000.1 & 51508 & 3.0 & 0.0100 & 32 & 11.0 & 11457 & 2 & 0 & 0 \\
45 & 30000.1 & 51508 & 3.0 & 0.0200 & 18 & 10.7 & 10637 & 0 & 0 & 0 \\
46 & 50000.0 & 85661 & 1.0 & 0.0002 & 77 & 7.9 & 51743 & 0 & 3 & 0 \\
47 & 50000.0 & 85660 & 1.0 & 0.0010 & 69 & 5.6 & 61924 & 0 & 1 & 1 \\
48 & 50000.0 & 85662 & 1.0 & 0.0050 & 66 & 4.0 & 68446 & 0 & 0 & 0 \\
49 & 50000.0 & 85662 & 1.0 & 0.0100 & 52 & 3.6 & 69981 & 1 & 3 & 0 \\
50 & 50000.0 & 85662 & 1.0 & 0.0200 & 32 & 1.9 & 77665 & 0 & 2 & 0 \\
51 & 50000.0 & 85662 & 2.0 & 0.0002 & 76 & 11.0 & 38615 & 0 & 0 & 0 \\
52 & 50000.0 & 85662 & 2.0 & 0.0010 & 77 & 11.0 & 39803 & 2 & 1 & 0 \\
53 & 50000.0 & 85662 & 2.0 & 0.0050 & 65 & 9.3 & 47519 & 0 & 3 & 1 \\
54 & 50000.0 & 85662 & 2.0 & 0.0100 & 53 & 6.5 & 60200 & 0 & 0 & 0 \\
55 & 50000.0 & 85662 & 2.0 & 0.0200 & 41 & 2.9 & 75480 & 3 & 0 & 0 \\
56 & 50000.0 & 85662 & 3.0 & 0.0002 & 80 & 11.0 & 34963 & 3 & 0 & 0 \\
57 & 50000.0 & 85662 & 3.0 & 0.0010 & 75 & 11.0 & 36463 & 5 & 0 & 0 \\
58 & 50000.0 & 85662 & 3.0 & 0.0050 & 63 & 11.0 & 38416 & 1 & 1 & 0 \\
59 & 50000.0 & 85662 & 3.0 & 0.0100 & 60 & 11.0 & 38961 & 0 & 0 & 0 \\
60 & 50000.0 & 85662 & 3.0 & 0.0200 & 30 & 11.0 & 38037 & 0 & 1 & 0 \\
61 & 75000.0 & 128343 & 1.0 & 0.0002 & 124 & 5.4 & 104244 & 0 & 4 & 3 \\
62 & 75000.0 & 128343 & 1.0 & 0.0010 & 109 & 5.4 & 103420 & 0 & 4 & 3 \\
63 & 75000.0 & 128343 & 1.0 & 0.0050 & 109 & 4.0 & 110207 & 2 & 5 & 2 \\
64 & 75000.0 & 128343 & 1.0 & 0.0100 & 81 & 2.8 & 116295 & 0 & 4 & 0 \\
65 & 75000.0 & 128343 & 1.0 & 0.0200 & 48 & 2.3 & 117970 & 1 & 0 & 0 \\
66 & 75000.0 & 128343 & 2.0 & 0.0002 & 112 & 11.0 & 78032 & 4 & 5 & 1 \\
67 & 75000.0 & 128343 & 2.0 & 0.0010 & 108 & 11.0 & 81463 & 2 & 3 & 4 \\
68 & 75000.0 & 128343 & 2.0 & 0.0050 & 107 & 9.7 & 87377 & 1 & 3 & 1 \\
69 & 75000.0 & 128341 & 2.0 & 0.0100 & 75 & 8.1 & 94562 & 1 & 1 & 1 \\
70 & 75000.0 & 128343 & 2.0 & 0.0200 & 53 & 6.6 & 101019 & 0 & 0 & 0 \\
71 & 75000.0 & 128342 & 3.0 & 0.0002 & 117 & 9.9 & 81921 & 4 & 2 & 3 \\
72 & 75000.0 & 128343 & 3.0 & 0.0010 & 110 & 11.0 & 76965 & 5 & 2 & 1 \\
73 & 75000.0 & 128343 & 3.0 & 0.0050 & 106 & 9.7 & 84644 & 5 & 1 & 2 \\
74 & 75000.0 & 128343 & 3.0 & 0.0100 & 88 & 7.7 & 94552 & 3 & 2 & 0 \\
75 & 75000.0 & 128343 & 3.0 & 0.0200 & 49 & 11.0 & 78319 & 2 & 1 & 0 \\
76 & 100000.1 & 171019 & 1.0 & 0.0002 & 162 & 3.3 & 155252 & 8 & 4 & 1 \\
77 & 100000.1 & 171019 & 1.0 & 0.0010 & 160 & 2.7 & 158083 & 22 & 2 & 1 \\
78 & 100000.1 & 171017 & 1.0 & 0.0050 & 126 & 3.4 & 155810 & 2 & 3 & 3 \\
79 & 100000.1 & 171019 & 1.0 & 0.0100 & 117 & 4.1 & 151359 & 0 & 3 & 1 \\
80 & 100000.1 & 171018 & 1.0 & 0.0200 & 73 & 2.8 & 157772 & 0 & 0 & 0 \\
81 & 100000.1 & 171018 & 2.0 & 0.0002 & 166 & 10.4 & 125080 & 4 & 4 & 3 \\
82 & 100000.1 & 171018 & 2.0 & 0.0010 & 156 & 11.0 & 119026 & 7 & 4 & 1 \\
83 & 100000.1 & 171019 & 2.0 & 0.0050 & 131 & 7.5 & 139780 & 2 & 6 & 0 \\
84 & 100000.1 & 171019 & 2.0 & 0.0100 & 117 & 3.5 & 157352 & 8 & 1 & 1 \\
85 & 100000.1 & 171019 & 2.0 & 0.0200 & 76 & 0.7 & 167527 & 23 & 0 & 1 \\
86 & 100000.1 & 171019 & 3.0 & 0.0002 & 162 & 10.7 & 114885 & 13 & 4 & 1 \\
87 & 100000.1 & 171019 & 3.0 & 0.0010 & 147 & 10.1 & 120540 & 15 & 2 & 1 \\
88 & 100000.1 & 171017 & 3.0 & 0.0050 & 137 & 10.3 & 123356 & 3 & 1 & 1 \\
89 & 100000.1 & 171019 & 3.0 & 0.0100 & 116 & 4.8 & 149719 & 20 & 1 & 0 \\
90 & 100000.1 & 171019 & 3.0 & 0.0200 & 78 & 10.9 & 122221 & 2 & 3 & 0 \\
\end{longtable}

\endgroup

\FloatBarrier

\section{Merger rate density from model merger population}\label{rate}

\begin{figure*}[!h]
\centering
\includegraphics[width = \textwidth, angle=0.0]{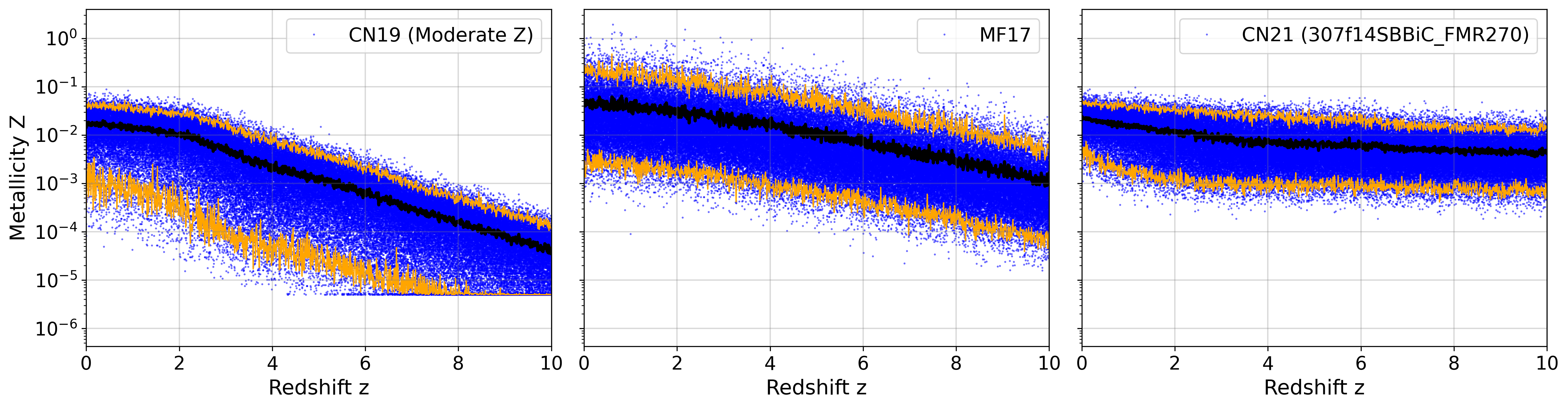}
	\caption{Redshift ($z$) - metallicity ($Z$) or $zZ$ sample as of Ref.~\cite{Madau_2017} (MF17; middle panel),
	Ref.~\cite{Chruslinska_2019} (CN19; left panel), and Ref.~\cite{Chruslinska_2021} (CN21; right panel).
	For the case of MF17 $zZ$ dependence, a spread in metallicity of 0.5 dex is assumed.
	For the case of CN19 $zZ$ dependence, the variant `Moderate-$Z$' is chosen.
	For the case of CN21 $zZ$ dependence, the variant ${\rm 307f14SBBiC\_FMR270}$
	is chosen. On each panel, the black line traces the mean and the orange
	lines enclose the 95\% CI of the $Z$-distribution at increasing $z$.}
\label{fig:zZsample}
\end{figure*}

To obtain the redshift-dependent merger rate density from YMC/OC,
a Model Universe is constructed out of the computed evolutionary model grid. The intrinsic merger
rate density, $\rate(z)$, at a redshift $z$ can be formally expressed as
\begin{equation}
\begin{split}
       & \rate(z) = \\
       & \frac{\ncl}{\delage(\delz(z))}
	 \int_{\zfhigh}^{\zf=z}\int_{Z=\metlow}^{\methigh}
	 \int_{\mcl=\mcllow}^{\mclhigh}\int_{\rh=\rhlow}^{\rhhigh}
	 \mrgfrac(z,\zf,\mcl,\rh,Z)\clmf(\mcl)\phir(\rh)\phizz(\zf,Z)\sfh(\zf) d\rh d\mcl dZ d\zf
\label{eq:rate0}
\end{split}
\end{equation}

Here, the functions $\clmf(\mcl)$, $\phir(\rh)$, $\phizz(\zf,Z)$, $\sfh(\zf)$ are, respectively, the
probability distribution of a star cluster's initial mass, $\mcl$, initial size, $\rh$\footnote{For the
convenience of notation, in this subsection, the initial mass and size of a cluster are denoted by
$\mcl$ and $\rh$, respectively, instead of $\mcl(0)$ and $\rh(0)$ as used in other places in this paper.},
metallicity, $Z$, and formation redshift, $\zf$, in the Universe. The function $\mrgfrac(z,\zf,\mcl,\rh,Z)$
is the number of GR mergers that a cluster, formed at a redshift $\zf$ with parameters $\mcl$, $\rh$, and $Z$,
produces at the redshift $z$. That way, $\mrgfrac$ allows for taking the mergers' delay time,
$\tmrg$, into account: in this function,
$z={\mathcal Z}({\mathcal Z}^{-1}(\zf)+\tmrg)$, where the function ${\mathcal Z}(t_{\rm age})$
converts the age of the Universe, $t_{\rm age}$, into redshift.
In this study, the age-redshift relation
corresponding to the standard $\Lambda$CDM cosmological framework \citep{Peebles_1993,Narlikar_2002}
is adopted, where the cosmological constants from the Planck mission \cite{Planck_2018}
(Hubble constant $H_0=67.4\kmps{\rm~Mpc}^{-1}$, matter content $\Omega_{\rm~m}=0.315$, flat Universe,
for which $\thub=13.79{\rm~Gyr}$) are applied. Note that $\mrgfrac$ effectively encodes the merger
efficiency $\etamrg$.

The quantity $\ncl$ is the number of star clusters formed
per unit comoving volume. The quantity $\delage(\delz(z))$ is the age range (or age bin) of the Universe
corresponding to the redshift bin $\delz(z)$ around the redshift of interest $z$.
As for the integral limits, $X_l$ and $X_u$ represent the lower and upper limits of the
integration variable $X$.

In practice, due to the discrete model grid points and a limited number of mergers (155 of them)
from them, a Monte-Carlo approach is adopted to tackle the multi-dimensional integral
in Eqn.~\ref{eq:rate0}. Here, $\nsamp$ star clusters are randomly selected such that
\begin{flalign}
\mcl \in \clmf(\mcl) \propto \mcl^{-2} & {\rm ~~~}10^4\Ms \leq \mcl \leq 10^5\Ms,\label{eq:mcldist}  \\
\rh  \in {\mathcal U}(0,1)(\rhhigh-\rhlow) + \rhlow & {\rm ~~~}\rhlow=1{\rm ~pc},{\rm ~}\rhhigh=3{\rm ~pc},\label{eq:rhdist}
\end{flalign}
where ${\mathcal U}(0,1)$ is the uniform distribution over $[0,1]$. The $\mcl^{-2}$ dependence
is supported by observations of young clusters in the Milky Way and nearby galaxies
\cite{Gieles_2006b,Larsen_2009,PortegiesZwart_2010,Bastian_2012}.
The formation redshift is chosen such that
\begin{equation}
\zf \in \sfh(\zf) \propto \frac{(1+\zf)^{2.6}}{1+[(1+\zf)/3.2]^{6.2}}, 
\label{eq:zfdist}
\end{equation}
which is the redshift dependence of star formation rate density as of Ref.~\cite{Madau_2017}.

As for the redshift-dependent metallicity distribution, a variety of observation-based $zZ$ dependence,
$\phi(zZ)$, have been proposed. A commonly used relation that is considered here is the observation-based
algebraic redshift-metallicity relation of
Ref.~\cite{Madau_2017} superimposed with a log-normal scatter in metallicity of 0.5 dex
(as in, \eg, Refs.~\cite{Belczynski_2016,Chatterjee_2017b}), \ie,
\begin{equation}
\log_{10}(Z(z)/\Zs) \in {\mathcal N}(\mu=0.153-0.074z^{1.34}, \sigma=0.5),
\label{eq:mdz}
\end{equation}
where ${\mathcal N}(\mu,\sigma)$ is the normal distribution with mean $\mu$ and dispersion $\sigma$.
The above spread in $Z$ is based on observations of damped Ly$\alpha$ systems up to high
redshifts \cite{Rafelski_2012}. This $zZ$ dependence is termed MF17 in this study.

Due to the importance of cosmic metallicity evolution 
on GW source formation and GW astrophysics (see Ref.~\cite{Chruslinska_2022} for a review),
more recent, observation-based $zZ$ dependences of Refs.~\cite{Chruslinska_2019}
and \cite{Chruslinska_2021} are also considered in this study. These latter $zZ$ dependences
explicitly provide observationally derived metallicity scatter in star
forming regions up to redshift 10. Among the wide variety of $\phizz(z,Z)$ provided
in those works, the variants {\tt Moderate-Z} (hereafter CN19) from Ref.~\cite{Chruslinska_2019}
and {\tt 307f14SBBiC\_FMR270} (hereafter CN21) from Ref.~\cite{Chruslinska_2021} 
are considered here due to their non-extreme nature and the inclusion
of the latest observations. Sample draws from these three $\phizz(z,Z)$ schemes
are shown in Fig.~\ref{fig:zZsample}\footnote{These $zZ$ samples are
generated utilising the scripts provided in Refs.~\cite{Chruslinska_2019,Chruslinska_2021}.}.
As seen in Fig.~\ref{fig:zZsample},
the metallicity declines with increasing redshift in all three cases. However,
the CN19, MF17, and CN21 schemes produce, in order, increasingly higher mean metallicity and
smaller scatter in metallicity. That way, these choices allow for some leeway
to explore potential variations in the redshift-metallicity dependence, consistently
with high-redshift observations. Since all the above $zZ$-schemes incorporate
observations beginning from the local Universe up to redshift 10, the lower and upper limits of the
cluster-formation redshift are set to be $\zflow=0$ and $\zfhigh=10$.

After each random draw of a star cluster's initial and formation parameters from the various distributions
as described above, the parameters $\mcl$, $\rh$, and $Z$ are discretised based on the available
grid points (Sec.~\ref{grid}, Figs.~\ref{fig:grid}), and the
GW coalescences (if any) from the corresponding model cluster are individually mapped at the appropriate
redshifts based on the events' $\tmrg$, the cluster's $\zf$, and the background
cosmological model (see above). The redshift-dependent intrinsic merger rate density is then
given by
\begin{equation}
\rate(z) \approx \left(\frac{\Delta\nmrg(\delz(z))}{\nsamp}\right)\left(\frac{1}{\delage(\delz(z))}\right)\\
\left[\frac{\int_{\mcllow}^{\mclhigh}\clmf(\mcl)d\mcl}{\int_{\mgclow}^{\mgchigh}\clmf(\mcl)d\mcl}\right]
\left[\frac{\int_{10}^{0}\sfh(\zf)d\zf}{\int_6^3\sfh(\zf)d\zf}\right]\ngc,
\label{eq:rate}
\end{equation}
where $\Delta\nmrg(\delz(z))$ is the number of events contained within the redshift bin $\delz(z)$.
The quantity $\ngc$ is the comoving spatial number density of GCs, with respect to which $\ncl$ in
Eqn.~\ref{eq:rate0} is scaled. In Eqn.~\ref{eq:rate}, this scaling is done in the same way as
in Ref.~\cite{Banerjee_2021} (see their Eqn.~5). To summarise, $\ngc$ is boosted to $\ncl$ due to
(a) newly formed star clusters extending to lower masses than the GCs and (b) cluster formation across
masses extending up to the present day, as opposed to the present-day GCs
that are old objects.

Due to the mass range of the model clusters with which the Model Universe is populated,  
$[\mcllow,\mclhigh]=[10^4\Ms,10^5\Ms]$ is set in Eqn.~\ref{eq:rate}. As in Ref.~\cite{Banerjee_2021},
the birth mass range of GCs, $[\mgclow,\mgchigh]$, is chosen based on the initial masses of the GC models
in the CMC Cluster Catalog \cite{Kremer_2020} (hereafter {\tt CCC}), which models reasonably represent the observed
present-day Milky Way GC population \cite{Ye_2025,Kremer_2025}.
In {\tt CCC}, the typical present-day GC masses were obtained by evolving from 
initial masses within $[\mgclow,\mgchigh]=[5\times10^5\Ms,1\times10^6\Ms]$, which
mass range is taken here for evaluating a `reference' $\rate(z)$ from Eqn.~\ref{eq:rate}. The entire
Milky Way GC mass range (\ie, including the least and most massive GCs) was obtained
from $[\mgclow,\mgchigh]=[1\times10^5\Ms,2\times10^6\Ms]$, which mass range
is used here for obtaining a `pessimistic' rate, $\rpess(z)$. The formation redshift range
of GCs in the denominator of Eqn.~\ref{eq:rate} is taken to be
$3.0\lesssim\zf\lesssim6.0$ based on cosmological structure formation modelling \citep{ElBadry_2019b}.
As for the spatial density of GCs, the observationally determined value of
$\ngc\approx8.4h^3{\rm Mpc}^{-3}\approx2.57{\rm Mpc}^{-3}$ \cite{PortegiesZwart_2000}
($h\equiv H_0/[100\kmps{\rm~Mpc}^{-1}]=0.674$ \cite{Planck_2018}) is applied for obtaining
the reference $\rate(z)$ and $\ngc\approx0.33{\rm Mpc}^{-3}$ \cite{Rodriguez_2015} is applied
for the pessimistic $\rpess(z)$.
While the above approach makes the merger rate estimation independent of explicitly assuming
a cluster formation efficiency, it implicitly assumes that (i) star cluster formation in the Universe follows
the cosmic star formation history and (ii) star clusters are formed at all redshifts with
the same universal mass distribution.
See Refs.~\cite{Banerjee_2021,Banerjee_2021b} for detailed discussions on these choices and assumptions.

If $p(A)|_z$ is the normalised density distribution of a GW observable, $A$, for the events
in the redshift bin $\delz(z)$, then the differential intrinsic merger rate density with respect to $A$ at
redshift $z$ is
\begin{equation}
\left.\frac{d\rate}{dA}(A)\right|_z = \rate(z)p(A)|_z 
\label{eq:diffrate0}
\end{equation}
If the number of events detected within the $i$-th $A$-bin of width $\Delta A_i$
around $A=A_i$ is $\Delta N_{A,i}|_z$ ($\sum_i\Delta N_{A,i}|_z=\Delta\nmrg(\delz(z))$) then
\begin{equation}
\left.\frac{d\rate}{dA}(A_i)\right|_z \approx
\rate(z) \frac{\Delta N_{A,i}|_z}{\Delta A_i \Delta\nmrg(\delz(z))}.
\label{eq:diffrate}
\end{equation}

\FloatBarrier

\section{Assigning BH spins and BBH spin-orbit tilts}\label{bhspin}

During the N-body simulations performed here, dimensionless spin magnitude (or Kerr parameter \cite{Kerr_1963})
is assigned runtime to all newly formed
or mass-gained stellar-remnant BHs. For a BH that is derived from a single star (that either is
evolved from ZAMS or is a rejuvenated star-star merger product) or from a non-interacting
stellar member of a binary or higher-multiplicity system inside the cluster, a small natal spin magnitude
is assigned based on hydrodynamic calculations of BH formation from single stars.
In particular, the `MESA' BH-spin model of Ref.~\cite{Belczynski_2020} is adopted in the
present N-body simulations, which prescription assigns a BH-spin within $0.05\lesssim a \lesssim0.15$
depending on the properties of the pre-core-collapse star and the metallicity. The small BH spin  
is due to the dynamo-driven efficient core-to-envelope angular momentum transport \cite{Spruit_2002}
in the stellar evolution models computed in Ref.~\cite{Belczynski_2020}.
In fact, Ref.~\cite{Fuller_2019a} modelled a rather extreme version of the dynamo-driven angular momentum
extraction, where the BHs were formed with nearly vanishing spin, $a\sim10^{-1}$. However,
the higher, standard-dynamo-based BH spin is preferred in this work since the to-date-observed GW events
tend to favour small but non-zero magnitudes of $a\sim10^{-1}$ \cite{Callister_2024a,GWTC4a_pop}.

In the current modelling, the spin magnitude of a BH can, however, be higher than this dynamo-driven
value, \ie, the BH can be `spun-up' due to
(a) tidal spin-up of the BH's WR-star progenitor in a close, tidally interacting binary
\cite{Qin_2018,Belczynski_2020,Bavera_2020b},
(b) matter accretion onto the BH due to a BH-star merger and/or mass transfer from a binary companion
\cite{Qin_2019,Kiroglu_2025b},
(c) GW coalescence with another BH \cite{Baker_2008,Hofmann_2016,JimenezForteza_2017}. In
the present $\nbseven$ version, no recipe or algorithm is yet implemented for runtime BH spin
alteration (boost) via channels (a) and (b), although these channels themselves are identified during a run.
The BHs that are outcomes of channels (a) and (b) are simply tagged runtime as maximally
spinning BHs, \ie, with $a=1$. At present, no Kerr parameter is assigned to NS members,
and any GR merger involving an NS is excluded from the calculation of spin-related parameters
such as $\xeff$. Improvements in the treatments are planned in the near future.

As for BBH mergers happening inside a model cluster (\ie, channel (c)), the GR recoil kick and the
final spin of the merged BH are
obtained based on analytical fits to NR BBH merger calculations as of Refs.~\cite{Lousto_2012,Hofmann_2016}.
The recoil velocity and final spin depend on both the BH components' spin magnitude and
direction relative to the BBH's orbital angular momentum.
Here, the spin-orbit tilt and azimuthal angles are drawn randomly from a full isotropic distribution if the merging
BBH is purely dynamically assembled. A partial alignment is applied with a truncated
isotropic tilt distribution if the BBH's components were members of the same primordial binary. Notably,
inside a cluster, a BH's spin can be modified by a combination of the above channels. The reader
is directed to Refs.~\cite{Banerjee_2020,Banerjee_2020c} for all the details and the 
formulae of the above implementations in the $\nbseven$ code.

However, the presence of by-construction maximally spinning BH members in a population of
merging BBHs can lead to artefacts in, \eg, the population's $\xeff$ distribution.
In this work, this is addressed by reassigning post-run the spins of those BHs that are
outcomes of channels (a) and (b). Spin alteration of a BH via mass accretion is rather poorly understood
(\eg, \cite{Berti_2008,Perego_2009,Cenci_2021});
hence for channel (b) the spin magnitude of the mass-accreted BH is reassigned tacitly from a uniform distribution
between $0.5\leq a \leq 1.0$. Refs.~\cite{Li_2024b,Banagiri_2025} provide indirect support to this range of spin for high-spin BHs. 

Those BHs that are derived via channel (a) are mainly primordially paired, since such close,
tidally interacting binaries (orbital period $\lesssim 1$ day) would rarely undergo close dynamical encounters \cite{Heggie_2003}.
Therefore, BH spins and spin-orbit tilts of all the primordially paired mergers in a BBH merger population
are replaced with those from a separate isolated binary population evolution calculation, where
each binary is evolved independently without undergoing any dynamical encounter. 
For consistency, this isolated binary evolution modelling is performed
with a standalone version of the same $\bse$ binary evolution code that is coupled with $\nbseven$ (Sec.~\ref{nbsims}).
However, this version, in addition, explicitly takes into account tidal spin-up of WR progenitors
while assigning spin magnitudes to newly formed BHs. Furthermore, the spin-orbit tilts are
determined based on the BHs' natal kicks. This version of $\bse$ is described in detail
in Ref.~\cite{Banerjee_2024}. In particular, the spin magnitudes are determined by their
Eqns.~4-6 and the tilts are given by their Eqn.~10.

The isolated binary evolution calculations are carried out with an initial massive
binary population whose distributions match those of the initial (primordial) massive binary population
in the N-body models computed here (Sec.~\ref{nbsims}). Some $2\times10^6$ binaries are
evolved for each metallicity in the N-body model grid, \ie, at $Z=0.0002$, 0.001, 0.005,
0.01, 0.02. Each set is evolved under identical model choices as for the N-body runs (Sec.~\ref{nbsims}), \ie,
delayed remnant mass model with PPSN/PSN, `MESA' (non-spun-up) BH spin model,
common envelope efficiency parameter $\ace=1.0$ \cite{Hurley_2002,Ivanova_2013},
$\bse$'s default binding energy parameter \cite{Hurley_2002}, fallback-modulated, momentum-conserving remnant
natal kick scheme \cite{Banerjee_2020}, and Eddington-limited mass accretion. The default binary mass transfer physics
of $\bse$ (\eg, mass transfer stability criteria, angular momentum transport, super-Eddington mass transfer treatment \cite{Hurley_2002})
is applied in these isolated binary evolution runs (although alternative treatments of
mass transfer physics were also demonstrated in Ref.~\cite{Banerjee_2024}), as for the N-body runs.

These isolated binary evolutionary models produce a population of merging BBHs whose members can potentially
be spun-up depending on the BBH's parent binary's evolution history. If not spun-up, then a merging BH member possesses the
appropriate dynamo-driven spin magnitude; see Ref.~\cite{Banerjee_2024} for the details. For every primordially
paired BBH merger originating from a cluster of metallicity $Z^\prime$, the member BHs' spin magnitudes and spin-orbit
tilts are reassigned by picking the spin pair, $a_1$, $a_2$, and the tilt pair, $\nu_1$, $\nu_2$, of a 
merging BBH chosen randomly out of the isolated-binary set of metallicity $Z^\prime$. That way, the reassigned
spin magnitudes statistically incorporate the effect of possible tidal spin-up of BH-progenitor WR stars.

Since the formation of the BH members, a primordially paired in-cluster or ejected merging BBH  
would generally experience additional spin-orbit tilt (which tilt is common to both members)
due to close dynamical encounters. This would make the BBH more misaligned than its isolated-binary
counterpart \cite{Trani_2021}. In this study, this effect is taken into account by vector-adding
a spin-orbit tilt, selected randomly from the dynamical tilt distribution of Ref.~\cite{Trani_2021} (corresponding
to the appropriate metallicity), to both members of a primordially paired BBH. The procedure followed
for this is detailed in Ref.~\cite{Banerjee_2023}. This reassignment
is done after the first set of reassignment from the isolated-binary models (see above). These spin and tilt
reassignments are done for both the BBH merger events extracted directly from the cluster models (Sec.~\ref{gwtc})
and the Model Universe BBH events across redshifts (Appendix~\ref{rate}). 

For all in-cluster or ejected BBH mergers that are dynamically paired, an isotropic spin-orbit tilt distribution is assumed
for both components independently, after reassigning the spins of the high-spin BH candidates (see above).
In other words, for all dynamically paired BBH inspirals, $(\cos\theta_1,\cos\theta_2) \in \unif(-1,1)$, where $\unif(-1,1)$ is
the uniform distribution over $[-1,1]$ (Eqn.~\ref{eq:xeffdef}). This is because members of
stellar entities can, in general, be taken to be uncorrelated over the length scale of a parsec-scale star cluster.

\FloatBarrier

\section{Parametric inference of the model merger primary-mass and
mass-ratio distributions}\label{massmodel}

The intrinsic primary-mass ($\mone$) and mass-ratio
($q\equiv\mtwo/\mone\leq1$) distributions of the model GR merger populations are
quantified with hierarchical, parametric fits of the type employed in the LVK
population analyses \cite{Abbott_GWTC3_prop,GWTC4a_pop}. Two data sets are
analysed: (i) the direct N-body merger sample, \ie, the 155 GR mergers
produced by the computed model clusters of this work
(Appendix~\ref{runlist}), and (ii) the population-synthesis sample, \ie, the
Model-Universe merger population (Appendix~\ref{rate}; CN21 $zZ$ variant), restricted to the redshift bin around
$z=0.2$. The latter comprises 4103 merger samples that repeat 91 unique
$(\mone,\mtwo)$ pairs; the
corresponding Kish effective sample size is $\approx8$.

A single global parametric form of the LVK \plp{} type is found to visibly smear
features of the model population distributions, \eg, the expected pile-up of primary
masses at the PPSN cap at $\mone=40.5\Ms$ (Sec.~\ref{nbsims}). The mass axis
is therefore divided into two regimes at $\msplit=45\Ms$ --- slightly above the
PPSN cap, so that the pile-up lies within the lower-mass regime --- and
independent mass models are inferred on either side. Since the joint population
density factorises as $p(\mone,q)=p_1(\mone)\,p(q|\mone)$ with disjoint
parameter sets, the regime-wise and mass-ratio inferences presented below are
equivalent to a joint fit.

\medskip
\noindent{\it Mass model.---}Each regime is described by a power law plus $K$
Gaussian peaks, a direct generalization of the LVK \plp{} form
\cite{Abbott_GWTC3_prop}:
\beq
p_1(\mone|\Lambda_K)\propto\left[\left(1-\sum_{k=1}^{K}\lambda_k\right)
C\,\mone^{-\alpha}
+\sum_{k=1}^{K}\lambda_k\,{\mathcal N}(\mone;\mu_k,\sigma_k)\right]
S(\mone|m_{\rm min},\delta_m),
\label{eq:plk}
\eeq
truncated at the regime limits and normalised therein numerically. Here $C$ is
the truncated-power-law normalisation and $S$ is the standard LVK low-mass
smoothing window (Appendix~B of Ref.~\cite{Abbott_GWTC3_prop}). The lower
regime is supported on $[m_{\rm min},\msplit]$ with $K=2$ or $3$; the upper
regime on $[\msplit,m_{\rm max}]$ with $K=1$ or $2$ and no smoothing at its
(hard, by construction) lower edge. The mass ratio follows the LVK conditional
power law,
\beq
p(q|\mone)\propto q^{\beta_q}\,S(q\mone|m_{{\rm min},q},\delta_{m,q}),
\qquad q\in[m_{{\rm min},q}/\mone,\,1],
\label{eq:pq}
\eeq
normalized per event, where the spectral index is allowed to
differ between the two mass regimes ($\betalow$ for $\mone\leq\msplit$,
$\betahigh$ for $\mone>\msplit$) (see also Ref.~\cite{Banagiri_2025}); the single-index model is recovered for $\betalow=\betahigh$. Since the
model mergers have exact masses and no detector selection is applied, the
hierarchical likelihood reduces to the point-estimate form
$\ln{\mathcal L}=\sum_i w_i \ln p(m_{1,i},q_i|\Lambda)$, with $w_i=1$ for the
direct N-body sample and $w_i=$ the template multiplicity for the
population-synthesis sample.

All priors are uniform (Table~\ref{tab:massmodel}). The peak-location boxes of
a given fit are disjoint, so the peaks are ordered by construction and no
ordering constraint is required; the only support restriction is the mixture
simplex $\sum_k\lambda_k\leq1$, whose exact prior fraction $1/K!$ is removed
from the reported evidences so that Bayes factors compare properly normalized
models. Posteriors and evidences are computed with the nested sampler
{\tt dynesty} \cite{Speagle_2020,Skilling_2006} through the {\tt bilby}
inference library \cite{Ashton_2019} (1000 live points, acceptance-walk
sampling), \ie, the same software stack as used in the LVK population analyses.
Non-parametric, penalised B-spline density fits (in the spirit of the LVK
\bspline{} model \cite{GWTC4a_pop}) are performed per regime as cross-checks
and corroborate all the parametric features discussed below.

\begin{figure*}[!h]
\centering
\includegraphics[width=\textwidth]{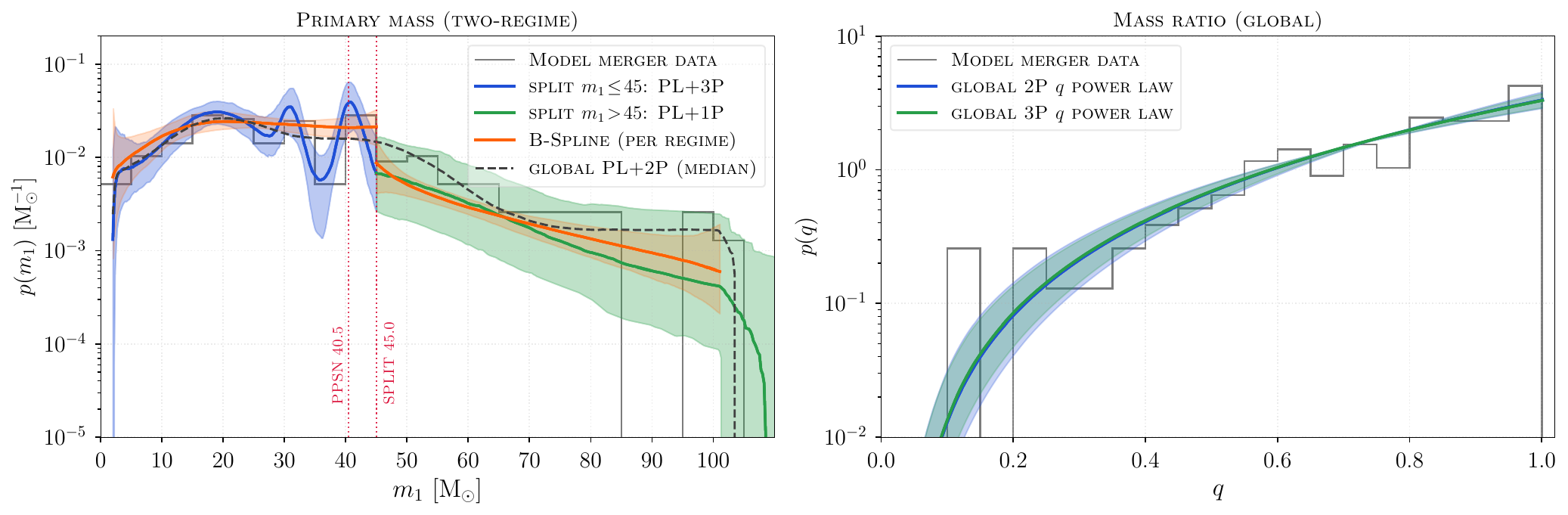}
\caption{Two-regime parametric inference for the direct N-body merger
sample (155 mergers). {\bf Left:} the inferred primary-mass distribution
(median and 90\% credible band) in the lower ($\mone\leq45\Ms$; \plp-type with
$K=3$ peaks) and upper ($\mone>45\Ms$; $K=1$) regimes, renormalised to be
continuous at $\msplit=45\Ms$. Also shown are the
B-spline cross-check (per regime), the global single-model fit, and the
merger-sample histogram. The PPSN cap at $40.5\Ms$ is marked (red dotted line).
{\bf Right:} the corresponding global mass-ratio distribution.}
\label{fig:massmodel_nbody}
\end{figure*}

\begin{figure*}[!h]
\centering
\includegraphics[width=\textwidth]{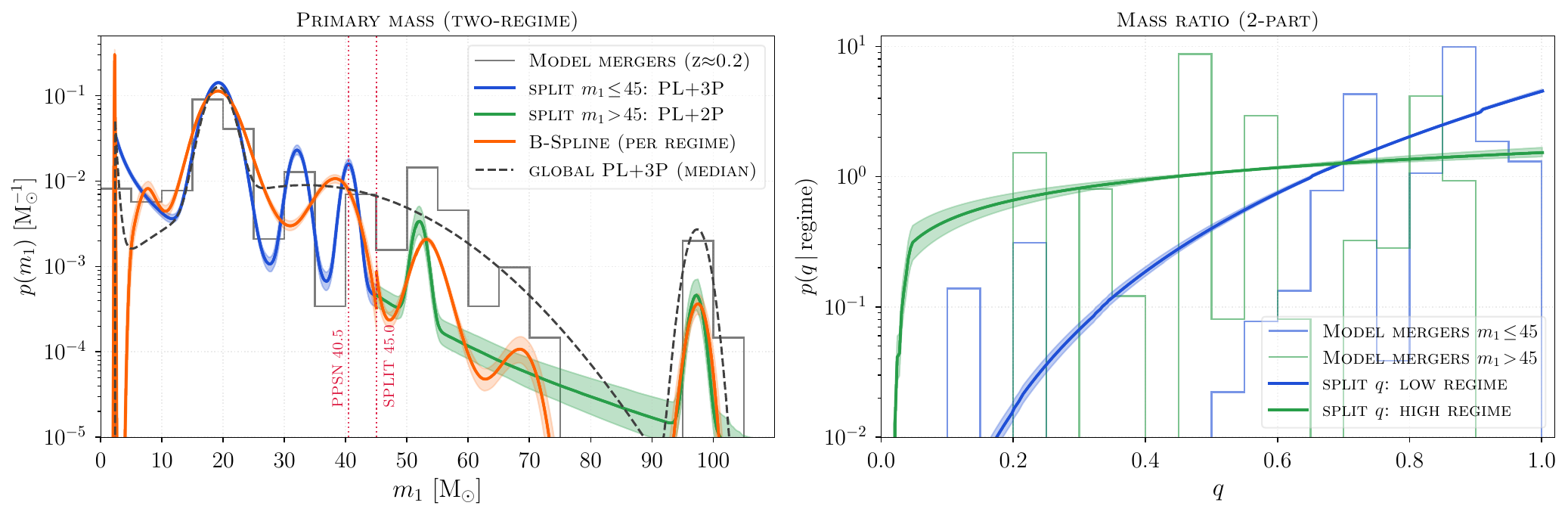}
\caption{Same as Fig.~\ref{fig:massmodel_nbody}, but for the
population-synthesis sample at $z\approx0.2$.
Here the right panel shows the regime-wise mass-ratio
distributions of the two-index model, Eq.~(\ref{eq:pq}).}
\label{fig:massmodel_cn21}
\end{figure*}

\begin{figure*}[!h]
\centering
\includegraphics[width=0.90\textwidth]{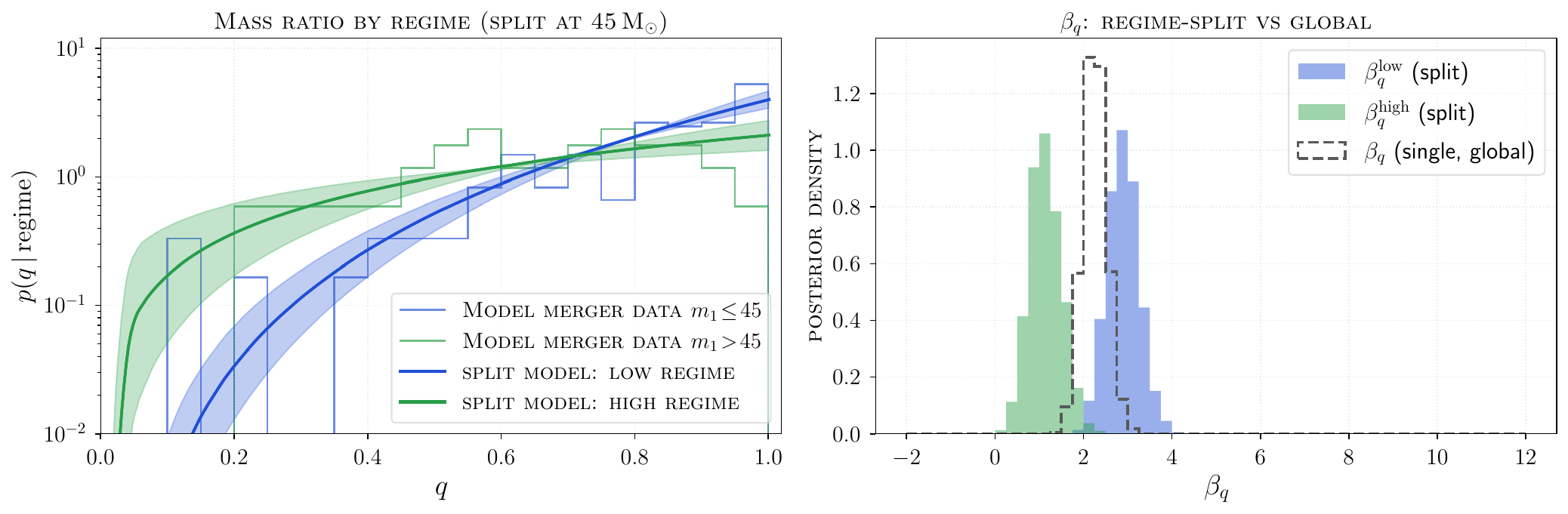}
\includegraphics[width=0.50\textwidth]{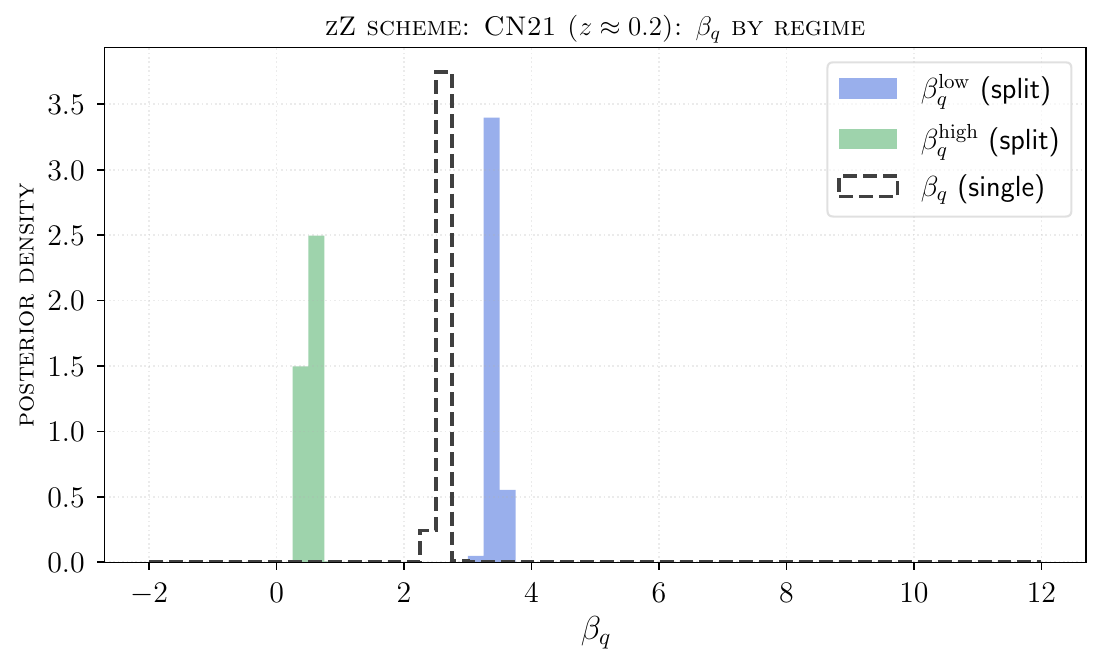}
\caption{Posterior distributions of the regime-wise mass-ratio spectral
indices, $\betalow$ ($\mone\leq45\Ms$) and $\betahigh$ ($\mone>45\Ms$), compared
with the single global index (dashed), for the direct N-body sample ({\bf
upper}; the left-hand sub-panel additionally shows the corresponding regime-wise
$q$ distributions) and the population-synthesis sample ({\bf lower}). In both
data sets, the upper mass regime prefers a distinctly flatter mass-ratio
distribution.}
\label{fig:qsplit}
\end{figure*}

\medskip
\noindent{\it Results.---}Figures~\ref{fig:massmodel_nbody} and
\ref{fig:massmodel_cn21} show the inferred two-regime $\mone$ distributions
(continuity-matched at \msplit; left panels) and the regime-wise $q$
distributions (right panels); Fig.~\ref{fig:qsplit} shows the corresponding
$\beta_q$ posteriors. Table~\ref{tab:massmodel} lists the priors and the
posteriors of the preferred models. For the direct N-body sample (121 and 34
mergers below and above \msplit, respectively), the lower
regime prefers three peaks with a (constraint-corrected) Bayes factor
$\ln{\mathcal B}_{3P/2P}=+4.3$: a broad $\mu_1\approx18.9\Ms$ bulk, a narrow
$\mu_2\approx30.9\Ms$ cluster, and a narrow peak at $\mu_3\approx40.9\Ms$ that
captures the PPSN pile-up. The upper regime is adequately described by a single
broad peak ($\mu_1\approx54\Ms$) over a declining power law
($\ln{\mathcal B}_{2P/1P}=-0.3$), with $m_{\rm max}$ constrained only from below
($\gtrsim101\Ms$, set by the single most massive merger). The mass-ratio index
differs between the regimes at high credibility:
$\betalow=2.90^{+0.60}_{-0.56}$ versus $\betahigh=1.12^{+0.65}_{-0.54}$
[$P(\betahigh<\betalow)=99.8\%$], with
$\ln{\mathcal B}_{\rm split/single}=+2.8$; \ie, the post-PPSN-cap or mass-gap mergers pair
distinctly more asymmetrically than the lower-mass population, as expected for
such over-massive BHs pairing (dynamically) with lower mass BHs derivable from single star evolution (see Sec.~\ref{nbsims}).
The population-synthesis slice (3607 and 496 weighted samples below and above
\msplit) yields the analogous decomposition --- peaks at
$19.2$, $32.1$, and $40.5\Ms$ below \msplit{} (the PPSN pile-up again), narrow
peaks at $52.0$ and $97.1\Ms$ above, and $\betalow=3.42^{+0.13}_{-0.12}$ versus
$\betahigh=0.52^{+0.12}_{-0.12}$ --- with counts-weighted Bayes factors
$\ln{\mathcal B}=+192$, $+68$, and $+294$ for the three comparisons which,
deflated by the per-regime information inflation $N_{\rm raw}/N_{\rm eff}$
($\approx590$, $110$, $540$), correspond to only $+0.3$ to $+0.6$ per effective
sample and are therefore descriptive.
Note that the low inferred $m_{\rm min}$ and $m_{{\rm min},q}$ (Table~\ref{tab:massmodel}) are driven by the
NS-bearing events: the NS-NS  ($2.28\Ms+2.07\Ms$) and the NS-BH ($2.41\Ms+10.30\Ms$) mergers.
The prior ranges of these parameters lie at or below the NS
mass range in the present computations (remnants of $\leq2.5\Ms$). Since no model merger populates masses below these
events, both parameters are bounded from above by them and are otherwise only weakly constrained: for the direct N-body
sample they remain broadly distributed within their prior, whereas for the population-synthesis sample $m_{\rm min}$ pins at the prior's upper edge and
$m_{{\rm min},q}$ tracks the lowest secondary mass of that sample. These parameters, therefore, describe the low-mass
turn-on of the fits and are not to be read as a minimum mass of the model BBH population.

\medskip
\noindent{\it Comparison with the LVK population.---}The inferred model
distributions are compared with the GWTC-4.0 (O4a) astrophysical population
\cite{GWTC4a_pop} through the JS divergence \cite{Lin_1991}
(base-2 logarithm; $0\leq{\rm JS}\leq1$~bit), computed between the normalised
model and the publicly released LVK \plp{}, \bplpp{},
and \bspline{} curves. Since the present model mass distribution is devoid of
the dominant $\mone\approx10\Ms$ peak of the LVK's mass distribution (Sec.~\ref{GWTC_pop}), the present $\mone$-distribution
comparison is restricted to  $\mone>45\Ms$ (renormalised on $[45,300]\Ms$), while the
$q$ comparison uses the full range. The results are listed in
Table~\ref{tab:jsd}. For the direct N-body sample, the high-mass conditional
$\mone$ distribution diverges from the LVK models by only $0.05-0.07$~bit ---
comparable to the $0.02-0.06$~bit spread among the LVK models themselves ---
and the $q$ distribution by $0.02-0.09$~bit, again within the LVK internal
spread (which reaches $0.21$~bit between the LVK \plp{} and \bspline{} $q$
curves). In other words, at the current measurement precision, the model
high-mass ($\mone>45\Ms$) primary-mass shape and the full-range mass-ratio
distribution are statistically consistent with the observed BBH population.
The larger $\mone$ divergences of the population-synthesis slice
($0.23-0.27$~bit) are dominated by its finite-template discreteness rather
than by its underlying shape, as indicated by the equally large divergence
($0.23$~bit) between the slice and its own parent N-body distribution,
alongside their near-identical $q$ predictions ($0.005$~bit).
All reported JS divergences are based on pairwise draws from
full posterior samples.

\begingroup
\setlength{\tabcolsep}{5.0pt}
\renewcommand{\arraystretch}{1.25}
\begin{table*}[!h]
\centering
\caption{Priors and marginalised posteriors (median and 90\% credible
interval) of the preferred two-regime mass models and of the two-index
mass-ratio model, for the direct N-body merger sample and the $z\approx0.2$
population-synthesis (CN21) sample. Masses are in $\Ms$. The $m_{\rm max}$ prior lower limit is $100.9\Ms$
($101.1\Ms$) for the N-body (CN21) sample, just above the respective most
massive merger; in both data sets $m_{\rm max}$ is constrained only from
below. Where two $\sigma_1$ priors are quoted, they refer to the $K=1$ and
$K=2$ upper-regime fits, respectively.}
\label{tab:massmodel}
\begin{tabular}{llcc}
\toprule
Parameter & Prior & N-body & CN21 $z\approx0.2$ \\
\midrule
\multicolumn{4}{l}{{\it Lower regime, $\mone\in[m_{\rm min},45\Ms]$: \plp-type, $K=3$ (preferred)}}\\
$\alpha$          & U$(-4,8)$        & $0.50^{+6.01}_{-3.39}$   & $1.47^{+0.13}_{-0.13}$ \\
$m_{\rm min}$     & U$(1,2.25)$      & $1.37^{+0.60}_{-0.34}$   & $2.24^{+0.01}_{-0.05}$ \\
$\delta_m$        & U$(0,10)$        & $1.85^{+3.77}_{-1.69}$   & $0.03^{+0.05}_{-0.02}$ \\
$\lambda_1$       & U$(0,1)$         & $0.47^{+0.18}_{-0.21}$   & $0.750^{+0.013}_{-0.015}$ \\
$\mu_1$           & U$(8,26)$        & $18.9^{+2.2}_{-2.3}$     & $19.23^{+0.07}_{-0.07}$ \\
$\sigma_1$        & U$(1,8)$         & $6.3^{+1.5}_{-2.3}$      & $2.11^{+0.07}_{-0.06}$ \\
$\lambda_2$       & U$(0,1)$         & $0.12^{+0.08}_{-0.07}$   & $0.072^{+0.008}_{-0.007}$ \\
$\mu_2$           & U$(26,36)$       & $30.9^{+0.9}_{-2.0}$     & $32.07^{+0.14}_{-0.15}$ \\
$\sigma_2$        & U$(1,6)$         & $1.4^{+2.9}_{-0.4}$      & $1.27^{+0.11}_{-0.10}$ \\
$\lambda_3$       & U$(0,1)$         & $0.17^{+0.09}_{-0.08}$   & $0.039^{+0.006}_{-0.005}$ \\
$\mu_3$           & U$(36,44.5)$     & $40.9^{+1.5}_{-0.8}$     & $40.47^{+0.15}_{-0.14}$ \\
$\sigma_3$        & U$(1,6)$         & $1.6^{+1.7}_{-0.6}$      & $1.007^{+0.024}_{-0.007}$ \\
\midrule
\multicolumn{4}{l}{{\it Upper regime, $\mone\in[45\Ms,m_{\rm max}]$: $K=1$ (N-body) / $K=2$ (CN21) preferred}}\\
$\alpha$          & U$(-4,8)$        & $3.1^{+1.8}_{-3.4}$      & $4.77^{+0.70}_{-0.65}$ \\
$m_{\rm max}$     & U$(100.9|101.1,150)$ & $108^{+33}_{-7}$ & $104^{+15}_{-3}$ \\
$\lambda_1$       & U$(0,1)$         & $0.23^{+0.49}_{-0.22}$   & $0.551^{+0.041}_{-0.041}$ \\
$\mu_1$           & U$(45.5,70)$     & $53.6^{+14.2}_{-7.1}$    & $51.99^{+0.10}_{-0.10}$ \\
$\sigma_1$        & U$(1,12)$/U$(1,10)$ & $8.3^{+3.4}_{-5.9}$   & $1.003^{+0.010}_{-0.003}$ \\
$\lambda_2$       & U$(0,1)$         & ---                      & $0.082^{+0.022}_{-0.019}$ \\
$\mu_2$           & U$(70,100.5)$    & ---                      & $97.14^{+0.28}_{-0.28}$ \\
$\sigma_2$        & U$(1,12)$        & ---                      & $1.02^{+0.08}_{-0.02}$ \\
\midrule
\multicolumn{4}{l}{{\it Mass ratio, Eq.~(\ref{eq:pq}), two-index (preferred)}}\\
$\betalow$        & U$(-4,12)$       & $2.90^{+0.60}_{-0.56}$   & $3.42^{+0.13}_{-0.12}$ \\
$\betahigh$       & U$(-4,12)$       & $1.12^{+0.65}_{-0.54}$   & $0.52^{+0.12}_{-0.12}$ \\
$m_{{\rm min},q}$ & U$(1,2.25)$      & $1.40^{+0.48}_{-0.37}$   & $2.066^{+0.003}_{-0.009}$ \\
$\delta_{m,q}$    & U$(0,10)$        & $1.15^{+1.81}_{-1.05}$   & $0.003^{+0.010}_{-0.003}$ \\
\bottomrule
\end{tabular}
\end{table*}
\endgroup

\begingroup
\setlength{\tabcolsep}{6.0pt}
\renewcommand{\arraystretch}{1.25}
\begin{table*}[!h]
\centering
\caption{Values of JS divergence (bits; median and 90\% interval over
posterior-draw pairs) between the inferred model distributions and the GWTC-4
population models \cite{GWTC4a_pop}. The $\mone$ comparison is between the
conditional distributions $p(\mone|\mone>45\Ms)$; the $q$ comparison uses the
full range. For comparison scale, the last row block gives the internal divergences among the three
LVK models.}
\label{tab:jsd}
\begin{tabular}{llccc}
\toprule
 & Model sample & vs \plp & vs \bplpp & vs \bspline \\
\midrule
\multirow{2}{*}{$\mone\,(>45\Ms)$}
 & N-body           & $0.065^{+0.042}_{-0.032}$ & $0.053^{+0.054}_{-0.037}$ & $0.066^{+0.092}_{-0.044}$ \\
 & CN21 $z\approx0.2$ & $0.266^{+0.034}_{-0.028}$ & $0.229^{+0.057}_{-0.031}$ & $0.249^{+0.094}_{-0.052}$ \\
\midrule
\multirow{2}{*}{$q$ (full range)}
 & N-body           & $0.043^{+0.041}_{-0.027}$ & $0.017^{+0.032}_{-0.010}$ & $0.093^{+0.065}_{-0.051}$ \\
 & CN21 $z\approx0.2$ & $0.023^{+0.027}_{-0.016}$ & $0.017^{+0.026}_{-0.010}$ & $0.125^{+0.071}_{-0.067}$ \\
\midrule
\multicolumn{5}{l}{{\it Reference (LVK internal), $\mone(>45\Ms)$\,/\,$q$:}\quad
\plp{} vs \bplpp: $0.024^{+0.068}_{-0.023}$\,/\,$0.032^{+0.068}_{-0.028}$;}\\
\multicolumn{5}{l}{\plp{} vs \bspline: $0.059^{+0.071}_{-0.033}$\,/\,$0.214^{+0.104}_{-0.094}$;\qquad
\bplpp{} vs \bspline: $0.044^{+0.092}_{-0.030}$\,/\,$0.103^{+0.141}_{-0.069}$}\\
\bottomrule
\end{tabular}
\end{table*}
\endgroup

\FloatBarrier

\section{Parametric inference of the model population $\xeff$ distribution}\label{xeffinfer}

In this appendix, parametric fitting is performed on the $\xeff$ distribution of the Model-Universe BBH merger population
(CN21 $zZ$ variant with dynamical tilt; see Secs.~\ref{zevol}, Fig.~\ref{fig:xeffz_cn21}, Appendix~\ref{bhspin}). Two samples are analysed: the
redshift bin nearest $z=0.2$ (the slice used in Appendix~\ref{massmodel}) and a random subsample across all redshifts
($N=5\times10^4$).

\begin{figure*}[!h]
\centering
\includegraphics[width=0.72\textwidth]{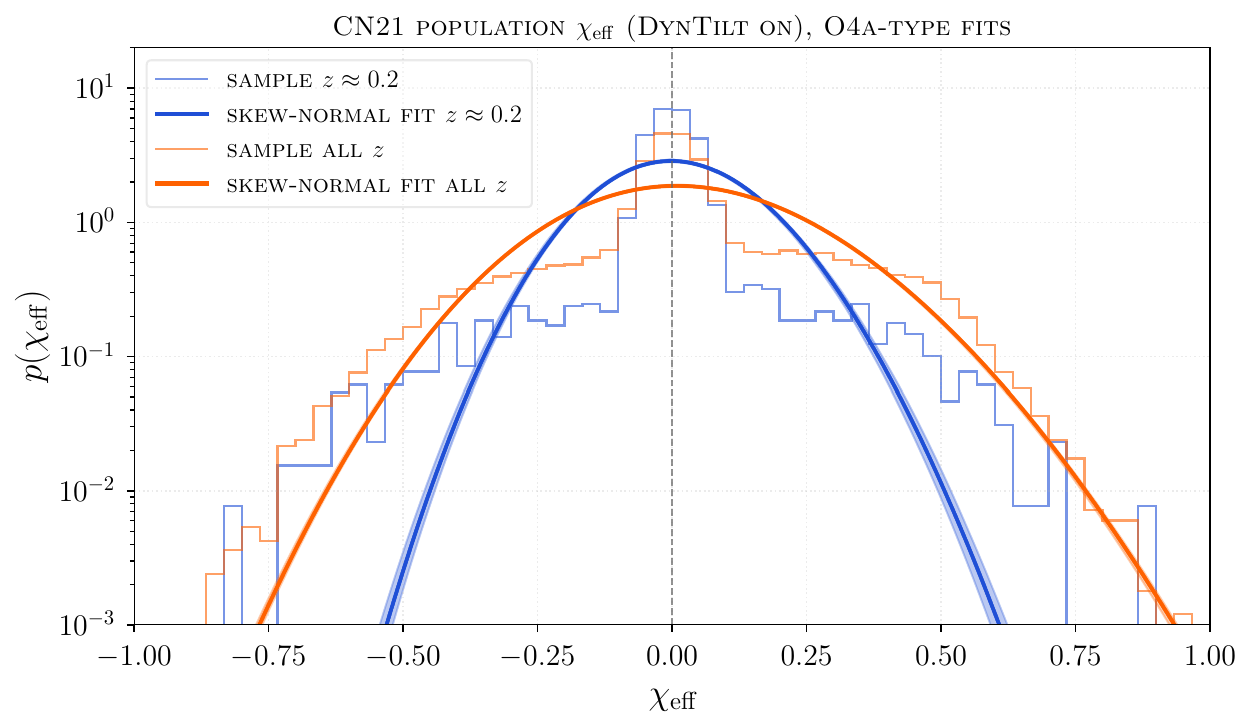}
\caption{Population $\xeff$ distribution of the CN21 Model Universe BBH mergers: shown are the $z\approx0.2$ slice and the all-redshift samples
(histograms) with the corresponding truncated-skew-normal posterior bands. Note the sharp
$\xeff\approx0$ core atop broad shoulders, and the broadening of the
distribution with redshift.}
\label{fig:xeffinfer}
\end{figure*}

Following the O4a population analysis \cite{GWTC4a_pop}, the $\xeff$
distribution is fit with the truncated-Gaussian and truncated-skew-normal
population models on $\xeff\in[-1,1]$,
\beq
p(\xeff)\propto\frac{2}{\sigma_\chi}\,
\varphi(t)\,\Phi(\eta_\chi t),\qquad t=\frac{\xeff-\mu_\chi}{\sigma_\chi},
\label{eq:skewnorm}
\eeq
[$\varphi$, $\Phi$ the unit-normal pdf and cdf,
$\Phi(x)=\frac{1}{2}\left\{1+{\rm erf}\left(x/\sqrt{2}\right)\right\}$;
$\eta_\chi=0$ recovers the Gaussian], using the point-estimate likelihood, uniform priors
[$\mu_\chi\sim{\rm U}(-1,1)$, $\sigma_\chi\sim{\rm U}(0.01,2)$,
$\eta_\chi\sim{\rm U}(-10,10)$], and the same {\tt bilby}/{\tt dynesty}
setup as Appendix~\ref{massmodel}. The inferred parameters of both models
are listed in Table~\ref{tab:xeffpars}. The skew-normal is strongly
preferred over the Gaussian ($\ln{\mathcal B}=+13.5$ and $+124$ for the
$z\approx0.2$ and all-$z$ samples, respectively): a slightly negative centered but
positively skewed distribution, as seen in Fig.~\ref{fig:xeffinfer} (\cf Fig.~\ref{fig:xeffz_cn21}). However, the model
population additionally exhibits a sharp $\xeff\approx0$ core (populated by the dominant low-natal-spin, isotropic,
dynamically paired mergers) atop the broad shoulders --- a two-component structure that no single
skew-normal can represent.

Table~\ref{tab:jsd_xeff} lists the JS divergences \cite{Lin_1991} between
the model and the public GWTC-4 $\xeff$ population posteriors
(\texttt{SkewNormal}, \texttt{EpsSkewNormal}, and \bspline{} variants;
conventions of Table~\ref{tab:jsd}). The fitted model diverges from the
LVK parametric variants by $0.10-0.21$~bit --- similar magnitude as the
internal disagreement between the LVK parametric and \bspline{} $\xeff$
inferences ($0.11-0.13$~bit). On the other hand, the agreement with the LVK
\bspline{} distribution is excellent: $0.02-0.03$~bit for
the fitted models and $0.055$~bit for the raw all-$z$ distribution. The profile of the
model-population $\xeff$ distribution is therefore consistent with the
observed O4a population at the level of the LVK analyses' own
systematics, agreeing best with the non-parametric inference.
The one exception (the raw distribution of the $z\approx0.2$ slice; JS $\approx0.14$ against
all LVK variants) traces the sharp $\xeff\approx0$ core, a
structure of the model $\xeff$ distribution that current GW observations may not resolve.

\begingroup
\setlength{\tabcolsep}{7.0pt}
\renewcommand{\arraystretch}{1.25}
\begin{table*}[!h]
\centering
\caption{Prior ranges and marginalised posteriors (median and 90\% credible
interval) of the $\xeff$ population models' ( Eq.~(\ref{eq:skewnorm}))
fit to the CN21 Model-Universe merger population: the truncated Gaussian
($\eta_\chi\equiv0$) and the truncated skew-normal, for the $z\approx0.2$
slice and the all-redshift sample.}
\label{tab:xeffpars}
\begin{tabular}{llcc}
\toprule
Parameter & Prior & $z\approx0.2$ & all $z$ \\
\midrule
\multicolumn{4}{l}{{\it Truncated Gaussian}}\\
$\mu_\chi$    & U$(-1,1)$    & $0.005^{+0.004}_{-0.004}$  & $0.024^{+0.002}_{-0.002}$ \\
$\sigma_\chi$ & U$(0.01,2)$  & $0.142^{+0.003}_{-0.003}$  & $0.217^{+0.001}_{-0.001}$ \\
\midrule
\multicolumn{4}{l}{{\it Truncated skew-normal (preferred)}}\\
$\mu_\chi$    & U$(-1,1)$    & $-0.085^{+0.010}_{-0.009}$ & $-0.127^{+0.005}_{-0.005}$ \\
$\sigma_\chi$ & U$(0.01,2)$  & $0.168^{+0.006}_{-0.006}$  & $0.264^{+0.003}_{-0.003}$ \\
$\eta_\chi$   & U$(-10,10)$  & $0.92^{+0.11}_{-0.12}$     & $1.03^{+0.05}_{-0.05}$ \\
\midrule
\multicolumn{2}{l}{$\ln{\mathcal B}$ (skew vs Gaussian)} & $+13.5$ & $+124.1$ \\
\bottomrule
\end{tabular}
\end{table*}
\endgroup

\begingroup
\setlength{\tabcolsep}{6.0pt}
\renewcommand{\arraystretch}{1.25}
\begin{table*}[!h]
\centering
\caption{Values of JS divergence (bits; median and 90\% interval over posterior-draw
pairs) between the CN21 model population $\xeff$ distributions and the GWTC-4
$\xeff$ populations \cite{GWTC4a_pop}. The `fitted' rows use the model
skew-normal posteriors, and the `raw' rows use the model
density distributions directly. For comparison, the last row block gives the internal spread among
the LVK models.}
\label{tab:jsd_xeff}
\begin{tabular}{lccc}
\toprule
Model sample & vs {\tt SkewNormal} & vs {\tt EpsSkewNormal} & vs \bspline \\
\midrule
Fitted, $z\approx0.2$ & $0.121^{+0.045}_{-0.053}$ & $0.102^{+0.051}_{-0.063}$ & $0.026^{+0.015}_{-0.011}$ \\
Fitted, all $z$       & $0.206^{+0.052}_{-0.053}$ & $0.188^{+0.062}_{-0.081}$ & $0.021^{+0.020}_{-0.012}$ \\
Raw, $z\approx0.2$    & $0.136^{+0.022}_{-0.019}$ & $0.145^{+0.023}_{-0.020}$ & $0.148^{+0.028}_{-0.029}$ \\
Raw, all $z$          & $0.148^{+0.027}_{-0.019}$ & $0.150^{+0.026}_{-0.026}$ & $0.055^{+0.011}_{-0.010}$ \\
\midrule
\multicolumn{4}{l}{{\it Reference (LVK internal):} {\tt SkewNormal} vs {\tt EpsSkewNormal}: $0.026^{+0.060}_{-0.021}$;}\\
\multicolumn{4}{l}{{\tt SkewNormal} vs \bspline: $0.125^{+0.057}_{-0.055}$;\quad
{\tt EpsSkewNormal} vs \bspline: $0.107^{+0.062}_{-0.053}$}\\
\bottomrule
\end{tabular}
\end{table*}
\endgroup

\FloatBarrier

\section{Additional illustrations}\label{more}

\begin{figure*}
\centering
\includegraphics[width = 0.9\textwidth, angle=0.0]{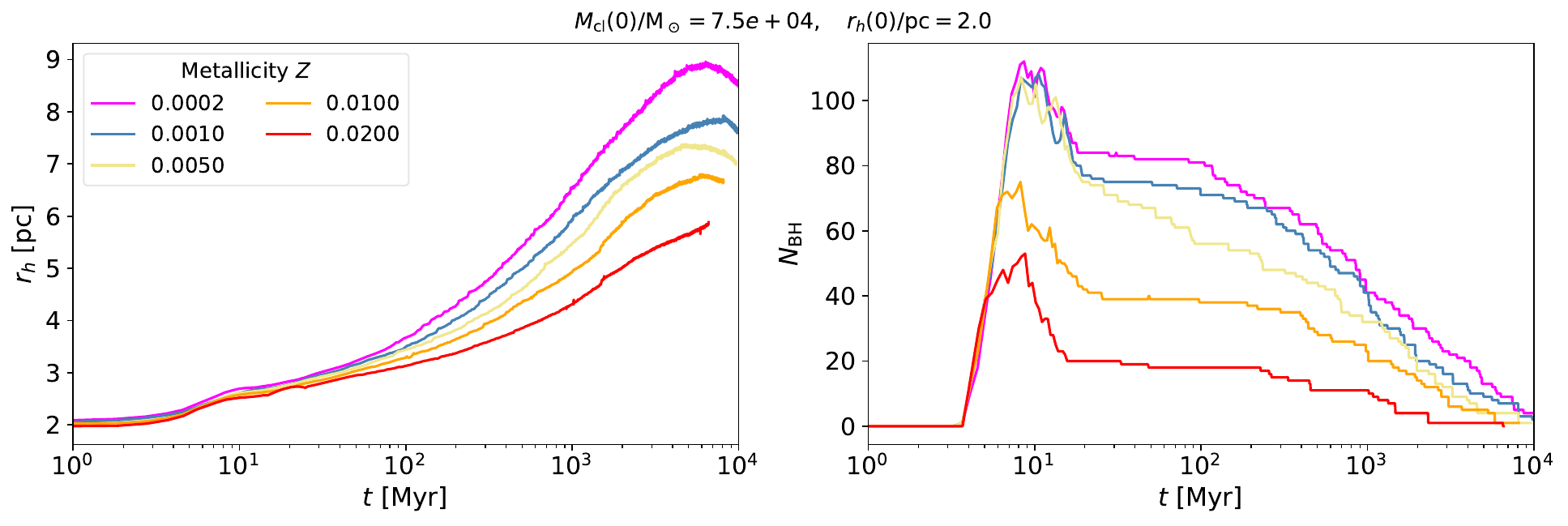}\\
\vspace{-0.22cm}
\includegraphics[width = 0.9\textwidth, angle=0.0]{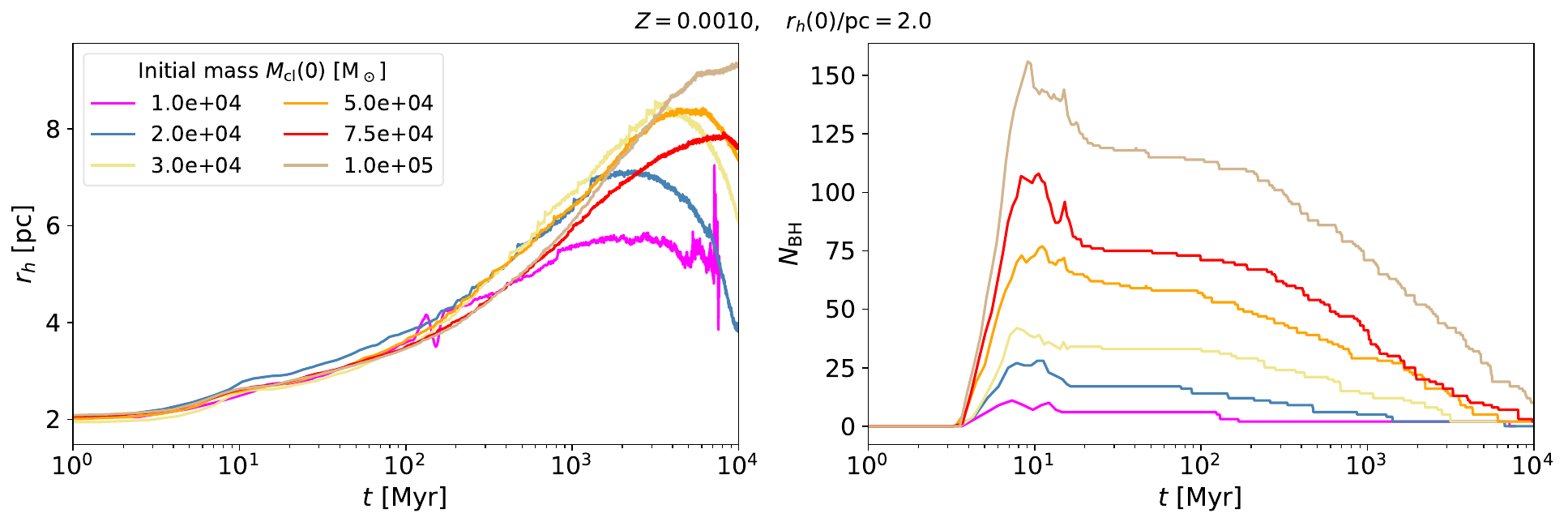}\\
\vspace{-0.22cm}
\includegraphics[width = 0.9\textwidth, angle=0.0]{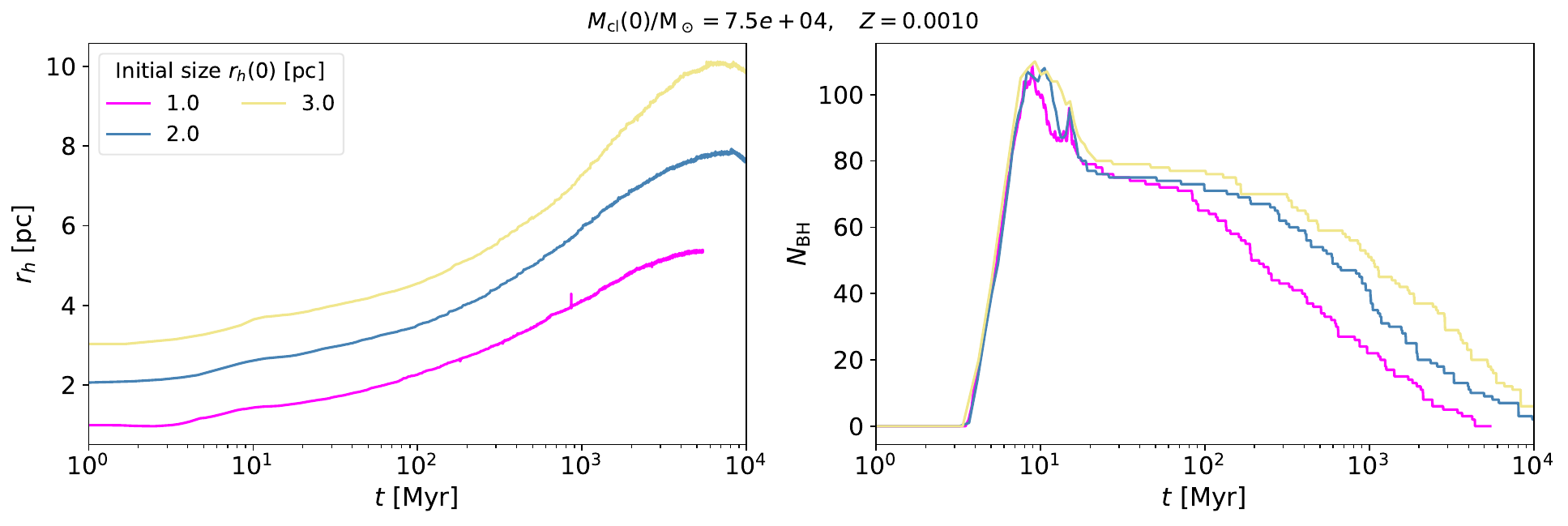}
	\caption{Dependence of model cluster evolution with variation of cluster metallicity, $Z$ (upper panel),
	initial mass, $\mcl(0)$ (middle panel), and initial size, $\rh(0)$ (lower panel). For each variation (legend),
	the other two grid coordinates are kept constant, as indicated in each row's title. Shown are the time
	evolution of the half-mass radius, $\rh$ (left column), and the number of BHs bound to the cluster,
	$\nbh$ (right column).}
\label{fig:evol}
\end{figure*}

\begin{figure*}
\centering
\includegraphics[width = \textwidth, angle=0.0]{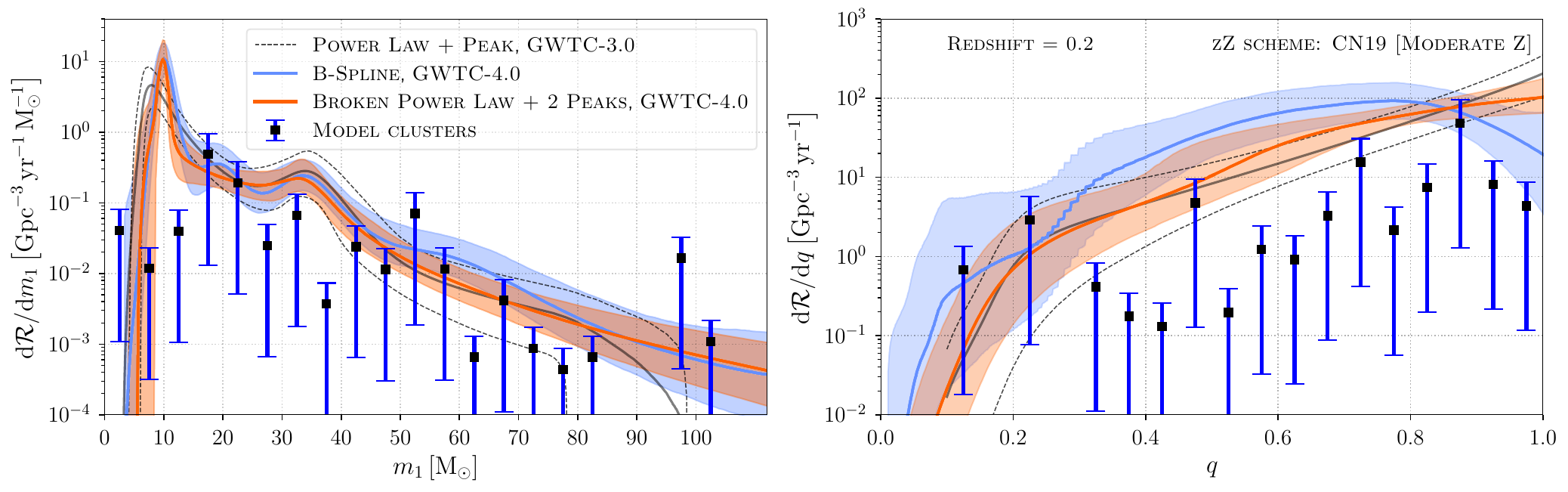}
\includegraphics[width = \textwidth, angle=0.0]{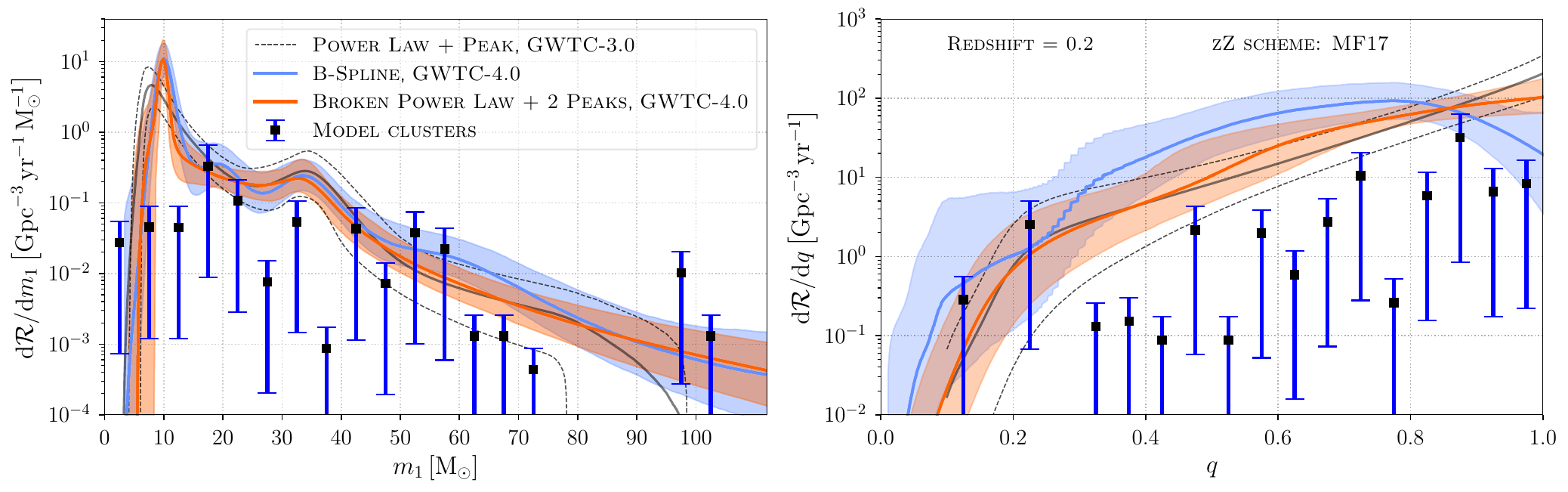}
	\caption{Same description as in Fig.~\ref{fig:m1_q_dist} applies. The model differential
	merger rate density distributions presented in the upper (lower) pair of panels correspond
	to the CN19 (MF17) $zZ$ scheme.}
\label{fig:m1_q_dist_xtra}
\end{figure*}

\begin{figure*}
\centering
\includegraphics[width = \textwidth, angle=0.0]{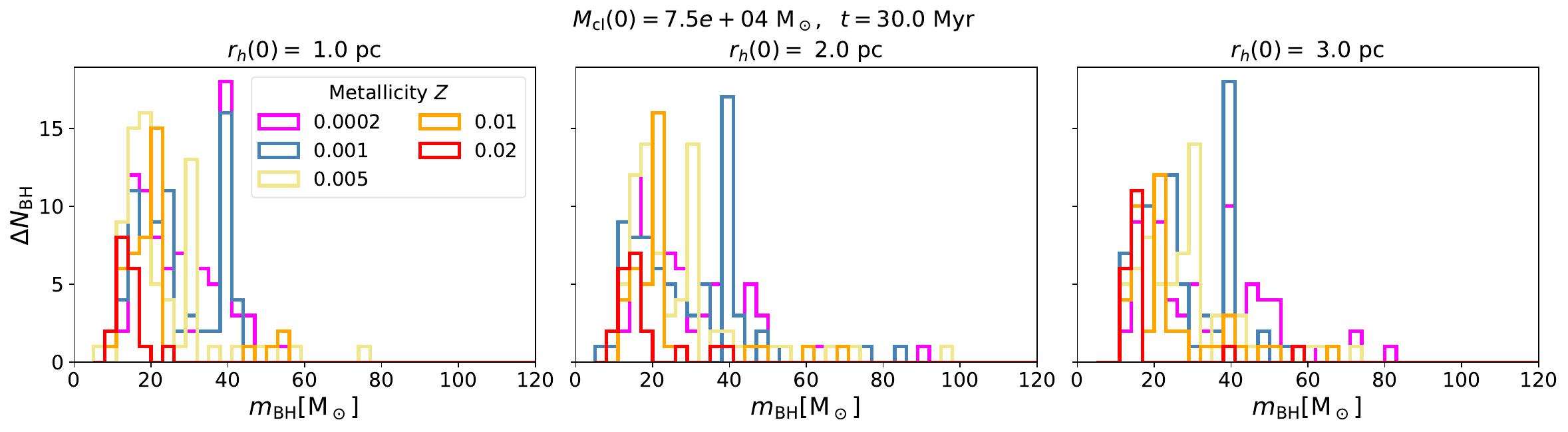}
\caption{Mass distribution of BHs that are retained in the cluster after formation. Here,
	the distributions for the $\mcl(0)=7.5\times10^4\Ms$ clusters are plotted.
	The left, middle, and right panels plot those for the $\rh(0)=1$ pc, 2 pc,
	and 3 pc model clusters, respectively, of all the five metallicities (legend).
	The distributions are plotted at an evolutionary time of 30 Myr, which is sufficiently
	beyond the time over which the BHs are formed inside the clusters (3-20 Myr) so
	that the unbound BHs with high natal kick have escaped from the cluster, but
	too early for the BHs to be dynamically active via central core formation
	(Sec.~\ref{evol}, Fig.~\ref{fig:evol}). That way, the birth mass distribution
	of the initially cluster-retained BHs is captured. The cluster-retained
	ESC-NSs are excluded from this plot.}
\label{fig:remdist}
\end{figure*}

\begin{figure*}
\centering
\includegraphics[width = 0.32\textwidth, angle=0.0]{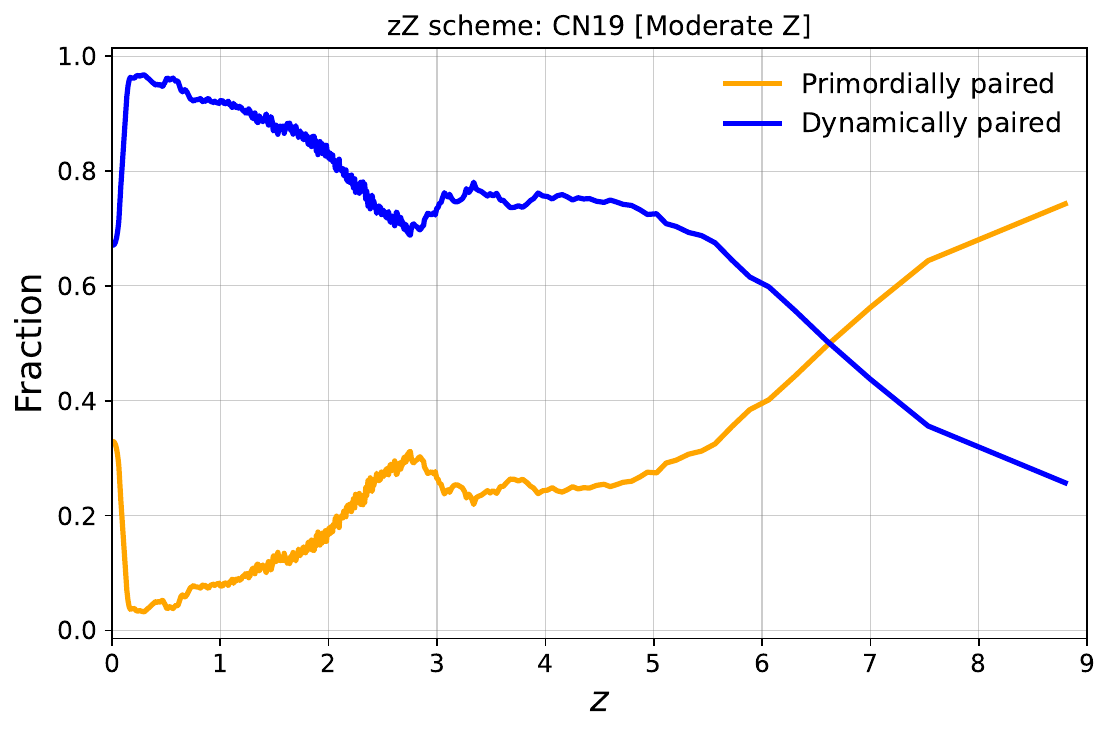}
\includegraphics[width = 0.32\textwidth, angle=0.0]{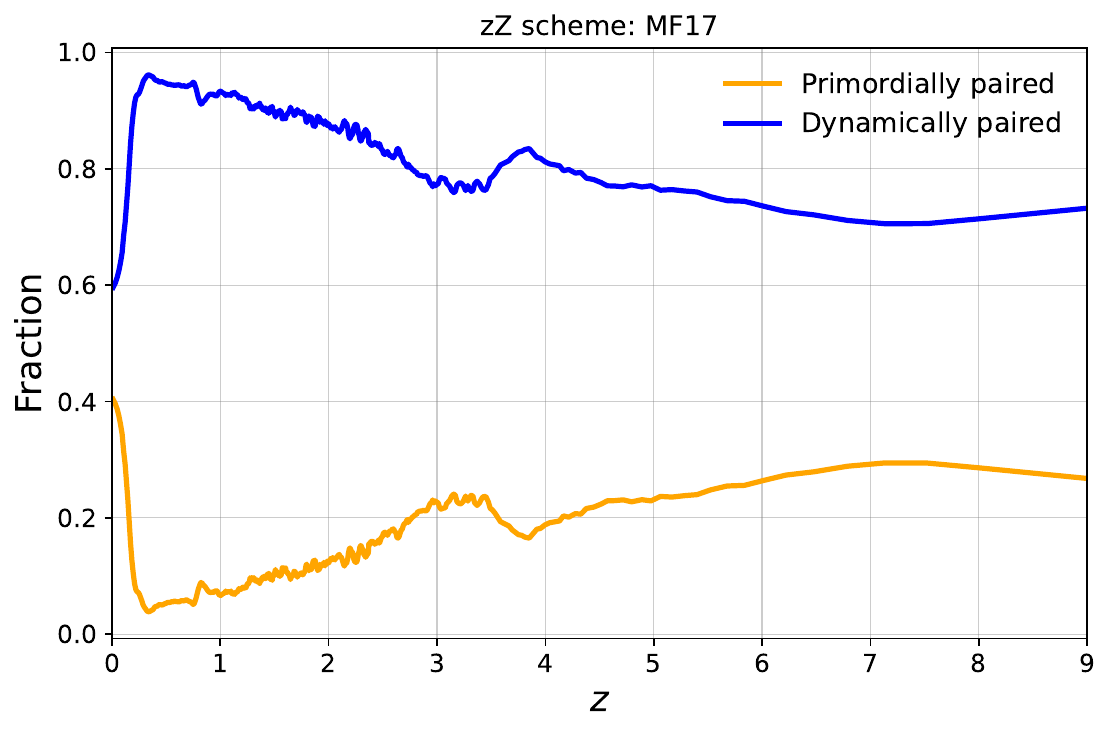}
\includegraphics[width = 0.32\textwidth, angle=0.0]{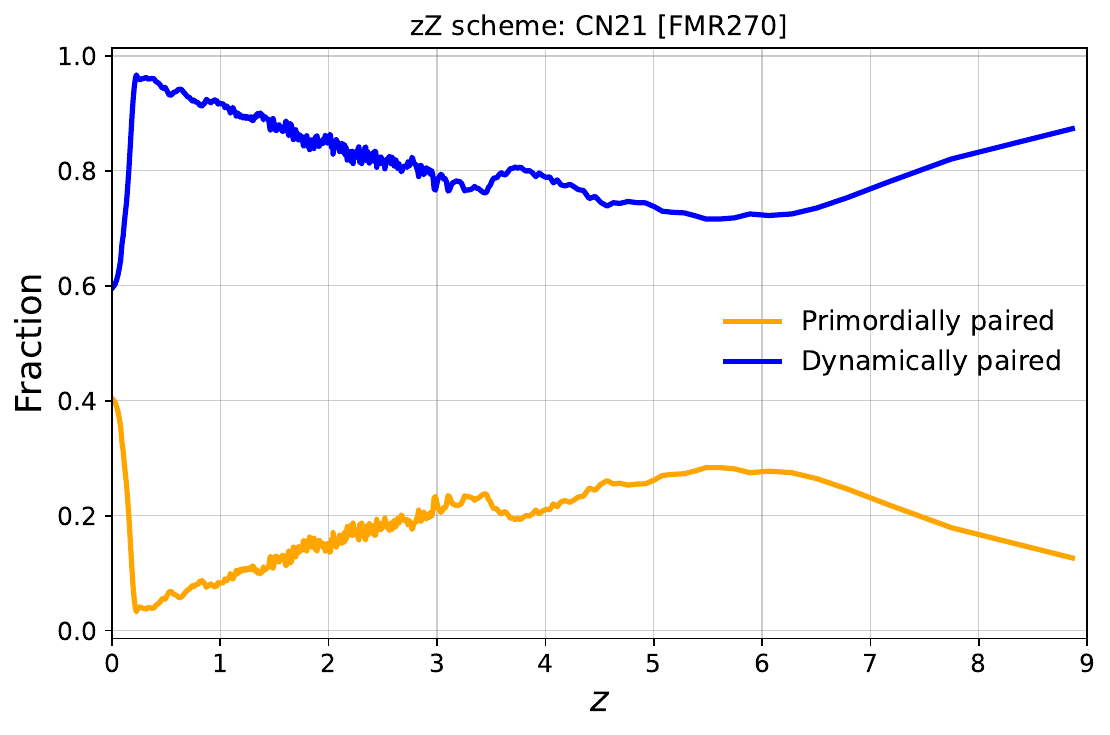}
	\caption{Evolution of the fractions of primordially paired and dynamically paired mergers (legend)
	with redshift, for the model merger population syntheses performed in this work.
	The three panels distinguish the adopted $zZ$ schemes, namely, MF17, CN19, and CN21 (panel subtitle).}
\label{fig:fracz}
\end{figure*}

\begin{figure*}
\centering
\includegraphics[width = \textwidth, angle=0.0]{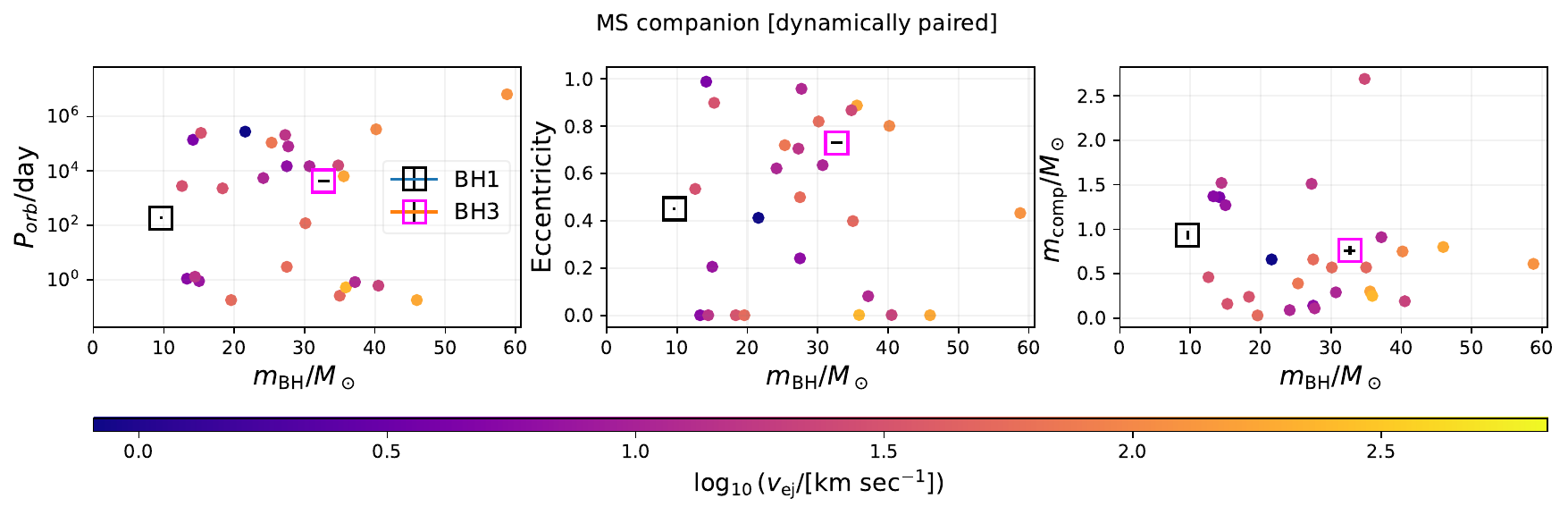}\\
\vspace{-1.3cm}
\includegraphics[width = \textwidth, angle=0.0]{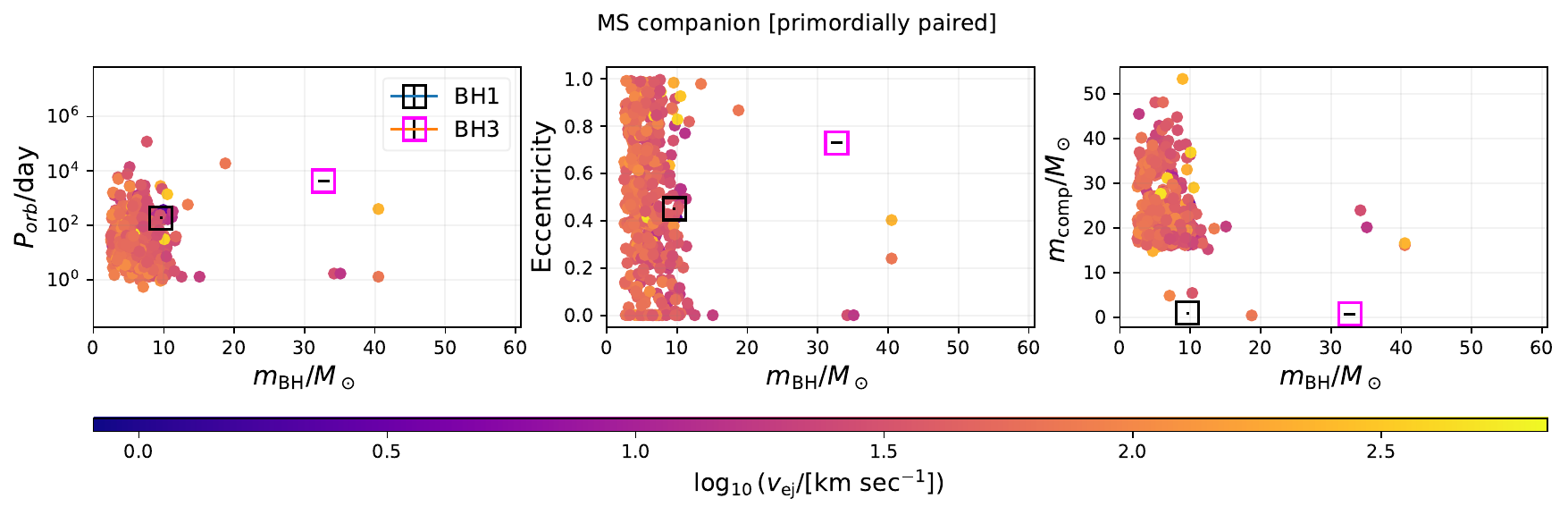}
\caption{Demographics of the BH-MS binaries that have escaped into the galactic field
from the evolutionary model star clusters in this study. For the filled circles, descriptions analogous to
Fig.~\ref{fig:GaiaNS} apply.
For comparison, the observed BH-MS binary candidates
Gaia BH1 and Gaia BH3 are shown in each panel (empty squares).
For these observed data points, the colour coding is not followed
and their uncertainties are marginally visible due to the scale of the figure axes. (The size
of the squares are chosen for legibility and does not represent
measurement uncertainties.)}
\label{fig:GaiaBH}
\end{figure*}

\twocolumngrid

\end{document}